\documentclass[english,prb,aps]{revtex4-2}
\pdfoutput=1
\usepackage[T1]{fontenc}
\usepackage[latin9]{inputenc}
\usepackage{geometry}
\usepackage{float}
\usepackage{dsfont}
\usepackage{amsmath}
\usepackage{amssymb}
\usepackage{graphicx}
\usepackage{esint}

\makeatletter
\usepackage{babel}
\usepackage{braket}
\usepackage{csquotes}

\@ifundefined{showcaptionsetup}{}{%
 \PassOptionsToPackage{caption=false}{subfig}}
\usepackage{subfig}
\makeatother

\usepackage{babel}
\begin{document}
\title{Two-level Systems Coupled to Graphene plasmons: A Lindblad equation
approach}
\author{T. V. C. Antão{\normalsize{}{}$^{1,\ast}$} and N. M. R. Peres$^{1,2,\dagger}$}
\address{$^{1}$Department and Centre of Physics, University of Minho, Campus
of Gualtar, 4710-057, Braga, Portugal}
\address{$^{2}$International Iberian Nanotechnology Laboratory (INL), Av. Mestre
José Veiga, 4715-330, Braga, Portugal}
\begin{abstract}
In this paper we review the theory of open quantum systems and macroscopic
quantum electrodynamics, providing a self-contained account of many
aspects of these two theories. The former is presented in the context
of a qubit coupled to a electromagnetic thermal bath, the latter is
presented in the context of a quantization scheme for surface-plasmon
polaritons (SPPs) in graphene based on Langevin noise currents. This
includes a calculation of the dyadic Green's function (in the electrostatic
limit) for a Graphene sheet between two semi-infinite linear dieletric
media, and its subsequent application to the construction of SPP creation
and annihilation operators. We then bring the two fields together
and discuss the entanglement of two qubits in the vicinity of a graphene
sheet which supports SPPs. The two qubits communicate with each other
via the emission and absorption of SPPs. We find that a Schödinger
cat state involving the two qubits can be partially protected from
decoherence by taking advantage of the dissipative dynamics in graphene.
A comparison is also drawn between the dynamics at zero temperature,
obtained via Schrodinger's equation, and at finite temperature, obtained
using the Lindblad equation. 
\end{abstract}
\maketitle
\tableofcontents{}

\section{Introduction}

\begin{figure}[H]
\begin{centering}
\includegraphics[scale=0.25]{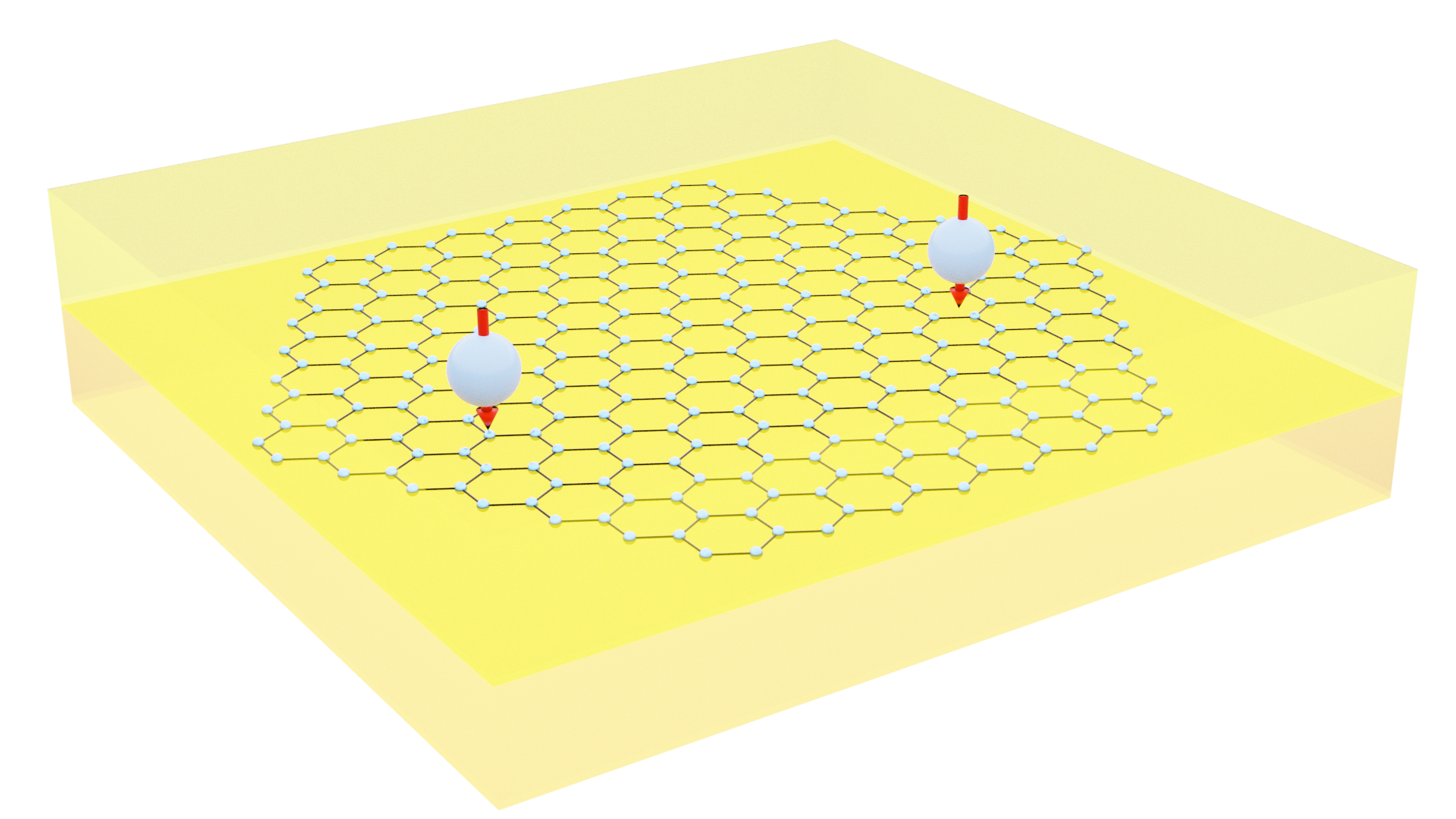} 
\par\end{centering}
\caption{Two qubits (blue spheres) placed near a graphene sheet (represented
by the hexagonal lattice) embedded between two semi-infinite dielectric
media with different electric permittivity (represented by the yellow
and orange slabs) \label{fig:System-we-will}}
\end{figure}

The following paper is concerned with studying the entanglement of
two two-level systems placed in close proximity to a graphene sheet.
More specifically, we wish to study the communication of these emitters
via so called surface plasmon polaritons (SPPs) that travel along
the graphene sheet (see Fig. \ref{fig:System-we-will}).

These SPPs are collective excitation of the electron gas that propagate
along interfaces, and are usually produced in metal-dielectric interfaces
with the aid of a prism, in setups such as the Kretschmann or Otto
configurations. However, it is also possible to generate these excitations
in a configuration where a graphene mono-layer is ``sandwiched''
between two dielectric media. This is because doped graphene acts
as a conductor of electric charge, characterized by a sheet conductivity,
which enables the existence of SPPs propagating on this material.
(see \citet{Goncalves2016Graphene}). In the configuration represented
in Fig. \ref{fig:System-we-will}, if two emitters are positioned
close to the graphene sheet, they will interact with the surface plasmon
bath and thus, emission and absorption of SPPs will provide a way
of communication between them. This opens the possibility of creation
of two (or more) entangled (in quantum mechanical sense) emitters.
So, the aim and motivation for our work is two fold. We will on one
hand describe the quantization of the SPPs for our configuration.
We will do so via the dyadic Green's function method, and this will
shine light onto the propagation of these SPPs and how they will be
used to establish a communication channel between the quantum emitters.

On the other hand, we will then focus on studying the dynamics of
the emitters, that is, their time evolution. The final objective of
our study is to show how dissipative currents in the graphene that
correspond to the production of quantized SPPs will, in fact, protect
the entanglement of the qubits and select a superposition of states
over any other, extending its decay time (see \citet{Gonzalez2011}).
To obtain the dynamics of the emitters, however, we must study their
interaction with the plasmon bath. This can be modeled as a system
interacting with a reservoir, and, to these kinds of problems, the
theory of open quantum systems is particularly well suited. Indeed,
at zero temperature, we can describe de dynamics of the emitters using
the Schrödinger formalism. However, at finite temperatures, this formalism
proves inadequate and the theory of open quantum systems has to be
used.

The literature on application of the Lindblad equation to graphene
is scarce to nonexistent, with a single paper, accordingly to a well
known scientific data basis, focused on the interaction of two-level
systems and graphene \cite{Nechepurenko2018} using the Lindblad equation.

The philosophy of writing of this paper was to keep the minimum of
technical detail in the main part of the text. Thus, going through
this part only, the reader will be able to capture the qualitative
aspects of the problem. However, to master the details the reader
has to work through the Appendices.

The paper is structured as follows: In section \ref{Sec:Steps-towards-the}
we will first introduce the theory of open quantum systems with a
focus on the Lindblad equation giving the main concepts toward its
derivation. In section \ref{Sec:Microscopic-Lindblad} a microscopic
derivation of the Lindblad equation is given. We will use this theory
to evaluate the dynamics of a quantum emitter coupled to a thermal
radiation bath, which is one of the most studied applications of open
quantum systems in quantum optics, and it will prove to be a good
starting point for the study of the dynamics of similar systems. In
section \ref{Sec:quantization_of_the_polar_field}, we will quantize
the SPP field using macroscopic Green's functions methods, known as
dyadic Green's functions; we will consider the electrostatic limit,
as a SPP is, for large wave vectors, insensitive to retardation effects.
This will allow us to quantize the SPP field in the presence of dissipation
and set the stage for the evaluation of the qubit dynamics. These
are finally described in section \ref{Sec:singleQubit}, where we
study a single qubit coupled to the bath. The entanglement between
two qubits mediated by SPPs as well as an investigation of the role
of dissipation in this quantum mechanical process are performed in
Sec. \ref{Sec:2qubits}. All the details of the calculations are given
in a set of Appendixes written using a didactic style.

\section{Steps towards the Lindblad equation \label{Sec:Steps-towards-the}}

The aim of this section is to approach the theory of open quantum
systems in general, providing a few significant steps toward the derivation
of the master equation in Lindblad form without making reference to
the particular properties of the system we wish to describe. Master
equations describe the dynamical evolution of a quantum system which
interacts with its environment (a so called \textit{open quantum system}).

As stated, we will start without making reference to any particular
system, following the approach of \citet{Manzano_2020}, and will,
a couple of sections ahead, specialize to actual applications of the
theory.

Note that decoupling a system from its environment is a very significant
endeavor both from a conceptual point of view, as well as from a practical
perspective. On the conceptual side, a closed quantum system which
does not interact with its environment, such as those treated in elementary
quantum mechanics is merely a useful idealization as in nature nothing
can be truly isolated. On the practical side, in some cases the environment
of a system plays an important and useful role in its dynamics, and
as such cannot be ignored. The goal of this section is therefore to
introduce the necessary steps and approximations used in order to
infer the evolution of the system of interest (which we will often
call the reduced system) from those of the total ``reduced system+environment"
ensemble. The end result will be the so called \textit{Lindblad equation}

\begin{equation}
\frac{d\hat{\rho}_{S}}{dt}=-\frac{i}{\hbar}[\hat{H}_{S}(t),\hat{\rho}_{S}(t)]+\sum_{k}\left(\hat{L}_{k}\hat{\rho}_{S}(t)\hat{L}_{k}^{\dagger}-\frac{1}{2}\left\{ \hat{L}_{k}^{\dagger}\hat{L}_{k},\hat{\rho}_{S}(t)\right\} \right).\label{eq:Lindbladgeneral}
\end{equation}

This equation gives the time evolution of the density operator $\hat{\rho}_{S}$,
which characterizes the state of the reduced system. The Lindblad
equation can be interpreted as involving two contributions: A coherent
evolution governed by the reduced system's Hamiltonian $\hat{H}_{S}(t)$
corresponding to the commutator present in equation (\ref{eq:Lindbladgeneral});
and also an incoherent evolution brought on by the reduced system's
interaction with the environment. In this second term of the Lindblad
equation there come into play a series of operators $\hat{L}_{k}$
which are called \textit{jump operators} that describe, for instance,
the excitation and relaxation of the system's state. The Lindblad
equation is the most general \textit{trace-preserving} dynamical equation
with \textit{Markovian} time-evolution (we shall clarify the meaning
of these terms later).

To start with, we make a few comments on the formalism we shall use
regarding composite systems. We may describe the reduced system $S$
by a \textit{wave-function} that lives in a \textit{Hilbert space}
$\mathcal{H}_{S}$ with an arbitrary complete basis $\left|\phi_{S}\right\rangle $.
The environment $R$ can be characterized similarly by a Hilbert space
$\mathcal{H}_{R}$ with basis $\left|\phi_{R}\right\rangle $ (we
use the $R$ index for reservoir. In practice along the work we use
the word \textit{reservoir} or \textit{bath} interchangeably with
the word \textit{environment}). The total system is characterized
by the tensor product space $\mathcal{H}_{S}\otimes\mathcal{H}_{R}$
with basis $\left|\phi_{S}\right\rangle \otimes\left|\phi_{R}\right\rangle $.
The total Hamiltonian which characterizes the dynamics of the reduced
system+environment ensemble can be constructed from the reduced system's
and environment's Hamiltonians (\textbf{$\hat{H}_{S}$} and $\hat{H}_{R}$
respectively) as well as a coupling or interaction term $\hat{V}$
as

\begin{equation}
\hat{H}=\hat{H}_{0}+\hat{V}=\hat{H}_{S}\otimes\mathds{1}_{R}\text{+}\mathds{1}_{S}\otimes\hat{H}_{R}+\hat{V}.\label{eq:Hamiltonian}
\end{equation}

\begin{figure}
\includegraphics[scale=0.3]{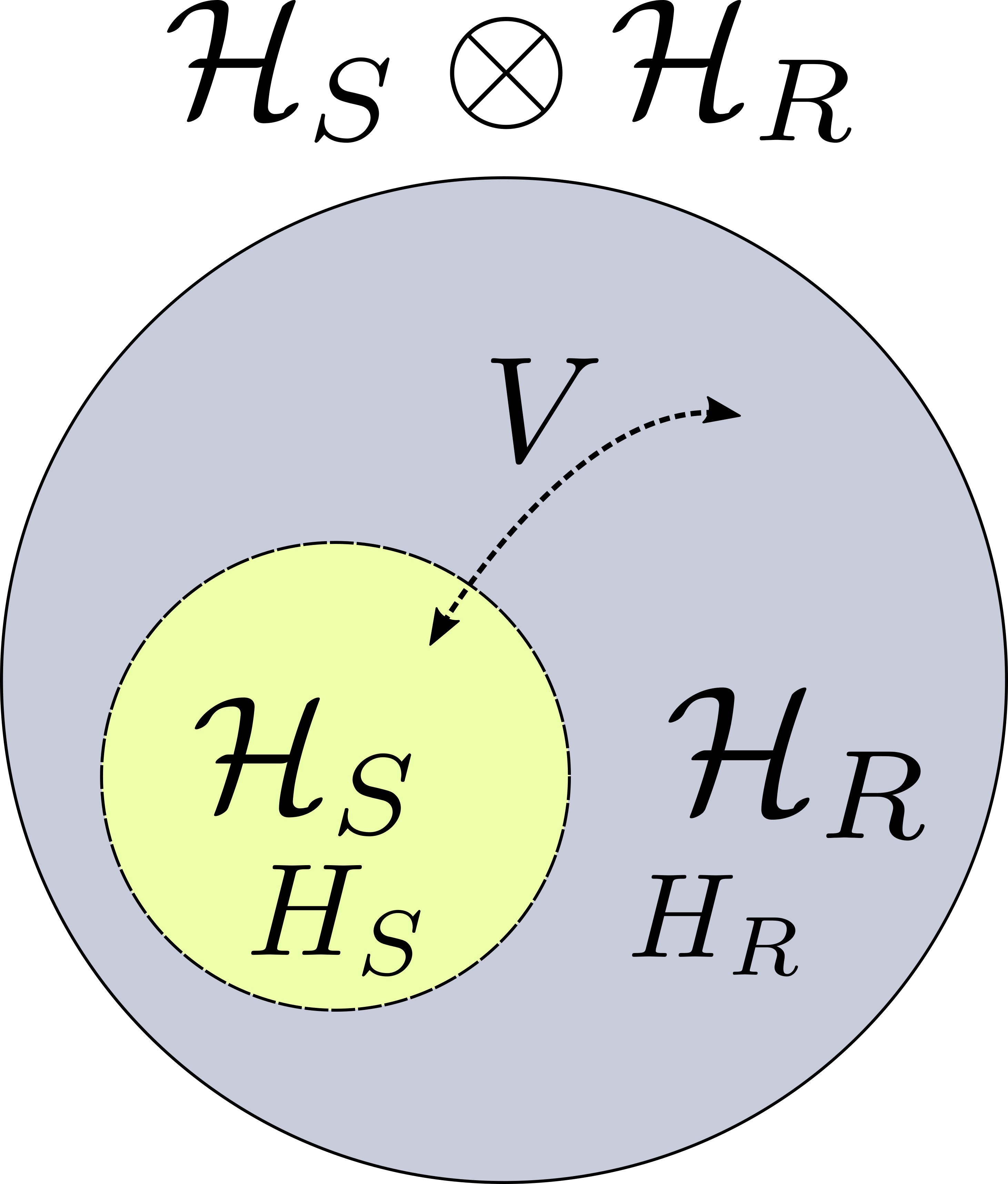} \caption{Graphical representation of the Hilbert space of an open quantum system
$\mathcal{H}_{S}\otimes\mathcal{H}_{R}$ and respective Hamiltonians.
In green we see the reduced system Hilbert space $\mathcal{H}_{S}$
to which corresponds a Hamiltonian $H_{S}$. In gray, encapsulating
the reduced system we have the environment or reservoir with Hilbert
space $\mathcal{H}_{R}$ with Hamiltonian $H_{R}$, and finally the
interaction Hamiltonian $V$ causing the correlations between reduced
system and environment.}
\end{figure}

As can be seen from the nature of the Lindblad equation (\ref{eq:Lindbladgeneral}),
we will be using the density operator formalism and thus describe
mixed states of the total system by a total density operator $\hat{w}(t)=\hat{\rho}_{S}(t)\otimes\hat{\rho}_{R}(t)$.
$\hat{w}(t)$ corresponds to a tensor product of the reduced system
density operator $\hat{\rho}_{S}$ and the reservoir density operator
$\hat{\rho}_{R}$ (see appendix \ref{Sec:rho_opperator} for relevant
definitions and calculations regarding the density operator and an
application to the dynamics of an optically driven two-level system).

The density operator evolves according to the von-Neumann equation
and its dynamics is governed by the Hamiltonian of equation (\ref{eq:Hamiltonian}).
If we wish to describe the evolution of a system in contact with an
environment in a quantitative and analytical manner we must proceed
in lowest order perturbation theory.

Our goal is to make clear how the interaction of the system with its
environment affects the dynamics and as such, we are really only interested
in the time evolution brought on by the coupling $\hat{V}$. The reduced
system itself can have non-trivial dynamics, so we want to separate
the time evolution due to the coupling by writing the equations of
motion in the interaction picture of quantum mechanics. In this representation,
the density operator is $\hat{w}_{I}(t)=e^{i\hat{H}_{0}t/\hbar}\hat{w}(t)e^{-i\hat{H}_{0}t/\hbar}$
and the coupling term in the Hamiltonian is $\hat{V}_{I}(t)=e^{i\hat{H}_{0}t/\hbar}\hat{V}e^{-i\hat{H}_{0}t/\hbar}$.
It is quite simple to show that the time evolution is given by the
von-Neumann equation but with the total Hamiltonian replaced by the
interaction term $\hat{V}_{I}(t)$ (see appendix \ref{subsec:Pictures-of-quantum}
for a discussion on the representations of quantum mechanics).

Another important tool which we shall employ often in the following
considerations is the \textit{partial trace}. It consists in a summation
over a select number of degrees of freedom of the total system. For
instance, one may take the partial trace over the reduced system degrees
of freedom, or over the environment degrees of freedom. Tracing over
the environment degrees of freedom, for instance, allows us to isolate
the density operator for the reduced system from the total density
operator. In symbols, we can write $\hat{\rho}_{S}=\text{tr}_{R}\left\{ \hat{w}\right\} $
and thus, tracing over the reservoir we can write the von-Neumann
equation for the system $S$.

On the other hand, the von-Neumann equation for the total density
operator $\hat{w}$ lends itself to an iterative solution. Iterating
once, and then tracing over the reservoir degrees of freedom, the
following integro-differential equation for the time evolution of
the system's density operator is easily computed

\begin{equation}
i\hbar\frac{d\hat{\rho}_{S}^{I}(t)}{dt}=\text{tr}_{R}\left\{ \left[\hat{V}_{I}(t),\hat{w}_{I}(t_{0})-\frac{i}{\hbar}\int_{t_{0}}^{t}\left[\hat{V}_{I}(t'),\hat{w}_{I}(t')\right]dt'\right]\right\} .
\end{equation}
The first term, containing $\hat{w}_{I}(t_{0})$ is often neglected
in the spirit of the molecular chaos hypothesis (see \citet{Hohenester2020NanoandQuantum}).
It contains information regarding the initial state of the system,
and we assume that initial correlations are meaningless in the evolution
of the system. In practice, this term can actually be shown to be
0 in many cases such as if the environment is thermal.

We also neglect the system-reservoir interaction to first order. This
is the so called \textit{Born approximation}, which allows us to write
$\hat{w}(t')\approx\hat{\rho}_{S}^{I}(t')\otimes\hat{\rho}_{R}^{I}(t')+\mathcal{O}\left(\hat{V}_{I}\right)$.
This approximation imposes that we work in the weak coupling regime,
and, as such, the evolution of the system occurs over a much larger
time scale $\tau_{sys}$ than the correlation and relaxation time
scales of the system $\tau_{corr}$ and $\tau_{rel}$, i.e. we have
$\tau_{sys}\gg\tau_{rel},\tau_{corr}$. We may assume that we start
from an uncorrelated state $\hat{w}_{I}(t_{0})=\hat{\rho}_{S}^{I}(t_{0})\otimes\hat{\rho}_{R}^{I}(t_{0})$,
for example by considering that the system and environment have not
interacted before $t=t_{0}$. After this initial point in time, we
assume that the system and reservoir stay uncorrelated, or that correlations
are sufficiently small to be ignored. This is a very strong assumption
which is the first of three key approximations we use to derive the
Lindblad equation, and is a necessary step to unravel the system from
its environment.

We also make the approximation $\hat{\rho}_{S}^{I}(t')\approx\hat{\rho}_{S}^{I}(t_{0})\approx\hat{\rho}_{S}^{I}(t)$
which allows us to obtain a time-local equation. This approximation
is called the \textit{first Markov approximation} since now the system
can be described by $\hat{\rho}(t)$ at the single time $t$ in a
manner which does not depend on its past history. This kind of evolution
is called \textit{Markovian}. This constitutes the second key approximation
in our journey toward the Lindblad equation. If we assume the case
of a thermal environment and reservoir is in thermal equilibrium,
or more generally that the environment has many more degrees of freedom
than the system and thus the interaction changes the environment in
a negligible manner, $\hat{\rho}_{R}^{I}(t)$ loses its time dependence
and the time-evolution of the density operator $\hat{\rho}_{S}(t)$
is given by

\begin{equation}
\frac{d\hat{\rho}_{S}^{I}(t)}{dt}=-\frac{1}{\hbar^{2}}\int_{0}^{t}\text{tr}_{R}\left\{ \left[\hat{V}_{I}(t),\left[\hat{V}_{I}(t'),\hat{\rho}_{S}^{I}(t)\otimes\hat{\rho}_{R}^{I}\right]\right]\right\} dt'+\mathcal{O}\left(\hat{V}_{I}^{2}\right),\label{eq:unuseful}
\end{equation}
where we have considered for simplicity that $t_{0}=0$. We have proceeded
in a quite general manner and the previous equation has many of the
characteristics we want, but we still need further work in order to
transform equation (\ref{eq:unuseful}) to a form that we can actually
compute any physical quantity. In many cases the interaction Hamiltonian
$\hat{V}_{I}(t)$ can be written as a tensor product of operators

\begin{equation}
\hat{V}_{I}(t)=\sum_{i}\beta_{i}^{S}(t)\otimes\Gamma_{i}^{R}(t),\label{eq:FactoredInteraction}
\end{equation}
where as indicated by the indices $\beta_{i}^{S}$ acts on the Hilbert
space of the system $\mathcal{H}_{S}$ and $\Gamma_{i}^{R}$ on the
Hilbert space of the reservoir $\mathcal{H}_{R}$. We can now perform
a series of simple expansions and calculations to write the master
equation in a more suitable form. Firstly, we substitute the interaction
Hamiltonian of the form (\ref{eq:FactoredInteraction}) into the master
equation and evaluate the commutators explicitly. Then, factoring
the expression, and cyclically permuting the operators $\Gamma_{i},\text{\ensuremath{\Gamma}}_{j}$
and $\hat{\rho}_{R}$ allows us to introduce several additional simplifications
(see appendix \ref{sec:Details-of-the-Derivation-of-the-Lindblad-Equation}
for details regarding these and following calculations). Finally,
we can introduce the \textit{correlation functions}

\begin{equation}
\begin{cases}
\left\langle \Gamma_{i}(t)\Gamma_{j}(t')\right\rangle _{R}\equiv\text{tr}_{R}\left\{ \hat{\rho}_{R}\Gamma_{i}(t)\Gamma_{j}(t')\right\} ,\\
\left\langle \Gamma_{j}(t)\Gamma_{i}(t')\right\rangle _{R}\equiv\text{tr}_{R}\left\{ \hat{\rho}_{R}\Gamma_{j}(t)\Gamma_{i}(t')\right\} ,
\end{cases}\label{eq:ReservoirCorr}
\end{equation}
where the super-index $R$ in the operators $\Gamma_{i}^{R}(t)$ has
been omitted for simplicity. The idea is that, given a concrete physical
system, we can approach the derivation by calculating these correlation
functions by hand first and then replacing them in the master equation
to obtain the time evolution. Note that the first Markov approximation
ceases to be an approximation and becomes exact only when these correlation
functions are proportional to $\delta(t-t')$. Such proportionality
would indeed allow us to replace $\hat{\rho}_{S}^{I}(t')$ by $\hat{\rho}_{S}^{I}(t)$.
The third, and final, key approximation we introduce is to assume
that the integrand decays fast enough with time, thus letting us extend
the integral to infinity. To do this we perform the change of variables
$t'=t-\tau$ and let the upper limit of integration go to infinity.
This procedure is the \textit{second Markov approximation} and it
results in the \textit{Redfield equation} written in terms of correlation
functions, which will allow us to give a derivation of the Lindblad
equation in physical applications. The Redfield equation reads

\begin{align}
\frac{d\hat{\rho}_{S}(t)}{dt} & =-\frac{1}{\hbar^{2}}\sum_{ij}\int_{0}^{\infty}\left(\beta_{i}(t)\beta_{j}(t-\tau)\hat{\rho}_{S}(t)-\beta_{j}(t-\tau)\hat{\rho}_{S}(t)\beta_{i}(t)\right)\left\langle \Gamma_{i}(t)\Gamma_{j}(t-\tau)\right\rangle \nonumber \\
 & +\left(\hat{\rho}_{S}(t)\beta_{j}(t-\tau)\beta_{i}(t)-\beta_{i}(t)\hat{\rho}_{S}(t)\beta_{j}(t-\tau)\right)\left\langle \Gamma_{j}(t)\Gamma_{i}(t-\tau)\right\rangle d\tau.\label{eq:Redfield}
\end{align}
For instance, we shall derive the Lindblad equation in the case of
a two-level system (qubit) coupled to the thermal E-M field in a later
section. We note that the procedure we are going over in this section
is often described as a microscopic derivation of the Lindblad equation,
since we explicitly construct the Lindblad jump operators from the
Hamiltonian, however there are many ways to derive this equation,
and in principle one could also derive it from macroscopic principles,
such as positivity of the density operator as well as linearity of
the time--evolution, along with the Markov approximations.

\section{Microscopic derivation of the optical master equation in Lindblad
form \label{Sec:Microscopic-Lindblad}}

\subsection{Thermal radiation - A short review \label{subSec:A-short-review}}

Before we concern ourselves with continuing the derivation of the
Lindblad equation, it is useful to first discuss the concept and some
of the mechanics of thermal radiation. In particular we aim to discuss
thermal radiation field states and derive the state distribution and
average photon number when a mode of the E-M field is in thermal equilibrium
with its environment at some temperature $T$. These kinds of states
are incoherent superpositions of Fock states. From thermodynamics,
it is well known that the probability of a field mode being in the
$n$th excited state, or in other words, that there are $n$ photons
occupying a certain mode is given by a\textit{ Boltzmann factor}

\begin{equation}
P_{n}=\frac{e^{-\frac{E_{n}}{k_{B}T}}}{Z},
\end{equation}
where $k_{B}$ is Boltzmann's constant and $Z=\sum_{n=0}^{\infty}e^{-E_{n}/k_{B}T}$
is the partition function of the E-M field. Using this we may calculate,
for instance, the average number of photons in such a thermal field
state, in a single mode characterized by a frequency $\omega$. This
amounts to summing the occupation numbers $n$ for each state, weighted
by the probability $P_{n}$. With the previous Boltzmann factor and
using the formula for the energy of the harmonic oscillator $E_{n}=\hbar\omega$
(note that rigorously the energy is written $E_{n}=\hbar\omega+1/2$,
however, since the partition function would then be written $Z'=e^{-1/(2k_{B}T)}Z$,
we can cancel this factor), this can be evaluated as a derivative
of the geometric series which yields the famous result

\begin{equation}
\bar{n}(\omega,T)=\frac{1}{e^{\hbar\omega/k_{B}T}-1}.
\end{equation}
This is the \textit{Bose-Einstein distribution for photons}. Thus
$\bar{n}(\omega,T)$ depends on the interplay between the mode energy
$\hbar\omega$ and the inverse thermal energy $1/k_{B}T$. Notice,
of course, that we can arrive at this same result using a more quantum
mechanical approach involving the density operator formalism. The
density operator is $\hat{\rho}=\sum_{n}P_{n}\left|n\right\rangle \left\langle n\right|$
where $\left|n\right\rangle $ is the Fock state with $n$ photons.
The distribution of states is the expectation value of the number
operator, which we can calculate by tracing over the degrees of freedom
of the field. This procedure equivalently yields the Bose-Einstein
distribution.

Another relevant quantity beyond the distribution of these thermal
states, is the density of states, or \textit{mode density}, of the
field, which corresponds to the number of allowed modes with wave-vector
between \textbf{$\boldsymbol{k}$} and $\boldsymbol{k}+d^{3}\boldsymbol{k}$.
To evaluate it, we need only count the number of allowed modes inside
a sphere of radius $k$ in $k$-space. If we let the E-M field be
quantized over a volume $V=L^{3}$ then the wave vector can only have
certain values (see appendix \ref{sec:Quantization-of-the-EM-Free}).
We thus integrate over the sphere in $k$-space and divide by the
effective volume occupied by each mode, which is $\left(2\pi/L\right)^{3}$.
The resulting integration kernel is the mode density

\begin{equation}
g(\boldsymbol{k})=\frac{\omega^{2}}{8\pi^{3}c^{3}}V,\label{eq:DOS}
\end{equation}
where $c$ is the speed of light.

\subsection{Derivation of the Lindblad equation for a two-level system coupled
to thermal radiation \label{subSec:Derivation-of-the}}

We now proceed with the derivation of the Lindblad equation by using
our previous considerations regarding thermal radiation to compute
the master equation for a two-level system coupled to the E-M field
when in thermal equilibrium at a temperature $T$. We follow the approach
outlined in \citet{carmichael1993open,carmichael2013statistical}.
By the end of this section, we will be able to write the master equation
in Lindblad form. Our starting point is the master equation for a
system coupled to a reservoir. We have derived this in section \ref{Sec:Steps-towards-the}
to lowest order perturbation theory, where upon introducing the Born
and first Markov approximations we were left with a Markovian dynamical
equation that gave the time evolution for the system. By assuming
that the interaction Hamiltonian could be written as a sum of tensor
products of operators which act on the space and reservoir separately,
we were able to write the master equation in terms of reservoir correlation
functions and thus we now actually apply this equation to a two-level
system interacting with thermal radiation.

To proceed, we need the Hamiltonians for the reduced system (two-level
atom), the reservoir (E-M field in thermal equilibrium at temperature
$T$), as well as the coupling Hamiltonian (responsible for the light-matter
interaction). The treatment of the two-level system is simple enough
(see appendix \ref{sec:A.-Two-level-systems}) as the Hamiltonian
can be described in terms of a Pauli matrix $\sigma_{z}$

\begin{equation}
H_{S}=\hbar\omega\sigma_{z}.
\end{equation}
The two-level system is coupled to the E-M field, which has the Hamiltonian

\begin{equation}
\hat{H}_{R}=\sum_{\boldsymbol{k}}\hbar\omega_{k}\hat{a}_{\boldsymbol{k}}^{\dagger}\hat{a}_{\boldsymbol{k}},
\end{equation}
where $\hat{a}_{\boldsymbol{k}}$ and $\hat{a}_{\boldsymbol{k}}^{\dagger}$
are the creation and annihilation operators for photons. Note that
we have chosen to neglect the $+1/2$ term that usually comes with
the Hamiltonian, since we are only interested in the evolution caused
by energy differences measured with respect to the vacuum state. The
coupling Hamiltonian is yet to be determined. It is well known that
in the dipole approximation the light-matter interaction Hamiltonian
has the form of a scalar product of the electric field with the dipole
moment, but this is a semi-classical description which does not take
into account the quantization of the E-M field. To arrive at the full
description of this operator we substitute the electric field $\boldsymbol{E}(\boldsymbol{r},t)$
for the electric field operator $\hat{\boldsymbol{E}}(\boldsymbol{r},t)$
in the Hesinberg picture, derived in the appendix \ref{sec:Quantization-of-the-EM-Free}.

For a single mode, the Hamiltonian is time independent and given in
terms of the number operator $\hat{n}_{\boldsymbol{k}}=\hat{a}_{\boldsymbol{k}}^{\dagger}\hat{a}_{\boldsymbol{k}}$
by $\hat{H}_{\boldsymbol{k}}=\hbar\omega_{k}\hat{n}_{\boldsymbol{k}}$.
The time evolution in the Heisenberg Picture of the creation and annihilation
operators ($a_{\boldsymbol{k}}$ and $a_{\boldsymbol{k}}^{\dagger}$)
is quite simple. In fact, we give a brief derivation, in appendix
\ref{sec:Quantization-of-the-EM-Free}, which shows they oscillate
in time with frequencies $-\omega_{\boldsymbol{k}}$ and $\omega_{\boldsymbol{k}}$
respectively. This makes it so we can construct the electric and magnetic
field operators from these operators in the Schrodinger picture quite
readily, from which the interaction Hamiltonian follows. The Hamiltonian
has, in principle, a complex structure, involving all combinations
of the two-level system raising and lowering operators ($\sigma_{\pm}=\sigma_{x}\pm i\sigma_{y}$,
where $\sigma_{x}$ and $\sigma_{y}$ are the second and third Pauli
matrices respectively) with creation and annihilation operators for
photons $(\hat{a}_{\boldsymbol{k}}^{\dagger}$ and $\hat{a}_{\boldsymbol{k}}$),
but we can simplify the dynamics by converting to the interaction
picture, performing the rotating wave approximation (RWA) and then
converting back to the Schrodinger picture (see appendix \ref{sec:Details-of-the-Derivation-of-the-Lindblad-Equation}).
In terms of the two-level system raising and lowering operators $\sigma_{\pm}$,
we can write the interaction Hamiltonian in the Schrodinger representation
as

\begin{equation}
\hat{V}=\hbar\sum_{\boldsymbol{k},\lambda}\kappa_{\boldsymbol{k},\lambda}^{*}\sigma_{-}\hat{a}_{\boldsymbol{k},\lambda}^{\dagger}+\kappa_{\boldsymbol{k},\lambda}\sigma_{+}\hat{a}_{\boldsymbol{k},\lambda},
\end{equation}
where the summation of the previous equation occurs over both the
wave-vectors $\boldsymbol{k}$ as well as the polarization components
$\lambda$. We have also introduced the coupling constants

\begin{equation}
\kappa_{\boldsymbol{k},\lambda}=-ie^{i\boldsymbol{k}\cdot\boldsymbol{r}}\sqrt{\frac{\omega_{k}}{2\hbar\varepsilon_{0}V}}\boldsymbol{e}_{\boldsymbol{k},\lambda}\cdot\boldsymbol{d}_{eg},\label{eq:Coupling}
\end{equation}
to group together the prefactors coming from our manipulations. Specifically,
in performing the several calculations, this coupling constant depends
on the inner product between the polarization vector $\boldsymbol{e}_{\boldsymbol{k},\lambda}$
and the dipole moment of the two-level system $\boldsymbol{d}_{eg}$
and on the permittivity of free space $\varepsilon_{0}$.

In the rotating wave approximation, the interaction Hamiltonian describes
photon absorption and emission by the two-state atom which leads to
its excitation and de-excitation respectively. We can now compare
the form of the interaction Hamiltonian to the master equation we
had written in terms of reservoir correlation functions. In the interaction
picture we can make the identifications

\begin{equation}
\begin{cases}
\tilde{\beta}_{1}(t)=\sigma_{-}e^{i\omega_{0}t} & \tilde{\beta}_{2}(t)=\sigma_{+}e^{i\omega_{0}t}\\
\tilde{\Gamma}_{1}(t)=\sum_{\boldsymbol{k},\lambda}\kappa_{\boldsymbol{k},\lambda}^{*}\hat{a}_{\boldsymbol{k},\lambda}^{\dagger}e^{i\omega_{k}t} & \tilde{\Gamma}_{2}(t)=\sum_{\boldsymbol{k},\lambda}\kappa_{\boldsymbol{k},\lambda}\hat{a}_{\boldsymbol{k},\lambda}e^{-i\omega_{k}t},
\end{cases}\label{eq:FactoringTwoLevelThermal}
\end{equation}
thus, in the master equation we sum over $i=1,2$ and $j=1,2$. The
reservoir correlation functions are, in this case, given by

\begin{equation}
\begin{cases}
\left\langle \tilde{\Gamma}_{1}(t)\tilde{\Gamma}_{1}(t')\right\rangle _{R}=\left\langle \tilde{\Gamma}_{2}(t)\tilde{\Gamma}_{2}(t')\right\rangle _{R}=0,\\
\left\langle \tilde{\Gamma}_{1}(t)\tilde{\Gamma}_{2}(t')\right\rangle _{R}=\sum_{\boldsymbol{k},\lambda}\left|\kappa_{\boldsymbol{k},\lambda}\right|^{2}e^{i\omega_{k}(t-\tau)}\bar{n}(\omega_{k},T),\\
\left\langle \tilde{\Gamma}_{2}(t)\tilde{\Gamma}_{1}(t')\right\rangle _{R}=\sum_{\boldsymbol{k},\lambda}\left|\kappa_{\boldsymbol{k},\lambda}\right|^{2}e^{i\omega_{k}(t-\tau)}\left[\bar{n}(\omega_{k},T)+1\right],
\end{cases}
\end{equation}
where we have used the orthogonality of the states $\left|n_{\boldsymbol{k},\lambda}\right\rangle $
to arrive at the results in the first line. We have also employed
the Bose-Einstein distribution for photons at temperature $T$ as
well as the relation $\hat{a}_{\boldsymbol{k},\lambda}\hat{a}_{\boldsymbol{k},\lambda}^{\dagger}=\hat{a}_{\boldsymbol{k},\lambda}^{\dagger}\hat{a}_{\boldsymbol{k},\lambda}+1$
to obtain the expectation values of the number operator needed for
the results of the second line. Making use of the density of states
$g(\boldsymbol{k})$, we replace the sum over the modes in the reservoir
correlation functions with an integral over the mode density. Then,
we can introduce the previously discussed first and second Markov
approximations, through which we write $\hat{\rho}_{S}^{I}(t-\tau)\approx\hat{\rho}_{S}^{I}(t)$
and let $t$ go to infinity. In this regime we are able to use the
Sokhotski-Plemelj theorem (see appendix \ref{sec:The-Sokhotski-Plemelj-identity})
and explicitly write the master equation as

\begin{align}
\frac{d\hat{\rho}_{S}(t)}{dt}= & \left(\sigma_{-}\hat{\rho}_{S}^{I}(t)\sigma_{+}-\sigma_{+}\sigma_{-}\hat{\rho}_{S}^{I}(t)\right)\left(\frac{\gamma}{2}\left[\bar{n}(\omega_{0},T)+1\right]+i(\Delta+\Delta')\right)\nonumber \\
 & +\left(\sigma_{+}\hat{\rho}_{S}^{I}(t)\sigma_{-}-\sigma_{-}\sigma_{+}\hat{\rho}_{S}^{I}(t)\right)\left(\frac{\gamma}{2}\bar{n}(\omega_{0},T)+i\Delta'\right)+\text{H.c.},\label{eq:DynamicswithConstants}
\end{align}
where H.c. stands for the Hermitian conjugate and we have introduced
the coefficients

\begin{equation}
\begin{cases}
\gamma_{A} & \equiv2\pi\sum_{\lambda}\int d^{3}\boldsymbol{k}g(\boldsymbol{k})\left|\kappa(\boldsymbol{k},\lambda)\right|^{2}\delta(kc-\omega_{0}),\\
\Delta_{L} & \equiv\sum_{\lambda}\mathcal{P}\int d^{3}\boldsymbol{k}\frac{\left|\kappa(\boldsymbol{k},\lambda)\right|^{2}g(\boldsymbol{k})}{\omega_{0}-kc},\\
\Delta'_{ac} & \equiv\sum_{\lambda}\mathcal{P}\int d^{3}\boldsymbol{k}\frac{\left|\kappa(\boldsymbol{k},\lambda)\right|^{2}g(\boldsymbol{k})}{\omega_{0}-kc}\bar{n}(kc,T).
\end{cases}\label{eq:AcoefShifts}
\end{equation}
where $\mathcal{P}$ denotes the Cauchy principal value of the integrals.
We wish now to introduce a more symmetric grouping of these terms,
such that the underlying physics becomes more apparent. Firstly we
write out all the terms of equation (\ref{eq:DynamicswithConstants})
making the Hermitian conjugate explicit and then grouping together
common factors in $\frac{\gamma_{A}}{2}\left[\bar{n}(\omega_{0},T)+1\right]$,
$\frac{\gamma_{A}}{2}\bar{n}(\omega_{0},T)$, $-i(\Delta_{L}+\Delta'_{L})$
and $i\Delta'_{L}$. We also make use the of the Pauli matrix identity
$\sigma_{\pm}\sigma_{\mp}=\left(\mathds{1}\pm\sigma_{z}\right)/2$.
Converting back to the Schrodinger picture we then obtain

\begin{align}
\frac{d\hat{\rho}_{S}(t)}{dt}= & -\frac{i}{2}\omega_{0}'\left[\sigma_{z},\hat{\rho}_{S}(t)\right]+\frac{\gamma_{A}}{2}\bar{n}(\omega_{0},T)\left(2\sigma_{+}\hat{\rho}_{S}(t)\sigma_{-}-\sigma_{-}\sigma_{+}\hat{\rho}_{S}(t)-\hat{\rho}_{S}(t)\sigma_{-}\sigma_{+}\right)\nonumber \\
 & +\frac{\gamma_{A}}{2}\left[\bar{n}(\omega_{0},T)+1\right]\left(2\sigma_{-}\hat{\rho}_{S}(t)\sigma_{+}-\sigma_{+}\sigma_{-}\hat{\rho}_{S}(t)-\hat{\rho}_{S}(t)\sigma_{+}\sigma_{-}\right),\label{eq:Mastereq}
\end{align}
where $\omega_{0}'=\omega_{0}+\Delta_{L}+2\Delta'_{ac}$. Note in
particular that this equation is easily written in Lindblad form

\begin{equation}
\frac{d\hat{\rho}_{S}(t)}{dt}=-\frac{i}{\hbar}\left[\hbar\omega_{0}'\sigma_{z}/2,\hat{\rho}_{S}(t)\right]+\sum_{k}\left(\hat{L}_{k}\hat{\rho}(t)\hat{L}_{k}^{\dagger}-\frac{1}{2}\left\{ \hat{L}_{k}^{\dagger}\hat{L}_{k},\hat{\rho}(t)\right\} \right).
\end{equation}
where, the sum occurs over $k=1,2$ and we have introduced the jump
operators $\hat{L}_{1}=\sqrt{\gamma_{A}\bar{n}(\omega_{0},T)}\sigma_{+}$
and $\hat{L}_{2}=\sqrt{\gamma_{A}\left[\bar{n}(\omega_{0},T)+1\right]}\sigma_{-}$.
In this form, we have made clear the existence of the Lamb shift $\Delta_{L}+2\Delta\text{'}_{ac}$
in the frequency corresponding to Hamiltonian of the two-level system.
The operators $\hat{L}_{1}=\sqrt{\Gamma_{+}}\sigma_{+}$, $\hat{L}_{2}=\sqrt{\Gamma_{-}}\sigma_{-}$
jump from the ground to excited state and vice-versa with decay rate
$\Gamma_{-}=\gamma_{A}\left[\bar{n}(\omega_{0},T)+1\right]$ and excitation
rate $\Gamma_{+}=\gamma_{A}\bar{n}(\omega_{0},T)$.

Note, in particular, that there are two contributions to the emission
rate. The first is temperature independent and given by $\gamma_{A}$,
while the second is temperature dependent and given by $\gamma_{A}\bar{n}(\omega_{0},T)$
. The former corresponds to a spontaneous decay rate and the latter
to stimulated transitions induced by thermal photons. The same goes
for the Lamb shift, since the term $\Delta_{L}$ is temperature independent
and corresponds to the normal Lamb shift and the term $\Delta'_{ac}$
gives the ac Stark shift associated with the thermal radiation field.
As for the excitation rate, there exists a single term given by $\gamma_{A}\bar{n}(\omega_{0},T)$.

We also note that these calculations give a correct prediction for
the \textit{Einstein A coefficient} ($\gamma_{A}$ in our notation),
associated with the spontaneous emission of light. With the density
of states given by equation (\ref{eq:DOS}) and the coupling constants
of equation (\ref{eq:Coupling}) and making a suitable choice of coordinates
in $k$-space it is simple enough to evaluate $\gamma_{A}$ as given
in equation (\ref{eq:AcoefShifts}). We obtain

\begin{equation}
\gamma_{A}=\frac{1}{4\pi\varepsilon_{0}}\frac{4\omega_{0}^{3}d_{eg}^{2}}{3\hbar c^{3}},\label{eq:EinsteinAcoef}
\end{equation}
and thus, we see that our open quantum systems approach gives the
correct result for the Einstein A coefficient and as such, a correct
prediction for the spontaneous decay rate for the two-level system.
This shows an agreement of the Wigner-Weisskopf theory prediction
of the Lamb shift and Einstein A coefficient (see \citet{Hecht2000})
and the open quantum systems approach. The latter also predicts an
ac Stark shift which is not taken into account by the former method.
This extension of the quantum mechanical description to a finite temperature
regime is something we will explore in more detail in the following
sections.

\section{Quantization of the surface plasmon polariton field via the dyadic
Green's function method \label{Sec:quantization_of_the_polar_field}}

\subsection{An introduction to dyadic Green's functions \label{Sec:dyadicGreen}}

In this paper, our final aim is to study the interaction of qubits
and the exchange of information via surface plasmon polaritons (SPPs)
in a graphene sheet. We thus now step away from the theory of open
quantum systems and step foot into the realm of macroscopic quantum
electrodynamics (macroscopic QED), which can be characterized as the
study of the quantized E-M field in the presence of macroscopic media
(see \citet{scheel2009macroscopic}, \citet{buhmann2013dispersion}).
The aim of this section is to introduce an important tool of macroscopic
QED, called the \textit{dyadic Green's function} or \textit{Green's
tensor}. This is quite a remarkable object because it allows for the
characterization of absorption and dispersion properties in media
characterized by a certain electric permittivity $\varepsilon(\boldsymbol{r},\omega)$
and a certain magnetic permeability $\mu(\boldsymbol{r},\omega)$.
This is especially noteworthy since the description it provides is
non-classical, despite the emergence of the Green's tensor from Maxwell's
equations, which are a phenomenological and classical description
of macroscopic systems. In general, the Green's tensor as a mathematical
object plays the role of the usual Green's function for partial differential
equations in the case of vector equations. In particular, we will
use it to find a solution of the inhomogeneous Helmholtz equation
for the electric field

\begin{equation}
\nabla\times\nabla\times\boldsymbol{E}(\boldsymbol{r},\omega)-\frac{\omega^{2}}{v^{2}}\boldsymbol{E}(\boldsymbol{r},\omega)=i\omega\mu\mu_{0}\boldsymbol{j}(\boldsymbol{r},\omega),\label{eq:Helmholtz}
\end{equation}
where again $\boldsymbol{E}(\boldsymbol{r},\omega)$ is the electric
field at a given point in space $\boldsymbol{r}$ and frequency $\omega$,
$\boldsymbol{j}(\boldsymbol{r},\omega)$ is the electrical current
at that point, $\mu_{0}$ is the magnetic permeability of free space
and $\mu$ is the relative magnetic permeability of the medium.

As stated, the idea of the mathematical application of the Green's
tensor is much the same as that of the usual Green's function (see
\citet{2016scalarandapplied} for a thorough introduction to dyadic
Green's functions), in the sense that we want to turn a partial differential
equation problem, into a linear algebra problem. The Green's tensor
will play the role of the inverse to the differential operator in
question. We can state the general second order ordinary differential
equation problem, up to transformations of the variable $x$, as

\begin{equation}
L[y(x)]=a_{0}(x)y''(x)+a_{1}(x)y'(x)+a_{2}(x)y(x)=-f(x),
\end{equation}
with $0<x<1$ and $y(0)=y(1)=0$. The aim of the Green's function
method is to present a solution in terms of an integral. To do so,
we find a function $G(x,t)$ for which

\begin{equation}
y(x)=\int_{0}^{1}G(x,t)f(t)\ dt.
\end{equation}
This function $G(x,t)$ is the namesake Green's function (see appendix
\ref{subsec:The-idea-behind}). There are usually several methods
at our disposal for finding such functions. We go over two of them
in full generality in the appendixes, \ref{subsec:Construction-of-Green's}
and \ref{subsec:Construction-of-Green's-1}, the first of which makes
use of the \textit{Dirac-$\delta$ function} and the second of \textit{eigenfunction
expansions}. Simply put, and as already stated, the general idea is
to proceed as in linear algebra, where if we want to solve a matrix
equation $A\boldsymbol{u}=\boldsymbol{f},$ we simply try to find
the inverse operator $A^{-1}$, which is possible as long as 0 is
not an eigenvalue of $A$. The same approach is valid for the dyadic
Green's function $\bar{\bar{G}}(\boldsymbol{r},\boldsymbol{r}',\omega)$,
which is constructed from a solution of the Helmholtz equation with
a Dirac-delta source

\begin{equation}
\nabla\times\nabla\times\bar{\bar{G}}(\boldsymbol{r},\boldsymbol{r}',\omega)-\frac{\omega^{2}}{c^{2}}\bar{\bar{G}}(\boldsymbol{r},\boldsymbol{r}',\omega)=\mathds{1}\delta(\boldsymbol{r}-\boldsymbol{r}'),
\end{equation}
where $\mathds{1}$ is the identity dyadic. Much like in linear algebra
an operator $B$ is the inverse of an operator $A$ if $AB=\mathds{1}$.
In the case of the Helmholtz equation, a solution for the electric
field can thus be constructed from as an integral

\begin{equation}
\boldsymbol{E}(\boldsymbol{r},\omega)=\boldsymbol{E}_{0}(\boldsymbol{r},\omega)+i\omega\mu_{0}\int d^{3}\boldsymbol{r}'\bar{\bar{G}}(\boldsymbol{r},\boldsymbol{r}',\omega)\cdot\boldsymbol{j}(\boldsymbol{r},\omega).
\end{equation}
Physically, the Green's tensor contains the geometry and physical
properties of the involved media, and will play a role in the microscopic
description of dissipation, as we shall later see. We will be able
to think of it as an electric response function that will carry the
interaction generated by the current $\boldsymbol{j}(\boldsymbol{r},\omega)$
from the point $\boldsymbol{r}'$ to the point $\boldsymbol{r}$.

We present in the appendix \ref{subsec:Dyadic-Green's-functions}
a derivation for the Green's tensor in free space for which we make
use of the Dirac-$\delta$ method, while in the following section
we will derive this quantity for the case of SPPs in graphene via
the eigenfunction expansion method, which will prove to be a more
challenging endeavor.

\subsection{Deriving the Green's tensor for our system}

A \textit{surface plasmon polariton} (SPP) is a two-dimensional electromagnetic
excitation that lives in dielectric-metal interfaces, or in more complicated
geometries such as the one we are considering, where a graphene sheet
is embedded or ``sandwiched'' between two semi-infinite dielectrics.
This configuration can be described by a dielectric function over
the whole space of the form (see \citet{Goncalves2016Graphene} for
discussions regarding the conductivity of graphene).

\begin{equation}
\varepsilon(z)=\varepsilon_{1}\Theta(z)+i\frac{\sigma(\omega)}{\varepsilon_{0}\omega}\delta(z)+\varepsilon_{2}\Theta(-z).\label{eq:dieletric}
\end{equation}
where $\sigma(\omega)$ is the conductivity of the graphene sheet.
For $z>0$ the dielectric function is a constant $\varepsilon_{1}$
while for $z<0$ the dielectric function is also constant and equal
to $\varepsilon_{2}$. This information is contained in the Heaviside
theta functions $\Theta(\pm z)$ written in equation (\ref{eq:dieletric}).
At $z=0$ we have a contribution from the graphene sheet. This dielectric
function defines the system we are interested in studying because
it contains, as a macroscopic quantity, all the microscopic information
about the media that make up the system.

\begin{figure}
\begin{centering}
\includegraphics[scale=0.45]{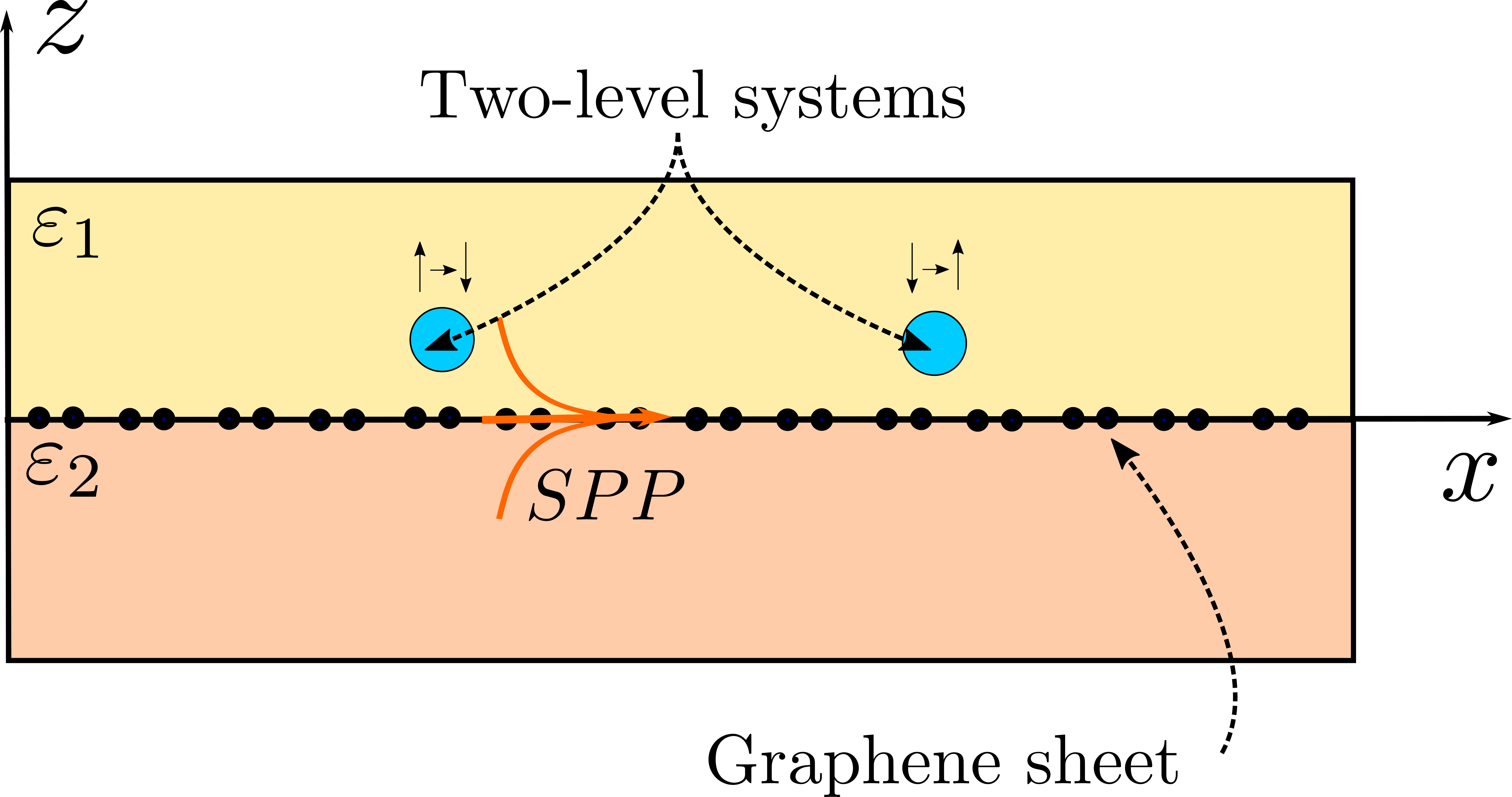} 
\par\end{centering}
\caption{Schematic for the system that concerns this study. It consists of
a graphene sheet (black dots) sandwiched between two dielectrics with
electric permittivity $\varepsilon_{1}$ (yellow) and $\varepsilon_{2}$
(salmon). Two two-level systems (qubits) represented by the blue circles,
interact via exchange of an SPP (in orange). Emission of an SPP causes
the first which was initially in the excited state to go to the ground
state, while the second one absorbs the SPP and goes from the ground
to the excited state.}
\end{figure}

Our discussion will now focus on essentially two topics. The first
is actually finding the Green's tensor for SPPs traveling along the
Graphene sheet, and the second is then using this Green's tensor to
follow through with a quantization scheme for the SPPs based on the
presence of Langevin noise currents in the Graphene (see \citet{Matloob1995}).

To find the Green's tensor we work in the Weyl or temporal gauge (see
\citet{Loffelholz2003,kurt1987,Creutz1979QuantumEI}) we have the
scalar potential $\phi=0$. In this gauge, the vector potential and
the electric field are related through $\boldsymbol{E}(\boldsymbol{r})=i\omega\boldsymbol{A}(\boldsymbol{r})$
and the magnetic field can also be derived from $\boldsymbol{B}(\boldsymbol{r})=\nabla\times\boldsymbol{A}(\boldsymbol{r})$.
The electric field relation allows for the Helmholtz equation to be
written in terms of the vector potential. Assuming that no currents
exist within the dielectric media on both sides of the graphene sheet,
we obtain

\begin{equation}
-\nabla\times\nabla\times\boldsymbol{A}(\boldsymbol{r})+\frac{\omega^{2}}{c^{2}}\varepsilon(z)\boldsymbol{A}(\boldsymbol{r})=0.
\end{equation}

Our approach will make use of an expansion in eigenmodes, from which
the Green's function will be calculated (see \citet{Sondergaard2001})
as

\begin{equation}
\bar{\bar{G}}(\boldsymbol{r},\boldsymbol{r}';\omega)=\int d^{2}\boldsymbol{q}\sum_{n}\frac{\boldsymbol{a}_{n}(\boldsymbol{r})\otimes\left[\tilde{\boldsymbol{a}}_{n}(\boldsymbol{r}')\right]^{*}}{N_{n}\lambda_{n}},\label{eq:GreensEigen}
\end{equation}
where the integral is over $\boldsymbol{q}$ corresponding to the
plasmon wave-vector, $\lambda_{n}$ are the eigenvalues, $N_{n}$
a normalization factor, and $\boldsymbol{a}_{n}$ the eigenmodes,
which are solutions to the equation

\begin{equation}
-\nabla\times\nabla\times\boldsymbol{a}_{n}(\boldsymbol{r})+\frac{\omega^{2}}{c^{2}}\varepsilon(z)\boldsymbol{a}_{n}(\boldsymbol{r})=\varepsilon(z)\lambda_{n}\boldsymbol{a}_{n}(\boldsymbol{r}).\label{eq:Eigenmodes}
\end{equation}
Note that $\boldsymbol{a}_{n}$ $\lambda_{n}$ and $N_{n}$ all depend
on $q$ but we omit this dependence for the sake of lightening up
the notation. Equation (\ref{eq:GreensEigen}) can be written as an
eigenvalue problem $H(\boldsymbol{r})\boldsymbol{a}_{n}(\boldsymbol{r})=\varepsilon(z)\lambda_{n}\boldsymbol{a}_{n}(\boldsymbol{r})$,
with the operator $H(\boldsymbol{r})=-\nabla\times\nabla\times+\omega^{2}\varepsilon(z)/c$.
This is still a complicated equation, but writing it in this form
makes it somewhat easier to handle. In fact, we already know all the
pieces to the puzzle, since all that we are looking for are solutions
in the form of SPPs. These correspond to evanescent waves along the
$z$ axis, with a small penetration depth along the dielectric media
(in fact this penetration depth can be remarkably small and thus graphene
plasmon polaritons are much more confined than usual SPPs, see \citet{Luo2013gp, Principi2018gp}),
and to traveling waves along the graphene sheet. This suitable choice
of solution conceptually closes the problem, since all we have to
do is substitute the form of the SPP eigenmodes

\begin{equation}
\begin{cases}
\boldsymbol{a}_{n}^{(1)}(\boldsymbol{r})=(a_{x}^{(1)},a_{y}^{(1)},a_{z}^{(1)})e^{i(q_{x}x+q_{y}y)}e^{-\kappa_{1}z}\Theta(z),\\
\boldsymbol{a}_{n}^{(2)}(\boldsymbol{r})=(a_{x}^{(2)},a_{y}^{(2)},a_{z}^{(2)})e^{i(q_{x}x+q_{y}y)}e^{\kappa_{2}z}\Theta(-z),
\end{cases}
\end{equation}
where the $(1)$ index shows that these propagate in the medium with
$\varepsilon_{1}$ while the index $(2)$ shows that these propagate
in the medium with $\varepsilon_{2}$, into equation (\ref{eq:Eigenmodes}).
We then write this as a matrix equation, solve for the eigenvalues
and normalization and then substitute this all into equation (\ref{eq:GreensEigen})
to find the Green's function. Of course things are not so easy in
practice since this procedure is mathematically challenging, but also,
several noteworthy physical insights can be derived by performing
it. We leave most mathematical details to the appendix \ref{subsec:Finding-the-eigenmodes}
and focus on the physical aspects of the derivation. The first of
these insights may be brought to the limelight when noticing that
in indexing the eigenmodes with the medium they propagate in, we are
actually ``splitting'' our solution in two, and thus we will have
two sets of eigenvalues $\left\{ \lambda_{n}^{(1)}\right\} $and $\left\{ \lambda_{n}^{(2)}\right\} $
which for each $n$ should actually correspond to the same SPP mode,
and therefore should agree with each other. This imposes a constraint
$\lambda_{n}^{(1)}=\lambda_{n}^{(2)}$ which allows us to connect
the solution across the graphene sheet as well as relate $\kappa_{1}$
with $\kappa_{2}$ and $q=\sqrt{q_{x}^{2}+q_{y}^{2}}$ through the
relation

\begin{equation}
\kappa_{2}^{2}=\frac{\varepsilon_{2}}{\varepsilon_{1}}\kappa_{1}^{2}+\left(\frac{\varepsilon_{1}-\varepsilon_{2}}{\varepsilon_{1}}\right)q^{2}.
\end{equation}
The second insight one gains comes from the form of the eigenvalues
themselves. In fact, upon examining the eigenvalues present in the
appendix \ref{subsec:Boundary-conditions--} we notice that they are
degenerate for $n=2,3$, which means they must indeed be superpositions
of $s-$ and $p-$polarization vectors $\hat{e}_{s}\propto\boldsymbol{k}_{1}\times(0,0,1)$
and $\hat{e}_{p}\propto\hat{e}_{s}\times\boldsymbol{k}_{1}$ with
$\boldsymbol{k}_{1}=(q_{x},q_{y},i\kappa_{1})$ (see \citet{Sipe1987}).
Since SPPs have only $p-$polarized modes, then we shall only retain
this $p-$polarization with eigenvalue $\varepsilon_{1}\lambda_{p}^{(1)}=\left(-q^{2}+\kappa_{1}^{2}+\varepsilon_{1}\frac{\omega^{2}}{c^{2}}\right)$.
To further connect the solution for $z>0$ and $z<0$ we look at the
boundary condition for the electric field at $z=0$. In particular,
we have $\boldsymbol{E}_{\parallel}^{(1)}=\boldsymbol{E}_{\parallel}^{(2)}$.
Noting the relation between the electric field and vector potential
that stems from the Weyl gauge, we can use it to relate the amplitudes
of the modes in both media, labeled $A_{1}$ and $A_{2}$. The boundary
condition yields the relation $A_{2}=-A_{1}\kappa_{1}/\kappa_{2}\sqrt{(q^{2}+\kappa_{2}^{2})/(q^{2}+\kappa_{1}^{2})}$.
We can also use the boundary condition for the magnetic field $\hat{\boldsymbol{z}}\times(\boldsymbol{B}_{\parallel}^{(1)}-\boldsymbol{B}_{\parallel}^{(2)})=\mu_{0}\boldsymbol{K}=\mu_{0}\sigma(\omega)\boldsymbol{E}_{\parallel}^{(1)}$
and Ohm's law in the graphene sheet to provide some additional constraints
into the wave-vector and allow us to relate it to the frequency $\omega$.
All this procedure results in a dispersion relation

\begin{equation}
\kappa_{2}\left(q^{2}-\kappa_{1}^{2}\right)+\kappa_{1}\left(q^{2}-\kappa_{2}^{2}\right)=-i\omega\mu_{0}\sigma(\omega)\kappa_{1}\kappa_{2},
\end{equation}
which, when assuming a conductivity computed with the Drude model,
and if we perform the electrostatic approximation $\kappa_{1}\approx\kappa_{2}\approx q$,
gives the final dispersion relation for SPPs which we will use throughout
the rest of this manuscript

\begin{equation}
\omega_{\text{spp}}^{2}(q)\equiv\frac{2\alpha cE_{F}}{\hbar}\frac{q}{\bar{\varepsilon}},\label{eq:dispersion}
\end{equation}
where $E_{F}$ is the Fermi energy, $\alpha$ is the fine structure
constant and $\bar{\varepsilon}=(\varepsilon_{1}+\varepsilon_{2})/2$
is the mean dielectric constant. The normalization constant can also
be found quite readily as described in the appendix \ref{subsec:Finding-the-normalization}.
The calculation yields, in the electrostatic approximation, $N_{p}\approx-\bar{\varepsilon}/q$.
Now, we have only to put these pieces together and crank the mathematical
engine (and by cranking the mathematical engine we mean calculating
16 integrals of Bessel functions over the complex plane, the details
of which are presented in the appendix \ref{subsec:Construction-of-the},
\ref{subsec:Evaluation-of-the}, and \ref{subsec:Putting-the-results}).
Making use of equation (\ref{eq:GreensEigen}) to construct the SPP
part of the Green's tensor. After toiling over the calculations, one
finds the dyadic Green's function

\begin{align}
\bar{\bar{G}}(\boldsymbol{r},\boldsymbol{r}';\omega)=c^{2}\frac{\pi^{2}}{\bar{\varepsilon}}e^{-q_{\text{spp}}(z+z')}\frac{q_{\text{spp}}^{2}}{A(\omega)}\bar{\bar{M}}(q_{\text{ssp}},R,\phi),\label{eq:GrapheneGreensFunction}
\end{align}
where $A(\omega)\approx\omega\sigma_{2}(\omega)/2\bar{\varepsilon}\varepsilon_{0}$
is a function of frequency such that the dispersion relation reads
$\omega_{\text{spp}}^{2}=A(\omega)q$, and $\phi$ is the angle that
the vector $\boldsymbol{\rho}-\boldsymbol{\rho}'$ corresponding to
the position at $z=0$ where the Green's function is evaluated, makes
with the $x-$axis. Finally the matrix $\bar{\bar{M}}(q,\phi)$ has
diagonal elements

\begin{equation}
\begin{cases}
M_{11}=\left[Y_{0}\left(q_{\text{spp}}R\right)-iJ_{0}\left(q_{\text{spp}}R\right)\right]\cos^{2}\phi-\frac{1}{Rq_{\text{ssp}}}\left[Y_{1}\left(q_{\text{spp}}R\right)-iJ_{1}\left(q_{\text{spp}}R\right)\right]\cos2\phi,\\
M_{22}=\left[Y_{0}\left(q_{\text{spp}}R\right)-iJ_{0}\left(q_{\text{spp}}R\right)\right]\sin^{2}\phi+\frac{1}{Rq_{\text{ssp}}}\left[Y_{1}\left(q_{\text{spp}}R\right)-iJ_{1}\left(q_{\text{spp}}R\right)\right]\cos2\phi,\\
M_{33}=\left[Y_{0}\left(q_{\text{spp}}R\right)-iJ_{0}\left(q_{\text{spp}}R\right)\right],
\end{cases}
\end{equation}
and off-diagonal elements

\begin{equation}
\begin{cases}
M_{12}=M_{21}=-\left[\left[Y_{2}\left(q_{\text{spp}}R\right)-iJ_{2}\left(q_{\text{spp}}R\right)\right]+\frac{2}{\pi q_{\text{spp}}R^{2}}e^{q_{\text{spp}}(z+z')}\frac{A(\omega)}{q_{\text{spp}}}\right]\cos\phi\sin\phi,\\
M_{13}=M_{31}=-\left[Y_{1}\left(q_{\text{spp}}R\right)-iJ_{1}\left(q_{\text{spp}}R\right)\right]\cos\phi,\\
M_{23}=M_{32}=-\left[Y_{1}\left(q_{\text{spp}}R\right)-iJ_{1}\left(q_{\text{spp}}R\right)\right]\sin\phi.
\end{cases}.
\end{equation}
We now conclude this section where we have found the SPP part of Green's
tensor for the specific geometry of a Graphene sheet embedded between
two dielectrics. We shall later see that it is this Green's tensor,
or rather, its imaginary part that plays a role in the interaction
between matter and the SPPs, determining specifically the decay rates
for two level systems.

\subsection{Quantization of surface plasmon polaritons using dyadic Green's functions}

In this section we take a step back from the specific form of our
Green's tensor and analyze a quantization scheme for SPPs utilizing
the method of dyadic Green's functions. Our approach will be based
on the introduction of \textit{Langevin-noise currents} $\boldsymbol{J}_{N}(\boldsymbol{\rho},\omega)$
in graphene to model dissipation. This procedure is outlined in \citet{allameh2015quantization}
in the case of a general dissipative medium. Here $\boldsymbol{\rho}$
is the longitudinal spatial coordinate, and $\omega$ is the frequency
of the noise current. These dissipative noise currents will then be
the basis upon which we construct the bosonic creation and annihilation
operators for the SPPs. To solve for the electric field with these
currents present, we will have to solve an non-homogeneous Helmholtz
equation of the form

\begin{equation}
-\nabla\times\nabla\times\boldsymbol{A}(\boldsymbol{r})+\frac{\omega^{2}}{c^{2}}\varepsilon(z)\boldsymbol{A}(\boldsymbol{r})=\mu_{0}\boldsymbol{J}_{N}(\boldsymbol{\rho},\omega)\delta(z),\label{eq:HelmholtzLangevin}
\end{equation}
to which, as we know from the previously discussed mathematics of
Green's tensors, a solution can be provided in the form of an integral

\begin{equation}
\boldsymbol{A}(\boldsymbol{r},\omega)=\mu_{0}\int d^{3}\boldsymbol{r}'\bar{\bar{G}}(\boldsymbol{r},\boldsymbol{r}';\omega)\cdot\boldsymbol{J}_{N}(\boldsymbol{\rho}',\omega)\delta(z').\label{eq:ANoise}
\end{equation}
In the Weyl gauge, since the Lagrangian density of the E-M field is
$\mathcal{L}=(\varepsilon_{0}\dot{\boldsymbol{A}}^{2}+(\nabla\times\boldsymbol{A})^{2}/\mu_{0})/2$,
we can find the canonically conjugate variable to $\boldsymbol{A}$
by calculating $\partial\mathcal{L}/\partial\dot{\boldsymbol{A}}=\varepsilon_{0}\dot{\boldsymbol{A}}(\boldsymbol{r})=-\varepsilon_{0}\boldsymbol{E}(\boldsymbol{r},\omega)$.
This means that when we promote the vector potential and the electric
field to operators, we have the canonical commutation relation $\left[\hat{A}_{i}(\boldsymbol{r},\omega),-\varepsilon_{0}\hat{E}_{j}(\boldsymbol{r}',\omega)\right]=i\hbar\delta_{ij}\delta(\boldsymbol{r}-\boldsymbol{r}')$.
This result holds in general, but since we are interested only in
the commutation relations for the SPP part of the fields, we instead
impose the commutation relation $\left[\hat{A}_{i}(\boldsymbol{r},\omega),-\varepsilon_{0}\hat{E}_{j}(\boldsymbol{r}',\omega)\right]=i\hbar\delta_{ij}\delta^{\parallel}(\boldsymbol{r}-\boldsymbol{r}')$,
where we have replaced the full Dirac-delta function with a \textit{longitudinal
delta function} (see the appendix \ref{subsec:Transverse-and-Longitudinal}
for a brief note regarding transverse and longitudinal delta functions).

In addition to this, we impose that the Langevin-noise currents satisfy
the commutation relations $\left[\hat{J}_{N,i}(\boldsymbol{\rho};\omega),\hat{J}_{N,j}^{\dagger}(\boldsymbol{\rho}';\omega')\right]=\mathcal{C}\alpha(\omega)\delta_{ij}\delta(\boldsymbol{\rho}-\boldsymbol{\rho}')\delta(\omega-\omega')$
where $\mathcal{C}$ and $\alpha(\omega)$ are a constant and a function
determined so that equation (\ref{eq:ANoise}) is consistent with
both commutation relations for the operators $\hat{\boldsymbol{A}}$,
$\hat{\boldsymbol{E}}$ as well as $\hat{\boldsymbol{J}}_{N}$. We
go over the derivation of $\mathcal{C}$ and $\alpha(\omega)$ in
detail in the appendix \ref{subsec:Quantizing-the-Plasmon}, but the
derivation is quite simple as we need only to substitute equation
(\ref{eq:ANoise}) into the commutator, bring the Green's tensors
outside it and then substitute the commutation relation for the noise
currents in. A few tricks are then necessary (see appendix \ref{sec:Details-of-the-calculation-of-the-dynamics-of-one-qubit}
for the details), but one can show that with the propper choice of
$\mathcal{C}=\bar{\varepsilon}\hbar/(2\pi)^{3}$ and $\alpha(\omega)=\omega\sigma_{1}(\omega)$,
it is possible to write the commutation relations for our system consistently
and therefore ``normalize'' the commutation relation for the noise
currents into a bosonic commutation relation which yields the SPP
creation and annihilation operators

\begin{equation}
\hat{\text{\textbf{f}}}(\boldsymbol{\rho},\omega)=\frac{\hat{\boldsymbol{J}}_{N}(\boldsymbol{\rho},\omega)}{\sqrt{\mathcal{\mathcal{C}}\alpha(\omega)}}.
\end{equation}

This result allows us to write the vector potential operator and the
electric field operator in terms of these $\hat{\text{\textbf{f}}}$-operators,
and therefore this concludes the section on the quantization of the
SPP field, as we can now write

\begin{equation}
\begin{cases}
\hat{\boldsymbol{A}}(\boldsymbol{r},\omega)=\mu_{0}\sqrt{\mathcal{\mathcal{C}}\alpha(\omega)}\int d^{3}\boldsymbol{r}'\bar{\bar{G}}(\boldsymbol{r},\boldsymbol{r}',\omega)\cdot\hat{\text{\textbf{f}}}(\boldsymbol{\rho}',\omega)\delta(z'),\\
\hat{\boldsymbol{E}}(\boldsymbol{r})=i\int_{0}^{\infty}d\omega\omega\hat{\boldsymbol{A}}(\boldsymbol{r},\omega)-i\int_{0}^{\infty}d\omega\omega\hat{\boldsymbol{A}}^{\dagger}(\boldsymbol{r},\omega).
\end{cases}
\end{equation}
where the electric field operator in the Schrodinger picture was obtained
by taking the Fourier transform of $\hat{\boldsymbol{E}}(\boldsymbol{r},\omega)$
at $t=0$. We see that the quantized field is connected deeply with
the Green's tensor, and in fact this macroscopic quantity appears
directly when the field is written in terms of the creation and annihilation
operators for the SPPs.

\section{Time evolution of a single qubit coupled to a plasmonic bath in thermal
equilibrium \label{Sec:singleQubit}}

In this section we bring together the two thematically disconnected
parts of this study we have so far discussed, and make use of the
theory of open quantum systems and dyadic Green's functions, with
which we arrived at the SPP field creation and destruction operators
$\hat{\text{\textbf{f}}}^{\dagger}(\boldsymbol{\rho},\omega)$ and
$\hat{\text{\textbf{f}}}(\boldsymbol{\rho},\omega)$ respectively,
to describe the dynamics of a two-level system coupled to a plasmonic
bath. The theory of open quantum systems will come into play, since
the SPP field will (at a finite temperature $T$) act as a thermal
reservoir of SPPs which will constitute an environment to which the
two-level system is coupled. Therefore, much like in the case of the
thermal E-M field, we proceed with a microscopic derivation of the
master equation in Lindblad form. We start with the study of the dynamics
of a single qubit coupled to a plasmonic bath in thermal equilibrium.

To develop our theory we need the Hamiltonians for the single qubit,
the reservoir (SPP field in thermal equilibrium at temperature $T$),
and the coupling Hamiltonian. We set the ground state of the two-level
system with zero energy, which amounts to the choice of a reduced
system Hamiltonian of the form

\begin{equation}
H_{S}=\hbar\omega_{0}\sigma_{+}\sigma_{-},
\end{equation}
instead of the usual $H_{S}=\hbar\omega_{0}\sigma_{z}$. The Hamiltonian
of the SPP field is constructed from the vector potential operator
and reads

\begin{equation}
H_{R}=\hbar\int d^{2}\boldsymbol{\rho}\int_{0}^{\infty}d\omega\omega\ \hat{\text{\textbf{f}}}^{\dagger}(\boldsymbol{\rho},\omega)\cdot\hat{\text{\textbf{f}}}(\boldsymbol{\rho},\omega).
\end{equation}
The interaction Hamiltonian can also be constructed from these operators
in the Schrodinger picture. It contains mixed terms involving the
creation and destruction SPP field operators, as well as the raising
and lowering operators for the qubit. The dyadic Green's function
will then carry the interaction from $\boldsymbol{r}$, which is the
position of the qubit to $\boldsymbol{\rho}$, where the SPP is created.
The summarized procedure leads to $\hat{V}$ written in the Schrödinger
picture

\begin{align}
\hat{V}= & -\sigma_{+}i\int_{0}^{\infty}d\omega\omega\beta(\omega)\int d^{2}\boldsymbol{\rho}\boldsymbol{d}_{eg}\cdot\bar{\bar{G}}(\boldsymbol{r},\boldsymbol{\rho},z=0;\omega)\cdot\hat{\text{\textbf{f}}}(\boldsymbol{\rho},\omega)\nonumber \\
 & +i\int_{0}^{\infty}d\omega\omega\beta(\omega)\int d^{2}\boldsymbol{\rho}\hat{\text{\textbf{f}}}^{\dagger}(\boldsymbol{\rho},\omega)\cdot\bar{\bar{G}}^{\dagger}(\boldsymbol{r},\boldsymbol{\rho},z=0;\omega)\cdot\boldsymbol{d}_{eg}\sigma_{-},
\end{align}
where, we have defined $\beta(\omega)=\mu_{0}\sqrt{\mathcal{C}\alpha(\omega)}$.
Although these Hamiltonians appear to be significantly more complicated
than that of simple thermal radiation, the procedure is essentially
the same as before, since we can separate the coupling Hamiltonian
into a tensor product of operators $\beta_{1,2}=\sigma_{\pm}$ and

\begin{equation}
\begin{cases}
\Gamma_{1}=i\int_{0}^{\infty}d\omega\omega\beta(\omega)\int d^{2}\boldsymbol{\rho}\hat{\text{\textbf{f}}}^{\dagger}(\boldsymbol{\rho},\omega)\cdot\bar{\bar{G}}^{\dagger}(\boldsymbol{r},\boldsymbol{\rho},z=0;\omega)\cdot\boldsymbol{d}_{eg},\\
\Gamma_{2}=i\int_{0}^{\infty}d\omega\omega\beta(\omega)\int d^{2}\boldsymbol{\rho}\boldsymbol{d}_{eg}\cdot\bar{\bar{G}}(\boldsymbol{r},\boldsymbol{\rho},z=0;\omega)\cdot\hat{\text{\textbf{f}}}(\boldsymbol{\rho},\omega).
\end{cases}\label{eq:GammaTensor}
\end{equation}
The aim is to calculate the reservoir correlation functions, which
in this case correspond to thermal averages of equation (\ref{eq:ReservoirCorr})
but with the $\beta_{12}$ and $\Gamma_{12}$ as defined in equation
(\ref{eq:GammaTensor}). This is no easy task and in fact conceals
several mathematical subtleties, because in order to calculate the
correlation functions it is necessary to evaluate the thermal averages
of the $\hat{\text{\textbf{f}}}$-operators. In the literature (see
\citet{Philbin2011}), it is often simply assumed that the following
relations hold, due to the mathematical subtlety of the evaluation
of the thermal averages

\begin{equation}
\begin{cases}
\left\langle \hat{\text{\textbf{f}}}(\boldsymbol{\rho},\omega)\hat{\text{\textbf{f}}}(\boldsymbol{\rho}',\omega')\right\rangle _{R}=0=\left\langle \hat{\text{\textbf{f}}}^{\dagger}(\boldsymbol{\rho},\omega)\hat{\text{\textbf{f}}}^{\dagger}(\boldsymbol{\rho}',\omega')\right\rangle _{R},\\
\left\langle \hat{\text{\textbf{f}}}^{\dagger}(\boldsymbol{\rho},\omega)\hat{\text{\textbf{f}}}(\boldsymbol{\rho}',\omega')\right\rangle _{R}=\bar{n}_{\text{spp}}(\omega)\delta(\boldsymbol{\rho}-\boldsymbol{\rho}')\delta(\omega-\omega'),\\
\left\langle \hat{\text{\textbf{f}}}(\boldsymbol{\rho},\omega)\hat{\text{\textbf{f}}}^{\dagger}(\boldsymbol{\rho}',\omega')\right\rangle _{R}=\left[\bar{n}_{\text{spp}}(\omega)+1\right]\delta(\boldsymbol{\rho}-\boldsymbol{\rho}')\delta(\omega-\omega').
\end{cases}
\end{equation}
We include in the appendix \ref{sec:Thermal-averages-of}, however,
a derivation based on discretizing the integrals in frequency and
position that come up in the Hamiltonian. We notice still, that the
reservoir correlation functions present in equation (\ref{eq:ReservoirCorr})
are written in the interaction picture, however, a simple derivation
included in the appendix \ref{sec:Details-of-the-calculation-of-the-dynamics-of-one-qubit}
shows that their representation in this picture is entirely analogous
to the representation of usual creation and annihilation operators.
After substituting the reservoir correlation functions back into the
Redfield equation and performing both the Born and Markov approximations,
we obtain a master equation of entirely the same form as equation
(\ref{eq:Mastereq}) which we repeat here for convenience

\begin{align}
\frac{d\hat{\rho}_{S}(t)}{dt}= & -\frac{i}{2}\omega_{0}'\left[\sigma_{z},\hat{\rho}_{S}(t)\right]+\frac{\gamma}{2}\bar{n}(\omega_{0},T)\left(2\sigma_{+}\hat{\rho}_{S}(t)\sigma_{-}-\sigma_{-}\sigma_{+}\hat{\rho}_{S}(t)-\hat{\rho}_{S}(t)\sigma_{-}\sigma_{+}\right)\nonumber \\
 & +\frac{\gamma}{2}\left[\bar{n}(\omega_{0},T)+1\right]\left(2\sigma_{-}\hat{\rho}_{S}(t)\sigma_{+}-\sigma_{+}\sigma_{-}\hat{\rho}_{S}(t)-\hat{\rho}_{S}(t)\sigma_{+}\sigma_{-}\right).
\end{align}

with $\omega_{0}'=\omega_{0}+\Delta+2\Delta'$. Here, however, the
decay rate $\gamma$ as well as the normal Lamb shift $\Delta$ and
thermal shift $\Delta'$ due to the SPP field are dependent on the
imaginary part of the Green's tensor which is obtained via an integral
of a product of Green's tensors (see appendix \ref{sec:Integrals-of-products}).
Their form can be written as

\begin{equation}
\begin{cases}
\gamma\equiv2\pi\int_{0}^{\infty}d\omega\boldsymbol{d}_{eg}\cdot\bar{\mathcal{K}}(\boldsymbol{r},\omega)\cdot\boldsymbol{d}_{eg}\delta(\omega-\omega_{0}),\\
\Delta\equiv\mathcal{P}\int_{0}^{\infty}d\omega\frac{\boldsymbol{d}_{eg}\cdot\bar{\mathcal{K}}(\boldsymbol{r},\omega)\cdot\boldsymbol{d}_{eg}}{\omega_{0}-\omega},\\
\Delta'\equiv\mathcal{P}\int_{0}^{\infty}d\omega\frac{\boldsymbol{d}_{eg}\cdot\bar{\mathcal{K}}(\boldsymbol{r},\omega)\cdot\boldsymbol{d}_{eg}\bar{n}_{\text{spp}}(\omega)}{\omega_{0}-\omega}.
\end{cases}\label{eq:ratesandshifts}
\end{equation}

where the kernel $\bar{\mathcal{K}}(\boldsymbol{r},\omega)$ is defined
as

\begin{equation}
\bar{\mathcal{K}}(\boldsymbol{r},\omega)=\frac{\bar{\varepsilon}\mu_{0}}{(2\pi)^{3}\hbar}\omega^{2}\text{Im}[\bar{\bar{G}}(\boldsymbol{r},\boldsymbol{r};\omega)],
\end{equation}

The previous equation can also be brought into Lindblad form by defining
the jump operators

\begin{equation}
\begin{cases}
\hat{L}_{1}=\sqrt{\frac{\bar{\varepsilon}\mu_{0}}{(2\pi)^{3}\hbar}\omega_{0}^{2}\boldsymbol{d}_{eg}\cdot\text{Im}[\bar{\bar{G}}(\boldsymbol{r},\boldsymbol{r};\omega_{0})]\cdot\boldsymbol{d}_{eg}\bar{n}_{\text{spp}}(\omega_{0})}\sigma_{\text{+}},\\
\hat{L}_{2}=\sqrt{\frac{\bar{\varepsilon}\mu_{0}}{(2\pi)^{3}\hbar}\omega_{0}^{2}\boldsymbol{d}_{eg}\cdot\text{Im}[\bar{\bar{G}}(\boldsymbol{r},\boldsymbol{r};\omega_{0})]\cdot\boldsymbol{d}_{eg}\left[\bar{n}_{\text{spp}}(\omega_{0})+1\right]}\sigma_{-}.
\end{cases}
\end{equation}

Note that we can explicitly evaluate the imaginary part of the Green's
function for our system, which yields a decay rate

\begin{equation}
\gamma=\frac{1}{8\hbar\varepsilon_{0}}\frac{(2\pi)^{3}}{\lambda_{\text{spp}}^{3}}e^{-2q_{\text{spp}}z}(d_{eg}^{\parallel2}+2d_{eg}^{\perp2}),
\end{equation}

where $d_{eg}^{\parallel2}=d_{eg}^{x2}+d_{eg}^{y2}$ and $d_{eg}^{\perp2}=d_{eg}^{z2}$
are the transverse and longitudinal parts of the two-level system
dipole moment $\boldsymbol{d}_{eg}$, $q_{\text{spp}}$ is the SPP
wave-vector, and $\lambda_{\text{spp}}=2\pi/q_{\text{spp}}$ is the
SPP wavelength. This result is seen to match others found in the literature,
for instance those by \citet{Ferreira2020quantization,Koppens2011}.
$\gamma$ of course corresponds to the spontaneous decay rate from
the excited to the ground state of the qubit coupled to the SPP field.
In fact, expanding the density matrix in the Lindblad equation and
solving for each probability we obtain the coupled rate equations

\begin{equation}
\begin{cases}
\dot{P}_{e}(t)=-\gamma\left[\bar{n}_{\text{spp}}(\omega_{0})+1\right]P_{e}(t)+\gamma\bar{n}_{\text{spp}}(\omega_{0})P_{g}(t),\\
\dot{P}_{g}(t)=\gamma\left[\bar{n}_{\text{spp}}(\omega_{0})+1\right]P_{e}(t)-\gamma\bar{n}_{\text{spp}}(\omega_{0})P_{g}(t).
\end{cases}
\end{equation}

\begin{figure}[H]
\begin{centering}
\subfloat[]{\begin{centering}
\includegraphics[scale=0.5]{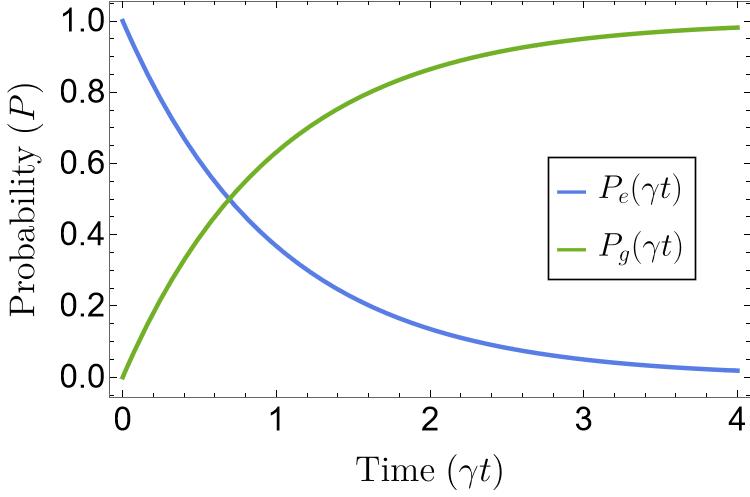} 
\par\end{centering}
}\subfloat[]{\begin{centering}
\includegraphics[scale=0.5]{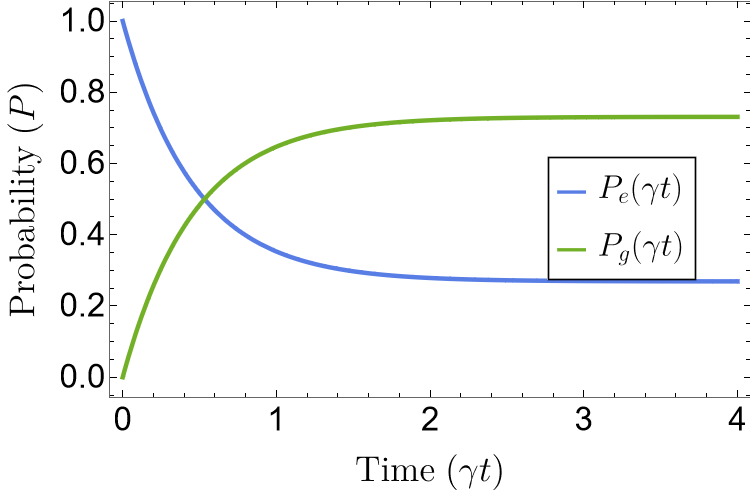} 
\par\end{centering}
}
\par\end{centering}
\begin{centering}
\subfloat[]{\begin{centering}
\includegraphics[scale=0.5]{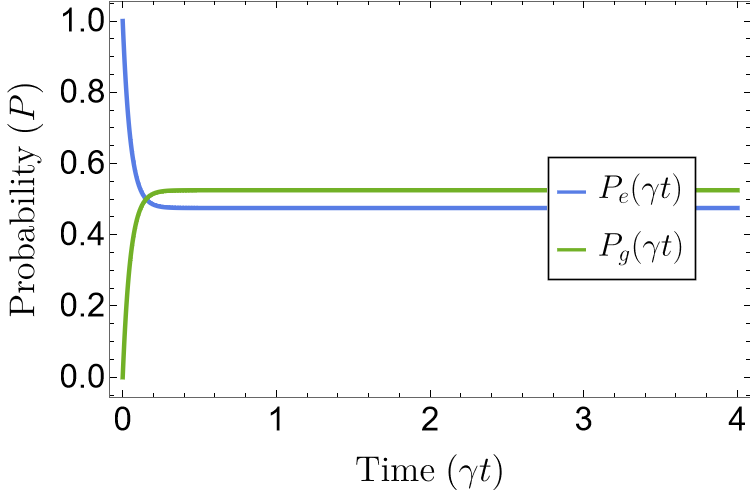} 
\par\end{centering}
}
\par\end{centering}
\caption{Plots of a numerical evaluation of the dynamics of a single qubit
interacting with a Plasmon field at several temperatures. The time
is adimensionalized by mutliplying with $\gamma$ which has S.I. units
of $s^{-1}$. We plot the probability of measuring the excited state
$\left|e\right\rangle $ in blue and ground state $\left|g\right\rangle $
in green. In these plots, for simplicity, we plot the time evolution
for a small temperature $\hbar\omega_{0}/k_{B}T=100$ in figure $(a)$,
for an intermediate temperature $\hbar\omega_{0}/k_{B}T=1$ in figure
$(b)$ and for a somewhat larger temperature $\hbar\omega_{0}/k_{B}T=0.1$
in figure $(c)$. We see from these plots that as the temperature
increases, the qubit achieves equilibrium with probability of being
found in the excited and ground state both approaching $1/2$, and
for temperatures very close to zero, the dynamics are well described
by the Schrodinger equation which predicts an exponential decay to
0 of the probability of finding the qubit in an excited state.\label{fig:Plots-of-a}}
\end{figure}

Note in particular that when the temperature becomes very small, in
the sense that $\hbar\omega_{0}\gg k_{B}T$, then the Bose distribution
function vanishes $\bar{n}_{\text{spp}}(\omega_{0})\to0$ and we are
left with the dynamics

\begin{equation}
\begin{cases}
\dot{P}_{e}=-\gamma P_{e},\\
\dot{P}_{g}=\gamma P_{e},
\end{cases}
\end{equation}
which can also be obtained via Shcrodinger's equation (see appendix
\ref{subsec:Obtaining-the-dynamics-1}). Thus, we see clearly that
the coupling between the SPP field and the qubit is described via
the Lindblad equation in a manner which is consistent with the zero
temperature description characteristic of the Schrodinger equation,
where initially the qubit is excited and there is no SPP present in
the graphene because there is no thermal excitation of the plasmon
gas. The open quantum systems approach, however, allows for a finite
temperature description as well. The effect of temperature is visible
if we make some plots as in Fig. \ref{fig:Plots-of-a}.

\section{Time evolution of two qubits coupled to a plasmonic bath in thermal
equilibrium \label{Sec:2qubits}}

We can now extend the previous discussion to the case we really want
to study, in particular the case of two qubits close to the graphene
sheet (the case of two color centers near a transition metal dichalcogenide
single layer has been studied already by \citet{Henriques2qubits}).
What we find, when we evaluate the coherent part of the dynamics (equivalently
the Schrodinger dynamics of the system with a zero temperature SPP
field) is that, as expected, the system will oscillate between the
states denoted $\left|1\right\rangle $ and $\left|2\right\rangle $,
where each qubit is excited, however there will also be an overall
decay due to the losses in the graphene sheet. This oscillation is
caused by the introduction of a coupling rate between the qubits $\gamma_{12}$
and an additional energy shift $g_{12}$ coming from their interaction.
We shall later provide their definition explicitly. A physical description
of this phenomenon may be obtained when we note that the Green's tensor
will not only propagate the interaction from the position of each
qubit $\boldsymbol{r}_{\alpha}$ to itself when written $\bar{\bar{G}}(\boldsymbol{r}_{\alpha},\boldsymbol{r}_{\alpha};\omega)$,
but also propagate the interaction from one qubit to the other when
written $\bar{\bar{G}}(\boldsymbol{r}_{1},\boldsymbol{r}_{2};\omega)$,
giving rise to the additional rates and shifts.

\begin{figure}[h]
\begin{centering}
\subfloat[\label{fig:Schrodinger-a}]{\begin{centering}
\includegraphics[scale=0.5]{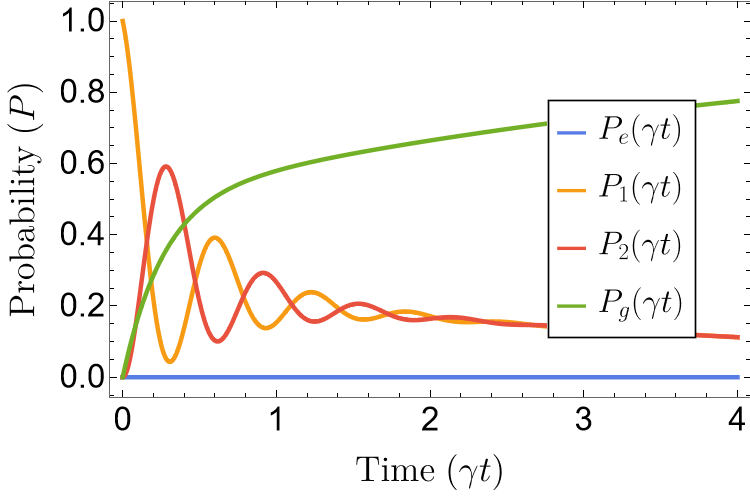} 
\par\end{centering}
}\subfloat[\label{fig:Schrodinger-b}]{\begin{centering}
\includegraphics[scale=0.5]{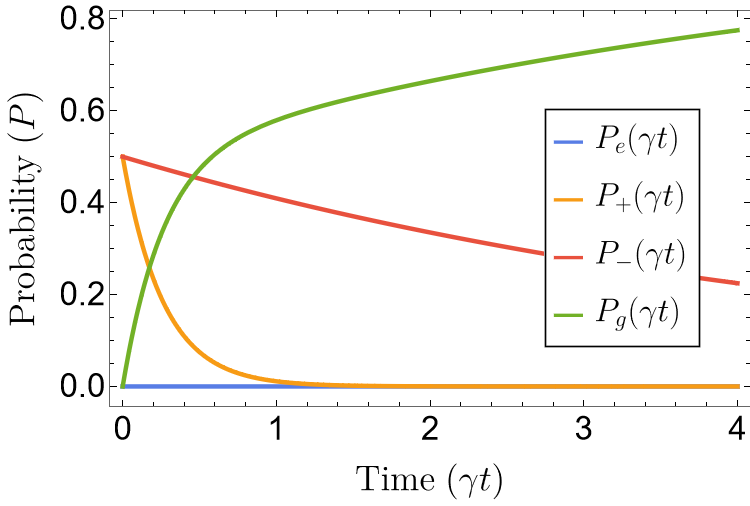} 
\par\end{centering}
}
\par\end{centering}

\caption{Schrodinger dynamics of two qubits, labeled $1$ and $2$, coupled
to an SPP field. The time is adimensionalized by multiplying with
$\gamma$ which has S.I. units of $s^{-1}$, and to illustrate/exaggerate
the possible behaviour we choose the parameters as $\gamma_{12}=0.9\gamma,$
$\Delta/\hbar\gamma=10,$ $g_{12}/\hbar\gamma=5$. Figure $(a)$ shows
the probability that atom 1 and 2 are in the excited state ($P_{1}$
in blue and $P_{2}$ in orange respectively), while figure $(b)$
shows the probability of measuring the Schrodinger cat states $\left|+\right\rangle =(\left|e\right\rangle \otimes\left|g\right\rangle +\left|g\right\rangle \otimes\left|e\right\rangle )/\sqrt{2}\text{ and }\left|-\right\rangle =(\left|e\right\rangle \otimes\left|g\right\rangle -\left|g\right\rangle \otimes\left|e\right\rangle )/\sqrt{2}$
($P_{+}$ in orange and $P_{-}$ in red respectively). In both figures
we also plot the probability of measuring the ground state in green
and of measuring the state where both states are excited in blue.
Inital conditions are $P_{1}(0)=1,$ $P_{2}(0)=0$, $P_{g}(0)=0$.
\label{fig:Dynamics-of-two-Schrodinger}}
\end{figure}

We find that the oscillations occur at a rate that depends on $g_{12}$
and the total populations decay slower as $\gamma_{12}\to\gamma$.
When $\gamma_{12}=\gamma$, the probability of finding the state in
the states $\left|1\right\rangle =\left|e\right\rangle \otimes\left|g\right\rangle $
and $\left|2\right\rangle =\left|g\right\rangle \otimes\left|e\right\rangle $
oscillate around $1/4$. Due to the dissipation, the amplitude of
these oscillations will diminish in time, and both states will eventually
tend to an equilibrium situation with equal probability $P_{1}(t\to\infty)=P_{2}(t\to\infty)=1/\sqrt{2}$.
Details regarding the derivation of the dynamics can be found in the
appendix \ref{subsec:Obtaining-the-dynamics-1}. We can see the dynamics
of the qubits in Fig. \ref{fig:Schrodinger-a}.

An alternative and perhaps clearer picture can be obtained when we
repeat this analysis for the ``Schrodinger cat'' states $\left|+\right\rangle =(\left|e\right\rangle \otimes\left|g\right\rangle +\left|g\right\rangle \otimes\left|e\right\rangle )/\sqrt{2}$
and $\left|-\right\rangle =(\left|e\right\rangle \otimes\left|g\right\rangle -\left|g\right\rangle \otimes\left|e\right\rangle )/\sqrt{2}$.
We see that the for $\gamma_{12}$ close enough to $\gamma$, the
state $\left|+\right\rangle $ decays very quickly, while the state
$\left|-\right\rangle $ is protected from decoherence via the dissipative
dynamics. In fact for $\gamma=\gamma_{12}$ the state $\left|-\right\rangle $
is stationary with probability $P_{-}(t)=1/2$ and the only interchange
that occurs is between the states $\left|-\right\rangle $ and $\left|0\right\rangle $.
We can plot the probability of finding the superposition states explicitly
as in Fig. \ref{fig:Schrodinger-b}. The moral we can extract from
this discussion is that dissipative dynamics in graphene allow us
to isolate one Schrodinger cat state over the other.

\begin{figure}[H]
\begin{centering}
\includegraphics[scale=0.5]{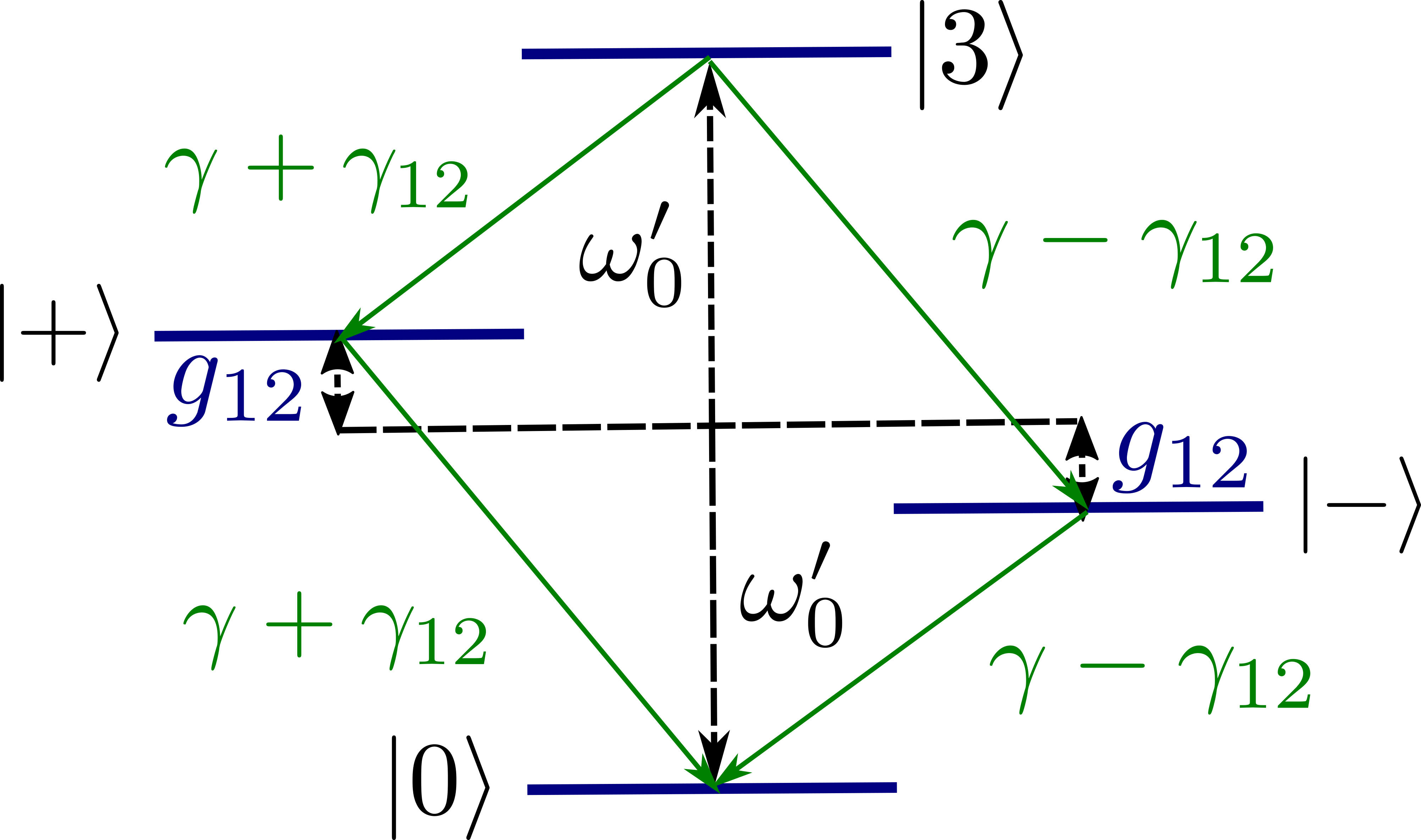} 
\par\end{centering}
\caption{Schematic for the dynamics of the two qubit system at zero temperature.
The state $\left|3\right\rangle $ decays with total rate $2\gamma$,
with a contribution of $\gamma+\gamma_{12}$ to the state $\left|+\right\rangle $
and a contribution $\gamma-\gamma_{12}$ to the state $\left|-\right\rangle $.
These decays are represented by top two the green arrows. These intermediate
states then decay to the state $\left|0\right\rangle $, with the
respective rates. Note that the rate $\gamma+\gamma_{12}$ is much
higher than the rate $\gamma-\gamma_{12}$, which suggests that the
Schrodinger cat state $\left|-\right\rangle $ can be isolated. In
the figure, the shifts $g_{12}$ and $g_{12}'$ are also represented
as shifts from the frequency $\omega_{0}'=\omega_{0}+\Delta$ of the
qubits. Note that the energy separation between the state $\left|0\right\rangle $
and $\left|3\right\rangle $ is $2\omega_{0}'$\label{fig:Schematic-for-the}.}
\end{figure}

The dynamics at zero temperature are summarized in Fig. \ref{fig:Schematic-for-the}
(see also \citet{Gonzalez2011}). The state $\left|+\right\rangle $
will decay quickly with rate $\gamma+\gamma_{12}$, while the state
$\left|-\right\rangle $ will be protected from decoherence as it
decays much slower with rate $\gamma-\gamma_{12}$. The state $\left|3\right\rangle =\left|e\right\rangle $
where both qubits are excited and $\left|0\right\rangle =\left|g\right\rangle $
where both are in the ground state have more straightforward dynamics
as their populations either decay or increase with overall rates $2\gamma$
going to/coming from the states $\left|\text{\ensuremath{\pm}}\right\rangle $.

An extension of this discussion can be performed in the case of finite
temperature by making use of Lindblad dynamics. A derivation of the
Lindblad equation for a two qubit system coupled to the plasmon field
proceeds in a manner which is very similar to the case of a single
qubit, with a few notable distinctions. We go over this derivation
in detail in the appendix \ref{subsec:Deriving-the-Lindblad-1} but
it suffices to say that because the summation of the jump operators
now occurs not only over the states of a single qubit, but rather
of both of them, there come into play terms which describe the emission
of a plasmon by one of the qubits followed by absorption by the other.
These terms lead to additional shifts in energy $g_{12}$ which must
be included in the coherent evolution of the system, as well as additional
rates $\gamma_{12}$ that set a time-scale for the communication between
the qubits. These rates are given by the same expression as for the
spontaneous decay rate for a single qubit coupled to the plasmon bath,
but with the dyadic Green's function evaluated at each position of
the two qubits thus leading to a kernel

\begin{equation}
\bar{\mathcal{K}}(\boldsymbol{r}_{1},\boldsymbol{r}_{2},\omega)=\frac{\bar{\varepsilon}\mu_{0}}{(2\pi)^{3}\hbar}\omega^{2}\text{Im}[\bar{\bar{G}}(\boldsymbol{r}_{1},\boldsymbol{r}_{2};\omega)],
\end{equation}
where $\bar{\varepsilon}$ is the average dielectric constant of the
mediums between which the graphene sheet is immersed, and $\bar{\bar{G}}(\boldsymbol{r}_{1},\boldsymbol{r}_{2};\omega)$
is the Green's tensor evaluated at the position of both qubits. This
kernel leads, by means of similar definitions to those of equation
(\ref{eq:ratesandshifts}), to the rates and shifts

\[
\begin{cases}
\gamma_{12}\equiv2\pi\int_{0}^{\infty}d\omega\boldsymbol{d}_{1}\cdot\bar{\mathcal{K}}(\boldsymbol{r}_{1},\boldsymbol{r}_{2},\omega)\cdot\boldsymbol{d}_{2}\delta(\omega-\omega_{0}),\\
g_{12}\equiv\mathcal{P}\int_{0}^{\infty}d\omega\frac{\boldsymbol{d}_{1}\cdot\bar{\mathcal{K}}(\boldsymbol{r}_{1},\boldsymbol{r}_{2},\omega)\cdot\boldsymbol{d}_{2}}{\omega_{0}-\omega},
\end{cases}
\]
or explicitly for the additional decay rate

\begin{equation}
\gamma_{12}=\frac{\bar{\varepsilon}\mu_{0}}{(2\pi)^{3}\hbar}\omega^{2}\boldsymbol{d}_{1}\cdot\text{Im}[\bar{\bar{G}}(\boldsymbol{r}_{1},\boldsymbol{r}_{2};\omega)]\cdot\boldsymbol{d}_{2}.
\end{equation}
Notice that we can interpret this in a somewhat physical manner by
noting that the Green's tensor is propagating the interaction between
them. The Lindblad equation can thus be written as

\begin{align}
\frac{d\hat{\rho}_{S}}{dt}= & -\frac{i}{\hbar}\left[\hbar\omega_{12}\sum_{\alpha\neq\beta}\sigma_{\alpha}^{+}\sigma_{\beta}^{-},\hat{\rho}_{S}(t)\right]-\frac{i}{\hbar}\left[\hbar\omega_{0}'\sum_{\alpha}\sigma_{\alpha}^{+}\sigma_{\alpha}^{-},\hat{\rho}_{S}(t)\right]\nonumber \\
 & +\sum_{\alpha}\left(\sum_{k}2\hat{L}_{\alpha,k}\hat{\rho}_{S}(t)\hat{L}_{\alpha,k}^{\dagger}-\left\{ \hat{L}_{\alpha,k}^{\dagger}\hat{L}_{\alpha,k},\hat{\rho}_{S}(t)\right\} \right)\nonumber \\
 & +\sum_{\alpha\neq\beta}\left(\sum_{k}2\hat{K}_{\alpha,k}\hat{\rho}_{S}(t)\hat{K}_{\beta,k}^{\dagger}-\left\{ \hat{K}_{\alpha,k}^{\dagger}\hat{K}_{\beta,k},\hat{\rho}_{S}(t)\right\} \right),
\end{align}
where the jump operators $\hat{L}_{\alpha,k}$ match those defined
for the single qubit except for the index $\alpha$ running over the
two qubits $1$ and $2$. The $\hat{K}_{\alpha,k}$ jump operators
also match these but give the interaction between the qubits and so
are constructed in a similar manner but with the rate $\gamma_{12}$
instead of $\gamma$. In the case of finite temperature, working in
the basis $\left\{ \left|3\right\rangle ,\left|+\right\rangle ,\left|-\right\rangle ,\left|0\right\rangle \right\} $
has some additional value, since the Lindblad equation becomes self-contained
within the diagonal elements. In other words, one would in principle
have to calculate the solution for a coupled set of 16 differential
equations to evaluate the dynamics (one corresponding to each element
of the density matrix). Hermiticiy of the density matrix brings this
number down to 10, but in this basis, the diagonal elements, which
correspond to the populations depend via the Lindblad equation only
on each other (see appendix \ref{subsec:Deriving-the-Lindblad-1}),
and therefore we only have only to solve a set of 4 coupled differential
equations to obtain the population dynamics.

\begin{figure}
\begin{centering}
\subfloat[\label{fig:Lindblad-a}]{\begin{centering}
\includegraphics[scale=0.5]{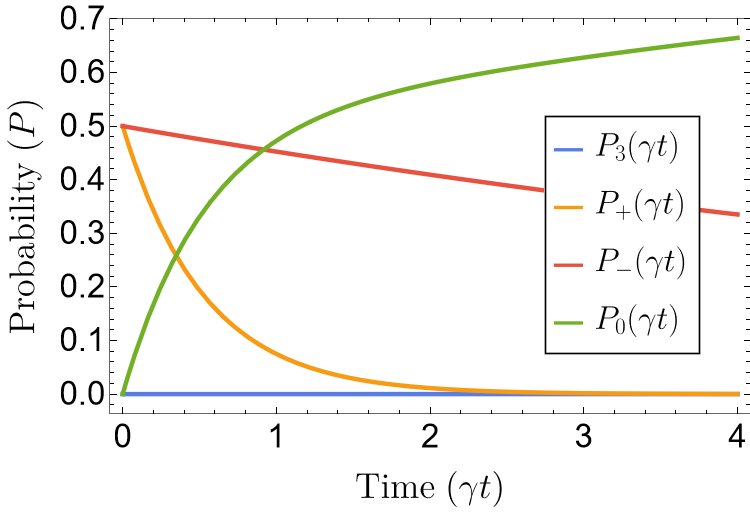} 
\par\end{centering}
}\subfloat[\label{fig:Lindblad-b}]{\begin{centering}
\includegraphics[scale=0.5]{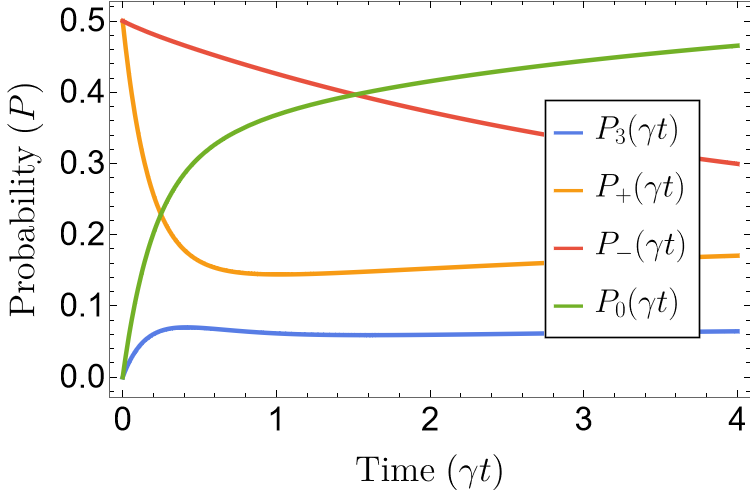} 
\par\end{centering}
}
\par\end{centering}
\begin{centering}
\subfloat[\label{fig:Lindblad-c}]{\begin{centering}
\includegraphics[scale=0.5]{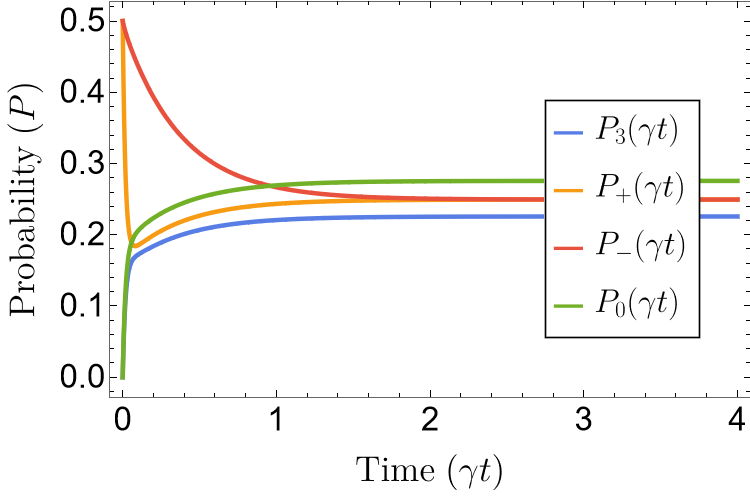} 
\par\end{centering}
}
\par\end{centering}
\caption{Lindblad dynamics of two qubits coupled to an SPP field. Each figure
shows the probability of measuring the Schrodinger cat states $\left|+\right\rangle =(\left|e\right\rangle \otimes\left|g\right\rangle +\left|g\right\rangle \otimes\left|e\right\rangle )/\sqrt{2}\text{ and }\left|-\right\rangle =(\left|e\right\rangle \otimes\left|g\right\rangle -\left|g\right\rangle \otimes\left|e\right\rangle )/\sqrt{2}$
($P_{+}$ in orange and $P_{-}$ in red respectively). The time is
adimentionalized by multiplying with $\gamma$ which has S.I. units
of $s^{-1}$. In all figures we also plot the probability of measuring
the ground state in green and the state where both qubits are in the
excited state in blue. Initial conditions are $P_{3}(0)=0,$$P_{+}(0)=P_{-}(0)=1/2$,
which corresponds to $P_{1}(0)=1,$$P_{2}(0)=0$ and$P_{0}(0)=0$.
Figure $(a)$ shows a small temperature $\hbar\omega_{0}/k_{B}T=100$,
which matches the Shcrodinger dynamics, while figure $(b)$ shows
an intermediate temperature $\hbar\omega_{0}/k_{B}T=1$, and figure
$(c)$ shows the temperature $\hbar\omega_{0}/k_{B}T=0.01$. We see
that as temperature increases several things happen. Firstly, the
excited sate $\left|3\right\rangle $ begins to play a role in the
dynamics, as it is capable of being excited by thermal plasmons. In
addition, the states tend to an equilibrium around $P=1/4$, as temperature
increases. For infinite temperature the decay is to happen much quicker
and all states will tend to the same probability. \label{fig:Dynamics-of-two-Lindblad}}
\end{figure}

A simple numerical evaluation of the Lindblad equation yields Fig.
\ref{fig:Dynamics-of-two-Lindblad}. 

Once again, from Fig. \ref{fig:Dynamics-of-two-Lindblad}, we see
that the zero temperature dynamics predicted by the approach based
on the Schrodinger equation are reproduced by the Lindblad equation
(compare Fig. \ref{fig:Dynamics-of-two-Schrodinger} with Fig. \ref{fig:Lindblad-a}),
which is really astonishing considering the differences in formalism.
As temperature increases some of the coherence is lost for longer
times, as all states, including the ground state tend to equilibrium
at probability $1/4$.

\section{Conclusions and outlook \label{Sec:conclusion}}

In the present study we aimed to investigate a physical system where
two two-level systems (qubits) interact via emission and absorption
of surface plasmon-polaritons (SPPs) in a graphene sheet placed between
two semi-infinite dielectrics. The evaluation of the Green's tensor
from the macroscopic Maxwell's equations allowed us to quantize the
SPP field and somewhat remarkably derive a quantized description of
the SPP field with dissipation, by arriving at creation and annihilation
operators of the SPPs. We studied in detail the derivation of the
master equation in Lindblad form using the Born and Markov approximations
for a single qubit coupled to a thermal electromagnetic field, and
used this as a stepping stone for the following analysis. This consisted
in studying the dynamics of one and two qubits coupled to the SPP
reservoir, via the Schrodinger equation as well as the Lindblad equation.
From a practical perspective, the results obtained using the Lindblad
equation matched those of the Schrodinger equation when we considered
the temperature to be absolute zero. Otherwise, the Lindblad equation
allowed the extension of these results to a finite temperature regime,
where the SPP reservoir plays an active role in the qubit dynamics.

These calculations revealed some interesting behavior in the qubit
dynamics as the dissipation introduces an asymmetry between the time
evolution of what we called ``Schrodinger cat states'' corresponding
to superpositions of the states where one qubit is excited while the
other is in the ground state. Effectively, this asymmetry manifests
in a larger decay rate for one Schrodinger cat state over the other.
This in turns leads to a protection from decoherence, where entanglement
is maintained and a superposition is isolated over a longer time-frame
than would otherwise be possible. We find this quite a beautiful result,
where we are able to take advantage of dissipation in macroscopic
media to maintain coherence in a quantum mechanical system. Not only
is this an interesting result in itself, but we can envision a continuation
of this work, where the usage of graphene-based metamaterials allows
for the consideration of an anisotropic conductivity and thus the
manipulation of SPP using wave-guides (see \citet{Billow2003,Alu2015,Gomez-Diaz:15})

\section*{Acknowledgments }

$^{\ast}$TVCA performed all the calculations and produced the first
written version of the paper.

$^{\dagger}$NMRP conceived the idea, supervised the work, and contributed
to the writing in the final stage of the work.

N.M.R.P. acknowledges support from the European Commission through
the project ``Graphene-Driven Revolutions in ICT and Beyond'' (Ref.
No. 881603, CORE 3) and COMPETE 2020, PORTUGAL 2020, FEDER and the
Portuguese Foundation for Science and Technology (FCT) through project
POCI-01- 0145-FEDER-028114. The Portuguese Foundation for Science
and Technology (FCT) is acknowledged in the framework of the Strategic
Funding UIDB/04650/2020. Both authors acknowledge Dr. Bruno Amorim
for providing the derivation found in Appendix J. N.M.R.P. Dr. Bruno
Amorim and João Carlos Henriques for discussions on the Lindblad equation.

\appendix

\section{The density operator\label{Sec:rho_opperator}}

\subsection{General definitions\label{subsec:General-definitions}}

In this appendix we follow \citet{Hohenester2020NanoandQuantum} for
a brief description of the density matrix formalism. We start by considering
a quantum system to be in a state $\left|\psi\right\rangle $ and
an hermitian operator $\hat{A}$ to be acting on the system's Hilbert
space, we define its expectation value as

\begin{equation}
\left\langle \hat{A}\right\rangle =\left\langle \psi\right|\hat{A}\left|\psi\right\rangle .
\end{equation}
This observable provides information as to the outcome of an experiment
where we have absolute certainty that the system is in the state $\left|\psi\right\rangle $.
We might however find some kind of statistical uncertainty in the
systems state before measurement. That is to say that to know that
the exact state of the system is often difficult to know in practice
since we have only imperfect information. To make this observation
quantitative we may think about an ensemble of almost identical systems,
where the state is not $\left|\psi\right\rangle $, possibly due to
the system's interaction with its environment. Rather, we may find
that the system exists in a superposition of states $\left|\psi^{(\mu)}\right\rangle $
with probability distribution $p_{\mu}$. In this case, the expectation
value is given by the weighted sum

\begin{equation}
\left\langle \hat{A}\right\rangle =\sum_{\mu}p_{\mu}\left\langle \psi^{(\mu)}\right|\hat{A}\left|\psi^{(\mu)}\right\rangle .
\end{equation}
We note the difference between the \textit{statistical uncertainty}
in the system's state and the \textit{quantum mechanical uncertainty}
which is accounted for by the quantum mechanical expectation value.
To account for these different uncertainties and corresponding averages
we develop a formalism based on the \textit{density operator}. This
is the mathematical tool that describes our knowledge of the system.
If we consider a complete basis of the system's Hilbert space made
up by the states $\left|i\right\rangle $, then the trace of an operator
is defined by

\begin{equation}
\text{tr}(\hat{A})=\sum_{i}\left\langle i\right|\hat{A}\left|i\right\rangle .
\end{equation}
This is in analogy to the trace of a matrix as the sum of its diagonal
elements. The cyclic property of the trace reads

\begin{equation}
\text{tr}(\hat{A}_{1}\hat{A}_{2}\dots\hat{A}_{n})=\text{tr}(\hat{A}_{n}\hat{A}_{1}\dots\hat{A}_{n-1}).\label{eq:cyclic}
\end{equation}
By decomposing the states $\left|\psi^{(\mu)}\right\rangle $ in the
complete basis of the Hilbert space with states $\left|i\right\rangle $,
we can write the expectation value in $\left|\psi^{(\mu)}\right\rangle $
as

\begin{align}
\left\langle \hat{A}\right\rangle  & =\sum_{\mu}p_{\mu}\left\langle \psi^{(\mu)}\right|\hat{A}\left|\psi^{(\mu)}\right\rangle \nonumber \\
 & =\sum_{\mu}p_{\mu}\sum_{ij}\braket{\psi^{(\mu)}|i}\left\langle i\right|\hat{A}\left|j\right\rangle \braket{j|\psi^{(\mu)}}\nonumber \\
 & =\sum_{\mu}p_{\mu}\sum_{ij}\braket{j|\psi^{(\mu)}}\braket{\psi^{(\mu)}|i}\left\langle i\right|\hat{A}\left|j\right\rangle .
\end{align}
Note that the states $\left|\psi^{(\mu)}\right\rangle $ may not form
a complete basis. As such, for convenience we introduce a definition
for a \textit{density operator}

\begin{equation}
\hat{\rho}=\sum_{\mu}p_{\mu}\left|\psi^{(\mu)}\right\rangle \left\langle \psi^{(\mu)}\right|.\label{eq:densityDefinition}
\end{equation}
Following these considerations we introduce the concept of a \textit{pure
state }as a unit vector in the Hilbert space of a certain system.
This is opposed to the concept of a \textit{mixed state}, which is
represented by a positive unit trace operator $\hat{\rho}$ corresponding
to the density operator. Using this operator, the expectation value
can be succinctly written as

\begin{align}
\left\langle \hat{A}\right\rangle  & =\sum_{\mu}p_{\mu}\sum_{ij}\braket{j|\psi^{(\mu)}}\braket{\psi^{(\mu)}|i}\left\langle i\right|\hat{A}\left|j\right\rangle \nonumber \\
 & =\sum_{ij}\left\langle i\right|\left(\sum_{\mu}p_{\mu}\left|\psi^{(\mu)}\right\rangle \left\langle \psi^{(\mu)}\right|\right)\left|j\right\rangle \left\langle i\right|\hat{A}\left|j\right\rangle \nonumber \\
 & =\sum_{ij}\left\langle j\right|\hat{\rho}\left|i\right\rangle \left\langle i\right|\hat{A}\left|j\right\rangle \nonumber \\
 & =\sum_{j}\left\langle j\right|\hat{\rho}\hat{A}\left|j\right\rangle \nonumber \\
 & =\text{tr}(\hat{\rho}\hat{A}).\label{eq:Expectation}
\end{align}
This definition for expectation value is thus more general as it accounts
for two kinds of averaging 
\begin{itemize}
\item The quantum mechanical averaging over the eigenvalues of a given hermitian
operator $\hat{A}$ when acting on a pure state. 
\item The statistical averaging over the probability distribution $p_{\mu}$
of the state of the system itself $\left|\psi^{(\mu)}\right\rangle $. 
\end{itemize}
We now note some of the properties of the density operator as defined
above. Its trace is

\begin{align}
\text{tr}(\hat{\rho}) & =\sum_{i}\left\langle i\right|\left(\sum_{\mu}p_{\mu}\left|\psi^{(\mu)}\right\rangle \left\langle \psi^{(\mu)}\right|\right)\left|i\right\rangle \nonumber \\
 & =\sum_{\mu}p_{\mu}\sum_{i}\braket{i|\psi^{(\mu)}}\braket{\psi^{(\mu)}|i}\nonumber \\
 & =\sum_{\mu}p_{\mu}\sum_{i}\braket{\psi^{(\mu)}|i}\braket{i|\psi^{(\mu)}}\nonumber \\
 & =\sum_{\mu}p_{\mu}\braket{\psi^{(\mu)}|\psi^{(\mu)}}\nonumber \\
 & =\sum_{\mu}p_{\mu}=1,
\end{align}
where we have used both the fact that $\braket{\psi^{(\mu)}|\psi^{(\mu)}}=1$
and $\sum_{\mu}p_{\mu}=1$. Note also that for a pure state $\left|\psi\right\rangle $,
that is a state in which there exists no statistical uncertainty,
we may compute the density operator as

\begin{equation}
\hat{\rho}_{pure}=\left|\psi\right\rangle \left\langle \psi\right|,
\end{equation}
which is easily seen to obey the projector relation

\begin{equation}
\hat{\rho}_{pure}^{2}=\hat{\rho}_{pure}.\label{eq:pure}
\end{equation}
In essence, this imposes an upper limit on the trace of $\hat{\rho}^{2}$.
Since, in general

\begin{align}
\hat{\rho}^{2} & =\sum_{\mu}p_{\mu}\left|\psi^{(\mu)}\right\rangle \left\langle \psi^{(\mu)}\right|\sum_{\mu}p_{\nu}\left|\psi^{(\nu)}\right\rangle \left\langle \psi^{(\nu)}\right|\nonumber \\
 & =\sum_{\mu\nu}p_{\mu}p_{\nu}\braket{\psi^{(\mu)}|\psi^{(\nu)}}\left|\psi^{(\mu)}\right\rangle \left\langle \psi^{(\nu)}\right|.
\end{align}
The trace is, thus

\begin{align}
\text{tr}(\hat{\rho}^{2}) & =\sum_{i}\left\langle i\right|\left(\sum_{\mu\nu}p_{\mu}p_{\nu}\braket{\psi^{(\mu)}|\psi^{(\nu)}}\left|\psi^{(\mu)}\right\rangle \left\langle \psi^{(\nu)}\right|\right)\left|i\right\rangle \nonumber \\
 & =\sum_{\mu\nu}p_{\mu}p_{\nu}\braket{\psi^{(\mu)}|\psi^{(\nu)}}\sum_{i}\braket{\psi^{(\nu)}|i}\braket{i|\psi^{(\mu)}}\nonumber \\
 & =\sum_{\mu\nu}p_{\mu}p_{\nu}\braket{\psi^{(\mu)}|\psi^{(\nu)}}\braket{\psi^{(\nu)}|\psi^{(\mu)}}\nonumber \\
 & =\sum_{\mu\nu}p_{\mu}p_{\nu}\left|\braket{\psi^{(\mu)}|\psi^{(\nu)}}\right|^{2}.
\end{align}
By the Cauchy-Schwartz inequality, we get

\begin{align}
\text{tr}(\hat{\rho}^{2}) & =\sum_{\mu\nu}p_{\mu}p_{\nu}\left|\braket{\psi^{(\mu)}|\psi^{(\nu)}}\right|^{2}\nonumber \\
 & \leq\sum_{\mu\nu}p_{\mu}p_{\nu}\left|\braket{\psi^{(\mu)}|\psi^{(\mu)}}\right|^{2}\left|\braket{\psi^{(\nu)}|\psi^{(\nu)}}\right|^{2}\nonumber \\
 & =\left(\sum_{\mu}p_{\mu}\left|\braket{\psi^{(\mu)}|\psi^{(\mu)}}\right|^{2}\right)\left(\sum_{\nu}p_{\nu}\left|\braket{\psi^{(\nu)}|\psi^{(\nu)}}\right|^{2}\right)\nonumber \\
 & =1,
\end{align}
where the equality in the second step holds only if $\nu=\mu$ for
all $\mu$ and $\nu$, which can only happen if the state is pure
and equation (\ref{eq:pure}) is satisfied. Thus we can identify a
valid density operator by two abstract properties 
\begin{itemize}
\item $\text{tr}(\hat{\rho})=1$ 
\item $\text{tr}(\hat{\rho}^{2})\leq1$ 
\end{itemize}
These properties do indeed have important physical meaning. The fact
that the trace of the density operator is unity represents the normalization
of mixed states. A system must be in some state after all, and as
such, the total probability of finding it in any state should add
up to one. On the other hand, the trace of the square of the density
operator measures the \textit{purity }(and indeed this quantity often
bares that name) of the state. If it is a pure state then $\text{tr}\left\{ \rho^{2}\right\} =\text{tr}\left\{ \rho\right\} =1$.
On the other hand, for a mixed state, $\text{tr}\left\{ \rho^{2}\right\} <1$.
In fact the purity is bounded by $1/d\leq\text{tr}\left\{ \rho^{2}\right\} <1$
where $d$ is the Hilbert space dimension. The lower bound is obtained
for the maximally mixed state, where all populations are $1/d$ (see
\citet{jaeger2006quantum}). A practical note on the density operator
may be that of course, in a finite dimensional Hilbert space it can
be represented by a matrix in an arbitrary basis $\left|i\right\rangle $.
In this basis we have $\rho=\sum_{ij}\rho_{ij}\left|i\right\rangle \left\langle j\right|$
or

\begin{equation}
\rho=\left(\begin{array}{cccc}
\rho_{00} & \rho_{01} & \cdots & \rho_{0N}\\
\rho_{10} & \rho_{11} & \dots & \rho_{1N}\\
\vdots & \vdots & \ddots & \vdots\\
\rho_{N0} & \rho_{N1} & \cdots & \rho_{NN}
\end{array}\right),
\end{equation}
The diagonal elements of this matrix are often called the \textit{populations},
whereas the off-diagonal elements are called the \textit{coherences}.

\subsection{Pictures of quantum mechanics\label{subsec:Pictures-of-quantum}}

In this appendix we give a brief overview of the pictures of quantum
mechanics. For a thorough overview see \citet{jishi_2013}.

\subsubsection{Schrodinger Representation}

Unitary transformations play an important role in quantum mechanics,
and in fact, the time evolution operator $\hat{U}(t,t_{0})$ is unitary.
This operator takes a wave-function at time $t_{0}$ and propagates
it to time $t$.

\begin{equation}
\left|\psi(t)\right\rangle =\hat{U}(t,t_{0})\left|\psi(t_{0})\right\rangle .
\end{equation}
If we insert this expression into the Schrodinger equation, we may
write an operator equality

\begin{equation}
i\hbar\frac{d}{dt}U(t,t_{0})=\hat{H}(t)U(t,t_{0}),
\end{equation}
which can also be written in integral form as

\begin{equation}
\hat{U}(t,t_{0})=\mathds{1}-\frac{i}{\hbar}\int_{t_{0}}^{t}\hat{H}(t')\hat{U}(t',t_{0})\ dt'.\label{eq:IntTimeEv}
\end{equation}
We can now look at time evolution from different perspectives, and
as such come up with several different pictures or representations
of quantum mechanics. Firstly, in the Schrodinger picture, the wave-functions
are time dependent, and their evolution is given by

\begin{equation}
\left|\psi_{S}(t)\right\rangle =\hat{U}(t,0)\left|\psi_{0}\right\rangle .
\end{equation}
\textit{As time evolves, only the wave-functions change and the operators
stay unchanged.}

\subsubsection{Heisenberg Representation}

In contrast, in the Heisenberg picture, the wave-functions are time
independent, and only the operators evolve in time, according to

\begin{equation}
\hat{A}_{H}(t)=\hat{U}^{\dagger}(t,0)\hat{A}_{S}\hat{U}(t,0).
\end{equation}
Note that the observable corresponding to the expectation value in
both pictures is the same. We may prove this succinctly

\begin{align}
\left\langle \hat{A}\right\rangle _{S} & =\left\langle \psi_{S}(t)\right|\hat{A}_{S}\left|\psi_{S}(t)\right\rangle =\left\langle \psi_{0}\right|\hat{U}^{\dagger}(t,0)A_{S}\hat{U}(t,0)\left|\psi_{0}\right\rangle \nonumber \\
 & =\left\langle \psi_{0}\right|\hat{A}_{H}(t)\left|\psi_{0}\right\rangle =\left\langle \hat{A}\right\rangle _{H}.
\end{align}
We may calculate the time evolution of an operator in the Heisenberg
picture taking the derivative of an operator $\hat{A}$ in the Heisenberg
picture $\hat{A}_{H}(t)$

\begin{equation}
\frac{d}{dt}\hat{A}_{H}(t)=\frac{d}{dt}\left\{ \hat{U}^{\dagger}(t,0)\hat{A}_{S}\hat{U}(t,0)\right\} =\frac{d\hat{U}^{\dagger}(t,0)}{dt}\hat{A}_{S}\hat{U}(t,0)+\hat{U}^{\dagger}(t,0)\hat{A}_{S}\frac{d\hat{U}(t,0)}{dt}.\label{eq:DerivU}
\end{equation}
By making use of the integral equation (\ref{eq:IntTimeEv}) we can
simplify the previous equation by calculating

\begin{equation}
\frac{d\hat{U}(t,0)}{dt}=\frac{d}{dt}\left(\mathds{1}-\frac{i}{\hbar}\int_{0}^{t}\hat{H}(t')\hat{U}(t',0)\ dt'\right),
\end{equation}
which by the fundamental theorem of calculus gives

\begin{equation}
\frac{d\hat{U}(t,0)}{dt}=-\frac{i}{\hbar}\hat{H}(t)\hat{U}(t,0).
\end{equation}
We then get, by substituting into equation (\ref{eq:DerivU})

\begin{equation}
\frac{d}{dt}\hat{A}_{H}(t)=+\frac{i}{\hbar}\hat{H}(t)\hat{U}^{\dagger}(t,0)\hat{A}_{S}\hat{U}(t,0)-\frac{i}{\hbar}\hat{U}^{\dagger}(t,0)\hat{A}_{S}\hat{H}(t)\hat{U}(t,0).
\end{equation}
By making use of the definition of the time evolution of an operator
in the Heisenberg picture, we write

\begin{equation}
\frac{d}{dt}\hat{A}_{H}(t)=+\frac{i}{\hbar}\hat{H}(t)\hat{A}_{H}(t)-\frac{i}{\hbar}\hat{A}_{H}(t)\hat{H}(t),
\end{equation}
which can be written in a form known as the Heisenberg equation of
motion

\begin{equation}
i\hbar\frac{d}{dt}\hat{A}_{H}(t)=[\hat{H}(t),\hat{A}_{H}(t)]\label{eq:HeisenbergEq}
\end{equation}
\textit{As time evolves, only the operators change and the wave-functions
stay unchanged.}

\subsubsection{Interaction or Dirac Representation}

There exists still another picture called the interaction or Dirac
picture. This is useful when considering perturbed Hamiltonians of
the form

\begin{equation}
\hat{H}(t)=\hat{H}_{0}(t)+\hat{V}(t),
\end{equation}
where $\hat{V}(t)$ is treated as a perturbation. We introduce a time
evolution operator $\hat{U}_{0}(t,0)$, which is associated only with
the evolution governed by the Hamiltonian $\hat{H}_{0}(t)$. In this
picture, both the wave-functions and operators evolve in time according
to

\begin{equation}
\begin{cases}
\left|\psi_{I}(t)\right\rangle =\hat{U}_{0}^{\dagger}(t,0)\left|\psi_{S}(t)\right\rangle ,\\
\hat{A}_{I}(t)=\hat{U}_{0}^{\dagger}(t,0)\hat{A}_{S}\hat{U}_{0}(t,0).
\end{cases}
\end{equation}
We can prove, yet again that this picture produces the same results
for the expectation value

\begin{align}
\left\langle \hat{A}\right\rangle _{I} & =\left\langle \psi_{I}(t)\right|\hat{A}_{I}(t)\left|\psi_{I}(t)\right\rangle =\left\langle \psi_{S}(t)\right|\hat{U}_{0}(t,0)\hat{U}_{0}^{\dagger}(t,0)\hat{A}_{S}\hat{U}_{0}(t,0)\hat{U}_{0}^{\dagger}(t,0)\left|\psi_{S}(t)\right\rangle \nonumber \\
 & =\left\langle \psi_{S}(t)\right|\hat{A}_{S}\left|\psi_{S}(t)\right\rangle =\left\langle \hat{A}\right\rangle _{S}.
\end{align}
We can still define the evolution operator of the interaction picture
as

\begin{equation}
\hat{U}_{I}(t,0)=\hat{U}_{0}^{\dagger}(t,0)\hat{U}(t,0).\label{eq:EvolutionInteraction}
\end{equation}
Taking its time derivative, and invoking once again the fundamental
theorem of calculus when using equation (\ref{eq:IntTimeEv}), write

\begin{align}
i\hbar\frac{d}{dt}\hat{U}_{I}(t,0) & =i\hbar\left[\left(\frac{d}{dt}\hat{U}_{0}^{\dagger}(t,0)\right)\hat{U}+\hat{U}_{0}^{\dagger}(t,0)\left(\frac{d}{dt}\hat{U}(t,0)\right)\right]\nonumber \\
 & =\left(-\hat{U}_{0}^{\dagger}(t,0)\hat{H}_{0}(t)\right)\hat{U}(t,0)+\hat{U}_{0}^{\dagger}(t,0)\left(\hat{H}_{0}(t)+\hat{V}(t)\right)\hat{U}(t,0)\nonumber \\
 & =\hat{U}_{0}^{\dagger}(t,0)\hat{V}(t)\hat{U}(t,0)\nonumber \\
 & =\hat{V}_{I}(t)\hat{U}_{I}(t,0),
\end{align}
where the perturbation in the interaction representation is defined
as $\hat{V}_{I}=\hat{U}_{0}^{\dagger}(t,0)\hat{V}\hat{U}_{0}(t,0)$.
As such, in the interaction picture, the time evolution operator in
the interaction picture is given by the solution of the equation

\begin{equation}
i\hbar\frac{d}{dt}\hat{U}_{I}(t,0)=\hat{V}_{I}(t)\hat{U}_{I}(t,0).
\end{equation}
By noting that $\hat{U}_{I}(0,0)=\mathds{1}$ this last equation can
be written in integral form as

\begin{equation}
\hat{U}_{I}(t,0)=\mathds{1}-\frac{i}{\hbar}\int_{0}^{t}\hat{V}_{I}(t')\hat{U}_{I}(t',0)\ dt'.\label{eq:UInteractionIter}
\end{equation}
The time evolution operator $\hat{U}_{I}(t,0)$ depends only on this
perturbation, and as such the dynamics of the system can be computed
by making use only of $\hat{V}$ since the zeroth order Hamiltonian
has been completely absorbed into the computation of $U_{0}^{\dagger}(t,0)$
and therefore into the time evolution operator and wave function in
the interaction picture. The integral of equation (\ref{eq:UInteractionIter})
can be solved iteratively

\begin{align}
\hat{U}_{I}(t,0) & =\mathds{1}-\frac{i}{\hbar}\int_{0}^{t}\hat{V}_{I}(t')\left(\mathds{1}-\frac{i}{\hbar}\int_{0}^{t'}\hat{V}_{I}(t'')\hat{U}_{I}(t',0)\ dt''\right)\ dt'\nonumber \\
 & =\mathds{1}-\frac{i}{\hbar}\int_{0}^{t}\hat{V}_{I}(t')\ dt'+\mathcal{O}\left(\hat{V}_{I}^{2}\right),
\end{align}
and thus, to first order in the perturbation $\hat{V_{I}}(t)$, we
have

\begin{equation}
\hat{U}_{I}(t,0)=\mathds{1}-\frac{i}{\hbar}\int_{0}^{t}\hat{V}_{I}(t')\ dt'.
\end{equation}
As such, the time evolution of an operator in the interaction picture
is calculated from differentiating $\hat{A}_{I}(t)=\hat{U}_{I}^{\dagger}(t,0)\hat{A}_{I}(0)\hat{U}_{I}(t,0)$
from the general expression (\ref{eq:UInteractionIter}) with respect
to $t$ on both sides. This gives

\begin{equation}
\frac{d}{dt}\hat{A}_{I}(t)=\frac{i}{\hbar}\hat{V}_{I}(t)\hat{U}_{I}^{\dagger}(t',0)\hat{A}_{I}(0)\hat{U}_{I}(t,0)-\hat{U}_{I}^{\dagger}(t,0)\hat{A}_{I}(0)\frac{i}{\hbar}\hat{V}_{I}(t)\hat{U}_{I}(t,0),
\end{equation}
or simplifying the expression using the commutator

\begin{equation}
i\hbar\frac{d}{dt}\hat{A}_{I}(t)=[\hat{A}_{I}(t),\hat{V}_{I}(t)],
\end{equation}
and the time evolution of a wave-function is

\begin{equation}
i\hbar\frac{d}{dt}\left|\psi_{I}(t)\right\rangle =\hat{V}_{I}(t)\left|\psi_{I}(t)\right\rangle .
\end{equation}
\textit{As time evolves, both the wave-functions and operators change.
This is an intermediate representation between the Schrodinger and
Dirac pictures.}

\subsection{Time evolution of the density operator\label{subsec:Time-evolution-of-the-density-operator}}

We note that we can write the time evolution of a given state ket
$\left|\psi^{(\mu)}\right\rangle $ as well as the corresponding bra
$\left\langle \psi^{(\mu)}\right|$ by making use of the time dependent
Schrodinger equation with a time dependent Hamiltonian $\hat{H}(t)$

\begin{equation}
\begin{cases}
i\hbar\frac{d}{dt}\left|\psi^{(\mu)}\right\rangle =\hat{H}(t)\left|\psi^{(\mu)}\right\rangle ,\\
-i\hbar\frac{d}{dt}\left\langle \psi^{(\mu)}\right|=\left\langle \psi^{(\mu)}\right|\hat{H}(t).
\end{cases}
\end{equation}
As such, we can construct the time evolution of the density operator
by using its definition as stated in equation (\ref{eq:densityDefinition}).
Thus, computing the derivative of the density operator, we get

\begin{align}
i\hbar\frac{d\hat{\rho}}{dt} & =\sum_{\mu}p_{\mu}i\hbar\frac{d}{dt}\left(\left|\psi^{(\mu)}\right\rangle \left\langle \psi^{(\mu)}\right|\right)\nonumber \\
 & =\sum_{\mu}p_{\mu}\left[i\hbar\frac{d}{dt}\left(\left|\psi^{(\mu)}\right\rangle \right)\left\langle \psi^{(\mu)}\right|+\left|\psi^{(\mu)}\right\rangle i\hbar\frac{d}{dt}\left(\left\langle \psi^{(\mu)}\right|\right)\right]\nonumber \\
 & =\sum_{\mu}p_{\mu}\left[\hat{H}(t)\left|\psi^{(\mu)}\right\rangle \left\langle \psi^{(\mu)}\right|-\left|\psi^{(\mu)}\right\rangle \left\langle \psi^{(\mu)}\right|\hat{H}(t)\right].
\end{align}
By pulling the Hamiltonian out of the sum and simplifying the notation
by recognizing the density operator, we may still write

\begin{align}
i\hbar\frac{d\hat{\rho}}{dt} & =\hat{H}(t)\sum_{\mu}p_{\mu}\left|\psi^{(\mu)}\right\rangle \left\langle \psi^{(\mu)}\right|-\sum_{\mu}p_{\mu}\left|\psi^{(\mu)}\right\rangle \left\langle \psi^{(\mu)}\right|\hat{H}(t)\nonumber \\
 & =\hat{H}(t)\hat{\rho}-\hat{\rho}\hat{H}(t)=[\hat{H}(t),\hat{\rho}].
\end{align}
We have thus arrived at the \textit{von-Neumann Equation for the Time
Evolution} of $\hat{\rho}:$

\begin{equation}
i\hbar\frac{d\hat{\rho}}{dt}=[\hat{H}(t),\hat{\rho}]\label{eq:vonNeumann}
\end{equation}
This equation can be solved in the interaction picture, by making
use of the time evolution operator of equation (\ref{eq:EvolutionInteraction}).
The time evolution is thus governed by

\begin{equation}
\hat{\rho}(t)=\hat{U}(t,0)\hat{\rho}_{0}\hat{U}^{\dagger}(t,0),
\end{equation}
where $\hat{\rho}_{0}$ corresponds to the density operator at time
$t=0$. Unitary time evolution is, as discussed, one of the postulates
of quantum mechanics. It is in fact, a very natural postulate, which
is realized in the conservation of the norm of states and means that
normalized states stay normalized.

\subsection{Time evolution of the Bloch vector in a two-level system \label{subsec:Time-evolution-of}}

In this section we analyze a two-level system (see appendix \ref{sec:A.-Two-level-systems})
form the density operator perspective and give the time evolution
of the Bloch vector. The density operator of a two-level system is
written as

\begin{equation}
\hat{\rho}=\frac{1}{2}(\mathds{1}+\boldsymbol{u}\cdot\boldsymbol{\sigma}),\label{eq:twoLevelDensity}
\end{equation}
where $\mathds{1}$ is the identity, $\boldsymbol{u}$ is an arbitrary
Bloch vector and $\boldsymbol{\sigma}$ is the vector containing the
Pauli Matrices. We note immediately that:

\begin{equation}
\begin{cases}
\text{tr(\ensuremath{\mathds{1)})=2}},\\
\text{tr}(\sigma_{i})=0 & \forall i.
\end{cases}
\end{equation}
This implies that

\begin{equation}
\text{tr}(\hat{\rho})=\frac{1}{2}(2)=1,\label{eq:Prop1}
\end{equation}
and we also note that

\begin{align}
\hat{\rho}^{2} & =\frac{1}{4}(\mathds{1}+\boldsymbol{u}\cdot\boldsymbol{\sigma})(\mathds{1}+\boldsymbol{u}\cdot\boldsymbol{\sigma})\nonumber \\
 & =\frac{1}{4}\left(\mathds{1}+2\boldsymbol{u}\cdot\boldsymbol{\sigma}+\sum_{ij}u_{i}u_{j}\sigma_{i}\sigma_{j}\right)\nonumber \\
 & =\frac{1}{4}\left[\mathds{1}+2\boldsymbol{u}\cdot\boldsymbol{\sigma}+\sum_{ij}u_{i}u_{j}\frac{1}{2}\left([\sigma_{i},\sigma_{j}]+\left\{ \sigma_{i},\sigma_{j}\right\} \right)\right]\nonumber \\
 & =\frac{1}{4}\left[\mathds{1}+2\boldsymbol{u}\cdot\boldsymbol{\sigma}+\sum_{ij}u_{i}u_{j}\left(i\varepsilon_{ijk}\sigma_{k}+\delta_{ij}\mathds{1}\right)\right],
\end{align}
where taking the trace, we get

\begin{align}
\text{tr}(\hat{\rho}^{2}) & =\frac{1}{4}\text{tr}\left[\mathds{1}+2\boldsymbol{u}\cdot\boldsymbol{\sigma}+\sum_{ij}u_{i}u_{j}\left(i\varepsilon_{ijk}\sigma_{k}+\delta_{ij}\mathds{1}\right)\right]\nonumber \\
 & =\frac{1}{4}\left[2+0+\sum_{ij}u_{i}u_{j}\text{tr}\left(i\varepsilon_{ijk}\sigma_{k}+\delta_{ij}\mathds{1}\right)\right]\nonumber \\
 & =\frac{1}{4}\left[2+2\sum_{ij}u_{i}u_{j}\right]\nonumber \\
 & =\frac{1}{2}\left(1+\boldsymbol{u}\cdot\boldsymbol{u}\right).
\end{align}
Noting that in a two level system the Bloch vector $\boldsymbol{u}$
has norm such that $\boldsymbol{u}\cdot\boldsymbol{u}\leq$1, then
the operator $\hat{\rho}$ of equation (\ref{eq:twoLevelDensity})
verifies the second condition of a density operator, and thus, since
it also verifies the first by virtue of the result in equation (\ref{eq:Prop1}),
it is indeed a valid density operator. The expectation value of $\sigma_{k}$
can thus be calculated utilizing

\begin{align}
\text{tr}(\hat{\rho}\sigma_{k}) & =\frac{1}{2}\text{tr}\left(\sigma_{k}+\sum_{j}u_{j}\sigma_{j}\sigma_{k}\right)\nonumber \\
 & =\frac{1}{2}\text{tr}\left\{ \sigma_{k}+\sum_{j}u_{j}(\delta_{jk}\mathds{1}+\varepsilon_{jkl}\sigma_{l})\right\} \nonumber \\
 & =\frac{1}{2}\text{tr}\left(u_{k}\mathds{1}\right)\nonumber \\
 & =u_{k},\label{eq:BlochElements}
\end{align}
which corresponds to a way of calculating the elements of the Bloch
vector. We can now aim to calculate the time evolution of a Bloch
vector in a two-level system. By employing the result of equation
(\ref{eq:BlochElements}) we can write

\begin{equation}
i\hbar\frac{du_{k}}{dt}=\frac{d}{dt}\text{tr}(\hat{\rho}\sigma_{k})=\text{tr}\left(\frac{d\hat{\rho}}{dt}\sigma_{k}\right),
\end{equation}
and by now making use of the von-Neumann time evolution, we write

\begin{equation}
i\hbar\frac{du_{k}}{dt}=\text{tr}\left([\hat{H}(t),\hat{\rho}(t)]\sigma_{k}\right).
\end{equation}
By cyclically permuting the operators under the trace, which is allowed
by equation (\ref{eq:cyclic}), we finally write

\begin{align}
i\hbar\frac{du_{k}}{dt} & =\text{tr}\left\{ \left(\hat{H}\hat{\rho}-\hat{\rho}\hat{H}\right)\sigma_{k}\right\} =\text{tr}\left\{ \sigma_{k}\hat{H}\hat{\rho}-\hat{\rho}\hat{H}\sigma_{k}\right\} =\text{tr}\left\{ \hat{\rho}\sigma_{k}\hat{H}-\hat{\rho}\hat{H}\sigma_{k}\right\} \nonumber \\
 & =\text{tr}\left\{ \hat{\rho}[\sigma_{k},\hat{H}]\right\} =\left\langle [\sigma_{k},\hat{H}]\right\rangle .\label{eq:BlochEvolution}
\end{align}
This result based on the density operator formalism reproduces the
well known result for the evolution of the components of the Bloch
vector in a two-level system but is in fact more general since it
also applies to mixed states, where the additional statistical averaging
is needed.

\subsection{An application of the density matrix formalism in the interaction
picture - Optically driven two-level system \label{subsec:An-application-of}}

We have so far described two-level systems in terms of the evolution
of the components of the Bloch vector, but we might as well pick a
basis made up by a ground state $\left|g\right\rangle $ and an excited
state $\left|e\right\rangle $, which we do following \citet{Lavine2018twolevel}.
In this basis our time dependent wave-function in the Schrodinger
picture can be written as

\begin{equation}
\left|\psi(t)\right\rangle =C_{g}(t)\left|g\right\rangle +C_{e}(t)\left|e\right\rangle .
\end{equation}
Of course $C_{e}$ and $C_{g}$ are in principle complex and as such
amount to four time-dependent functions. Of course, if this is to
be consistent with the Bloch formalism, we know there must be a constraint,
which is simply the normalization condition

\begin{equation}
|C_{e}(t)|^{2}+|C_{g}(t)|^{2}=1.
\end{equation}
To explicitly find the matrix elements of the density operator in
this basis, we just need to calculate

\begin{align}
\left\langle g\right|\hat{\rho}\left|g\right\rangle  & =\left\langle g\right|\left(C_{g}(t)\left|g\right\rangle +C_{e}(t)\left|e\right\rangle \right)\left(C_{g}^{*}(t)\left\langle g\right|+C_{e}^{*}(t)\left\langle e\right|\right)\left|g\right\rangle \nonumber \\
 & =C_{g}^{*}C_{g}=\left|C_{g}(t)\right|^{2},
\end{align}

\begin{align}
\left\langle e\right|\hat{\rho}\left|g\right\rangle  & =\left\langle e\right|\left(C_{g}(t)\left|g\right\rangle +C_{e}(t)\left|e\right\rangle \right)\left(C_{g}^{*}(t)\left\langle g\right|+C_{e}^{*}(t)\left\langle e\right|\right)\left|g\right\rangle \nonumber \\
 & =C_{e}C_{g}^{*},
\end{align}

\begin{align}
\left\langle g\right|\hat{\rho}\left|e\right\rangle  & =\left\langle g\right|\left(C_{g}(t)\left|g\right\rangle +C_{e}(t)\left|e\right\rangle \right)\left(C_{g}^{*}(t)\left\langle g\right|+C_{e}^{*}(t)\left\langle e\right|\right)\left|e\right\rangle \nonumber \\
 & =C_{e}^{*}C_{g},
\end{align}

\begin{align}
\left\langle e\right|\hat{\rho}\left|e\right\rangle  & =\left\langle g\right|\left(C_{g}(t)\left|g\right\rangle +C_{e}(t)\left|e\right\rangle \right)\left(C_{g}^{*}(t)\left\langle g\right|+C_{e}^{*}(t)\left\langle e\right|\right)\left|e\right\rangle \nonumber \\
 & =C_{e}^{*}C_{e}=\left|C_{e}(t)\right|^{2},
\end{align}
which along with the von-Neumann equation for the time evolution of
the density matrix allows us to calculate explicitly the time evolution
of the density operator for a given Hamiltonian. We aim now, however,
to solve this in the case of a perturbed system, for example in the
case of a time dependent harmonic electric field. We first approach
this problem using the von-Neumann equation in the Schrodinger picture
but later we shall see that the usage of the interaction picture may
provide much simpler results. We star with the general form of a perturbed
Hamiltonian

\begin{equation}
\hat{H}=\hat{H}_{0}+\hat{V}(t).
\end{equation}
where in the basis $\left\{ \left|g\right\rangle ,\left|e\right\rangle \right\} $
we have

\begin{equation}
\hat{H}_{0}\left|i\right\rangle =E_{i}\left|i\right\rangle ,
\end{equation}
and the matrix elements of our perturbed Hamiltonian are therefore

\begin{equation}
\left\langle i\right|\hat{H}_{0}+\hat{V}(t)\left|j\right\rangle =E_{j}\delta_{ij}+\left\langle i\right|\hat{V}(t)\left|j\right\rangle .
\end{equation}
The time evolution of the density matrix elements is given in the
Schrodinger representation by the von-Neumann equation (\ref{eq:vonNeumann})

\begin{align}
i\hbar\frac{d\rho_{ij}}{dt} & =\sum_{k}\left(E_{k}\delta_{ik}+\left\langle i\right|\hat{V}(t)\left|k\right\rangle \right)\rho_{kj}(t)\nonumber \\
 & -\rho_{ik}(t)\sum_{k}\left(E_{k}\delta_{kj}+\left\langle k\right|\hat{V}(t)\left|j\right\rangle \right),
\end{align}
and noting that only the terms that obey the conditions imposed by
the Kronecker-delta survive, we write

\begin{equation}
i\hbar\frac{d\rho_{ij}}{dt}=(E_{i}-E_{j})\rho_{ij}(t)+\sum_{k}\left\{ \left\langle i\right|\hat{V}(t)\left|k\right\rangle \rho_{kj}(t)-\rho_{ik}(t)\left\langle k\right|\hat{V}(t)\left|j\right\rangle \right\} ,
\end{equation}
or more succinctly, in terms of the commutator

\begin{equation}
i\hbar\frac{d\rho_{ij}}{dt}=(E_{i}-E_{j})\rho_{ij}(t)+\left\langle i\right|\left[\hat{V}(t),\rho(t)\right]\left|j\right\rangle ,
\end{equation}
which corresponds to the von-Neumann equation in the Schrodinger representation
for two-level systems. We now proceed to convert this result to the
interaction picture, which will simplify our calculations. Specifically
the interaction picture removes the time dependence due to $\hat{H}_{0}.$
To start, we write the interaction picture wave-function as

\begin{equation}
\left|\psi_{I}(t)\right\rangle =\hat{U}_{0}^{\dagger}(t,0)\left|\psi_{S}(t)\right\rangle =e^{i\hat{H}_{0}t/\hbar}\left|\psi_{S}(t)\right\rangle .
\end{equation}
We follow this up by calculating $\hat{V}(t)$ in the interaction
picture. This is simply

\begin{equation}
\hat{V}_{I}(t)=\hat{U}_{0}^{\dagger}(t,0)\hat{V}(t)\hat{U}_{0}(t,0)=e^{i\hat{H}_{0}t/\hbar}\hat{V}(t)e^{-i\hat{H}_{0}t/\hbar}.
\end{equation}
The density operator in the interaction picture can be constructed
from the wave-functions. It is given by

\begin{equation}
\hat{\rho}_{I}(t)=e^{i\hat{H}_{0}t/\hbar}\left|\psi_{S}(t)\right\rangle \left\langle \psi_{S}(t)\right|e^{-i\hat{H}_{0}t/\hbar}=e^{i\hat{H}_{0}t/\hbar}\hat{\rho}(t)e^{-i\hat{H}_{0}t/\hbar},\label{eq:rhoInteraction}
\end{equation}
and noting that the von-Neumann equation in the Schrodinger representation
for a perturbed Hamiltonian is

\begin{equation}
i\hbar\frac{d\hat{\rho}(t)}{dt}=\hat{H}_{0}\hat{\rho}(t)-\hat{\rho}(t)\hat{H}_{0}+\hat{V}(t)\hat{\rho}(t)-\hat{\rho}(t)\hat{V}(t),\label{eq:NeumannPerturbed}
\end{equation}
if we make use of equation (\ref{eq:rhoInteraction}) and multiply
on the right by $e^{i\hat{H}_{0}t/\hbar}$ and on the left by $e^{-i\hat{H}_{0}t/\hbar}$
on both sides, we can write $\hat{\rho}(t)=e^{-i\hat{H}_{0}t/\hbar}\hat{\rho}_{I}(t)e^{i\hat{H}_{0}t/\hbar}$
as well as differentiate both sides with respect to time and multiply
by $i\hbar$, we have

\begin{align}
i\hbar\frac{d\hat{\rho}(t)}{dt} & =i\hbar\left(-\frac{i}{\hbar}\hat{H}_{0}t\right)e^{-i\hat{H}_{0}t/\hbar}\hat{\rho}_{I}(t)e^{i\hat{H}_{0}t/\hbar}+i\hbar e^{-i\hat{H}_{0}t/\hbar}\left(\frac{d\hat{\rho_{I}}(t)}{dt}\right)e^{i\hat{H}_{0}t/\hbar}\nonumber \\
 & +i\hbar e^{-i\hat{H}_{0}t/\hbar}\hat{\rho}_{I}(t)\left(\frac{i}{\hbar}\hat{H}_{0}t\right)e^{i\hat{H}_{0}t/\hbar}.
\end{align}
Comparing with the von-Neumann equation in the Schrodinger picture,
and noting that the Hamiltonian commutes with its own exponential,
we have that

\begin{equation}
\hat{V}(t)\hat{\rho}(t)-\hat{\rho}(t)\hat{V}(t)=i\hbar e^{-i\hat{H}_{0}t/\hbar}\left(\frac{d\hat{\rho}_{I}(t)}{dt}\right)e^{i\hat{H}_{0}t/\hbar},
\end{equation}
and, multiplying on both sides by $i\hbar$ before solving for $i\hbar d\hat{\rho}/dt$

\begin{align}
i\hbar\left(\frac{d\hat{\rho}_{I}(t)}{dt}\right) & =e^{i\hat{H}_{0}t/\hbar}\left[\hat{V}(t)\hat{\rho}(t)-\hat{\rho}(t)\hat{V}(t)\right]e^{-i\hat{H}_{0}t/\hbar}\nonumber \\
 & =e^{i\hat{H}_{0}t/\hbar}\hat{V}(t)e^{-i\hat{H}_{0}t/\hbar}\hat{\rho}_{I}(t)e^{i\hat{H}_{0}t/\hbar}e^{-i\hat{H}_{0}t/\hbar}\nonumber \\
 & -e^{i\hat{H}_{0}t/\hbar}e^{-i\hat{H}_{0}t/\hbar}\hat{\rho}_{I}(t)e^{i\hat{H}_{0}t/\hbar}\hat{V}(t)e^{-i\hat{H}_{0}t/\hbar}\nonumber \\
 & =V_{I}(t)\hat{\rho}_{I}(t)-\hat{\rho}_{I}(t)\hat{V}_{I}(t)\nonumber \\
 & =\left[V_{I}(t),\hat{\rho}_{I}(t)\right].
\end{align}
This result is the von-Neumann equation in the interaction picture.
We shall now use this to calculate the time evolution of an optically
driven two-level system. The first step is to specify the form of
the perturbation or interaction Hamiltonian. We consider this to be
that of a time dependent electric field in the dipole approximation

\begin{equation}
\hat{V}(t)=\hat{H}_{\text{op}}(t)=-q\boldsymbol{r}\cdot\boldsymbol{E}(t),\label{eq:HopSimple}
\end{equation}
and in particular, we will consider a harmonic field, oscillating
with a positive and negative frequency. That is to say

\[
\boldsymbol{E}(t)=\boldsymbol{E}_{0}\left(e^{i\omega t}+e^{-i\omega t}\right).
\]
We now calculate, as in the general case, the matrix elements of this
perturbation

\begin{equation}
\left\langle i\right|\hat{H}_{\text{op}}(t)\left|j\right\rangle =-\left\langle i\right|q\boldsymbol{r}\left|j\right\rangle \cdot\boldsymbol{E}_{0}(e^{i\omega t}+e^{-i\omega t}).
\end{equation}
Since $\boldsymbol{r}$ is odd, then for $i=j$ the bra-ket vanishes.
As for the remaining matrix elements, we calculate them in the interaction
picture using $\hat{V}_{I}(t)=e^{i\hat{H}_{0}t/\hbar}\hat{H}_{\text{op}}(t)e^{-i\hat{H}_{0}t/\hbar}$.
Since for the eigenstates of the zeroth order Hamiltonian, the evolution
is trivial, then

\begin{align}
\hat{V}_{eg}^{I}(t)=\left\langle e\right|\hat{V}_{I}(t)\left|g\right\rangle  & =\left\langle e\right|e^{i\hat{H}_{0}t/\hbar}\hat{H}_{\text{op}}(t)e^{-i\hat{H}_{0}t/\hbar}\left|g\right\rangle \nonumber \\
 & =\left\langle e\right|e^{iE_{e}t/\hbar}q\boldsymbol{r}e^{-iE_{g}t/\hbar}\left|g\right\rangle \cdot\boldsymbol{E}_{0}(e^{i\omega t}+e^{-i\omega t})\nonumber \\
 & =\left\langle e\right|q\boldsymbol{r}\left|g\right\rangle \cdot\boldsymbol{E}_{0}(e^{i\omega t}+e^{-i\omega t})e^{i(E_{e}-E_{g})t/\hbar},
\end{align}
and similarly for $\hat{V}_{ge}^{I}$

\begin{equation}
\hat{V}_{ge}^{I}(t)=\left\langle e\right|q\boldsymbol{r}\left|g\right\rangle \cdot\boldsymbol{E}_{0}(e^{i\omega t}+e^{-i\omega t})e^{i(E_{g}-E_{e})t/\hbar}.
\end{equation}
The density operator has thus only non-diagonal elements. To simplify
our notation we introduce the Rabi energies $\hbar\Omega_{ij}$ defined
through

\begin{equation}
\Omega_{ij}=\left\langle i\right|q\boldsymbol{r}\left|j\right\rangle \cdot\boldsymbol{E}_{0},
\end{equation}
and a frequency $\omega_{0}=(E_{e}-E_{g})/\hbar$. For simplicity,
if we write $\Omega=\Omega_{12}$, we have a perturbation of the form

\begin{equation}
\hat{V}_{I}=\begin{bmatrix}0 & \Omega(e^{i\omega t}+e^{-i\omega t})e^{-i\omega_{0}t}\\
\Omega(e^{i\omega t}+e^{-i\omega t})e^{i\omega_{0}t} & 0
\end{bmatrix}.
\end{equation}
With the intent of using the von-Neumann equation in the interaction
picture we calculate the commutator

\begin{align}
\left[\hat{V}_{I}(t)/\hbar,\hat{\rho}_{I}(t)\right] & =\begin{bmatrix}0 & \Omega(e^{i\omega t}+e^{-i\omega t})e^{-i\omega_{0}t}\\
\Omega(e^{i\omega t}+e^{-i\omega t})e^{i\omega_{0}t} & 0
\end{bmatrix}\begin{bmatrix}\rho_{ee}^{I} & \rho_{eg}^{I}\\
\rho_{ge}^{I} & \rho_{gg}^{I}
\end{bmatrix}\nonumber \\
 & -\begin{bmatrix}\rho_{ee}^{I} & \rho_{eg}^{I}\\
\rho_{ge}^{I} & \rho_{gg}^{I}
\end{bmatrix}\begin{bmatrix}0 & \Omega(e^{i\omega t}+e^{-i\omega t})e^{-i\omega_{0}t}\\
\Omega(e^{i\omega t}+e^{-i\omega t})e^{i\omega_{0}t} & 0
\end{bmatrix}\nonumber \\
 & =\Omega(e^{i\omega t}+e^{-i\omega t})\begin{bmatrix}\rho_{ge}^{I}e^{-i\omega_{0}t}-\rho_{eg}^{I}e^{i\omega_{0}t} & \left(\rho_{gg}^{I}-\rho_{ee}^{I}\right)e^{-i\omega_{0}t}\\
\left(\rho_{ee}^{I}-\rho_{gg}^{I}\right)e^{i\omega_{0}t} & \rho_{eg}^{I}e^{i\omega_{0}t}-\rho_{ge}^{I}e^{-i\omega_{0}t}
\end{bmatrix},
\end{align}
and we note that this equality is equivalent to a system of equations
$\hat{\rho}$

\begin{equation}
\begin{cases}
i\frac{d\rho_{ee}^{I}}{dt} & =\Omega(e^{i\omega t}+e^{-i\omega t})\left[\rho_{ge}^{I}e^{-i\omega_{0}t}-\rho_{eg}^{I}e^{i\omega_{0}t}\right],\\
i\frac{d\rho_{eg}^{I}}{dt} & =\Omega(e^{i\omega t}+e^{-i\omega t})\left(\rho_{gg}^{I}-\rho_{ee}^{I}\right)e^{-i\omega_{0}t},\\
i\frac{d\rho_{ge}^{I}}{dt} & =\Omega(e^{i\omega t}+e^{-i\omega t})\left(\rho_{ee}^{I}-\rho_{gg}^{I}\right)e^{i\omega_{0}t},\\
i\frac{d\rho_{gg}^{I}}{dt} & =\Omega(e^{i\omega t}+e^{-i\omega t})\left[\rho_{eg}^{I}e^{i\omega_{0}t}-\rho_{ge}^{I}e^{-i\omega_{0}t}\right].
\end{cases}
\end{equation}
We now apply the rotating wave approximation and keep only the resonant
terms containing $\omega-\omega_{0}$ or $\omega_{0}-\omega$. We
are left with

\begin{equation}
\begin{cases}
i\frac{d\rho_{ee}^{I}}{dt} & =\Omega\left(\rho_{ge}^{I}e^{i(\omega-\omega_{0})t}+\rho_{eg}^{I}e^{i(\omega_{0}-\omega)t}\right),\\
i\frac{d\rho_{eg}^{I}}{dt} & =\Omega\left(\rho_{gg}^{I}-\rho_{ee}^{I}\right)e^{i(\omega-\omega_{0})t},\\
i\frac{d\rho_{ge}^{I}}{dt} & =\Omega\left(\rho_{ee}^{I}-\rho_{gg}^{I}\right)e^{i(\omega_{0}-\omega)t},\\
i\frac{d\rho_{gg}^{I}}{dt} & =\Omega(\rho_{ge}^{I}e^{i(\omega-\omega_{0})t}+\rho_{eg}^{I}e^{i(\omega-\omega_{0})t}).
\end{cases}
\end{equation}
If we introduce the parameter $\Delta=\omega-\omega_{0}$ and $\bar{\Omega}=\sqrt{\Omega^{2}+\Delta^{2}}$
and admit that the system is originally in the ground state, which
corresponds to $\rho_{gg}=1$ and $\rho_{eg}=\rho_{ge}=\rho_{ee}=0$
it is possible to solve this system of equations. This solution is
difficult to obtain, however we can simplify the calculations assuming
that the system is in resonance and $\Delta=0$. This removes the
time dependence and we obtain a much simpler system of equations of
the form

\begin{equation}
\begin{cases}
i\frac{d\rho_{ee}^{I}}{dt} & =\Omega\left(\rho_{ge}^{I}+\rho_{eg}^{I}\right)=\Omega\text{Re\ensuremath{\left(\rho_{ge}^{I}\right)}},\\
i\frac{d\rho_{eg}^{I}}{dt} & =\Omega\left(\rho_{gg}^{I}-\rho_{ee}^{I}\right),\\
i\frac{d\rho_{ge}^{I}}{dt} & =\Omega\left(\rho_{ee}^{I}-\rho_{gg}^{I}\right),\\
i\frac{d\rho_{gg}^{I}}{dt} & =\Omega\left(\rho_{ge}^{I}+\rho_{eg}^{I}\right)=\Omega\text{Re\ensuremath{\left(\rho_{ge}^{I}\right)}}.
\end{cases}\label{eq:systeminresonance}
\end{equation}
If we take the time derivative of the first equation, we obtain the
second order differential equation

\begin{equation}
i\frac{d^{2}\rho_{ee}^{I}}{dt^{2}}=\Omega\left(\frac{d\rho_{ge}^{I}}{dt}+\frac{d\rho_{eg}^{I}}{dt}\right),
\end{equation}
and we now employ the equations for the time derivatives of the relevant
matrix elements in equation (\ref{eq:systeminresonance}). We obtain

\begin{align}
i\frac{d^{2}\rho_{ee}^{I}}{dt^{2}} & =-i\Omega^{2}\left(\rho_{ee}^{I}-\rho_{gg}^{I}-\rho_{gg}^{I}+\rho_{ee}^{I}\right)\nonumber \\
\Leftrightarrow & \frac{d^{2}\rho_{ee}^{I}}{dt^{2}}=-2\Omega^{2}(\rho_{ee}^{I}-\rho_{gg}^{I}),
\end{align}
and remembering that the trace of the density matrix is 1, we write

\begin{equation}
\frac{d^{2}\rho_{ee}^{I}}{dt^{2}}=-2\Omega^{2}(2\rho_{ee}^{I}-1).
\end{equation}
We propose a solution of the form $\rho_{ee}^{I}(t)=A\cos\left(2\Omega t\right)+B\sin\left(2\Omega t\right)+C$
and see if it solves the differential equation

\begin{equation}
-4\Omega^{2}A\cos\left(2\Omega t\right)-4\Omega^{2}B\sin\left(2\Omega t\right)=-4\Omega^{2}A\cos\left(2\Omega t\right)-4\Omega^{2}B\cos\left(2\Omega t\right)-4\Omega^{2}C+2\Omega^{2}.
\end{equation}
Indeed, canceling all the terms proportional to sines and cosines,
we get that the equality holds if $C=1/2$. We also make use of the
initial conditions

\begin{equation}
\begin{cases}
\rho_{ee}^{I}(t=0)=0,\\
\frac{d\rho_{ee}^{I}}{dt}(t=0)=\Omega\left(\rho_{ge}^{I}+\rho_{eg}^{I}\right)=0,
\end{cases}
\end{equation}
which mean that the two-level system is initially in the ground state.
Using these, we can find

\begin{equation}
\begin{cases}
A+1/2=0 & \Rightarrow A=-1/2,\\
B=0,
\end{cases}
\end{equation}
and as such, we have a solution for the matrix element $\rho_{ee}^{I}(t)$
in resonance, which reads

\begin{align}
\rho_{ee}^{I}(t) & =\frac{1}{2}\left[1-\cos\left(2\Omega t\right)\right]\nonumber \\
 & =\frac{1}{2}\left[1-\cos^{2}(\Omega t)+\sin^{2}(\Omega t)\right]\nonumber \\
 & =\sin^{2}(\Omega t).
\end{align}
As for $\rho_{gg}^{I}(t)$ we need not calculate its evolution in
the same manner. We can simply invoke that the trace of the density
matrix should be one. This leads us to

\begin{equation}
\rho_{gg}^{I}(t)=\cos^{2}(\Omega t).
\end{equation}
As for the non-diagonal elements of the density matrix, we can now
substitute these last few results into equation (\ref{eq:systeminresonance})
and obtain

\begin{equation}
i\frac{d\rho_{eg}^{I}}{dt}=\Omega\left(\cos^{2}(\Omega t)-\sin^{2}(\Omega t)\right)=\Omega\cos\left(2\Omega t\right),
\end{equation}
which can easily be solved to yield

\begin{equation}
\rho_{eg}^{I}(t)=-\frac{i}{2}\sin(2\Omega t).
\end{equation}
Finally, noting that $\rho_{ge}^{I}(t)=\rho_{eg}^{I*}(t)$ we have
the full solution to the time evolution of the density operator

\begin{equation}
\begin{cases}
\rho_{ee}^{I}(t) & =\sin^{2}(\Omega t),\\
\rho_{eg}^{I}(t) & =-\frac{i}{2}\sin(2\Omega t),\\
\rho_{ge}^{I}(t) & =\frac{i}{2}\sin(2\Omega t),\\
\rho_{ee}^{I}(t) & =\cos^{2}(\Omega t).
\end{cases}
\end{equation}
This example serves as a somewhat pedagogical introduction to calculations
of the time evolution of the density matrix, which is useful when
more complex cases arise ahead.

\section{Details of the derivation of the Lindblad equation\label{sec:Details-of-the-Derivation-of-the-Lindblad-Equation}}

\subsection{Considerations regarding the general derivation of the Lindblad equation\label{subsec:Considerations-regarding-the}}

In this appendix we provide a few discussions and steps which were
skipped over or just mentioned in the main text along the general
considerations made toward the Lindblad equation as well as its derivation
in the case of the two-level system in contact with the thermal reservoir.

We make a few remarks regarding the procedure of simplification of
the made in the main text more explicit. Starting from the integro-differential
equation for the von-Neumann equation it was stated that the commutators
were explicitly calculated for a interaction Hamiltonian of the form
given in equation (\ref{eq:FactoredInteraction}). Indeed the result
of this procedure is:

\begin{align}
\frac{d\hat{\rho}_{S}^{I}(t)}{dt}= & -\frac{1}{\hbar^{2}}\sum_{ij}\int_{t_{0}}^{t}\text{tr}_{R}\left\{ \beta_{i}^{S}(t)\otimes\Gamma_{i}^{R}(t)\left(\beta_{j}^{S}(\tau)\hat{\rho}_{S}^{I}(t)\otimes\Gamma_{j}^{R}(\tau)\hat{\rho}_{R}^{I}-\hat{\rho}_{S}^{I}(t)\beta_{j}^{S}(\tau)\otimes\hat{\rho}_{R}^{I}\Gamma_{j}^{R}(\tau)\right)\right.\nonumber \\
 & -\left.\left(\beta_{j}^{S}(\tau)\hat{\rho}_{S}^{I}(t)\otimes\Gamma_{j}^{R}(\tau)\hat{\rho}_{R}^{I}-\hat{\rho}_{S}^{I}(t)\beta_{j}^{S}(\tau)\otimes\hat{\rho}_{R}^{I}\Gamma_{j}^{R}(\tau)\right)\beta_{i}^{S}(t)\otimes\Gamma_{i}^{R}(t)\right\} d\tau.
\end{align}
Dropping the tensor products and $S$ and $R$ indices for the operators
which make up the interaction as well as the $I$ index, and then
factoring the expression, we are able to write:

\begin{align}
\frac{d\hat{\rho}_{S}(t)}{dt}= & -\frac{1}{\hbar^{2}}\sum_{ij}\int_{t_{0}}^{t}\text{tr}_{R}\left\{ \beta_{i}(t)\Gamma_{i}(t)\left(\beta_{j}(\tau)\hat{\rho}_{S}(t)\Gamma_{j}(\tau)\hat{\rho}_{R}-\hat{\rho}_{S}(t)\beta_{j}(\tau)\hat{\rho}_{R}\Gamma_{j}(\tau)\right)\right.\nonumber \\
 & -\left.\left(\beta_{j}(\tau)\hat{\rho}_{S}(t)\Gamma_{j}(\tau)\hat{\rho}_{R}-\hat{\rho}_{S}(t)\beta_{j}(\tau)\hat{\rho}_{R}\Gamma_{j}(\tau)\right)\beta_{i}(t)\Gamma_{i}(t)\right\} d\tau\nonumber \\
= & -\frac{1}{\hbar^{2}}\sum_{ij}\int_{t_{0}}^{t}\text{tr}_{R}\left\{ \left(\beta_{i}(t)\beta_{j}(\tau)\hat{\rho}_{S}(\tau)\right)\left(\Gamma_{i}(t)\Gamma_{j}(\tau)\hat{\rho}_{R}\right)\right.\nonumber \\
 & -\left(\beta_{i}(t)\hat{\rho}_{S}(t)\beta_{j}(\tau)\right)\left(\Gamma_{i}(t)\hat{\rho}_{R}\Gamma_{j}(\tau)\right)\nonumber \\
 & -\left(\beta_{j}(\tau)\hat{\rho}_{S}(t)\beta_{i}(t)\right)\left(\Gamma_{j}(\tau)\hat{\rho}_{R}\Gamma_{i}(t)\right)\nonumber \\
 & \left.+\left(\hat{\rho}_{S}(t)\beta_{j}(\tau)\beta_{i}(t)\right)\left(\hat{\rho}_{R}\Gamma_{j}(\tau)\Gamma_{i}(t)\right)\right\} d\tau.
\end{align}
Under the trace over the reservoir degrees of freedom we can cyclically
permute the operators $\Gamma_{i},\text{\ensuremath{\Gamma}}_{j}$
and $\hat{\rho}_{R}$. Doing this allows us to further simplify the
previous equation:

\begin{align}
\frac{d\hat{\rho}_{S}(t)}{dt} & =-\frac{1}{\hbar^{2}}\sum_{ij}\int_{t_{0}}^{t}\text{tr}_{R}\left\{ \left(\beta_{i}(t)\beta_{j}(\tau)\hat{\rho}_{S}(t)-\beta_{j}(\tau)\hat{\rho}_{S}(t)\beta_{i}(t)\right)\left(\hat{\rho}_{R}\Gamma_{i}(t)\Gamma_{j}(\tau)\right)\right.\nonumber \\
 & \left.+\left(\hat{\rho}_{S}(t)\beta_{j}(\tau)\beta_{i}(t)-\beta_{i}(t)\hat{\rho}_{S}(t)\beta_{j}(\tau)\right)\left(\hat{\rho}_{R}\Gamma_{j}(\tau)\Gamma_{i}(t)\right)\right\} d\tau,
\end{align}
and finally, introducing the correlation functions, we write:

\begin{align}
\frac{d\hat{\rho}_{S}(t)}{dt}= & -\frac{1}{\hbar^{2}}\sum_{ij}\int_{t_{0}}^{t}\left(\beta_{i}(t)\beta_{j}(\tau)\hat{\rho}_{S}(t)-\beta_{j}(\tau)\hat{\rho}_{S}(t)\beta_{i}(t)\right)\left\langle \Gamma_{i}(t)\Gamma_{j}(\tau)\right\rangle \nonumber \\
 & +\left(\hat{\rho}_{S}(t)\beta_{j}(\tau)\beta_{i}(t)-\beta_{i}(t)\hat{\rho}_{S}(t)\beta_{j}(\tau)\right)\left\langle \Gamma_{j}(t)\Gamma_{i}(\tau)\right\rangle d\tau.\label{eq:Correlation}
\end{align}

This is the equation (\ref{eq:Redfield}) presented in the main text
except for the use of the first and second Markov approximations.

\subsection{Detailed derivation of the light-matter interaction Hamiltonian in
the rotating wave approximation\label{subsec:Detailed-derivation-of}}

Here we provide a detailed derivation of the light-matter coupling
Hamiltonian in the rotating wave approximation. We start with the
semi-classical light-matter Hamiltonian.

\begin{align}
\hat{V}_{I} & =-q\hat{\boldsymbol{r}}\cdot\boldsymbol{E}\nonumber \\
 & =-\left(\boldsymbol{d}_{ge}\sigma_{-}+\boldsymbol{d}_{ge}^{*}\sigma_{+}\right)\cdot\boldsymbol{E}.
\end{align}
To arrive at the full description of this operator we want to substitute
the electric field $\boldsymbol{E}$ for the electric field operator
$\hat{\boldsymbol{E}}(\boldsymbol{r},t)$. Note that this operator
depends on time since the quantization of the E-M field was done in
the Heisenberg representation. We must therefore convert $\hat{\boldsymbol{E}}(\boldsymbol{r},t)$
to the Schrodinger picture. In the Heisenberg Picture we have

\begin{equation}
\hat{\boldsymbol{E}}_{\boldsymbol{k}}(\boldsymbol{r},t)=i\boldsymbol{e}_{\boldsymbol{k}}\sqrt{\frac{\hbar\omega_{k}}{2\varepsilon_{0}V}}\left[\hat{a}_{\boldsymbol{k}}e^{i\left(\boldsymbol{k}\cdot\boldsymbol{r}-\omega_{k}t\right)}-\hat{a}_{\boldsymbol{k}}^{\dagger}e^{-i\left(\boldsymbol{k}\cdot\boldsymbol{r}-\omega_{k}t\right)}\right].
\end{equation}
For a single mode, the Hamiltonian is time independent

\begin{equation}
\hat{H}_{\boldsymbol{k}}=\hbar\omega_{k}\left(\hat{n}_{\boldsymbol{k}}+1/2\right),
\end{equation}
and therefore, the time evolution of the ladder operators in the Heisenberg
Picture is given by

\begin{align}
i\hbar\frac{d}{dt}\hat{a}_{\boldsymbol{k}} & =\left[\hat{H}_{\boldsymbol{k}},\hat{a}_{\boldsymbol{k}}\right]=\left[\hbar\omega\left(\hat{a}_{\boldsymbol{k}}\hat{a}_{\boldsymbol{k}}^{\dagger}+1/2\right),\hat{a}_{\boldsymbol{k}}\right]\nonumber \\
 & =\left[\hbar\omega\hat{a}_{\boldsymbol{k}}\hat{a}_{\boldsymbol{k}}^{\dagger},\hat{a}_{\boldsymbol{k}}\right]\nonumber \\
 & =\left[\hbar\omega\hat{a}_{\boldsymbol{k}}\hat{a}_{\boldsymbol{k}}^{\dagger},\hat{a}_{\boldsymbol{k}}\right]\nonumber \\
 & =\hbar\omega\hat{a}_{\boldsymbol{k}}\left[\hat{a}_{\boldsymbol{k}}^{\dagger},\hat{a}_{\boldsymbol{k}}\right]+\hbar\omega\left[\hat{a}_{\boldsymbol{k}},\hat{a}_{\boldsymbol{k}}\right]\hat{a}_{\boldsymbol{k}}^{\dagger}\nonumber \\
 & =-\hbar\omega\hat{a}_{\boldsymbol{k}},
\end{align}
where we have used in the next to last step the commutator property
$\left[AB,C\right]=\left[A,B\right]C+B\left[A,C\right]$. Thus, by
taking the complex conjugate, we also obtain the equation of motion
for the creation operator

\begin{equation}
\begin{cases}
i\hbar\frac{d}{dt}\hat{a}_{\boldsymbol{k}}=-\hbar\omega\hat{a}_{\boldsymbol{k}},\\
i\hbar\frac{d}{dt}\hat{a}_{\boldsymbol{k}}^{\dagger}=\hbar\omega\hat{a}_{\boldsymbol{k}}^{\dagger}.
\end{cases}
\end{equation}
They therefore have solutions of the form

\begin{equation}
\begin{cases}
\hat{a}_{\boldsymbol{k}}(t)=\hat{a}_{\boldsymbol{k}}e^{-i\omega t},\\
\hat{a}_{\boldsymbol{k}}^{\dagger}(t)=\hat{a}_{\boldsymbol{k}}^{\dagger}e^{i\omega t}.
\end{cases}
\end{equation}
We can identify these in the operator for the electric field and make
the inverse correspondence in order to obtain the operator in the
Schrodinger picture

\begin{equation}
\hat{\boldsymbol{E}}_{\boldsymbol{k},\lambda}(\boldsymbol{r})=i\boldsymbol{e}_{\boldsymbol{k},\lambda}\sqrt{\frac{\hbar\omega_{k}}{2\varepsilon_{0}V}}\left[\hat{a}_{\boldsymbol{k},\lambda}e^{i\boldsymbol{k}\cdot\boldsymbol{r}}-\hat{a}_{\boldsymbol{k},\lambda}^{\dagger}e^{-i\boldsymbol{k}\cdot\boldsymbol{r}}\right],
\end{equation}
and thus, the interaction Hamiltonian is

\begin{equation}
\hat{H}_{I}=-i\boldsymbol{e}_{\boldsymbol{k},\lambda}\sqrt{\frac{\hbar\omega_{k}}{2\varepsilon_{0}V}}\cdot\left(\boldsymbol{d}_{ge}\sigma_{-}+\boldsymbol{d}_{ge}^{*}\sigma_{+}\right)\left[\hat{a}_{\boldsymbol{k},\lambda}e^{i\boldsymbol{k}\cdot\boldsymbol{r}}-\hat{a}_{\boldsymbol{k},\lambda}^{\dagger}e^{-i\boldsymbol{k}\cdot\boldsymbol{r}}\right].
\end{equation}
We write the $\sigma_{\pm}$ operators in the interaction picture,
using

\begin{equation}
\sigma_{\pm}^{I}(t)=e^{iH_{0}t/\hbar}\sigma_{\pm}e^{-iH_{0}t/\hbar}.
\end{equation}
Since the harmonic oscillator Hamiltonian does nothing to the Pauli
matrices, then

\begin{equation}
\sigma_{\pm}^{I}(t)=e^{i\omega_{0}\sigma_{z}t/2}\sigma_{\pm}e^{-i\omega_{0}\sigma_{z}t/2},
\end{equation}
and using the fact that $\hat{A}e^{\hat{B}}=e^{c}e^{\hat{B}}\hat{A}$,
where $\left[\hat{A},\hat{B}\right]=c\hat{A}$, we have $\left[\sigma_{\pm},-i\omega_{0}t/2\sigma_{z}\right]=\pm2i\omega_{0}t/2\sigma_{\pm}=\pm i\omega_{0}t\sigma_{\pm}$
and therefore

\begin{align}
\sigma_{\pm}^{I}(t) & =e^{i\omega_{0}\sigma_{z}}e^{-i\omega_{0}\sigma_{z}}e^{\pm i\omega_{0}t}\sigma_{\pm}\nonumber \\
 & =\sigma_{\pm}e^{\pm i\omega_{0}t}.
\end{align}
In a similar manner, the raising and lowering operators in the interaction
picture as

\begin{equation}
\hat{a}_{\boldsymbol{k}.\lambda}^{I}(t)=e^{iH_{0}t/\hbar}\hat{a}_{\boldsymbol{k}.\lambda}^{I}(t)e^{-iH_{0}t/\hbar}.
\end{equation}
Following the same procedure and using the previously calculated commutator,
we have

\begin{equation}
\begin{cases}
\hat{a}_{\boldsymbol{k}.\lambda}^{I}(t)=\hat{a}_{\boldsymbol{k},\lambda}e^{-i\omega_{k}t},\\
\hat{a}_{\boldsymbol{k}.\lambda}^{\dagger I}(t)=\hat{a}_{\boldsymbol{k},\lambda}^{\dagger}e^{i\omega_{k}t},
\end{cases}\label{eq:EvolutionLadder}
\end{equation}
and therefore, if we switch to the interaction picture and write the
interaction Hamiltonian, we have

\begin{equation}
\hat{H}_{I,\boldsymbol{k},\lambda}=-i\boldsymbol{e}_{\boldsymbol{k},\lambda}\sqrt{\frac{\hbar\omega_{k}}{2\varepsilon_{0}V}}\cdot\left(\boldsymbol{d}_{ge}\sigma_{-}e^{-i\omega_{0}t}+\boldsymbol{d}_{ge}^{*}\sigma_{+}e^{i\omega_{0}t}\right)\left[\hat{a}_{\boldsymbol{k},\lambda}e^{i\boldsymbol{k}\cdot\boldsymbol{r}-i\omega_{k}t}-\hat{a}_{\boldsymbol{k},\lambda}^{\dagger}e^{-i\boldsymbol{k}\cdot\boldsymbol{r}+i\omega_{k}t}\right].
\end{equation}
The simplification we introduce is the rotating wave approximation
in which we discard terms in high frequency and retain only those
that can resonate. Thus we have

\begin{equation}
\hat{H}_{I,\boldsymbol{k},\lambda}=-i\boldsymbol{e}_{\boldsymbol{k},\lambda}\sqrt{\frac{\hbar\omega_{k}}{2\varepsilon_{0}V}}\cdot\left(\boldsymbol{d}_{ge}\sigma_{-}\hat{a}_{\boldsymbol{k},\lambda}^{\dagger}e^{-i\boldsymbol{k}\cdot\boldsymbol{r}-i(\omega_{0}-\omega_{k})t}+\boldsymbol{d}_{ge}^{*}\sigma_{+}\hat{a}_{\boldsymbol{k},\lambda}e^{i\boldsymbol{k}\cdot\boldsymbol{r}+i\left(\omega_{0}-\omega_{k}\right)t}\right),
\end{equation}
which when converting back to the Schrodinger picture, becomes

\begin{equation}
\hat{H}_{I,\boldsymbol{k},\lambda}=-i\boldsymbol{e}_{\boldsymbol{k},\lambda}\sqrt{\frac{\hbar\omega_{k}}{2\varepsilon_{0}V}}\cdot\left(\boldsymbol{d}_{ge}\sigma_{-}\hat{a}_{\boldsymbol{k},\lambda}^{\dagger}e^{-i\boldsymbol{k}\cdot\boldsymbol{r}}+\boldsymbol{d}_{ge}^{*}\sigma_{+}\hat{a}_{\boldsymbol{k},\lambda}e^{i\boldsymbol{k}\cdot\boldsymbol{r}}\right).
\end{equation}
We see that the rotating wave-approximation retains only the terms
which correspond to simple photon absorption and emission. We may
further simplify notation by introducing the coupling constants $\kappa_{\boldsymbol{k},\lambda}$
defined as

\begin{equation}
\kappa_{\boldsymbol{k},\lambda}=-ie^{i\boldsymbol{k}\cdot\boldsymbol{r}}\sqrt{\frac{\omega_{k}}{2\hbar\varepsilon_{0}V}}\boldsymbol{e}_{\boldsymbol{k},\lambda}\cdot\boldsymbol{d}_{eg},\label{eq:Coupling-1}
\end{equation}
such that the interaction Hamiltonian for a single mode $\boldsymbol{k},\lambda$
is written as

\begin{equation}
\hat{H}_{I,\boldsymbol{k},\lambda}=\hbar\left(\kappa_{\boldsymbol{k},\lambda}^{*}\sigma_{-}\hat{a}_{\boldsymbol{k},\lambda}^{\dagger}+\kappa_{\boldsymbol{k},\lambda}\sigma_{+}\hat{a}_{\boldsymbol{k},\lambda}\right).
\end{equation}
Thus, the full Hamiltonian describing the light-matter interaction
is a sum over the modes

\begin{equation}
\hat{H}_{I}=\sum_{\boldsymbol{k},\lambda}\hat{H}_{I,\boldsymbol{k},\lambda}=\hbar\sum_{\boldsymbol{k},\lambda}\kappa_{\boldsymbol{k},\lambda}^{*}\sigma_{-}\hat{a}_{\boldsymbol{k},\lambda}^{\dagger}+\kappa_{\boldsymbol{k},\lambda}\sigma_{+}\hat{a}_{\boldsymbol{k},\lambda},
\end{equation}
which is the desired form for the light-matter Hamiltonian in the
Schrodinger picture and rotating wave approximation.

\subsection{Writing the Lindblad equation in terms of correlation functions and
their calculation\label{subsec:Writing-the-Lindblad}}

In this section we give more detail on the identifications made to
write the master equation of the two-level system coupled to the thermal
radiation field. In the Schrodinger picture, comparing the light-matter
Hamiltonian with the general form given in equation (\ref{eq:FactoredInteraction}),
we write

\begin{equation}
\begin{cases}
\beta_{1}=\beta & =\sigma_{-},\\
\beta_{2}=\beta^{\dagger} & =\sigma_{+},\\
\Gamma_{1}=\Gamma^{\dagger} & =\sum_{\boldsymbol{k},\lambda}\kappa_{\boldsymbol{k},\lambda}^{*}\hat{a}_{\boldsymbol{k},\lambda}^{\dagger},\\
\Gamma_{2}=\Gamma & =\sum_{\boldsymbol{k},\lambda}\kappa_{\boldsymbol{k},\lambda}\hat{a}_{\boldsymbol{k},\lambda},
\end{cases}
\end{equation}
with which the Hamiltonian is written

\begin{equation}
\hat{H}_{I}=\sum_{\boldsymbol{k},\lambda}\hat{H}_{I,\boldsymbol{k},\lambda}=\hbar\left(\sigma_{-}\Gamma^{\dagger}+\sigma_{+}\Gamma\right).\label{eq:B25}
\end{equation}
Of course, these operators can be converted to the interaction picture
using the previously derived identities for $\sigma_{\pm}$ and the
ladder operators of the E-M field is given as in the equation (\ref{eq:EvolutionLadder}).
As such, the sum of equation (\ref{eq:B25}) is over $i=1,2$ and
$j=1,2$, and we can evaluate the reservoir correlation functions
as

\begin{align}
\left\langle \tilde{\Gamma}_{1}(t)\tilde{\Gamma}_{1}(\tau)\right\rangle  & =\text{tr}_{R}\left\{ \hat{\rho}_{R}\tilde{\Gamma}(t)\tilde{\Gamma}(\tau)\right\} \nonumber \\
 & =\text{tr}_{R}\left\{ \hat{\rho}_{R}\sum_{\boldsymbol{k},\lambda}\kappa_{\boldsymbol{k},\lambda}^{*2}\hat{a}_{\boldsymbol{k},\lambda}^{\dagger}\hat{a}_{\boldsymbol{k},\lambda}^{\dagger}e^{i\omega_{k}(t+\tau)}\right\} \nonumber \\
 & =\sum_{\boldsymbol{k},\lambda}\text{tr}_{R}\left\{ \hat{\rho}_{R}\kappa_{\boldsymbol{k},\lambda}^{*2}\hat{a}_{\boldsymbol{k},\lambda}^{\dagger}\hat{a}_{\boldsymbol{k},\lambda}^{\dagger}e^{i\omega_{k}(t+\tau)}\right\} \nonumber \\
 & =0,
\end{align}
and a similar calculation yields

\begin{equation}
\left\langle \tilde{\Gamma}_{2}(t)\tilde{\Gamma}_{2}(\tau)\right\rangle =0.
\end{equation}
Finally, the cross terms evaluate to

\begin{align}
\left\langle \tilde{\Gamma}_{1}(t)\tilde{\Gamma}_{2}(\tau)\right\rangle  & =\text{tr}_{R}\left\{ \hat{\rho}_{R}\tilde{\Gamma}^{\dagger}(t)\tilde{\Gamma}(\tau)\right\} \nonumber \\
 & =\text{tr}_{R}\left\{ \hat{\rho}_{R}\sum_{\boldsymbol{k},\lambda}\left|\kappa_{\boldsymbol{k},\lambda}\right|^{2}e^{i\omega_{k}(t-\tau)}\hat{n}_{\boldsymbol{k},\lambda}\right\} \nonumber \\
 & =\sum_{\boldsymbol{k},\lambda}\left|\kappa_{\boldsymbol{k},\lambda}\right|^{2}e^{i\omega_{k}(t-\tau)}\text{tr}_{R}\left\{ \hat{\rho}_{R}\hat{n}_{\boldsymbol{k},\lambda}\right\} \nonumber \\
 & =\sum_{\boldsymbol{k},\lambda}\left|\kappa_{\boldsymbol{k},\lambda}\right|^{2}e^{i\omega_{k}(t-\tau)}\bar{n}(\omega_{k},T),
\end{align}
and since $\hat{n}_{\boldsymbol{k},\lambda}^{\dagger}=\hat{n}_{\boldsymbol{k},\lambda}+1$,
we have

\begin{align}
\left\langle \tilde{\Gamma}_{1}(t)\tilde{\Gamma}_{2}(\tau)\right\rangle  & =\text{tr}_{R}\left\{ \hat{\rho}_{R}\tilde{\Gamma}^{\dagger}(t)\tilde{\Gamma}(\tau)\right\} \nonumber \\
 & =\text{tr}_{R}\left\{ \hat{\rho}_{R}\sum_{\boldsymbol{k},\lambda}\left|\kappa_{\boldsymbol{k},\lambda}\right|^{2}e^{-i\omega_{k}(t-\tau)}\hat{n}_{\boldsymbol{k},\lambda}^{\dagger}\right\} \nonumber \\
 & =\sum_{\boldsymbol{k},\lambda}\left|\kappa_{\boldsymbol{k},\lambda}\right|^{2}e^{-i\omega_{k}(t-\tau)}\text{tr}_{R}\left\{ \hat{\rho}_{R}\left(\hat{n}_{\boldsymbol{k},\lambda}+1\right)\right\} \nonumber \\
 & =\sum_{\boldsymbol{k},\lambda}\left|\kappa_{\boldsymbol{k},\lambda}\right|^{2}e^{-i\omega_{k}(t-\tau)}\left[\bar{n}(\omega_{k},T)+1\right],
\end{align}
where we have used in the last step that the density operator has
unity trace. In these considerations we have also used the mean photon
number we had previously derived for the E-M field in thermal equilibrium
at temperature $T$. Using these correlation functions we can write
the master equation as

\begin{align}
\frac{d\hat{\rho}_{S}^{I}(t)}{dt} & =-\int_{t_{0}}^{t}\left(\sigma_{-}\sigma_{-}\hat{\rho}_{S}^{I}(\tau)-\sigma_{-}\hat{\rho}_{S}^{I}(\tau)\sigma_{-}\right)\left\langle \tilde{\Gamma}_{1}(t)\tilde{\Gamma}_{1}(\tau)\right\rangle e^{-i\omega_{0}(t+\tau)}+\text{H.c.}\nonumber \\
 & +\left(\sigma_{+}\sigma_{+}\hat{\rho}_{S}^{I}(\tau)-\sigma_{+}\hat{\rho}_{S}^{I}(\tau)\sigma_{+}\right)\left\langle \tilde{\Gamma}_{2}(t)\tilde{\Gamma}_{2}(\tau)\right\rangle e^{i\omega_{0}(t+\tau)}+\text{H.c.}\nonumber \\
 & +\left(\sigma_{-}\sigma_{+}\hat{\rho}_{S}^{I}(\tau)-\sigma_{+}\hat{\rho}_{S}^{I}(\tau)\sigma_{-}\right)\left\langle \tilde{\Gamma}_{1}(t)\tilde{\Gamma}_{2}(\tau)\right\rangle e^{-i\omega_{0}(t-\tau)}+\text{H.c.}\nonumber \\
 & +\left(\sigma_{+}\sigma_{-}\hat{\rho}_{S}^{I}(\tau)-\sigma_{-}\hat{\rho}_{S}^{I}(\tau)\sigma_{+}\right)\left\langle \tilde{\Gamma}_{2}(t)\tilde{\Gamma}_{1}(\tau)\right\rangle e^{i\omega_{0}(t-\tau)}+\text{H.c.}\nonumber \\
 & d\tau.
\end{align}
We have seen that two of these terms vanish since the reservoir correlation
functions evaluate to 0. Additionally, for simplicity we take $t_{0}=0$.
Thus, the master equation reduces to

\begin{align}
\frac{d\hat{\rho}_{S}^{I}(t)}{dt} & =-\int_{0}^{t}\left(\sigma_{-}\sigma_{+}\hat{\rho}_{S}^{I}(\tau)-\sigma_{+}\hat{\rho}_{S}^{I}(\tau)\sigma_{-}\right)\left\langle \tilde{\Gamma}_{1}(t)\tilde{\Gamma}_{2}(\tau)\right\rangle e^{-i\omega_{0}(t-\tau)}+\text{H.c.}\nonumber \\
 & +\left(\sigma_{+}\sigma_{-}\hat{\rho}_{S}^{I}(\tau)-\sigma_{-}\hat{\rho}_{S}^{I}(\tau)\sigma_{+}\right)\left\langle \tilde{\Gamma}_{2}(t)\tilde{\Gamma}_{1}(\tau)\right\rangle e^{i\omega_{0}(t-\tau)}+\text{H.c.}\nonumber \\
 & d\tau.
\end{align}
We can also introduce the change of variables $\tau\to t-\tau$. Thus,
we write

\begin{align}
\frac{d\hat{\rho}_{S}^{I}(t)}{dt} & =-\frac{1}{\hbar^{2}}\int_{0}^{t}\left(\sigma_{-}\sigma_{+}\hat{\rho}_{S}^{I}(t-\tau)-\sigma_{+}\hat{\rho}_{S}^{I}(t-\tau)\sigma_{-}\right)\left\langle \tilde{\Gamma}_{1}(t)\tilde{\Gamma}_{2}(t-\tau)\right\rangle e^{-i\omega_{0}\tau}+\text{H.c.}\nonumber \\
 & +\left(\sigma_{+}\sigma_{-}\hat{\rho}_{S}^{I}(t-\tau)-\sigma_{-}\hat{\rho}_{S}^{I}(t-\tau)\sigma_{+}\right)\left\langle \tilde{\Gamma}_{2}(t)\tilde{\Gamma}_{1}(t-\tau)\right\rangle e^{i\omega_{0}\tau}+\text{H.c.}\ d\tau,
\end{align}
where again, $\text{H.c.}$ stands for the Hermitian conjugate and
finally we make use of the density of states $g(\boldsymbol{k})$
of equation (\ref{eq:DOS}) to evaluate the reservoir correlation
functions. We replace the sum over the modes with an integral over
the mode density

\begin{align}
\left\langle \tilde{\Gamma}_{1}(t)\tilde{\Gamma}_{2}(t-\tau)\right\rangle  & =\sum_{\lambda}\int\left|\kappa(\boldsymbol{k},\lambda)\right|^{2}e^{i\omega\tau}\bar{n}(\omega,T)g(\boldsymbol{k})d^{3}\boldsymbol{k},\label{eq:corrfunctionsint-1}\\
\left\langle \tilde{\Gamma}_{2}(t)\tilde{\Gamma}_{1}(t-\tau)\right\rangle  & =\sum_{\lambda}\int\left|\kappa(\boldsymbol{k},\lambda)\right|^{2}e^{-i\omega\tau}\left[\bar{n}(\omega,T)+1\right]g(\boldsymbol{k})d^{3}\boldsymbol{k}.\label{eq:corrfunctionsint-2}
\end{align}
We conclude this section of the appendix remarking that with the reservoir
correlation calculated, we may now substitute them into the Redfield
equation. This next section is devoted to finishing the calculation
and writing the master equation in Lindblad form.

\subsection{Markov approximations and writing the master equation in Lindblad
form\label{subsec:Markov-approximations-and}}

If we introduce the Markov approximation, through which we write $\hat{\rho}_{S}^{I}(t-\tau)\approx\hat{\rho}_{S}^{I}(t)$
the only dependence on $\tau$ comes from the correlation functions.
We also let $t$ go to infinity (second Markov Approximation), regime
in which we are able to evaluate via the Sokhotski-Plemelj identity
of appendix \ref{sec:The-Sokhotski-Plemelj-identity}

\begin{equation}
\mathcal{I}_{SP}(\omega)=\lim_{t\to\infty}\int_{0}^{t}e^{i\left(\omega-\omega_{0}\right)\tau}=\pi\delta(\omega-\omega_{0})+\frac{i\mathcal{P}}{\omega_{0}-\omega},
\end{equation}
where $\mathcal{P}$ is the Cauchy principal value. Thus we introduce
the definitions

\begin{equation}
\begin{cases}
\gamma_{A}\equiv2\pi\sum_{\lambda}\int d^{3}\boldsymbol{k}\ g(\boldsymbol{k})\left|\kappa(\boldsymbol{k},\lambda)\right|^{2}\delta(kc-\omega_{0}),\\
\Delta\equiv\sum_{\lambda}\mathcal{P}\int d^{3}\boldsymbol{k}\frac{\left|\kappa(\boldsymbol{k},\lambda)\right|^{2}g(\boldsymbol{k})}{\omega_{0}-kc},\\
\Delta'\equiv\sum_{\lambda}\mathcal{P}\int d^{3}\boldsymbol{k}\frac{\left|\kappa(\boldsymbol{k},\lambda)\right|^{2}g(\boldsymbol{k})}{\omega_{0}-kc}\bar{n}(kc,T).
\end{cases}
\end{equation}
As such, we have

\begin{align}
\lim_{t\to\infty}\int_{0}^{t}\left\langle \tilde{\Gamma}_{1}(t)\tilde{\Gamma}_{2}(t-\tau)\right\rangle e^{-i\omega_{0}\tau}d\tau & =\frac{\gamma_{A}}{2}\bar{n}(\omega_{0},T)+i\Delta,\\
\lim_{t\to\infty}\int_{0}^{t}\left\langle \tilde{\Gamma}_{2}(t)\tilde{\Gamma}_{1}(t-\tau)\right\rangle e^{i\omega_{0}\tau}d\tau & =\frac{\gamma_{A}}{2}\left[\bar{n}(\omega_{0},T)+1\right]+i(\Delta+\Delta').
\end{align}
These simplifications allow us to write the master equation as

\begin{align}
\frac{d\hat{\rho}_{S}^{I}(t)}{dt}= & \left(\sigma_{-}\hat{\rho}_{S}^{I}(t)\sigma_{+}-\sigma_{+}\sigma_{-}\hat{\rho}_{S}^{I}(t)\right)\left(\frac{\gamma_{A}}{2}\left[\bar{n}(\omega_{0},T)+1\right]+i(\Delta+\Delta')\right)\nonumber \\
 & +\left(\sigma_{+}\hat{\rho}_{S}^{I}(t)\sigma_{-}-\sigma_{-}\sigma_{+}\hat{\rho}_{S}^{I}(t)\right)\left(\frac{\gamma_{A}}{2}\bar{n}(\omega_{0},T)+i\Delta'\right)+h.c.,\label{eq:DynamicswithConstants-1}
\end{align}
but, as stated in the main text, we wish now to introduce a more symmetric
grouping of these terms, such that the underlying physics becomes
more apparent. Firstly we write out all the terms of equation (\ref{eq:DynamicswithConstants-1})
making the Hermitian conjugate explicit

\begin{align}
\frac{d\hat{\rho}_{S}^{I}(t)}{dt}= & \left(\sigma_{-}\hat{\rho}_{S}^{I}(t)\sigma_{+}-\sigma_{+}\sigma_{-}\hat{\rho}_{S}^{I}(t)\right)\left(\frac{\gamma_{A}}{2}\left[\bar{n}(\omega_{0},T)+1\right]+i(\Delta+\Delta')\right)\nonumber \\
 & +\left(\sigma_{+}\hat{\rho}_{S}^{I}(t)\sigma_{-}-\sigma_{-}\sigma_{+}\hat{\rho}_{S}^{I}(t)\right)\left(\frac{\gamma_{A}}{2}\bar{n}(\omega_{0},T)+i\Delta'\right)\nonumber \\
 & +\left(\sigma_{-}\hat{\rho}_{S}^{I}(t)\sigma_{+}-\hat{\rho}_{S}^{I}(t)\sigma_{+}\sigma_{-}\right)\left(\frac{\gamma_{A}}{2}\left[\bar{n}(\omega_{0},T)+1\right]-i(\Delta+\Delta')\right)\nonumber \\
 & +\left(\sigma_{+}\hat{\rho}_{S}^{I}(t)\sigma_{-}-\hat{\rho}_{S}^{I}(t)\sigma_{-}\sigma_{+}\right)\left(\frac{\gamma_{A}}{2}\bar{n}(\omega_{0},T)-i\Delta'\right).
\end{align}
We then group together common factors

\begin{align}
\frac{d\hat{\rho}_{S}^{I}(t)}{dt}= & \frac{\gamma_{A}}{2}\left[\bar{n}(\omega_{0},T)+1\right]\left(2\sigma_{-}\hat{\rho}_{S}^{I}(t)\sigma_{+}-\sigma_{+}\sigma_{-}\hat{\rho}_{S}^{I}(t)-\hat{\rho}_{S}^{I}(t)\sigma_{+}\sigma_{-}\right)\nonumber \\
 & -i(\Delta+\Delta')\left[\sigma_{+}\sigma_{-},\hat{\rho}_{S}^{I}(t)\right]\nonumber \\
 & +\frac{\gamma_{A}}{2}\bar{n}(\omega_{0},T)\left(2\sigma_{+}\hat{\rho}_{S}^{I}(t)\sigma_{-}-\sigma_{-}\sigma_{+}\hat{\rho}_{S}^{I}(t)-\hat{\rho}_{S}^{I}(t)\sigma_{-}\sigma_{+}\right)\nonumber \\
 & +i\Delta'\left[\sigma_{-}\sigma_{+},\hat{\rho}_{S}^{I}(t)\right].
\end{align}
Making use of the identities of equation (\ref{eq:Commutators+-}),
we write the previous equation as

\begin{align}
\frac{d\hat{\rho}_{S}^{I}(t)}{dt}= & \frac{\gamma_{A}}{2}\left[\bar{n}(\omega_{0},T)+1\right]\left(2\sigma_{-}\hat{\rho}_{S}^{I}(t)\sigma_{+}-\sigma_{+}\sigma_{-}\hat{\rho}_{S}^{I}(t)-\hat{\rho}_{S}^{I}(t)\sigma_{+}\sigma_{-}\right)\nonumber \\
 & -\frac{i}{2}(\Delta+\Delta')\left[\mathds{1}+\sigma_{z},\hat{\rho}_{S}^{I}(t)\right]\nonumber \\
 & +\frac{\gamma_{A}}{2}\bar{n}(\omega_{0},T)\left(2\sigma_{+}\hat{\rho}_{S}^{I}(t)\sigma_{-}-\sigma_{-}\sigma_{+}\hat{\rho}_{S}^{I}(t)-\hat{\rho}_{S}^{I}(t)\sigma_{-}\sigma_{+}\right)\nonumber \\
 & +\frac{i}{2}\Delta'\left[\mathds{1}-\sigma_{z},\hat{\rho}_{S}^{I}(t)\right],
\end{align}
which further reduces it to

\begin{align}
\frac{d\hat{\rho}_{S}^{I}(t)}{dt}= & -\frac{i}{2}(\Delta+2\Delta')\left[\sigma_{z},\hat{\rho}_{S}^{I}(t)\right]\nonumber \\
 & +\frac{\gamma_{A}}{2}\bar{n}(\omega_{0},T)\left(2\sigma_{+}\hat{\rho}_{S}^{I}(t)\sigma_{-}-\sigma_{-}\sigma_{+}\hat{\rho}_{S}^{I}(t)-\hat{\rho}_{S}^{I}(t)\sigma_{-}\sigma_{+}\right)\nonumber \\
 & +\frac{\gamma_{A}}{2}\left[\bar{n}(\omega_{0},T)+1\right]\left(2\sigma_{-}\hat{\rho}_{S}^{I}(t)\sigma_{+}-\sigma_{+}\sigma_{-}\hat{\rho}_{S}^{I}(t)-\hat{\rho}_{S}^{I}(t)\sigma_{+}\sigma_{-}\right).
\end{align}
The final step in the derivation of the master equation is to convert
back to the Schrodinger picture. We can use the following relation

\begin{align}
\frac{d\hat{\rho}_{S}^{I}(t)}{dt}= & i\hat{H}_{0}/\hbar e^{i\hat{H}_{0}t/\hbar}\hat{\rho}_{S}e^{-i\hat{H}_{0}t/\hbar}-e^{i\hat{H}_{0}t/\hbar}\hat{\rho}_{S}i\hat{H}_{0}/\hbar e^{-i\hat{H}_{0}t/\hbar}\nonumber \\
 & +e^{i\hat{H}_{0}t/\hbar}\frac{d\hat{\rho}_{S}}{dt}e^{-i\hat{H}_{0}t/\hbar}\nonumber \\
= & \frac{i}{\hbar}e^{i\hat{H}_{0}t/\hbar}\left[\hat{\rho}_{S},\hat{H}_{0}\right]e^{-i\hat{H}_{0}t/\hbar}+e^{i\hat{H}_{0}t/\hbar}\frac{d\hat{\rho}_{S}}{dt}e^{-i\hat{H}_{0}t/\hbar},
\end{align}
and invert it to write

\begin{equation}
\frac{d\hat{\rho}_{S}}{dt}=-\frac{i}{\hbar}\left[\hat{\rho}_{S},\hat{H}_{0}\right]+e^{-i\hat{H}_{0}t/\hbar}\frac{d\hat{\rho}_{S}^{I}(t)}{dt}e^{i\hat{H}_{0}t/\hbar}.
\end{equation}
From here, we can use the substitute the master equation and obtain

\begin{align}
\frac{d\hat{\rho}_{S}}{dt} & =-\frac{i}{\hbar}\left[\hat{\rho}_{S},\hat{H}_{0}\right]\nonumber \\
 & -\frac{i}{2}(\Delta+2\Delta')\left[\sigma_{z},e^{-i\hat{H}_{0}t/\hbar}\hat{\rho}_{S}^{I}(t)e^{i\hat{H}_{0}t/\hbar}\right]\nonumber \\
 & +\frac{\gamma_{A}}{2}\bar{n}(\omega_{0},T)e^{-i\hat{H}_{0}t/\hbar}\left(2\sigma_{+}\hat{\rho}_{S}^{I}(t)\sigma_{-}-\sigma_{-}\sigma_{+}\hat{\rho}_{S}^{I}(t)-\hat{\rho}_{S}^{I}(t)\sigma_{-}\sigma_{+}\right)e^{i\hat{H}_{0}t/\hbar}\nonumber \\
 & +\frac{\gamma_{A}}{2}\left[\bar{n}(\omega_{0},T)+1\right]e^{-i\hat{H}_{0}t/\hbar}\left(2\sigma_{-}\hat{\rho}_{S}^{I}(t)\sigma_{+}-\sigma_{+}\sigma_{-}\hat{\rho}_{S}^{I}(t)-\hat{\rho}_{S}^{I}(t)\sigma_{+}\sigma_{-}\right)e^{i\hat{H}_{0}t/\hbar}.
\end{align}
We rearrange the previous expression using the commutator relations
of $\sigma_{z}$ and $\sigma_{\pm}$ to bring the exponentials to
either side of the density operator in the interaction picture. The
contributions from the commutators arising from rearranging the operator
products cancel each other out and thus, we are left with

\begin{align}
\frac{d\hat{\rho}_{S}}{dt}= & -\frac{i}{2}\omega_{0}'\left[\sigma_{z},\hat{\rho}_{S}(t)\right]\nonumber \\
 & +\frac{\gamma_{A}}{2}\bar{n}(\omega_{0},T)\left(2\sigma_{+}\hat{\rho}_{S}(t)\sigma_{-}-\sigma_{-}\sigma_{+}\hat{\rho}_{S}(t)-\hat{\rho}_{S}(t)\sigma_{-}\sigma_{+}\right)\nonumber \\
 & +\frac{\gamma_{A}}{2}\left[\bar{n}(\omega_{0},T)+1\right]\left(2\sigma_{-}\hat{\rho}_{S}^{I}(t)\sigma_{+}-\sigma_{+}\sigma_{-}\hat{\rho}_{S}^{I}(t)-\hat{\rho}_{S}^{I}(t)\sigma_{+}\sigma_{-}\right),
\end{align}
where $\omega_{0}'=\omega+\Delta+2\Delta'$. We note that we can define

\begin{equation}
\begin{cases}
\hat{L}_{1}=\sqrt{\gamma_{A}\bar{n}(\omega_{0},T)}\sigma_{+},\\
\hat{L}_{2}=\sqrt{\gamma_{A}\left[\bar{n}(\omega_{0},T)+1\right]}\sigma_{-}.
\end{cases}
\end{equation}
With these operators, we have

\begin{align}
\frac{d\hat{\rho}_{S}(t)}{dt}= & -\frac{i}{2}\omega_{0}'\left[\sigma_{z},\hat{\rho}_{S}(t)\right]+\hat{L}_{1}\hat{\rho}_{S}(t)\hat{L}_{1}^{\text{\ensuremath{\dagger}}}-\frac{1}{2}\left(\hat{L}_{1}\hat{L}_{1}^{\dagger}\hat{\rho}_{S}(t)-\hat{\rho}_{S}(t)\hat{L}_{1}\hat{L}_{1}^{\dagger}\right)\nonumber \\
 & +\hat{L}_{2}\hat{\rho}_{S}(t)\hat{L}_{2}^{\dagger}-\frac{1}{2}\left(\hat{L}_{2}^{\dagger}\hat{L}_{2}\hat{\rho}_{S}(t)-\hat{\rho}_{S}(t)\hat{L}_{2}^{\dagger}\hat{L}_{2}\right),
\end{align}
and with some further rewriting, we are able to bring the master equation
into the Lindblad form presented in the main text.

\subsection{Calculation of the Einstein A coefficient\label{subsec:Calculation-of-the}}

To provide a check on the open quantum systems approach, we calculate
the coefficient $\gamma_{A}$ which in fact corresponds to the Einstein
A coefficient as it gives the rate for spontaneous emission of photons.
It was originally defined as

\begin{equation}
\gamma_{A}\equiv2\pi\sum_{\lambda}\int d^{3}\boldsymbol{k}\ g(\boldsymbol{k})\left|\kappa(\boldsymbol{k},\lambda)\right|^{2}\delta(\omega-\omega_{0}).
\end{equation}
The density of states in a single direction is given in equation (\ref{eq:DOS})
and the coupling constants in equation (\ref{eq:Coupling-1}). We
thus have to solve

\begin{align}
\gamma_{A} & =-2\pi\sum_{\lambda}\int_{0}^{\infty}d\omega\frac{\omega^{2}}{8\pi^{3}c^{3}}V\int_{0}^{\pi}\sin\theta d\theta\int_{0}^{2\pi}d\phi\ e^{2i\boldsymbol{k}\cdot\boldsymbol{r}}\frac{\omega}{2\hbar\varepsilon_{0}V}\left(\boldsymbol{e}_{\boldsymbol{k},\lambda}\cdot\boldsymbol{d}_{eg}\right)^{2}\delta(\omega-\omega_{0})\nonumber \\
 & =\frac{\omega_{0}^{3}}{8\pi^{2}\hbar\varepsilon_{0}c^{3}}\sum_{\lambda}\int_{0}^{\pi}\sin\theta d\theta\int_{0}^{2\pi}d\phi\left(\boldsymbol{e}_{\boldsymbol{k},\lambda}\cdot\boldsymbol{d}_{eg}\right)^{2},
\end{align}
where we have considered the atom to be located at the origin and
thus $\boldsymbol{r}=0$. To evaluate the angular integral we chose
the first polarization component to be such that $\boldsymbol{d}_{eg}\cdot\boldsymbol{e}_{\boldsymbol{k},\lambda_{1}}=0$.
The second polarization component will make an angle $\alpha$ with
$\boldsymbol{k}$. Thus

\begin{equation}
\left(\boldsymbol{e}_{\boldsymbol{k},\lambda_{2}}\cdot\boldsymbol{d}_{eg}\right)^{2}=d_{eg}^{2}(1-\cos^{2}\alpha)=d_{eg}^{2}\left(1-\left(\hat{d}_{eg}\cdot\hat{k}\right)^{2}\right),
\end{equation}
if we choose $\hat{k}_{z}$ such that it is aligned with $\hat{d}_{eg}$
we have

\begin{equation}
\left(\boldsymbol{e}_{\boldsymbol{k},\lambda_{2}}\cdot\boldsymbol{d}_{eg}\right)^{2}=d_{eg}^{2}\left(1-\cos^{2}\theta\right),
\end{equation}
and thus

\begin{equation}
\gamma_{A}=\frac{\omega_{0}^{3}}{4\pi\hbar\varepsilon_{0}c^{3}}\int_{0}^{\pi}d\theta d_{eg}^{2}\sin\theta\left(1-\cos^{2}\theta\right),
\end{equation}
which can be easily shown to yield

\begin{equation}
\gamma_{A}=\frac{\omega_{0}^{3}}{4\pi\hbar\varepsilon_{0}c^{3}}\frac{4}{3}d_{eg}^{2}=\frac{1}{4\pi\varepsilon_{0}}\frac{4\omega_{0}^{3}d_{eg}^{2}}{3\hbar c^{3}},
\end{equation}
which is the correct result for the Einstein A coefficient and gives
the spontaneous decay rate for the two-level system.

\section{Quantization of the electromagnetic field in free space\label{sec:Quantization-of-the-EM-Free}}

In this appendix we quantize the E-M field in free space in a macroscopic
manner. We follow the most common approach, which can for instance
be found in \citet{Milonni2019}. Our starting point for this section
is the Maxwell's equations in the absence of currents and charges.
These read 
\begin{equation}
\left\{ \begin{array}{ll}
\nabla\cdot\boldsymbol{E}=0, & \nabla\times\boldsymbol{E}=-\frac{\partial B}{\partial t},\\
\nabla\cdot\boldsymbol{B}=0, & \nabla\times\boldsymbol{B}=\frac{1}{c^{2}}\frac{\partial E}{\partial t}.
\end{array}\right.
\end{equation}
These can be rewritten in terms of the scalar potential $\phi$ and
vector potential $\boldsymbol{A}$. In particular, if we work in the
coulomb gauge, we have that $\nabla\cdot\boldsymbol{A}=0$. In this
gauge the vector potential contains all the information about the
magnetic and electric fields, since

\begin{equation}
\left\{ \begin{array}{l}
\boldsymbol{B}=\nabla\times\boldsymbol{A}(\boldsymbol{r},t),\\
\boldsymbol{E}=-\frac{\partial\boldsymbol{A}(\boldsymbol{r},t)}{\partial t}.
\end{array}\right.\label{eq:FieldsfromA}
\end{equation}
Calculating the time evolution for the vector potential therefore
allows us to obtain the time evolution of the E-M fields. This is
done with the use of Ampère-Maxwell law, which reads 
\begin{equation}
\nabla\times\boldsymbol{B}=\frac{1}{c^{2}}\frac{\partial\boldsymbol{E}}{\partial t}.
\end{equation}
If we substitute here the fields by the vector potential it is easy
to arrive at 
\begin{equation}
\nabla\times(\nabla\times\boldsymbol{A})=-\frac{1}{c^{2}}\frac{\partial^{2}}{\partial t^{2}}\boldsymbol{A},
\end{equation}
which we can simplify a lot using the identity$\nabla\times(\nabla\times\boldsymbol{A})=\nabla(\nabla\cdot\boldsymbol{A})-\nabla^{2}\boldsymbol{A}$.
Since we are working in the Coulomb gauge the first term will be $0,$
and therefore we are left with 
\begin{equation}
\nabla^{2}\boldsymbol{A}-\frac{1}{c^{2}}\frac{\partial^{2}}{\partial t^{2}}\boldsymbol{A}=0.
\end{equation}
This is just a wave equation for the vector potential. Therefore the
dynamics of the E-M field in a vacuum can be obtained by solving a
wave equation for the vector potential. The solutions of the equation
may for instance be simply plane waves of the form 
\begin{equation}
\boldsymbol{A}_{\boldsymbol{k},\lambda}=\boldsymbol{e}_{\boldsymbol{k},\lambda}A_{\boldsymbol{k},\lambda}e^{i\left(\boldsymbol{k}\cdot\boldsymbol{r}-\omega_{k}t\right)},
\end{equation}
where in this equation $\boldsymbol{e}_{\boldsymbol{k}}$ is the polarization
vector, $A_{\boldsymbol{k}}$ a complex amplitude, $\boldsymbol{k}$
is the wave-vector, $\lambda$ is the polarization state and $\omega_{k}=ck=c|\boldsymbol{k}|$
the frequency associated with the wave. If we assume that we are solving
the wave equation over the whole space then $k$ is unrestricted and
can have any value. If however, for simplicity we confine the field
in a finite volume $V$ or impose periodic boundary conditions, then
$\boldsymbol{k}$ will only be able to have certain values. If we
consider for instance a box of side $L$ such that $V=L^{3},$ then
in the left and right as well as top and bottom and front and back
of the box the phase factors must coincide. This of course implies
that only discrete wave vectors are allowed, namely those such that
\begin{equation}
k_{x}=\frac{2\pi}{L}n_{x},\quad k_{y}=\frac{2\pi}{L}n_{y},\quad k_{z}=\frac{2\pi}{L}n_{z}.\label{eq:QuantizedWaveVec}
\end{equation}
These discrete wave-vectors together with their polarizations constitute
the allowed modes of the field. In terms of these modes we can express
a general field through Fourier decomposition. This amounts to summing
over the modes and their complex conjugates such that the electromagnetic
vector potential is real 
\begin{equation}
\boldsymbol{A}(\boldsymbol{r},t)=\sum_{\boldsymbol{k},\lambda}\boldsymbol{e}_{\boldsymbol{k},\lambda}\left(A_{\boldsymbol{k},\lambda}e^{i\left(\boldsymbol{k}\cdot\boldsymbol{r}-\omega_{k}t\right)}+A_{\boldsymbol{k},\lambda}^{*}e^{-i\left(\boldsymbol{k}\cdot\boldsymbol{r}-\omega_{k}t\right)}\right).
\end{equation}
If we want to derive the electric field from this vector potential
we simply do as in equation (\ref{eq:FieldsfromA}) and take the time
derivative. We arrive at 
\begin{equation}
\boldsymbol{E}(\boldsymbol{r},t)=-\frac{\partial}{\partial t}\boldsymbol{A}(\boldsymbol{r},t)=\sum_{\boldsymbol{k},\lambda}\boldsymbol{e}_{\boldsymbol{k},\lambda}i\omega_{k}\left(A_{\boldsymbol{k},\lambda}e^{i\left(\boldsymbol{k}\cdot\boldsymbol{r}-\omega_{k}t\right)}-A_{\boldsymbol{k},\lambda}^{*}e^{-i\left(\boldsymbol{k}\cdot\boldsymbol{r}-\omega_{k}t\right)}\right).
\end{equation}
Similarly if we take the curl of $\boldsymbol{A}$ we might write
the magnetic field as 
\begin{equation}
\boldsymbol{B}(\boldsymbol{r},t)=\sum_{\boldsymbol{k},\lambda}i\left(\boldsymbol{k}\times\boldsymbol{e}_{\boldsymbol{k},\lambda}\right)\left[A_{\boldsymbol{k},\lambda}e^{i\left(\boldsymbol{k}\cdot\boldsymbol{r}-\omega_{k}t\right)}-A_{\boldsymbol{k},\lambda}^{*}e^{-i\left(\boldsymbol{k}\cdot\boldsymbol{r}-\omega_{k}t\right)}\right].
\end{equation}
Since we have the expression for the E-M fields, we may now calculate
the energy stored in the volume $V$. Classically, this is simply
the integral over the energy density 
\begin{equation}
H=\frac{1}{2}\int_{V}\left[\varepsilon_{0}\boldsymbol{E}(\boldsymbol{r},t)^{2}+\frac{1}{\mu_{0}}\boldsymbol{B}(\boldsymbol{r},t)^{2}\right]d^{3}\boldsymbol{r},
\end{equation}
where we have used the suggestive notation $H$ for the energy. Note
that this can be simplified to 
\begin{equation}
H=\frac{\varepsilon_{0}}{2}\int_{V}\left[\boldsymbol{E}(\boldsymbol{r},t)^{2}+c^{2}\boldsymbol{B}(\boldsymbol{r},t)^{2}\right]d^{3}\boldsymbol{r}.
\end{equation}
We first calculate the magnetic contribution to the energy 
\begin{equation}
\begin{aligned}\int_{V}c^{2}\boldsymbol{B}^{2}d^{3}\boldsymbol{r}= & -c^{2}\int_{V}\sum_{k,\lambda,k^{\prime},\lambda^{\prime}}\left(\boldsymbol{k}\times\boldsymbol{e}_{\boldsymbol{k},\lambda}\right)\cdot\left(\boldsymbol{k}^{\prime}\times\boldsymbol{e}_{\boldsymbol{k}^{\prime},\lambda^{\prime}}\right)\left[A_{\boldsymbol{k},\lambda^{\prime}}A_{\boldsymbol{k}^{\prime}\lambda^{\prime}}e^{i\left(\boldsymbol{k}+\boldsymbol{k}^{\prime}\right)\cdot\boldsymbol{r}-2i\omega_{k}t}\right.\\
 & +A_{\boldsymbol{k},\lambda}^{*}A_{\boldsymbol{k}^{\prime},\lambda^{\prime}}^{*}e^{-i\left(\boldsymbol{k}+\boldsymbol{k}^{\prime}\right)\cdot\boldsymbol{r}+2i\omega_{k}t}\\
 & -A_{\boldsymbol{k},\lambda}A_{\boldsymbol{k}^{\prime},\lambda^{\prime}}^{*}e^{i\left(\boldsymbol{k}-\boldsymbol{k}^{\prime}\right)\cdot\boldsymbol{r}}\\
 & \left.-A_{\boldsymbol{k},\lambda}^{*}A_{\boldsymbol{k}^{\prime},\lambda^{\prime}}e^{-i\left(\boldsymbol{k}-\boldsymbol{k}^{\prime}\right)\cdot\boldsymbol{r}}\right]d^{3}\boldsymbol{r},
\end{aligned}
\end{equation}
where noting that 
\begin{equation}
\int_{V}e^{i\left(\boldsymbol{k}-\boldsymbol{k}^{\prime}\right)\cdot\boldsymbol{r}}d^{3}\boldsymbol{r}=V\delta_{\boldsymbol{k}\boldsymbol{k}^{\prime}},
\end{equation}
the whole expression simplifies and using that $\omega_{k}=ck$ we
have 
\begin{equation}
\int_{V}c^{2}\boldsymbol{B}^{2}d^{3}\boldsymbol{r}=\sum_{\boldsymbol{k},\lambda}\omega_{k}^{2}V\left[A_{\boldsymbol{k},\lambda}A_{\boldsymbol{k},\lambda}^{*}+A_{\boldsymbol{k},\lambda}^{*}A_{\boldsymbol{k},\lambda}\right].
\end{equation}
Doing the exact same calculation for the electric contribution, we
have 
\begin{equation}
\int_{V}\boldsymbol{E}^{2}d^{3}\boldsymbol{r}=\sum_{\boldsymbol{k},\lambda}\omega_{k}^{2}V\left[A_{\boldsymbol{k},\lambda}A_{\boldsymbol{k},\lambda}^{*}+A_{\boldsymbol{k},\lambda}^{*}A_{\boldsymbol{k},\lambda}\right],
\end{equation}
and thus, the total energy is

\begin{equation}
H=\sum_{\boldsymbol{k}}\varepsilon_{0}V\omega_{k}^{2}\left[A_{\boldsymbol{k},\lambda}A_{\boldsymbol{k},\lambda}^{*}+A_{\boldsymbol{k},\lambda}^{*}A_{\boldsymbol{k},\lambda}\right].\label{eq:ClassicalHamiltonian}
\end{equation}
Which corresponds to a sum of the energy over the several modes of
the system. Note that we did not commute and add together the complex
numbers $A_{k}^{*}A_{k}$ precisely because this brings out a fortuitous
comparison that we can make with the energy levels of a harmonic oscillator.
This gives us a hint toward the following considerations we can make
for the quantization of the E-M fields. The fundamental idea is that
to each mode of the radiation field, we associate a harmonic oscillator.
The degree of excitation of this harmonic oscillator gives us the
number of photons that occupy a mode. Thus, the term photon acquires
a precise definition in this context. A photon is an excitation of
the harmonic oscillator associated with a mode of the E-M fields.
The Hamiltonian for a harmonic oscillator is 
\begin{equation}
\hat{H}_{k}=\frac{1}{2}\hbar\omega_{k}\left(\hat{a}_{\boldsymbol{k},\lambda}\hat{a}_{\boldsymbol{k},\lambda}^{\dagger}+\hat{a}_{\boldsymbol{k},\lambda}^{\dagger}\hat{a}_{\boldsymbol{k},\lambda}\right).
\end{equation}
If we associate one such Hamiltonian to a mode, then the total Hamiltonian
will of course be the sum over the modes of each individual Hamiltonian
\begin{equation}
\hat{H}=\sum_{\boldsymbol{k}}\hat{H}_{\boldsymbol{k}}=\sum_{\boldsymbol{k}}\frac{1}{2}\hbar\omega_{k}\left(\hat{a}_{\boldsymbol{k},\lambda}\hat{a}_{\boldsymbol{k},\lambda}^{\dagger}+\hat{a}_{\boldsymbol{k},\lambda}^{\dagger}\hat{a}_{\boldsymbol{k},\lambda}\right).\label{eq:QuantumHamiltonian}
\end{equation}
Note the striking similarity with the equation (\ref{eq:ClassicalHamiltonian})
for the classical energy stored in the E-M field in the volume $V$.
We can in fact make a simple correspondence between equations (\ref{eq:QuantumHamiltonian})
and (\ref{eq:ClassicalHamiltonian}) by introducing the quantization
rules 
\begin{equation}
\left\{ \begin{array}{l}
A_{\boldsymbol{k},\lambda}\rightarrow\sqrt{\frac{\hbar}{2\varepsilon_{0}V\omega_{k}}}\hat{a}_{\boldsymbol{k},\lambda},\\
A_{\boldsymbol{k},\lambda}^{*}\rightarrow\sqrt{\frac{\hbar}{2\varepsilon_{0}V\omega_{k}}}\hat{a}_{\boldsymbol{k},\lambda}^{\dagger}.
\end{array}\right.
\end{equation}
For notation's sake we compress both indices $\boldsymbol{k}$ and
$\lambda$ into a single $\boldsymbol{k}$. Thus, for each mode, the
creation operator$\hat{a}_{\boldsymbol{k}}^{\dagger}$ acts by creating
a photon in this mode, while the destruction operator acts by removing
a photon from the mode

\begin{align}
\hat{a}_{\boldsymbol{k}}^{\dagger}\left|n_{\boldsymbol{k}}\right\rangle  & =\sqrt{n_{\boldsymbol{k}}+1}\left|n_{\boldsymbol{k}}+1\right\rangle ,\\
\hat{a}_{\boldsymbol{k}}\left|n_{\boldsymbol{k}}\right\rangle  & =\sqrt{n_{\boldsymbol{k}}}\left|n_{\boldsymbol{k}}-1\right\rangle .
\end{align}

Also, it is possible to measure how many photons occupy each mode
by acting with the number operator $\hat{n}_{\boldsymbol{k}}=\hat{a}_{\boldsymbol{k}}^{\dagger}\hat{a}_{\boldsymbol{k}},$
since 
\begin{equation}
\hat{n}_{\boldsymbol{k}}\left|n_{\boldsymbol{k}}\right\rangle =n_{\boldsymbol{k}}\left|n_{\boldsymbol{k}}\right\rangle .
\end{equation}
These states $\left|n_{k}\right\rangle $ with a precisely defined
photon number are called Fock states but we may also consider states
which are superpositions of these states an thus do not have a defined
number of photons.

Since we have determined the quantization rules which give us the
creation and annihilation operators for the radiation field, we may
now construct the operator which corresponds to the vector potential
\begin{equation}
\hat{\boldsymbol{A}}_{\boldsymbol{k}}(\boldsymbol{r},t)=\boldsymbol{e}_{k}\sqrt{\frac{\hbar}{2\varepsilon_{0}V\omega_{k}}}\left[\hat{a}_{\boldsymbol{k}}e^{i\left(\boldsymbol{k}\cdot\boldsymbol{r}-\omega_{k}t\right)}+\hat{a}_{\boldsymbol{k}}^{\dagger}e^{-i\left(\boldsymbol{k}\cdot\boldsymbol{r}-\omega_{k}t\right)}\right].
\end{equation}
From this operator we may also construct the field operators that
correspond to the electric and magnetic field by using equations (\ref{eq:FieldsfromA}).
We have for a single mode 
\begin{equation}
\left\{ \begin{array}{l}
\hat{\boldsymbol{E}}_{\boldsymbol{k}}(\boldsymbol{r},t)=-\frac{\partial}{\partial t}\hat{\boldsymbol{A}}(\boldsymbol{r},t)=i\boldsymbol{e}_{\boldsymbol{k}}\sqrt{\frac{\hbar\omega_{k}}{2\varepsilon_{0}V}}\left[\hat{a}_{\boldsymbol{k}}e^{i\left(\boldsymbol{k}\cdot\boldsymbol{r}-\omega_{k}t\right)}-\hat{a}_{\boldsymbol{k}}^{\dagger}e^{-i\left(\boldsymbol{k}\cdot\boldsymbol{r}-\omega_{k}t\right)}\right],\\
\hat{\boldsymbol{B}}_{\boldsymbol{k}}(\boldsymbol{r},t)=\nabla\times\hat{\boldsymbol{A}}(\boldsymbol{r},t)=i\left(\boldsymbol{k}\times\boldsymbol{e}_{\boldsymbol{k}}\right)\sqrt{\frac{\hbar}{2\varepsilon_{0}V\omega_{k}}}\left[\hat{a}_{\boldsymbol{k}}e^{i\left(\boldsymbol{k}\cdot\boldsymbol{r}-\omega_{k}t\right)}-\hat{a}_{\boldsymbol{k}}^{\dagger}e^{-i\left(\boldsymbol{k}\cdot\boldsymbol{r}-\omega_{k}t\right)}\right].
\end{array}\right.
\end{equation}
Of course, the total operators are given by a sum over the modes 
\begin{equation}
\hat{\boldsymbol{E}}=\sum_{\boldsymbol{k}}\boldsymbol{E}_{\boldsymbol{k}},
\end{equation}
and note also, that quite generally, these field operators can be
decomposed into parts which oscillate with positive and negative frequencies
as 
\begin{equation}
\hat{\boldsymbol{E}}(\boldsymbol{r},t)=\hat{\boldsymbol{E}}^{+}(\boldsymbol{r},t)+\hat{\boldsymbol{E}}^{-}(\boldsymbol{r},t).
\end{equation}
Using the quantization rules, one has that the photon Hamiltonian
is given by 
\begin{equation}
\hat{H}=\sum_{\boldsymbol{k}}\hbar\omega_{k}\left(\hat{a}_{\boldsymbol{k}}^{\dagger}\hat{a}_{\boldsymbol{k}}+1/2\right),
\end{equation}
which acts on a general Fock state with many modes (multi-mode Fock
state) as

\begin{equation}
\hat{H}\left|n_{\boldsymbol{k}_{1}},n_{\boldsymbol{k}_{2}},\ldots\right\rangle =\left[\hbar\omega_{\boldsymbol{k}_{1}}\left(n_{\boldsymbol{k}_{1}}+1/2\right)+\hbar\omega_{\boldsymbol{k}_{2}}\left(n_{\boldsymbol{k}_{2}}+1/2\right)+\cdots\right]\left|n_{\boldsymbol{k}_{1}},n_{\boldsymbol{k}_{2}},\ldots\right\rangle 
\end{equation}
In general we shall, in the main text and following appendices work
with thermal states, which are indeed multi-mode Fock states with
a probability distribution given by the canonical distribution.

\section{Two-level systems\label{sec:A.-Two-level-systems} }

In this appendix we go over a few useful properties of a two-level
system (two-state atom or qubit), with states $|e\rangle,|g\rangle$
characterized by energies $E_{e}$ and $E_{g}$ respectively. The
energy $E_{e}$ corresponds to the excited state while $E_{g}$ corresponds
to the ground-state. Of course the states are labeled such that $E_{e}>E_{g}$.
We will make use of the Pauli matrices as in particular, the system
Hamiltonian for a two-level system is of course 
\begin{equation}
H_{S}=E_{e}|e\rangle\langle e|+E_{g}|g\rangle\langle g|,
\end{equation}
which can be written in a more convenient form as 
\begin{equation}
H_{S}=\frac{1}{2}\left(E_{e}+E_{g}\right)+\frac{1}{2}\left(E_{e}-E_{g}\right)\sigma_{z},
\end{equation}
where the Pauli matrix $\sigma_{z}$ corresponds to 
\begin{equation}
\sigma_{z}=|e\rangle\langle e|-|g\rangle\langle g|.
\end{equation}
Choosing the energy reference so that the zero is in the middle of
the energies $E_{e}$ and $E_{g}.$ In this case the Hamiltonian is
just 
\begin{equation}
H_{S}=\frac{1}{2}\left(E_{e}-E_{g}\right)\sigma_{z},
\end{equation}
which we can write as 
\begin{equation}
H_{S}=\frac{1}{2}\hbar\omega_{0}\sigma_{z}.
\end{equation}
We may want to describe the light-matter interaction for this two-level
system. To do so we need to write the dipole operator $q\boldsymbol{r}$
conveniently, and as such, if we introduce the frequency $\omega_{0}=\left(E_{e}-E_{g}\right)/\hbar$,
we note that it can be expanded in terms of a complete basis of the
system's Hilbert space $\mathcal{H}_{S}$ 
\begin{equation}
q\hat{\boldsymbol{r}}=q\sum_{ij}\langle i|\hat{\boldsymbol{r}}|j\rangle|i\rangle\langle j|.
\end{equation}
In the case of the two-level system this reduces to 
\begin{equation}
q\hat{\boldsymbol{r}}=q(\langle e|\hat{\boldsymbol{r}}|g\rangle|e\rangle\langle g|+\langle g|\hat{\boldsymbol{r}}|e\rangle|g\rangle\langle e|),
\end{equation}
and thus can define the dipole operators 
\begin{equation}
\left\{ \begin{array}{l}
\boldsymbol{d}_{ge}=q\langle g|\hat{\boldsymbol{r}}|e\rangle,\\
\boldsymbol{d}_{eg}=q\langle e|\hat{\boldsymbol{r}}|g\rangle=\boldsymbol{d}_{ge}^{*}.
\end{array}\right.
\end{equation}
With this notation we write 
\begin{equation}
q\hat{\boldsymbol{r}}=\boldsymbol{d}_{ge}\sigma_{-}+\boldsymbol{d}_{ge}^{*}\sigma_{+}.
\end{equation}
It may be also worth noting that the expectation value of the dipole
operator has a particular simple realization in the density operator
formalism. It is given by

\begin{equation}
\begin{aligned}\langle q\hat{r}\rangle & =\text{tr}\{\hat{\rho}q\hat{\boldsymbol{r}}\}=\boldsymbol{d}_{ge}\text{tr}\left\{ \hat{\rho}\sigma_{-}\right\} +\boldsymbol{d}_{ge}^{*}\text{tr}\left\{ \hat{\rho}\sigma_{+}\right\} \\
 & =\boldsymbol{d}_{ge}\langle e|\hat{\rho}|g\rangle+\boldsymbol{d}_{ge}^{*}\langle g|\hat{\rho}|e\rangle\\
 & =\boldsymbol{d}_{ge}\rho_{eg}+\boldsymbol{d}_{ge}^{*}\rho_{ge}.
\end{aligned}
\end{equation}
Of use in our considerations are also the identities

\begin{equation}
\left[\sigma_{+},\sigma_{-}\right]=\sigma_{z},
\end{equation}

\begin{equation}
\left[\sigma_{\pm},\sigma_{z}\right]=\mp2\sigma_{\pm},
\end{equation}

\begin{equation}
\sigma_{\pm}\sigma_{\mp}=\frac{1}{2}\left(\mathds{1}\pm\sigma_{z}\right).\label{eq:Commutators+-}
\end{equation}
These will be used throughout the text very often, for instance the
last equation (\ref{eq:Commutators+-}) represents in the case of
a two-level system, a freedom in the choice of the zero of the energy.
If we choose the zero to be exactly in the middle between the excited
and ground states, the Hamiltonian will read $H=\hbar\omega_{0}\sigma_{z}/2$,
where $\hbar\omega_{0}$ is the energy gap between excited and ground
state. If we make the zero of the energy coincide with the ground
state, then we can write $H=\hbar\omega_{0}\sigma_{+}\sigma_{-}/2$.
Both these representations for two-level atoms are used throughout
the paper, and equation (\ref{eq:Commutators+-}) is useful as a transition
between them.

\section{The Sokhotski-Plemelj identity\label{sec:The-Sokhotski-Plemelj-identity}}

This appendix is dedicated to obtaining a result that is used again
and again throughout the paper. It is sometimes referred to as the
Sokhotski-Plemelj theorem or identity, and states that

\begin{align}
\lim_{t\to\infty}\int_{0}^{t}dt_{1}e^{-i(\omega_{0}-\omega)\left(t_{1}-t\right)} & =\pi\delta(\omega-\omega_{0})-i\mathcal{P}\frac{1}{\omega-\omega_{0}}.
\end{align}
For completeness, we prove this identity here, noting again that we
will often employ it. We start by performing a change of variables
$\tau=t-t_{1}$ and write

\begin{align}
\lim_{t\to\infty}\int_{0}^{t}dt_{1}e^{-i(\omega_{0}-\omega)\left(t_{1}-t\right)} & =-\lim_{t\to\infty}\int_{t}^{0}d\tau e^{i(\omega_{0}-\omega)\tau}\nonumber \\
 & =\lim_{t\to\infty}\int_{0}^{t}d\tau e^{i(\omega_{0}-\omega)\tau}\nonumber \\
 & =\int_{0}^{\infty}d\tau e^{i(\omega_{0}-\omega)\tau}.
\end{align}
This integral can be regularized by adding a small imaginary part
$-i\varepsilon$ to the frequency $\omega$, and solving the integral
before letting it go to 0.

\begin{align}
\lim_{\varepsilon\to0^{+}}\int_{0}^{t}d\tau e^{i(\omega_{0}-\omega+i\varepsilon)\tau} & =\lim_{\varepsilon\to0^{+}}\frac{-1}{i(\omega_{0}-\omega+i\varepsilon)}\nonumber \\
 & =\frac{i}{\omega_{0}-\omega+i\eta},
\end{align}
where $\eta=0^{+}$ is written so as to omit the limit. This form
of writing the integral is often utilized, but note that it is merely
a formal expression. It is in fact a distribution, and makes sense
only inside an integral. We can manipulate it as

\begin{align}
\frac{i}{\omega_{0}-\omega+i\eta} & =i\frac{1}{\omega_{0}-\omega+i\eta}\nonumber \\
 & =i\left(\frac{\omega_{0}-\omega}{(\omega_{0}-\omega)^{2}+\eta^{2}}-\frac{i\eta}{(\omega_{0}-\omega)^{2}+\eta^{2}}\right)\nonumber \\
 & =\pi\frac{\eta}{\pi\left[(\omega_{0}-\omega)^{2}+\eta^{2}\right]}+i\frac{\omega_{0}-\omega}{(\omega_{0}-\omega)^{2}+\eta^{2}}.
\end{align}
We now can calculate the following quantity

\begin{align}
\int_{-\infty}^{\infty}d\omega f(\omega)\frac{i}{\omega_{0}-\omega+i\eta} & =\pi\int_{-\infty}^{\infty}d\omega\frac{\eta}{\pi\left[(\omega_{0}-\omega)^{2}+\eta^{2}\right]}f(\omega)+i\int_{-\infty}^{\infty}d\omega\frac{\left(\omega_{0}-\omega\right)^{2}}{(\omega_{0}-\omega)^{2}+\eta^{2}}\frac{f(\omega)}{\omega_{0}-\omega}.
\end{align}
For the first integral we find, when $\omega\to\omega_{0}$ it approaches
$\infty$ and when $\omega\neq\omega_{0}$, approaches $0$. Additionally,
since

\[
\int_{-\infty}^{\infty}d\omega\frac{\eta}{\pi\left[(\omega_{0}-\omega)^{2}+\eta^{2}\right]}=1,
\]
this is a nascent $\delta-$function, which means

\begin{equation}
\int_{-\infty}^{\infty}d\omega f(\omega)\frac{i}{\omega_{0}-\omega+i\eta}=\pi\int_{-\infty}^{\infty}d\omega f(\omega)\delta(\omega-\omega_{0}).
\end{equation}
For the second integral, when $\omega\to\omega_{0}$, that when $\omega_{0}-\omega\gg\eta$,
we have

\begin{align}
\int_{-\infty}^{\infty}d\omega\frac{\left(\omega_{0}-\omega\right)^{2}}{(\omega_{0}-\omega)^{2}+\eta^{2}}\frac{f(\omega)}{\omega_{0}-\omega} & \approx\int_{-\infty}^{\infty}d\omega\frac{\left(\omega_{0}-\omega\right)^{2}}{(\omega_{0}-\omega)^{2}}\frac{f(\omega)}{\omega_{0}-\omega}\nonumber \\
 & =\int_{-\infty}^{\infty}d\omega\frac{f(\omega)}{\omega_{0}-\omega},
\end{align}
but when $\omega_{0}-\omega\ll\eta$, we have

\begin{align}
\int_{-\infty}^{\infty}d\omega\frac{\left(\omega_{0}-\omega\right)^{2}}{(\omega_{0}-\omega)^{2}+\eta^{2}}\frac{f(\omega)}{\omega_{0}-\omega} & \approx\int_{-\infty}^{\infty}d\omega\frac{\left(\omega_{0}-\omega\right)}{\eta^{2}}f(\omega)\nonumber \\
 & \to0.\label{eq:nearpole}
\end{align}
where the last line of equation (\ref{eq:nearpole}) follows from
the fact that $(\omega_{0}-\omega)/\eta^{2}\to0$. As such, we have

\begin{equation}
\int_{-\infty}^{\infty}d\omega\frac{\left(\omega_{0}-\omega\right)^{2}}{(\omega_{0}-\omega)^{2}+\eta^{2}}\frac{f(\omega)}{\omega_{0}-\omega}=\mathcal{P}\int_{-\infty}^{\infty}d\omega\frac{f(\omega)}{\omega_{0}-\omega},
\end{equation}
and by putting the results together we have proved the Sokhotski-Plemelj
theorem, that states

\begin{align}
\lim_{t\to\infty}\int_{0}^{t}dt_{1}e^{-i(\omega_{0}-\omega)\left(t_{1}-t\right)} & =\pi\delta(\omega_{0}-\omega)+i\mathcal{P}\frac{1}{\omega_{0}-\omega}\nonumber \\
 & =\pi\delta(\omega-\omega_{0})-i\mathcal{P}\frac{1}{\omega-\omega_{0}}.
\end{align}
Thus, we conclude this section, introducing the notation

\begin{equation}
\mathcal{I}_{SP}(\omega)=\lim_{t\to\infty}\int_{0}^{t}dt_{1}e^{-i(\omega_{0}-\omega)\left(t_{1}-t\right)}=\pi\delta(\omega-\omega_{0})-i\mathcal{P}\frac{1}{\omega-\omega_{0}}.\label{eq:Sokhotski-Plemelj}
\end{equation}
and noting that we shall often call upon this result.

\section{Details on the calculation of the Green's function\label{sec:Details-on-calculating}}

\subsection{The idea behind Green's functions\label{subsec:The-idea-behind}}

We start this appendix with a few considerations and a refresher on
the Green's function method for ODEs following the approach of \citet{kirkwood2011mathematical}.
We discuss how to solve a non homogeneous linear second-order ordinary
differential equation with given boundary conditions by presenting
the solution in terms of an integral. More precisely we find a function
$G(x,t)$ for which

\begin{equation}
L[y(x)]=a_{0}(x)y''(x)+a_{1}(x)y'(x)+a_{2}(x)y(x)=-f(x),
\end{equation}
with $0<x<1$ and $y(0)=y(1)=0$ and

\begin{equation}
y(x)=\int_{0}^{1}G(x,t)f(t)\ dt.
\end{equation}
This function $G(x,t)$ is the namesake Green's function. We shall
obtaining via two usual methods. The first will make use of the Dirac-delta
function and the second of eigenfunction expansions.The general idea
is to proceed as in linear algebra, where if we want to solve a matrix
equation

\begin{equation}
A\boldsymbol{u}=\boldsymbol{f},
\end{equation}
we simply try to find the inverse operator $A^{-1}$, which is possible
as long as 0 is not an eigenvalue of $A$. We shall also see that
if $0$ is not an eigenvalue of $L$, we shall also be able to construct
the Green's function that provides the unique solution to $L[y]=f(x)$,
with appropriate boundary conditions.

\subsection{Construction of Green's function using the Dirac-delta function\label{subsec:Construction-of-Green's}}

We provide first some intuitive motivation for this approach. If we
let

\begin{equation}
L[y(x)]=a_{0}(x)y''(x)+a_{1}(x)y'(x)+a_{2}(x)y(x)
\end{equation}
We are trying to solve

\begin{equation}
L[y(x)]=-f(x),
\end{equation}
for $t$ such that $0<t<1$. Suppose there is a function $G(x,t)$
for which

\begin{equation}
L[G(x,t)]=-\delta(x-t),
\end{equation}
we can multiply both sides by $f(t)$ and integrate with respect to
$t$. Doing so, we have

\begin{equation}
\int_{0}^{1}\left\{ L[G(x,t)]\right\} f(t)\ dt=-\int_{0}^{1}\delta(x-t)f(t)\ dt=-f(x).
\end{equation}
If we can pull the operator $L$ outside the integral, we have

\begin{equation}
L\left[\int_{0}^{1}G(x,t)f(t)\ dt\right]=-f(x),
\end{equation}
and thus

\begin{equation}
y(x)=\int_{0}^{1}G(x,t)f(t)\ dt,
\end{equation}
is a solution to the equation

\begin{equation}
L[y(x)]=-f(x).
\end{equation}
Note that the previous considerations are sound, however we need a
way to construct the Green's function. We first find $G(x,t)$ such
that it satisfies the homogeneous equation for $x\neq t$.

\[
L[G(x,t)]=0.
\]
The solution will have two parts, one for $x<t$ and $x>t$. If we
suppose that these solutions are $y_{1}(x)$ and $y_{2}(x)$ respectively.
Then the Green's function can be written as

\begin{equation}
G(x,t)=\begin{cases}
c_{1}(t)y_{1}(x)+c_{2}(t)y_{2}(x), & 0\leq x<t\\
c_{3}(t)y_{1}(x)+c_{4}(t)y_{2}(x). & t<x\leq1
\end{cases}
\end{equation}
We then need to find each $c_{i}(t)$ and the Green's function becomes
fully determined. To do this we use the boundary conditions

\begin{equation}
G(0,t)=G(1,t)=y(0)=y(1)=0,
\end{equation}
along with continuity at $x=t$. This means that

\begin{equation}
\lim_{x\to t^{+}}G(x,t)=\lim_{x\to t^{-}}G(x,t).
\end{equation}
Additionally, the derivative of the Green's function can be shown
to have a discontinuity at $x=t$:

\begin{equation}
\lim_{x\to t^{+}}\frac{dG(x,t)}{dx}-\lim_{x\to t^{-}}\frac{dG(x,t)}{dx}=-\frac{1}{a_{0}(t)}.
\end{equation}
The final condition is that it is symmetric and thus

\begin{equation}
G(x,t)=G(t,x).
\end{equation}
From these boundary conditions one can obtain the coefficients $c_{i}(t)$
and thus determine the Green's function.

\subsection{Construction of Green's function from eigenfunctions\label{subsec:Construction-of-Green's-1}}

We continue to consider the operator

\[
L[y(x)]=[p(x)y'(x)]'+q(x)y(x),
\]
and want to solve the equation

\[
L[y(x)]=-f(x),
\]
with boundary conditions

\[
y(0)=0,\ \ y(1)=0.
\]
Suppose that $\left\{ \phi_{n}\right\} $ is a complete orthonormal
basis for the vector space consisting of eigenvectors of $L$ that
satisfy the boundary conditions and that $\lambda_{n}$is the eigenvalue
of $\phi_{n}$. If we equip the vector space with the inner product

\[
\left\langle f(x),g(x)\right\rangle =\int_{0}^{1}f(x)g(x)\ dx.
\]
We then have

\[
f(x)=\sum_{n}f_{n}\text{\ensuremath{\phi_{n}(x),\ \ y(x)=\sum_{n}y_{n}\phi_{n}(x)},}
\]
where

\begin{equation}
f_{n}=\left\langle f(x),\phi_{n}(x)\right\rangle =\int_{0}^{1}f(x)\phi_{n}(x)\ dx,
\end{equation}
and therefore we want to find the $y_{n}$'s to construct the solution.
Since $L$ is a linear operator and $\phi_{n}(x)$ is an eigenvector
of $L$ with eigenvalue $\lambda_{n}$ for each $n$ we have that

\[
L[y(x)]=L\left(\sum_{n}y_{n}\phi_{n}(x)\right)=\sum_{n}y_{n}\lambda_{n}\phi_{n}(x).
\]
We get a relation for the coefficients since $L[y(x)]=-f(x)$ and
therefore

\begin{equation}
\sum_{n}y_{n}\lambda_{n}\phi_{n}(x)=-\sum_{n}f_{n}\phi_{n}(x),
\end{equation}
and since $\phi_{n}(x)$ is a basis, we can equate the previous equation
term by term

\begin{equation}
y_{n}\lambda_{n}=-f_{n}.
\end{equation}
Since $0$ is not an eigenvalue, we can write

\begin{equation}
y_{n}=-\frac{f_{n}}{\lambda_{n}},
\end{equation}
and thus

\begin{equation}
y(x)=-\sum_{n}\frac{f_{n}}{\lambda_{n}}\phi_{n}(x).
\end{equation}
If we recall that $f_{n}=\left\langle f(x),\phi_{n}(x)\right\rangle $,
we have

\begin{align}
y(x) & =-\sum_{n}\frac{\phi_{n}(x)}{\lambda_{n}}f_{n}=-\sum_{n}\frac{\phi_{n}(x)}{\lambda_{n}}\int_{0}^{1}f(t)\phi_{n}(t)\ dt\\
 & =-\sum_{n}\int_{0}^{1}\frac{\phi_{n}(x)\phi_{n}(t)}{\lambda_{n}}f(t)\ dt,\nonumber 
\end{align}
which, if we can exchange the summation and integral due to proper
convergence properties of the functions, we can write as

\begin{equation}
y(x)=-\int_{0}^{1}\left(\sum_{n}\frac{\phi_{n}(x)\phi_{n}(t)}{\lambda_{n}}\right)f(t)\ dt.
\end{equation}
Thus, the Green's function is

\begin{equation}
G(x,t)=-\sum_{n}\frac{\phi_{n}(x)\phi_{n}(t)}{\lambda_{n}}.
\end{equation}
The crucial step to construct the Green's function using this method
is thus to find the eigenvalues and eigenfunctions for $L$ that satisfy
the initial conditions. Note that an alternative is to consider the
problem

\begin{equation}
\tilde{L}[y(x)]+\mu y(x)=\left[p(x)y'(x)\right]'+\tilde{q}(x)y(x)+\mu y(x)=-f(x),
\end{equation}
where $\mu$ is not an eigenvalue of $\tilde{L}$ and thus, one obtains
the Green's function

\begin{equation}
\tilde{G}(x,t)=-\sum_{n}\frac{\phi_{n}(x)\phi_{n}(t)}{\lambda_{n}-\mu}.
\end{equation}
One can easily show, however that one form can be converted into another
by suitable renaming of $q(x)$ and $\tilde{q}(x)$.

\subsection{Dyadic Green's functions using the Dirac-$\delta$ - Example of a
single dipole\label{subsec:Dyadic-Green's-functions}}

In this appendix we introduce the dyadic Green's function as a natural
extension of the Green's function discussed in the previous section
for the case of a vector equation. We focus on a critical and instructive
example of the wave equation for the electric field

\begin{equation}
\nabla\times\nabla\times\boldsymbol{E}(\boldsymbol{r},\omega)-\frac{\omega^{2}}{v^{2}}\boldsymbol{E}(\boldsymbol{r},\omega)=i\omega\mu\mu_{0}\boldsymbol{j}_{t}(\boldsymbol{r},\omega).
\end{equation}
We introduce the wave number $k^{2}=\omega^{2}/v^{2}$. The Green's
tensor, or dyadic Green's function $\bar{\bar{G}}(\boldsymbol{r},\boldsymbol{r}',\omega)$
is defined through

\begin{equation}
\nabla\times\nabla\times\bar{\bar{G}}(\boldsymbol{r},\boldsymbol{r}',\omega)-\frac{\omega^{2}}{c^{2}}\bar{\bar{G}}(\boldsymbol{r},\boldsymbol{r}',\omega)=\mathds{1}\delta(\boldsymbol{r}-\boldsymbol{r}'),
\end{equation}
where $\mathds{1}$ is the $3\times3$ unit matrix. Similarly to the
procedure followed in the previous section, we can multiply on both
sides by \textbf{$i\omega\mu\mu_{0}\boldsymbol{j}_{t}(\boldsymbol{r}',\omega)$}
and integrate over the whole space.

\begin{equation}
\left[-\frac{\omega^{2}}{v^{2}}+\nabla\times\nabla\times\right]\left\{ i\omega\mu\mu_{0}\int d^{3}\boldsymbol{r}'\bar{\bar{G}}(\boldsymbol{r},\boldsymbol{r}',\omega)\boldsymbol{j}_{t}(\boldsymbol{r}',\omega)\right\} =i\omega\mu\mu_{0}\boldsymbol{j}_{t}(\boldsymbol{r}',\omega),
\end{equation}
and thus this is a particular solution to the non-homogeneous equation.
If $\boldsymbol{E}_{0}(\boldsymbol{r},\omega)$ is a solution to the
homogeneous equation, the general solution is of the form

\begin{equation}
\boldsymbol{E}(\boldsymbol{r},\omega)=\boldsymbol{E}_{0}(\boldsymbol{r},\omega)+i\omega\mu\mu_{0}\int d^{3}\boldsymbol{r}'\bar{\bar{G}}(\boldsymbol{r},\boldsymbol{r}',\omega)\boldsymbol{j}_{t}(\boldsymbol{r}',\omega).
\end{equation}
To find the Green's function we assume that the source of current
is a dipole at $\boldsymbol{r}=\boldsymbol{r}_{0}$, which means that
$\boldsymbol{j}_{t}(\boldsymbol{r},\omega)=-i\omega\boldsymbol{d}_{0}\delta(\boldsymbol{r}-\boldsymbol{r}_{0})$.
The integral of the previous equation is thus immediate, and reads

\begin{equation}
\boldsymbol{E}(\boldsymbol{r},\omega)=\boldsymbol{E}_{0}(\boldsymbol{r},\omega)+\omega^{2}\mu\mu_{0}\bar{\bar{G}}(\boldsymbol{r},\boldsymbol{r}',\omega)\boldsymbol{d}_{0}.
\end{equation}
In the Lorentz gauge, we can write the electric and magnetic fields
in terms of a scalar and vector potential $\boldsymbol{A}(\boldsymbol{r},\omega)$
and $\phi(\boldsymbol{r},\omega)$. In this gauge we have

\begin{equation}
\begin{cases}
\boldsymbol{E}(\boldsymbol{r},\omega) & =i\omega\boldsymbol{A}(\boldsymbol{r},\omega)-\nabla\phi(\boldsymbol{r},\omega),\\
\boldsymbol{B}(\boldsymbol{r},\omega) & =\nabla\times\boldsymbol{A}(\boldsymbol{r},\omega).
\end{cases}
\end{equation}
If we substitute the electric field in the electric field equation,
we have

\begin{align}
\nabla\times\nabla\times i\omega\boldsymbol{A}(\boldsymbol{r},\omega)-\nabla\times\nabla\times\nabla\phi(\boldsymbol{r},\omega)\nonumber \\
-\frac{\omega^{2}}{v^{2}}\left[i\omega\boldsymbol{A}(\boldsymbol{r},\omega)-\nabla\phi(\boldsymbol{r},\omega)\right] & =i\omega\mu\mu_{0}\boldsymbol{j}_{t}(\boldsymbol{r},\omega),
\end{align}
or

\begin{align}
i\nabla\left[\nabla\cdot\boldsymbol{A}(\boldsymbol{r},\omega)\right]-i\nabla^{2}\boldsymbol{A}(\boldsymbol{r},\omega)\nonumber \\
-\frac{\omega}{v^{2}}\left[i\omega\boldsymbol{A}(\boldsymbol{r},\omega)-\nabla\phi(\boldsymbol{r},\omega)\right] & =i\mu\mu_{0}\boldsymbol{j}_{t}(\boldsymbol{r},\omega),
\end{align}
and using the fact that $\nabla\cdot\boldsymbol{A}(\boldsymbol{r},\omega)=i\omega\mu\mu_{0}\varepsilon\varepsilon_{0}\phi(\boldsymbol{r},\omega)=i\omega\phi(\boldsymbol{r},\omega)/v^{2}$
we have

\begin{align}
-\frac{\omega}{v^{2}}\nabla\phi(\boldsymbol{r},\omega)-i\nabla^{2}\boldsymbol{A}(\boldsymbol{r},\omega)\nonumber \\
-i\frac{\omega^{2}}{v^{2}}\boldsymbol{A}(\boldsymbol{r},\omega)+\frac{\omega}{v^{2}}\nabla\phi(\boldsymbol{r},\omega) & =i\mu\mu_{0}\boldsymbol{j}_{t}(\boldsymbol{r},\omega),
\end{align}
where finally canceling the electric field term, we have

\begin{equation}
\left[\nabla^{2}+k^{2}\right]\boldsymbol{A}(\boldsymbol{r},\omega)=-\mu\mu_{0}\boldsymbol{j}_{t}(\boldsymbol{r},\omega),
\end{equation}
and if we replace the current with that produced by the dipole we
have

\begin{equation}
\left[\nabla^{2}+k^{2}\right]\boldsymbol{A}(\boldsymbol{r},\omega)=-\mu\mu_{0}\boldsymbol{d}_{0}\delta(\boldsymbol{r}-\boldsymbol{r}_{0}).
\end{equation}
We get a differential equation for the components of the Green's function
associated with the vector potential

\begin{equation}
\nabla^{2}g_{j}(\boldsymbol{r},\boldsymbol{r})+k^{2}g_{j}(\boldsymbol{r},\boldsymbol{r}',\omega)=i\omega\mu\mu_{0}d_{0}^{j}\delta(\boldsymbol{r}-\boldsymbol{r}').\label{eq:diffeqfeeEM}
\end{equation}
This equation can be solved using the Fourier transform. If we define

\begin{equation}
g_{j}(\boldsymbol{r},\boldsymbol{r}',\omega)=\int\frac{d^{3}\boldsymbol{p}}{(2\pi)^{3}}g_{j}(\boldsymbol{p},\omega)e^{i\boldsymbol{p}\cdot(\boldsymbol{r}-\boldsymbol{r}')},
\end{equation}
the differential equation (\ref{eq:diffeqfeeEM}) reduces to

\begin{equation}
g_{j}(\boldsymbol{p},\omega)=-\frac{i\omega\mu\mu_{0}d_{0}^{j}}{p^{2}-k^{2}}.
\end{equation}
We can then make an inverse Fourier transform arriving at

\begin{align}
g_{j}(\boldsymbol{r},\boldsymbol{r}',\omega) & =-i\omega\omega\mu\mu_{0}d_{0}^{j}\int\frac{d^{3}\boldsymbol{p}}{(2\pi)^{3}}\frac{e^{i\boldsymbol{p}\cdot(\boldsymbol{r}-\boldsymbol{r}')}}{p^{2}-k^{2}}.
\end{align}
We can perform the integral by contour integration noting that there
is a pole at $p=\pm k$. We have

\begin{align*}
\int\frac{d^{3}\boldsymbol{p}}{(2\pi)^{3}}\frac{e^{i\boldsymbol{p}\cdot(\boldsymbol{r}-\boldsymbol{r}')}}{p^{2}-k^{2}} & =\frac{1}{(2\pi)^{3}}\int_{0}^{2\pi}d\phi\int_{0}^{\pi}d\theta\sin\theta\int_{0}^{\infty}dp\ \frac{p^{2}e^{ip|\boldsymbol{r}-\boldsymbol{r}'|\cos\theta}}{p^{2}-k^{2}}\\
 & =\frac{1}{(2\pi)^{2}}\int_{0}^{\infty}dp\frac{p^{2}}{p^{2}-k^{2}}\int_{0}^{\pi}d\theta\sin\theta e^{ip|\boldsymbol{r}-\boldsymbol{r}'|\cos\theta}\\
 & =-\frac{1}{4\pi^{2}}\int_{0}^{\infty}dp\ \frac{ip\sin(p|\boldsymbol{r}-\boldsymbol{r}|)}{\left(p^{2}-k^{2}\right)|\boldsymbol{r}-\boldsymbol{r}'|}.
\end{align*}
The result allows us to write

\begin{align}
\boldsymbol{E}(\boldsymbol{r},\omega)= & \boldsymbol{E}_{0}(\boldsymbol{r},\omega)+\omega^{2}\mu\mu_{0}\boldsymbol{d}_{0}\frac{e^{ik|\boldsymbol{r}-\boldsymbol{r}'|}}{4\pi|\boldsymbol{r}-\boldsymbol{r}'|}\nonumber \\
 & +\frac{1}{\varepsilon\varepsilon_{0}}\nabla\left[\nabla\cdot\left(\boldsymbol{d}_{0}\frac{e^{ik|\boldsymbol{r}-\boldsymbol{r}'|}}{4\pi|\boldsymbol{r}-\boldsymbol{r}'|}\right)\right].
\end{align}
Factoring the above expression we have

\begin{equation}
\boldsymbol{E}(\boldsymbol{r},\omega)=\boldsymbol{E}_{0}(\boldsymbol{r},\omega)+\mu\mu_{0}\omega^{2}\left(\mathds{1}+k^{-2}\nabla\nabla\cdot\right)\boldsymbol{d}_{0}\frac{e^{ik|\boldsymbol{r}-\boldsymbol{r}'|}}{4\pi|\boldsymbol{r}-\boldsymbol{r}'|},
\end{equation}
which allows us to identify the dyadic Green's function in real space
as

\begin{equation}
\bar{\bar{G}}(\boldsymbol{r},\boldsymbol{r}',\omega)=(\mathds{1}+k^{-2}\nabla\nabla)\frac{e^{ik|\boldsymbol{r}-\boldsymbol{r}'|}}{4\pi|\boldsymbol{r}-\boldsymbol{r}'|},
\end{equation}
where $\nabla\nabla=\nabla\otimes\nabla$ is a dyadic differential
operator and $\otimes$ is the tensor product. This serves as an example
of the computation of a dyadic Green's function using from a Dirac-$\delta$
function.

\subsection{Finding the eigenmodes\label{subsec:Finding-the-eigenmodes}}

In this appendix we begin the quantization of surface plasmons via
dyadic Green's functions. Contrary to the previous section we shall
not use the construction from a delta function, but rather an expansion
of eigenfunctions (eigenmodes in this case).

The problem we are interested in solving consists of a graphene sheet
embedded between two semi-infinite dielectrics. As such, for $z>0$
the dielectric function is a constant $\varepsilon_{1}$ while for
$z<0$ the dielectric function is also constant and equal to $\varepsilon_{2}$.
At $z=0$ we have a contribution from the graphene and therefore the
dielectric function over the whole space can be written as

\begin{equation}
\varepsilon(z)=\varepsilon_{1}\Theta(z)+i\frac{\sigma(\omega)}{\varepsilon_{0}\omega}\delta(z)+\varepsilon_{2}\Theta(-z),
\end{equation}
where $\sigma(\omega)$ is the conductivity of the graphene sheet.
In the Weyl gauge (see \citet{Loffelholz2003,kurt1987,Creutz1979QuantumEI})
we have that $\boldsymbol{E}(\boldsymbol{r})=i\omega\boldsymbol{A}(\boldsymbol{r})$,
where $\boldsymbol{E}(\boldsymbol{r})$ is the electric field and
$\boldsymbol{A}(\boldsymbol{r})$ is the vector potential. We thus
have a wave equation reading

\begin{equation}
-\nabla\times\nabla\times\boldsymbol{A}(\boldsymbol{r})+\frac{\omega^{2}}{c^{2}}\varepsilon(z)\boldsymbol{A}(\boldsymbol{r})=0,
\end{equation}
where we have assumed that there are no currents in the dielectric.
The graphene sheet is a boundary condition for this wave equation.
We will solve this equation via a dyadic Green's function. It is such
that

\begin{equation}
-\nabla\times\nabla\times\bar{\bar{G}}(\boldsymbol{r},\boldsymbol{r}';\omega)+\frac{\omega^{2}}{c^{2}}\varepsilon(z)\bar{\bar{G}}(\boldsymbol{r},\boldsymbol{r}';\omega)=-\mathds{1}\delta(\boldsymbol{r}-\boldsymbol{r}'),
\end{equation}
and our first goal is to construct this dyadic Green's function. Our
approach will make use of an expansion in eigenmodes. We write the
Green's function as

\begin{equation}
\bar{\bar{G}}(\boldsymbol{r},\boldsymbol{r}';\omega)=\int d^{2}\boldsymbol{q}\sum_{n}\frac{\boldsymbol{a}_{n}(\boldsymbol{r})\otimes\left[\tilde{\boldsymbol{a}}_{n}(\boldsymbol{r}')\right]^{*}}{N_{n}\lambda_{n}},
\end{equation}
where $\lambda_{n}$ are the eigenvalues, $N_{n}$ a normalization
factor, and $\boldsymbol{a}_{n}$ the eigenmodes, solutions to the
equation

\begin{equation}
-\nabla\times\nabla\times\boldsymbol{a}_{n}(\boldsymbol{r})+\frac{\omega^{2}}{c^{2}}\varepsilon(z)\boldsymbol{a}_{n}(\boldsymbol{r})=\varepsilon(z)\lambda_{n}\boldsymbol{a}_{n}(\boldsymbol{r}),\label{eq:Eigenmodes-1}
\end{equation}
which we can write as

\begin{equation}
H(\boldsymbol{r})\boldsymbol{a}_{n}(\boldsymbol{r})=\varepsilon(z)\lambda_{n}\boldsymbol{a}_{n}(\boldsymbol{r}).
\end{equation}
Since the system has dissipation\textbf{ $H(\boldsymbol{r})$} is
not hermitian and the left and right eigenmodes are not the same.
The left eigenmodes are

\begin{equation}
\left[\tilde{\boldsymbol{a}}_{n}(\boldsymbol{r})\right]^{*}H(\boldsymbol{r})=\varepsilon(z)\lambda_{n}\left[\tilde{\boldsymbol{a}}_{n}(\boldsymbol{r})\right]^{*},
\end{equation}
or

\begin{equation}
\left[H(\boldsymbol{r})\right]^{\dagger}\tilde{\boldsymbol{a}}_{n}(\boldsymbol{r})=\varepsilon(z)^{*}\lambda_{n}^{*}\tilde{\boldsymbol{a}}_{n}(\boldsymbol{r}).
\end{equation}
The normalization factor can be found by evaluating

\begin{equation}
N_{n}=\int_{-\infty}^{\infty}dz\ \varepsilon(z)[\tilde{\boldsymbol{a}}_{n}^{(1)}(\boldsymbol{r})]^{*}\cdot\tilde{\boldsymbol{a}}_{n}^{(1)}(\boldsymbol{r}).
\end{equation}
We assume $\varepsilon_{1},\varepsilon_{2}\in\mathbb{R}$ and want
to solve for mode functions of the form

\begin{equation}
\begin{cases}
\boldsymbol{a}_{n}^{(1)}(\boldsymbol{r})=(a_{x}^{(1)},a_{y}^{(1)},a_{z}^{(1)})e^{i(q_{x}x+q_{y}y)}e^{-\kappa_{1}z}\Theta(z),\\
\boldsymbol{a}_{n}^{(2)}(\boldsymbol{r})=(a_{x}^{(2)},a_{y}^{(2)},a_{z}^{(2)})e^{i(q_{x}x+q_{y}y)}e^{\kappa_{2}z}\Theta(-z),
\end{cases}
\end{equation}
where the $(1)$ index shows that these propagate in the medium with
$\varepsilon_{1}$ while the index $(2)$ shows that these propagate
in the medium with $\varepsilon_{2}$. This is the desired form for
our problem since we are interested in solutions to the equation that
propagate like waves along the graphene sheet and decay exponentially
as we go along into the dielectric media corresponding to the surface
plasmon polariton solutions. Substituting this form into equation
(\ref{eq:Eigenmodes-1}) one can easily check that it can be written
as a matrix equation

\begin{equation}
\begin{bmatrix}-q_{y}^{2}+\text{\ensuremath{\kappa}}_{1}^{2}+\varepsilon_{1}\frac{\omega^{2}}{c^{2}} & q_{x}q_{y} & iq_{x}\kappa_{1}\\
q_{x}q_{y} & -q_{y}^{2}+\kappa_{1}^{2}+\varepsilon_{1}\frac{\omega^{2}}{c^{2}} & iq_{y}\kappa_{1}\\
iq_{x}\kappa_{1} & iq_{y}\kappa_{1} & -q_{x}^{2}-q_{y}^{2}+\varepsilon_{1}\frac{\omega^{2}}{c^{2}}
\end{bmatrix}\begin{bmatrix}a_{x}^{(1)}\\
a_{y}^{(1)}\\
a_{z}^{(1)}
\end{bmatrix}=\varepsilon_{1}\lambda_{n}\begin{bmatrix}a_{x}^{(1)}\\
a_{y}^{(1)}\\
a_{z}^{(1)}
\end{bmatrix},\label{eq:Matrix}
\end{equation}
for $z>0$. We can solve this system for the eigenmodes and corresponding
eigenvalues. These are

\begin{equation}
\begin{cases}
\varepsilon_{1}\lambda_{1}^{(1)}=\frac{\varepsilon_{1}\omega^{2}}{c^{2}} & \boldsymbol{a}_{1}^{(1)}=A_{1}e^{i\left(q_{x}x+q_{y}y\right)}e^{-\kappa_{1}z}(-iq_{x},-iq_{y},\kappa_{1})/\kappa_{1},\\
\varepsilon_{1}\lambda_{2}^{(1)}=\kappa_{1}^{2}-q^{2}+\frac{\varepsilon_{1}\omega^{2}}{c^{2}} & \boldsymbol{a}_{2}^{(1)}=A_{1}e^{i\left(q_{x}x+q_{y}y\right)}e^{-\kappa_{1}z}(-i\kappa_{1},0,q_{x})/q_{x},\\
\varepsilon_{1}\lambda_{3}^{(1)}=\kappa_{1}^{2}-q^{2}+\frac{\varepsilon_{1}\omega^{2}}{c^{2}} & \boldsymbol{a}_{3}^{(1)}=A_{1}e^{i\left(q_{x}x+q_{y}y\right)}e^{-\kappa_{1}z}(-q_{y},q_{x},0)/q_{x}.
\end{cases}
\end{equation}
The eigenvalues for the dielectric on the other side of the graphene
sheet is obtained by replacing $\kappa_{1}\to-\kappa_{2}$ and $\varepsilon_{1}\to\varepsilon_{2}$.
We have, for this case

\begin{equation}
\begin{cases}
\varepsilon_{2}\lambda_{1}^{(2)}=\frac{\varepsilon_{2}\omega^{2}}{c^{2}} & \boldsymbol{a}_{1}^{(2)}=A_{2}e^{i\left(q_{x}x+q_{y}y\right)}e^{\kappa_{2}z}(iq_{x},iq_{y},\kappa_{2})/\kappa_{2},\\
\varepsilon_{2}\lambda_{2}^{(2)}=\kappa_{2}^{2}-q^{2}+\frac{\varepsilon_{2}\omega^{2}}{c^{2}} & \boldsymbol{a}_{2}^{(2)}=A_{2}e^{i\left(q_{x}x+q_{y}y\right)}e^{\kappa_{2}z}(i\kappa_{2},0,q_{x})/q_{x},\\
\varepsilon_{2}\lambda_{3}^{(2)}=\kappa_{2}^{2}-q^{2}+\frac{\varepsilon_{2}\omega^{2}}{c^{2}} & \boldsymbol{a}_{3}^{(2)}=A_{2}e^{i\left(q_{x}x+q_{y}y\right)}e^{\kappa_{2}z}(-q_{y},q_{x},0)/q_{x}.
\end{cases}
\end{equation}
An important observation, however, is that each eigenvalue with the
same index $n$ on both sides of the graphene sheet must correspond
a single mode across the whole system. In symbols, this must mean
that $\lambda_{n}^{(1)}=\lambda_{n}^{(2)}$. This allows us to connect
the solution across the graphene sheet and relate $\kappa_{1}$ with
$\kappa_{2}$ and $q$. We can do this by equating

\begin{align}
 & \frac{\kappa_{1}^{2}}{\varepsilon_{1}}-\frac{q^{2}}{\varepsilon_{1}}+\frac{\omega^{2}}{c^{2}}=\frac{\kappa_{2}^{2}}{\varepsilon_{2}}-\frac{q^{2}}{\varepsilon_{2}}+\frac{\omega^{2}}{c^{2}}\nonumber \\
\Leftrightarrow & \kappa_{2}^{2}=q^{2}-\frac{\varepsilon_{2}}{\varepsilon_{1}}q^{2}+\frac{\varepsilon_{2}}{\varepsilon_{1}}\kappa_{1}^{2}\nonumber \\
 & \kappa_{2}^{2}=\frac{\varepsilon_{2}}{\varepsilon_{1}}\kappa_{1}^{2}+\left(\frac{\varepsilon_{1}-\varepsilon_{2}}{\varepsilon_{1}}\right)q^{2}.\label{eq:k2k1}
\end{align}
Note however that the eigenvalues are degenerated for $n=2,3$. They
must indeed be superpositions of the $s-$ and $p-$ polarization
vectors

\begin{equation}
\begin{cases}
\hat{e}_{s}=\frac{\boldsymbol{p}_{1}\times(0,0,1)}{q}=\left(-\frac{q_{x}}{q},-\frac{q_{y}}{q},0\right),\\
\hat{e}_{p}=\frac{\hat{e}_{s}\times\boldsymbol{p}_{1}}{\sqrt{\boldsymbol{p}_{1}^{*}\cdot\boldsymbol{p}_{1}}}=\frac{1}{\sqrt{q^{2}+\kappa_{1}^{2}}}\left(-i\frac{q_{x}\kappa_{1}}{q},-i\frac{q_{y}\kappa_{1}}{q},0\right),
\end{cases}
\end{equation}
where $\boldsymbol{p}_{1}=(q_{x},q_{y},i\kappa_{1})$. We can find
the $s-$ and $p-$ polarization components for $z<0$ by writing
$\kappa_{1}$ as a function $\kappa_{2}$ by using equation (\ref{eq:k2k1}).
Note that if we substitute these polarizations into the matrix equation
(\ref{eq:Matrix}) for the eigenmodes, we have

\begin{align}
\begin{bmatrix}-q_{y}^{2}+\text{\ensuremath{\kappa}}_{1}^{2}+\varepsilon_{1}\frac{\omega^{2}}{c^{2}} & q_{x}q_{y} & iq_{x}\kappa_{1}\\
q_{x}q_{y} & -q_{y}^{2}+\kappa_{1}^{2}+\varepsilon_{1}\frac{\omega^{2}}{c^{2}} & iq_{y}\kappa_{1}\\
iq_{x}\kappa_{1} & iq_{y}\kappa_{1} & -q_{x}^{2}-q_{y}^{2}+\varepsilon_{1}\frac{\omega^{2}}{c^{2}}
\end{bmatrix}\begin{bmatrix}-\frac{q_{x}}{q}\\
-\frac{q_{y}}{q}\\
0
\end{bmatrix}\nonumber \\
=\left(-q^{2}+\kappa_{1}^{2}+\varepsilon_{1}\frac{\omega^{2}}{c^{2}}\right)\begin{bmatrix}-\frac{q_{x}}{q}\\
-\frac{q_{y}}{q}\\
0
\end{bmatrix},
\end{align}
and similarly for the $z<0$ with the previously mentioned substitutions.
The same eigenvalue solves the equation for the $p-$polarization.
In particular, we want to solve only for plasmonic eigenmodes, which
only have $p-$ polarization.

\subsection{Boundary conditions - Dispersion relation in the electrostatic limit\label{subsec:Boundary-conditions--}}

We now proceed with connecting the solution for $z>0$ with that for
$z<0$. We have the usual boundary conditions for Maxwell's equations.
For the electric field we have

\begin{equation}
\boldsymbol{E}_{\parallel}^{(1)}=\boldsymbol{E}_{\parallel}^{(2)},
\end{equation}
which we can translate into an equation for the eigenmodes since $\boldsymbol{E}_{\parallel}^{(j)}=i\omega\boldsymbol{a}_{\parallel}^{(j)}$.
Using then the $p-$ polarization, which is parallel to the graphene
sheet at $z=0$, we have

\begin{align}
A_{1}e^{i\left(q_{x}x+q_{y}y\right)}\frac{1}{\sqrt{q^{2}+\kappa_{1}^{2}}}\left(-i\frac{q_{x}\kappa_{1}}{q},-i\frac{q_{y}\kappa_{1}}{q},0\right) & =\nonumber \\
A_{2}e^{i\left(q_{x}x+q_{y}y\right)}\frac{1}{\sqrt{q^{2}+\kappa_{2}^{2}}}\left(i\frac{q_{x}\kappa_{2}}{q},i\frac{q_{y}\kappa_{2}}{q},0\right),
\end{align}
from where follows

\begin{align}
 & -A_{1}\frac{1}{\sqrt{q^{2}+\kappa_{1}^{2}}}\kappa_{1}=A_{2}\frac{1}{\sqrt{q^{2}+\kappa_{2}^{2}}}\kappa_{2}\nonumber \\
\Leftrightarrow & A_{2}=-A_{1}\frac{\kappa_{1}}{\kappa_{2}}\sqrt{\frac{q^{2}+\kappa_{2}^{2}}{q^{2}+\kappa_{1}^{2}}}.
\end{align}
The boundary condition for the magnetic field will take into account
the surface current that propagates along the graphene sheet. We can
write it as

\begin{equation}
\hat{\boldsymbol{z}}\times(\boldsymbol{B}_{\parallel}^{(1)}-\boldsymbol{B}_{\parallel}^{(2)})=\mu_{0}\boldsymbol{K}=\mu_{0}\sigma(\omega)\boldsymbol{E}_{\parallel}^{(1)},
\end{equation}
where we have used Ohm's law. The magnetic field can be evaluated
as $\boldsymbol{B}_{\parallel}^{(j)}=\nabla\times\boldsymbol{a}_{\parallel}^{(j)}=i\boldsymbol{p}_{1}\times\boldsymbol{a}_{\parallel}^{(j)}$.
Explicitly, we have

\begin{align}
\boldsymbol{B}_{\parallel}^{(1)} & =A_{1}e^{i\left(q_{x}x+q_{y}y\right)}\frac{1}{\sqrt{q^{2}+\kappa_{1}^{2}}}i(q_{x},q_{y},i\kappa_{1})\times\left(\frac{q_{x}\kappa_{1}}{q},\frac{q_{y}\kappa_{1}}{q},0\right)\nonumber \\
 & =A_{1}e^{i\left(q_{x}x+q_{y}y\right)}\frac{1}{q\sqrt{q^{2}+\kappa_{1}^{2}}}\left(iq_{y}\left(q^{2}-\kappa_{1}^{2}\right),-iq_{x}(q^{2}-\kappa_{1}^{2}),0\right),
\end{align}
and thus making the usual substitutions we find $\boldsymbol{B}_{\parallel}^{(2)}$.
Thus

\begin{align}
\hat{\boldsymbol{z}}\times(\boldsymbol{B}_{\parallel}^{(1)}-\boldsymbol{B}_{\parallel}^{(2)})= & A_{1}e^{i(q_{x}x+q_{y}y)}\frac{1}{q\sqrt{q^{2}+\kappa_{1}^{2}}}\left[\left(q^{2}-\kappa_{1}^{2}\right)+\frac{\kappa_{1}}{\kappa_{2}}\left(q^{2}-\kappa_{2}^{2}\right)\right]\nonumber \\
 & \hat{\boldsymbol{z}}\times\left(iq_{y},-iq_{x},0\right)\nonumber \\
= & A_{1}e^{i(q_{x}x+q_{y}y)}\frac{\left(q^{2}-\kappa_{1}^{2}\right)+\frac{\kappa_{1}}{\kappa_{2}}\left(q^{2}-\kappa_{2}^{2}\right)}{q\sqrt{q^{2}+\kappa_{1}^{2}}}\left(iq_{x},iq_{y},0\right).
\end{align}
We have the additional constraints from the boundary condition

\begin{align}
\left[\left(q^{2}-\kappa_{1}^{2}\right)+\frac{\kappa_{1}}{\kappa_{2}}\left(q^{2}-\kappa_{2}^{2}\right)\right]\left(iq_{x},iq_{y},0\right) & =\nonumber \\
i\omega\mu_{0}\sigma(\omega)\left(-i\frac{q_{x}\kappa_{1}}{q},-i\frac{q_{y}\kappa_{1}}{q},0\right),
\end{align}
which yields

\begin{equation}
\left[\frac{1}{\kappa_{1}}\left(q^{2}-\kappa_{1}^{2}\right)+\frac{1}{\kappa_{2}}\left(q^{2}-\kappa_{2}^{2}\right)\right]=-i\omega\mu_{0}\sigma(\omega),
\end{equation}
or rewriting

\begin{equation}
\kappa_{2}\left(q^{2}-\kappa_{1}^{2}\right)+\kappa_{1}\left(q^{2}-\kappa_{2}^{2}\right)=-i\omega\mu_{0}\sigma(\omega)\kappa_{1}\kappa_{2}.
\end{equation}
and recalling the equation (\ref{eq:k2k1}) that related $\kappa_{1}$,
$\kappa_{2}$ and $q$, we can write

\begin{equation}
\kappa_{2}^{2}=\frac{\varepsilon_{2}}{\varepsilon_{1}}\kappa_{1}^{2}+\left(\frac{\varepsilon_{1}-\varepsilon_{2}}{\varepsilon_{1}}\right)q^{2},
\end{equation}
and solving for $q$, we have

\begin{equation}
q^{2}=\left(\frac{\varepsilon_{1}}{\varepsilon_{1}-\varepsilon_{2}}\right)\kappa_{2}^{2}-\left(\frac{\varepsilon_{2}}{\varepsilon_{1}-\varepsilon_{2}}\right)\kappa_{1}^{2},\label{eq:F70}
\end{equation}
and therefore

\begin{align}
\lambda_{p}^{(1)} & =-q^{2}/\varepsilon_{1}+\kappa_{1}^{2}/\varepsilon_{1}+\frac{\omega^{2}}{c^{2}}\nonumber \\
 & =-\frac{\kappa_{2}^{2}}{\varepsilon_{1}-\varepsilon_{2}}+\frac{\varepsilon_{2}}{\varepsilon_{1}}\frac{\kappa_{1}^{2}}{\varepsilon_{1}-\varepsilon_{2}}+\frac{\kappa_{1}^{2}}{\varepsilon_{1}}+\frac{\omega^{2}}{c^{2}}\nonumber \\
 & =-\frac{\kappa_{2}^{2}}{\varepsilon_{1}-\varepsilon_{2}}+\frac{\kappa_{1}^{2}}{\varepsilon_{1}}\left(\frac{\varepsilon_{2}}{\varepsilon_{1}-\varepsilon_{2}}+1\right)+\frac{\omega^{2}}{c^{2}}\nonumber \\
 & =-\frac{\kappa_{2}^{2}}{\varepsilon_{1}-\varepsilon_{2}}+\frac{\kappa_{1}^{2}}{\varepsilon_{1}}\left(\frac{\varepsilon_{1}}{\varepsilon_{1}-\varepsilon_{2}}\right)+\frac{\omega^{2}}{c^{2}}\nonumber \\
 & =\left(\frac{1}{\varepsilon_{1}-\varepsilon_{2}}\right)(\kappa_{1}^{2}-\kappa_{2}^{2})+\frac{\omega^{2}}{c^{2}}.
\end{align}
The conductivity can be computed using the Drude model, with which
we have

\begin{equation}
\sigma(\omega)=\frac{e^{2}}{4\pi\hbar}\frac{4E_{F}}{\Gamma-i\hbar\omega}\approx i\frac{e^{2}}{4\pi\hbar}\frac{4E_{F}}{\hbar\omega},
\end{equation}
where $E_{F}$ is the Fermi energy and $\hbar\Gamma$ is the momentum
relaxation rate. This corresponds to an approximation of the form
$\sigma(\omega)\equiv\sigma_{1}(\omega)+i\sigma_{2}(\omega)\approx i\sigma_{2}(\omega).$
The dispersion relation can be written as

\begin{equation}
\kappa_{2}\left(q^{2}-\kappa_{1}^{2}\right)+\kappa_{1}\left(q^{2}-\kappa_{2}^{2}\right)=\frac{e^{2}}{4\pi\varepsilon_{0}}\frac{4E_{F}}{c^{2}\hbar^{2}}\kappa_{1}\kappa_{2},
\end{equation}
and replacing $q^{2}$ with the result of equation (\ref{eq:F70})
in this equation we have

\begin{equation}
(\kappa_{1}^{2}-\kappa_{2}^{2})\frac{\kappa_{2}\varepsilon_{1}+\kappa_{1}\varepsilon_{2}}{\varepsilon_{2}-\varepsilon_{1}}=\frac{e^{2}}{4\pi\varepsilon_{0}}\frac{4E_{F}}{c^{2}\hbar^{2}}\kappa_{1}\kappa_{2},
\end{equation}
or in other words

\begin{equation}
(\kappa_{1}^{2}-\kappa_{2}^{2})\left(\frac{\varepsilon_{1}}{\kappa_{1}}+\frac{\varepsilon_{2}}{\kappa_{2}}\right)=\frac{e^{2}}{4\pi\varepsilon_{0}}\frac{4E_{F}}{c^{2}\hbar^{2}}\left(\varepsilon_{2}-\varepsilon_{1}\right),
\end{equation}
or yet

\begin{equation}
(\kappa_{1}^{2}-\kappa_{2}^{2})=\frac{e^{2}}{4\pi\varepsilon_{0}}\frac{4E_{F}}{c^{2}\hbar^{2}}\frac{\left(\varepsilon_{2}-\varepsilon_{1}\right)}{\varepsilon_{1}/\kappa_{1}+\varepsilon_{2}/\kappa_{2}}.
\end{equation}
This allows us to write the eigenvalues as

\begin{equation}
\lambda_{p}^{(1)}=\frac{\omega^{2}}{c^{2}}-\frac{e^{2}}{4\pi\varepsilon_{0}}\frac{4E_{F}}{c^{2}\hbar^{2}}\frac{1}{\varepsilon_{1}/\kappa_{1}+\varepsilon_{2}/\kappa_{2}},
\end{equation}
and in the electrostatic approximation

\begin{align}
\lambda_{p}^{(1)} & \approx\frac{\omega^{2}}{c^{2}}-\frac{e^{2}}{4\pi\varepsilon_{0}}\frac{4E_{F}}{\hbar^{2}c^{2}}\frac{q}{\varepsilon_{1}+\varepsilon_{2}}\nonumber \\
 & =\frac{\omega^{2}}{c^{2}}-\frac{4\alpha E_{F}c^{2}}{\hbar^{2}c^{2}}\frac{\hbar cq}{\varepsilon_{1}+\varepsilon_{2}}\nonumber \\
 & =\frac{\omega^{2}}{c^{2}}-\frac{2\alpha E_{F}}{\hbar c^{2}}\frac{cq}{\bar{\varepsilon}}\nonumber \\
 & =\frac{\omega^{2}}{c^{2}}-\frac{\omega_{\text{spp}}^{2}(q)}{c^{2}},
\end{align}
where $\bar{\varepsilon}=(\varepsilon_{1}+\varepsilon_{2})/2$ and
where we have introduced the surface plasmon polariton (SPP) frequency
as

\begin{equation}
\omega_{\text{spp}}^{2}(q)\equiv\frac{2\alpha cE_{F}}{\hbar}\frac{q}{\bar{\varepsilon}}=\frac{q\omega\sigma_{2}(\omega)}{2\bar{\varepsilon}\varepsilon_{0}}.\label{eq:wspp}
\end{equation}
The first form is valid for a Drude conductivity and the second in
general. The dispersion can therefore be found when $\omega=\omega_{\text{spp}},$or
in other words

\begin{equation}
\frac{q\omega\sigma_{2}(\omega)}{2\bar{\varepsilon}\varepsilon_{0}}=\omega^{2},
\end{equation}
or rather still

\begin{equation}
\frac{\sigma_{2}(\omega)}{\varepsilon_{0}\omega}=\frac{2\bar{\varepsilon}}{q},
\end{equation}
which also means that

\begin{equation}
i\frac{\sigma(\omega)}{\varepsilon_{0}\omega}\approx-\frac{2\bar{\varepsilon}}{q}.\label{eq:disper}
\end{equation}
This is an implicit equation for the spectrum of the SPPs, which may
be solved analytically or numerically depending on the complexity
of $\sigma(\omega)$.

\subsection{Finding the normalization constant\label{subsec:Finding-the-normalization}}

In the previous appendix we have found the eigenmodes and eigenvalues,
as well as written them in a convenient form, relating $q$, $\kappa_{1}$
and $\kappa_{2}$ through the boundary conditions of our problem.
To find the Green's function, in this appendix find the normalization
coefficient $N_{p}$. For the $p-$polarized modes we can evaluate
the normalization as

\begin{equation}
N_{p}=\int_{0}^{\infty}dz\varepsilon_{1}|A_{1}|^{2}e^{-2\kappa_{1}z}+\int_{-\infty}^{0}dz\varepsilon_{2}|A_{2}|^{2}e^{2\kappa_{2}z}+i\frac{\sigma(\omega)}{\varepsilon_{0}\omega},
\end{equation}
where the first term covers $z>0$, the second $z<0$ and the final
comes from the graphene boundary condition at $z=0$. We have, choosing
$A_{1}=1$

\begin{align}
N_{p} & =\int_{0}^{\infty}dz\varepsilon_{1}e^{-2\kappa_{1}z}+\int_{-\infty}^{0}dz\varepsilon_{2}\frac{\kappa_{1}^{2}}{\kappa_{2}^{2}}\frac{q^{2}+\kappa_{2}^{2}}{q^{2}+\kappa_{1}^{2}}e^{2\kappa_{2}z}+i\frac{\sigma(\omega)}{\varepsilon_{0}\omega}\nonumber \\
 & =\frac{\varepsilon_{1}}{2\kappa_{1}}+\frac{\varepsilon_{2}}{2\kappa_{2}}\frac{\kappa_{1}^{2}}{\kappa_{2}^{2}}\frac{q^{2}+\kappa_{2}^{2}}{q^{2}+\kappa_{1}^{2}}+\frac{i\sigma(\omega)}{\varepsilon_{0}\omega},
\end{align}
which if we use that $\varepsilon_{1}=\varepsilon_{2}$ we can write
more simply as

\begin{equation}
N_{p}=\frac{\varepsilon_{1}}{\kappa_{1}}+\frac{i\sigma(\omega)}{\varepsilon_{0}\omega}=\frac{\varepsilon_{1}}{\kappa_{1}}-\frac{e^{2}}{4\pi\varepsilon_{0}}\frac{4E_{F}}{\hbar^{2}\omega^{2}}.
\end{equation}
In the electrostatic limit, we have $\kappa_{1}\approx\kappa_{2}\approx q$
and thus we can write for the case when $\varepsilon_{1}\neq\varepsilon_{2}$

\begin{equation}
N_{p}\approx\frac{\varepsilon_{1}}{2q}+\frac{\varepsilon_{2}}{2q}+i\frac{\sigma(\omega)}{\varepsilon_{0}\omega}=\frac{\bar{\varepsilon}}{q}+i\frac{\sigma(\omega)}{\varepsilon_{0}\omega}.
\end{equation}
We can further write this result using the dispersion relation of
equation (\ref{eq:disper}), and this way obtain

\begin{equation}
N_{p}\approx\frac{\bar{\varepsilon}}{q}-\frac{2\bar{\varepsilon}}{q}=-\frac{\bar{\varepsilon}}{q}
\end{equation}

\subsection{Construction of the dyadic Green's function\label{subsec:Construction-of-the}}

In the present appendix we note that the dyadic Green's function can
then be computed taking only into account the mode with $p-$polarization.
It reads

\begin{equation}
\bar{\bar{G}}(\boldsymbol{r},\boldsymbol{r}';\omega)\approx\int d^{2}\boldsymbol{q}\frac{\left[\boldsymbol{a}_{p}^{(1)}(\boldsymbol{r})\right]^{T}\otimes\left[\boldsymbol{a}_{p}^{(1)}(\boldsymbol{r}')\right]^{*}}{N_{p}\lambda_{p}^{(1)}}.
\end{equation}
In the limit $\kappa_{1}\approx q$ the polarization vectors can be
written as

\begin{align}
\boldsymbol{a}_{p}^{(1)}(\boldsymbol{r}) & =\frac{1}{\sqrt{q^{2}+\kappa_{1}^{2}}}\left(-\frac{iq_{x}\kappa_{1}}{q},-\frac{iq_{y}\kappa_{1}}{q},q\right)e^{i\boldsymbol{q}\cdot\boldsymbol{\rho}}e^{-\kappa_{1}z}\nonumber \\
 & \approx\frac{1}{q\sqrt{2}}\left(-iq_{x},-iq_{y},q\right)e^{i\boldsymbol{q}\cdot\boldsymbol{\rho}}e^{-qz},
\end{align}
where $\boldsymbol{\rho}=(x,y)$ and thus, evaluating the tensor product
explicitly, and then substituting the forms of $N_{p}$ and $\lambda_{p}^{(1)}$,
we have

\begin{align}
\bar{\bar{G}}(\boldsymbol{r},\boldsymbol{r}';\omega) & =\int d^{2}\boldsymbol{q}\frac{e^{i\boldsymbol{q}\cdot(\boldsymbol{\rho}'-\boldsymbol{\rho})}e^{-q(z+z')}}{2q^{2}N_{p}\lambda_{p}^{(1)}}\begin{pmatrix}-q_{x}^{2} & -q_{x}q_{y} & -iqq_{x}\\
-q_{x}q_{y} & -q_{y}^{2} & -iqq_{y}\\
-iqq_{x} & -iqq_{y} & q^{2}
\end{pmatrix}\nonumber \\
 & =-\frac{1}{2\bar{\varepsilon}}\int d^{2}\boldsymbol{q}\frac{q}{q^{2}}\frac{c^{2}e^{i\boldsymbol{q}\cdot(\boldsymbol{\rho}'-\boldsymbol{\rho})}e^{-q(z+z')}}{\omega^{2}-\left[\omega_{\text{spp}}(\boldsymbol{q})\right]^{2}}\begin{pmatrix}-q_{x}^{2} & -q_{x}q_{y} & -iqq_{x}\\
-q_{x}q_{y} & -q_{y}^{2} & -iqq_{y}\\
-iqq_{x} & -iqq_{y} & q^{2}
\end{pmatrix}.
\end{align}
The prefactor in the integral can be simplified. We introduce the
notation

\begin{equation}
C(q)=\frac{q}{q^{2}}\frac{c^{2}e^{-q(z+z')}}{\omega^{2}-\left[\omega_{\text{spp}}(\boldsymbol{q})\right]^{2}}=\frac{1}{q}\frac{e^{-q(z+z')}}{\omega^{2}/c^{2}-\frac{2\alpha E_{F}}{\bar{\varepsilon}\hbar c}q},
\end{equation}
and this way, the integral can be written as

\begin{equation}
\bar{\bar{G}}(\boldsymbol{r},\boldsymbol{r}';\omega)=-\frac{1}{2\bar{\varepsilon}}\int_{0}^{\infty}qdq\int_{0}^{2\pi}d\theta\ C(q)e^{i\boldsymbol{q}\cdot(\boldsymbol{\rho}'-\boldsymbol{\rho})}\begin{pmatrix}-q_{x}^{2} & -q_{x}q_{y} & -iqq_{x}\\
-q_{x}q_{y} & -q_{y}^{2} & -iqq_{y}\\
-iqq_{x} & -iqq_{y} & q^{2}
\end{pmatrix}.
\end{equation}
And defining $R\equiv|\boldsymbol{\rho}'-\boldsymbol{\rho}|$, we
have

\begin{equation}
\bar{\bar{G}}(\boldsymbol{r},\boldsymbol{r}';\omega)=\frac{1}{2\bar{\varepsilon}}\int_{0}^{\infty}dq\int_{0}^{2\pi}d\theta\ q^{3}C(q)e^{iqR\cos\left(\theta-\phi\right)}\begin{pmatrix}\cos^{2}\theta & \cos\theta\sin\theta & i\cos\theta\\
\cos\theta\sin\theta & \sin^{2}\theta & i\sin\theta\\
i\cos\theta & i\sin\theta & 1
\end{pmatrix},
\end{equation}
where $\boldsymbol{\rho}'-\boldsymbol{\rho}$ makes an angle $\phi$
with the $x-$axis. We therefore can evaluate the angular integrals

\begin{equation}
\int_{0}^{2\pi}d\theta e^{iqR\cos\left(\theta-\phi\right)}\times\begin{cases}
1 & =2\pi J_{0}(qR),\\
\cos^{2}\theta & =2\pi\left[J_{0}(qR)\cos^{2}\phi-\frac{J_{1}(qR)}{qR}\cos2\phi\right],\\
\sin^{2}\theta & =2\pi\left[J_{0}(qR)\sin^{2}\phi+\frac{J_{1}(qR)}{qR}\cos2\phi\right],\\
\cos\theta\sin\theta & =-2\pi J_{2}(qR)\cos\phi\sin\phi,\\
\cos\theta & =2\pi iJ_{1}(qR)\cos\phi,\\
\sin\theta & =2\pi iJ_{1}(qR)\sin\phi,
\end{cases}
\end{equation}
and now we have left to evaluate a series of integrals over the coordinate
$q$.

\subsection{Evaluation of the radial integrals\label{subsec:Evaluation-of-the}}

\begin{figure}[H]
\begin{centering}
\includegraphics[scale=0.6]{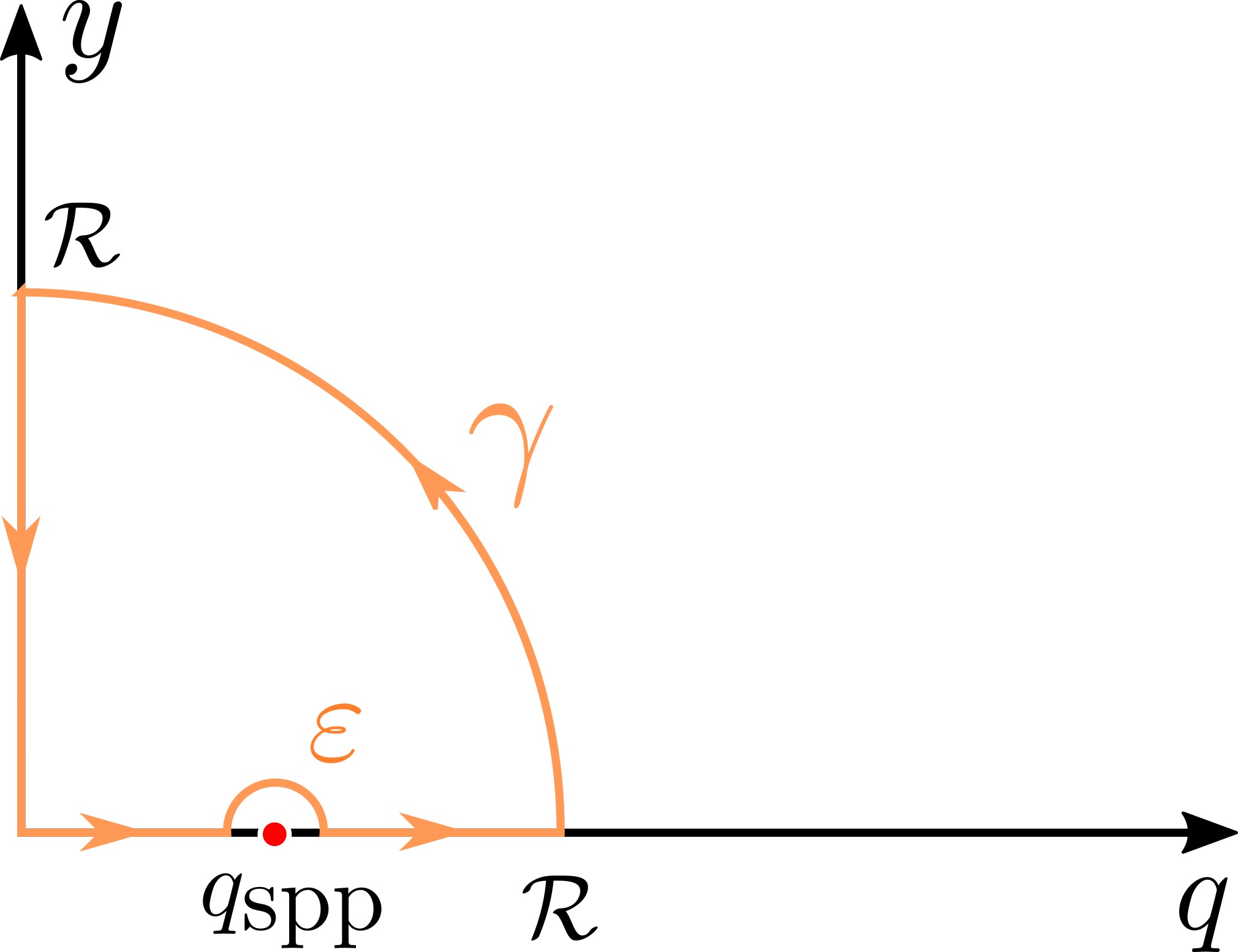} 
\par\end{centering}
\caption{Contour $\gamma$ over which we integrate equation (\ref{eq:Contour1})
for $i=1$ and $\nu=0,1$. We integrate over a large quarter circle
of radius $\mathcal{R}$ avoiding the pole with radius $\varepsilon$.
We will let $\mathcal{R}$ go to infinity to obtain the desired integral.\label{fig:Contour-over-1}}
\end{figure}

The angular integrals can be solved by adding a vanishing quantity
to the frequency, we will thus have, naming the matrix $\bar{\bar{\Theta}}(q,R,\phi)$
containing the previously calculated terms involving the Bessel functions,

\begin{equation}
\bar{\bar{G}}(\boldsymbol{r},\boldsymbol{r}';\omega)=\frac{c^{2}}{2\bar{\varepsilon}}\int_{0}^{\infty}dqq^{2}\frac{e^{-q(z+z')}}{\left(\omega+i0^{+}\right)^{2}-\left[\omega_{\text{spp}}(\boldsymbol{q})\right]^{2}}\bar{\bar{\Theta}}(q,R,\phi),
\end{equation}
where we have added a small part to the imaginary frequency, such
that we can use the Sokhotski--Plemelj identity (see appendix \ref{sec:The-Sokhotski-Plemelj-identity})
to regularize the integral

\begin{equation}
\frac{1}{\omega^{2}-\left[\omega_{\text{spp}}(\boldsymbol{q})\right]^{2}+i0^{+}}=\mathcal{P}\frac{1}{\omega^{2}-\left[\omega_{\text{spp}}(\boldsymbol{q})\right]^{2}}-i\pi\delta\left(\omega^{2}-\left[\omega_{\text{spp}}(\boldsymbol{q})\right]^{2}\right).
\end{equation}
We note that we have previously seen that in the case of a Drude conductivity,
$\omega_{\text{spp}}^{2}=A(\omega)q$. We now have to evaluate two
contributions to the integral, the first is the so called half-space
Hilbert transform, defined as

\begin{equation}
H(t)=\mathcal{P}\int_{0}^{\infty}d\tau\frac{u(\tau)}{t-\tau}.
\end{equation}
In our case, we will have integrals of the form

\begin{equation}
\mathcal{I}_{1}^{(1)}(\omega)=\mathcal{P}\int_{0}^{\infty}dq\frac{q^{2}e^{-q(z+z')}J_{\nu}(qR)}{\omega^{2}-A(\omega)q}.\label{eq:form1}
\end{equation}
It is useful to split this integral of a Bessel function into two
integrals involving the Hankel functions, such that

\begin{equation}
\mathcal{I}_{1}^{(1)}(\omega)=\frac{1}{2A(\omega)}\mathcal{P}\int_{0}^{\infty}dq\frac{q^{2}e^{-q(z+z')}\left[H_{\nu}^{(1)}(qR)+H_{\nu}^{(2)}(qR)\right]}{q_{\text{spp}}-q},
\end{equation}
where we have introduced the SSP wave vector $q_{\text{ssp}}=\omega^{2}/A(\omega)$.
A useful result is that for large $qR$ we can approximate the Hankel
functions as

\begin{equation}
\begin{cases}
H_{\nu}^{(1)}(qR)\approx\sqrt{\frac{2}{\pi qR}}e^{i(qR-\nu\pi/2-\pi/4)},\\
H_{\nu}^{(2)}(qR)\approx\sqrt{\frac{2}{\pi qR}}e^{-i(qR-\nu\pi/2-\pi/4)}.
\end{cases}
\end{equation}
The integral of the term in $H_{\nu}^{(i)}(qR)$ can be performed
in the complex plane, over a contour $\gamma$ such that

\begin{equation}
\oint_{\gamma}dz\frac{ze^{-z\zeta}H_{\nu}^{(i)}(zR)}{q_{\text{spp}}-z}=0,\label{eq:Contour1}
\end{equation}
and that the contour goes around the pole at $\omega^{2}/A(\omega)$.
Splitting the integral over the real line into the principal part
and the integral for $i=1$ and $\nu=0,1$ can be performed over the
small curve around the pole, and integrating over a quarter circle
of radius $\mathcal{R}$(see Fig. \ref{fig:Contour-over-1})

\begin{figure}[H]
\begin{centering}
\includegraphics[scale=0.6]{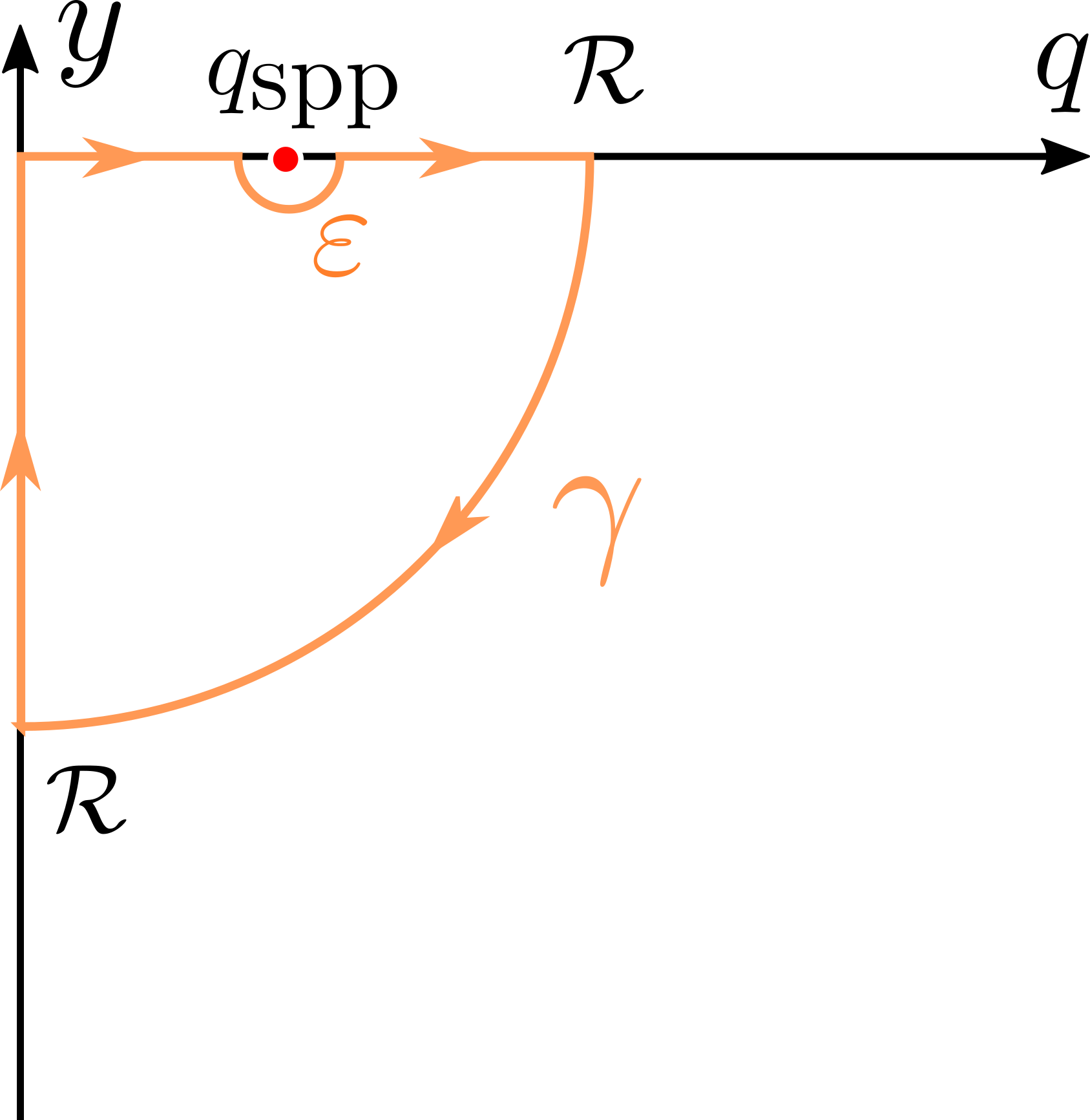} 
\par\end{centering}
\caption{Contour $\gamma$ over which we integrate equation (\ref{eq:Contour1})
for $i=2$ and $\nu=0,1$. We integrate over a large quarter circle
of radius $\mathcal{R}$ in the fourth quadrant, avoiding the pole
by moving along a semi-circle of radius $\varepsilon$ around it.
We will let $\mathcal{R}$ go to infinity to obtain the desired integral.\label{fig:Contour-over-2}}
\end{figure}

We have, for $i=1$

\begin{align}
\mathcal{P}\int_{0}^{\mathcal{R}}dq\frac{q^{2}e^{-q\zeta}H_{\nu}^{(1)}(qR)}{q_{\text{spp}}-q}+\int_{\text{pole}}\frac{dqq^{2}e^{-q\zeta}H_{\nu}^{(1)}(qR)}{q_{\text{spp}}-q}\nonumber \\
+\int_{0}^{\pi/2}d\theta\frac{-\mathcal{R}^{2}e^{2i\theta}\mathcal{R}e^{i\theta}e^{-\mathcal{R}e^{i\theta}\zeta}H_{\nu}^{(1)}(\mathcal{R}e^{i\theta}R)}{q_{\text{spp}}-\mathcal{R}e^{i\theta}}+\int_{\mathcal{R}}^{0}\frac{idy\ -y^{2}e^{-2iy\zeta}H_{\nu}^{(1)}(iyR)}{q_{\text{spp}}-q} & =0,
\end{align}
and thus, letting $\mathcal{R}\to\infty$, the integral over the arc
vanishes, and we are left with.

\begin{align}
\mathcal{P}\int_{0}^{\infty}dq\frac{q^{2}e^{-q\zeta}H_{\nu}^{(1)}(qR)}{q_{\text{spp}}-q}+\lim_{\varepsilon\to0}\int_{0}^{\pi}d\theta\frac{i\varepsilon e^{i\theta}}{\varepsilon e^{i\theta}}q_{\text{spp}}^{2}e^{-q_{\text{spp}}\zeta}H_{\nu}^{(1)}\left(q_{\text{spp}}R\right)\nonumber \\
-i\int_{0}^{\infty}dy\frac{y^{2}e^{-iy\zeta}H_{\nu}^{(1)}(iyR)}{q_{\text{spp}}-iy} & =0,
\end{align}
where we can now solve for the desired half-space Hilbert transform

\begin{align}
\mathcal{P}\int_{0}^{\infty}dq\frac{q^{2}e^{-q\zeta}H_{\nu}^{(1)}(qR)}{q_{\text{spp}}-q}= & -\int_{0}^{\pi}d\theta\frac{i\varepsilon e^{i\theta}}{\varepsilon e^{i\theta}}q_{\text{spp}}^{2}e^{-q_{\text{spp}}\zeta}H_{\nu}^{(1)}\left(q_{\text{spp}}R\right)\nonumber \\
 & +i\int_{0}^{\infty}dy\frac{y^{2}e^{-iy\zeta}H_{\nu}^{(1)}(iyR)}{q_{\text{spp}}-iy}\nonumber \\
= & -i\pi q_{\text{spp}}^{2}e^{-\omega^{2}\zeta/A}H_{\nu}^{(1)}\left(q_{\text{spp}}R\right)\nonumber \\
 & +i\int_{0}^{\infty}dy\frac{y^{2}e^{-iy\zeta}H_{\nu}^{(1)}(iyR)}{q_{\text{spp}}-iy}.
\end{align}
We have a part which we were able to calculate analytically for all
$R$, along with an integral we shall later approximate. For the second
Hankel function $H_{\nu}^{(2)}(qR)$, we must integrate over an analogous
contour in the fourth quadrant so that when we let the arc length
$\mathcal{R}$ go to infinity the corresponding integral vanishes
(see Fig. \ref{fig:Contour-over-2}).

\begin{figure}[H]
\begin{centering}
\includegraphics[scale=0.6]{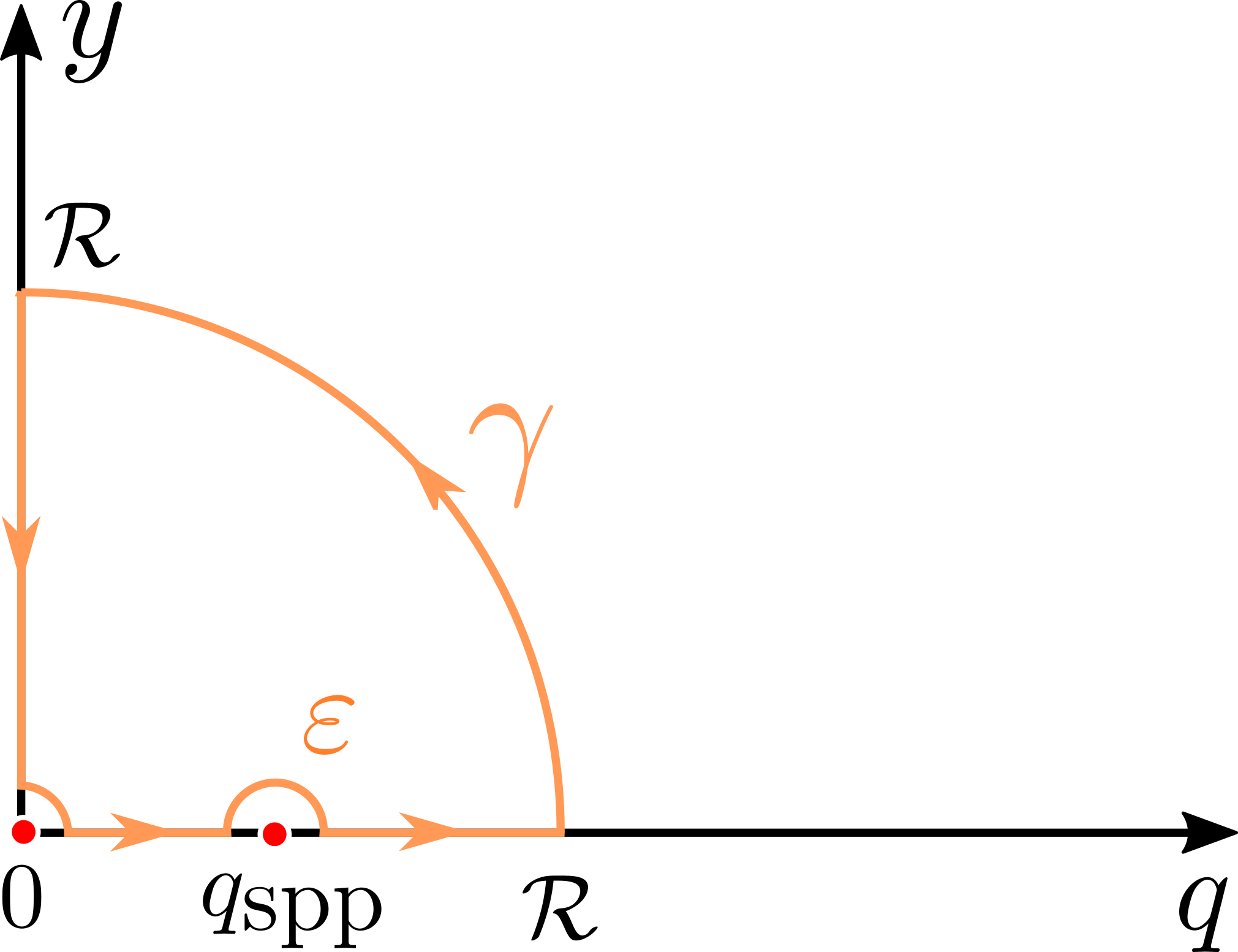} 
\par\end{centering}
\caption{Contour $\gamma$ over which we integrate equation (\ref{eq:Contour1})
for $i=1$ and $\nu=2$. We integrate over a large quarter circle
of radius $\mathcal{R}$ in the first quadrant, avoiding the pole
at $q_{\text{spp}}$ by moving along a semi-circle of radius $\varepsilon$
around it. We also avoid the pole at $z=0$, by moving another quarter
circle of radius $\varepsilon$ around it. We will let $\mathcal{R}$
go to infinity to obtain the desired integral.\label{fig:Contour-over-3}}
\end{figure}

We have

\begin{align*}
\mathcal{P}\int_{0}^{\infty}dq\frac{q^{2}e^{-q\zeta}H_{\nu}^{(2)}(qR)}{q_{\text{spp}}-q}+\lim_{\varepsilon\to0}\int_{\pi}^{0}d\theta\frac{i\varepsilon e^{i\theta}}{\varepsilon e^{i\theta}}q_{\text{spp}}^{2}e^{-q_{\text{spp}}\zeta}H_{\nu}^{(2)}\left(q_{\text{spp}}R\right)\\
+\int_{0}^{-\pi/2}d\theta\frac{-\mathcal{R}^{2}e^{2i\theta}\mathcal{R}e^{i\theta}e^{-\mathcal{R}e^{i\theta}\zeta}H_{\nu}^{(2)}(\mathcal{R}e^{i\theta}R)}{q_{\text{spp}}-\mathcal{R}e^{i\theta}}+\int_{-\infty}^{0}dy\frac{-iy^{2}e^{-iy\zeta}H_{\nu}^{(2)}(iyR)}{q_{\text{spp}}-q} & =0,
\end{align*}
which, proceeding as before, reduces to

\begin{align}
\mathcal{P}\int_{0}^{\infty}dq\frac{q^{2}e^{-q\zeta}H_{\nu}^{(2)}(qR)}{q_{\text{spp}}-q}= & \lim_{\varepsilon\to0}\int_{0}^{\pi}d\theta\frac{i\varepsilon e^{i\theta}}{\varepsilon e^{i\theta}}q_{\text{spp}}^{2}e^{-q_{\text{spp}}\zeta}H_{\nu}^{(2)}\left(q_{\text{spp}}R\right)\nonumber \\
 & +i\int_{0}^{\infty}dy\frac{y^{2}e^{-iy\zeta}H_{\nu}^{(2)}(iyR)}{q_{\text{spp}}-iy}\nonumber \\
= & i\pi q_{\text{spp}}^{2}e^{-q_{\text{spp}}\zeta}H_{\nu}^{(2)}\left(\frac{\omega^{2}}{A}R\right)\nonumber \\
 & +i\int_{-\infty}^{0}dy\frac{y^{2}e^{-iy\zeta}H_{\nu}^{(2)}(iyR)}{q_{\text{spp}}-iy}\nonumber \\
= & i\pi q_{\text{spp}}^{2}e^{-q_{\text{spp}}\zeta}H_{\nu}^{(2)}\left(q_{\text{spp}}R\right)\nonumber \\
 & +i\int_{0}^{\infty}dy\frac{y^{2}e^{iy\zeta}H_{\nu}^{(2)}(-iyR)}{q_{\text{spp}}+iy},
\end{align}
We now substitute the asymptotic form of the Hankel function for large
arguments in the integral that is left. We find, in the limit $q(z+z')\ll1$
and $qR\gg1$, the second integral vanishes and we are left with

\begin{equation}
\begin{cases}
\mathcal{P}\int_{0}^{\infty}dq\frac{q^{2}e^{-z\zeta}H_{\nu}^{(1)}(qR)}{q_{\text{spp}}-q}\approx-i\pi q_{\text{spp}}^{2}e^{-q_{\text{spp}}\zeta}H_{\nu}^{(1)}\left(q_{\text{spp}}R\right),\\
\mathcal{P}\int_{0}^{\infty}dq\frac{q^{2}e^{-z\zeta}H_{\nu}^{(2)}(qR)}{q_{\text{spp}}-q}\approx i\pi q_{\text{spp}}^{2}e^{-q_{\text{spp}}\zeta}H_{\nu}^{(2)}\left(q_{\text{spp}}R\right).
\end{cases}
\end{equation}
This means that we have the result

\begin{align}
\mathcal{I}_{1}^{(1)}(\omega)=\frac{1}{A(\omega)}\mathcal{P}\int_{0}^{\infty}dq\frac{q^{2}e^{-q(z+z')}J_{\nu}(qR)}{\omega^{2}-A(\omega)q} & \approx i\pi\frac{q_{\text{spp}}^{2}}{2A(\omega)}e^{-q_{\text{spp}}\zeta}\left[H_{\nu}^{(2)}\left(q_{\text{spp}}R\right)-H_{\nu}^{(1)}\left(q_{\text{spp}}R\right)\right]\nonumber \\
 & =\pi\frac{q_{\text{spp}}^{2}}{A(\omega)}e^{-q_{\text{spp}}\zeta}Y_{\nu}\left(q_{\text{spp}}R\right),
\end{align}
where $Y_{\nu}$ is a Bessel function of the second kind. These results
are valid for $\nu=0,1$. For $\nu=2$ we must also consider a pole
at the origin. The contour is thus a little bit different as we also
need to consider an additional small quarter turn to avoid the pole
at $z=0$ (see Fig. \ref{fig:Contour-over-3}).

Here, our integral will read

\begin{align}
\mathcal{P}\int_{0}^{\infty}dq\frac{q^{2}e^{-q\zeta}H_{\nu}^{(1)}(qR)}{q_{\text{spp}}-q}= & -i\pi e^{-q_{\text{spp}}\zeta}H_{\nu}^{(1)}\left(q_{\text{spp}}R\right)\nonumber \\
 & +i\int_{0}^{\infty}dy\frac{y^{2}e^{iy\zeta}H_{\nu}^{(1)}(iyR)}{q_{\text{spp}}-iy}\nonumber \\
 & -\int_{pole_{0}}dq\frac{q^{2}}{q_{\text{spp}}-q}\left(-\frac{4i}{\pi q^{2}R^{2}}-\frac{i}{\pi}\right),
\end{align}
where we have expanded the Hankel function for small $qR$ as

\begin{equation}
H_{2}^{(1)}(qR)\approx-\frac{4i}{\pi q^{2}R^{2}}-\frac{i}{\pi}.
\end{equation}
The final term can be rewritten as

\begin{align}
\int_{pole_{0}}dq\frac{q^{2}}{q_{\text{spp}}-q}\left(-\frac{4i}{\pi q^{2}R^{2}}-\frac{i}{\pi}\right) & =\int_{pole_{0}}dq\frac{1}{q_{\text{spp}}-q}\left(-\frac{4i}{\pi R^{2}}-q^{2}\frac{i}{\pi}\right)\nonumber \\
 & =\lim_{\varepsilon\to0}-\int_{0}^{\pi/2}i\varepsilon e^{i\theta}d\theta\frac{1}{q_{\text{spp}}-\varepsilon e^{i\theta}}\left(-\frac{4i}{\pi R^{2}}-\varepsilon e^{i\theta}\frac{i}{\pi}\right)\nonumber \\
 & =-\frac{\pi}{2}\frac{4}{\pi R^{2}}\nonumber \\
 & =-\frac{2}{R^{2}},
\end{align}
thus, for $H_{2}^{(1)}(qR)$ the integral contains an additional term.

\begin{equation}
\mathcal{P}\int_{0}^{\infty}dq\frac{q^{2}e^{-q\zeta}H_{2}^{(1)}(qR)}{q_{\text{spp}}-q}=-i\pi q_{\text{spp}}^{2}e^{-q_{\text{spp}}\zeta}H_{2}^{(1)}\left(q_{\text{spp}}R\right)+\frac{2}{R^{2}}.
\end{equation}
The analogous result holds for $H_{2}^{(1)}(qR)$ since we can expand
it as

\begin{equation}
H_{2}^{(2)}(qR)\approx+\frac{4i}{\pi q^{2}R^{2}}+\frac{i}{\pi},
\end{equation}
and therefore have

\begin{equation}
\mathcal{P}\int_{0}^{\infty}dq\frac{q^{2}e^{-q\zeta}H_{2}^{(2)}(qR)}{q_{\text{spp}}-q}=-i\pi q_{\text{spp}}^{2}e^{-q_{\text{spp}}\zeta}H_{2}^{(1)}\left(q_{\text{spp}}R\right)+\frac{2}{R^{2}}
\end{equation}
In this case we would have

\begin{align}
\mathcal{I}_{1}^{(1)}(\omega) & =\frac{1}{A(\omega)}\mathcal{P}\int_{0}^{\infty}dq\frac{q^{2}e^{-q(z+z')}J_{2}(qR)}{\omega^{2}-A(\omega)q},\nonumber \\
 & \approx i\pi\frac{q_{\text{spp}}^{2}}{2A(\omega)}e^{-q_{\text{spp}}\zeta}\left[H_{2}^{(2)}\left(q_{\text{spp}}R\right)+\frac{2}{R^{2}}-H_{2}^{(1)}\left(q_{\text{spp}}R\right)+\frac{2}{R^{2}}\right]\nonumber \\
 & =\pi\frac{q_{\text{spp}}^{2}}{A(\omega)}e^{-q_{\text{spp}}\zeta}Y_{2}\left(q_{\text{spp}}R\right)+\frac{2}{R^{2}}\nonumber \\
 & =\pi\frac{q_{\text{spp}}^{2}}{A(\omega)}e^{-q_{\text{spp}}\zeta}\left[Y_{2}\left(q_{\text{spp}}R\right)+e^{q_{\text{spp}}\zeta}\frac{2A(\omega)}{\pi q_{\text{spp}}R^{2}}\right].
\end{align}
We have thus evaluated the principal value of the Sokhotski--Plemelj
identity, we have still to derive the \textbf{$\delta-$}function
part. This part is much simpler, however, since we have

\begin{equation}
\delta(\omega^{2}-A(\omega)q)=\frac{1}{A(\omega)}\delta(q-q_{\text{spp}}),
\end{equation}
and thus

\begin{align}
\mathcal{I}_{2}^{(1)}(\omega) & =-i\pi\frac{1}{A(\omega)}\int_{0}^{\infty}dqq^{2}e^{-q\zeta}J_{\nu}(qR)\delta(q-q_{\text{spp}})\nonumber \\
 & =-i\pi\frac{q_{\text{spp}}^{2}}{A(\omega)}e^{-q_{\text{spp}}\zeta}J_{\nu}\left(q_{\text{spp}}R\right).
\end{align}
The total radial integral is therefore

\begin{align}
\mathcal{I}^{(1)}(\omega) & =\mathcal{I}_{1}^{(1)}(\omega)+\mathcal{I}_{2}^{(1)}(\omega)\nonumber \\
 & \approx\pi\frac{q_{\text{spp}}^{2}}{A(\omega)}e^{-\omega^{2}\zeta/A(\omega)}\left[Y_{\nu}\left(q_{\text{spp}}R\right)-iJ_{\nu}\left(q_{\text{spp}}R\right)\right].
\end{align}
This is an approximate solution to the integrals of the form of equation
(\ref{eq:form1}), but we still have to solve integrals of the form

\begin{align}
\mathcal{I}_{1}^{(2)}(\omega) & =\mathcal{P}\int_{0}^{\infty}dq\frac{q^{2}e^{-q\zeta}\left[J_{\nu}(qR)\right]/qR}{\omega^{2}-A(\omega)q}\nonumber \\
 & =\frac{1}{R}\mathcal{P}\int_{0}^{\infty}dq\frac{qe^{-q\zeta}J_{\nu}(qR)}{\omega^{2}-A(\omega)q}\nonumber \\
 & =\frac{1}{RA(\omega)}\mathcal{P}\int_{0}^{\infty}dq\frac{qe^{-q\zeta}J_{\nu}(qR)}{q_{\text{spp}}-q}.
\end{align}
We proceed in the same manner, splitting the Bessel function into
Hankel functions and integrating over contours in the complex plane.
We have, for the first Hankel function

\begin{align}
\mathcal{P}\int_{0}^{\infty}dq\frac{qe^{-q\zeta}H_{\nu}^{(1)}(qR)}{q_{\text{spp}}-q}+\int_{\text{pole}}dq\frac{qe^{-q\zeta}H_{\nu}^{(1)}(qR)}{q_{\text{spp}}-q}\nonumber \\
+\int_{0}^{\pi/2}d\theta\frac{i\mathcal{R}e^{i\theta}\mathcal{R}e^{i\theta}e^{-\mathcal{R}e^{i\theta}\zeta}H_{\nu}^{(1)}(\mathcal{R}e^{i\theta}R)}{q_{\text{spp}}-\mathcal{R}e^{i\theta}}-\int_{\infty}^{0}dy\frac{ye^{-iy}H_{\nu}^{(1)}(iyR)}{q_{\text{spp}}-q} & =0.
\end{align}
Like before, the integral over the arc vanishes and we are left with

\begin{align}
\mathcal{P}\int_{0}^{\infty}dq\frac{qe^{-q\zeta}H_{\nu}^{(1)}(qR)}{q_{\text{spp}}-q}=- & \lim_{\varepsilon\to0}\int_{0}^{\pi}d\theta\frac{i\varepsilon e^{i\theta}}{\varepsilon e^{i\theta}}q_{\text{spp}}e^{-q_{\text{spp}}\zeta}H_{\nu}^{(1)}\left(q_{\text{spp}}R\right)\nonumber \\
 & -\int_{0}^{\infty}dy\frac{ye^{-iy\zeta}H_{\nu}^{(1)}(iyR)}{q_{\text{spp}}-iy}\nonumber \\
= & -i\pi q_{\text{spp}}e^{-\omega^{2}\zeta/A(\omega)}H_{\nu}^{(1)}\left(q_{\text{spp}}R\right)\nonumber \\
 & -\int_{\infty}^{0}dy\frac{ye^{-iy\zeta}H_{\nu}^{(1)}(iyR)}{q_{\text{spp}}-iy}\nonumber \\
= & -i\pi q_{\text{spp}}e^{-\omega^{2}\zeta/A(\omega)}H_{\nu}^{(1)}\left(q_{\text{spp}}R\right)\nonumber \\
 & +\int_{0}^{\infty}dy\frac{ye^{iy\zeta}H_{\nu}^{(1)}(iyR)}{q_{\text{spp}}-iy}.
\end{align}
Using the same considerations as in the first case, we are left with
the approximate result

\begin{equation}
\mathcal{P}\int_{0}^{\infty}dq\frac{e^{-q\zeta}H_{\nu}^{(1)}(qR)}{q_{\text{spp}}-q}\approx-i\pi q_{\text{spp}}e^{-q_{\text{spp}}\zeta}H_{\nu}^{(1)}\left(q_{\text{spp}}R\right),
\end{equation}
and similarly for $H_{\nu}^{(2)}(qR)$,

\begin{equation}
\mathcal{P}\int_{0}^{\infty}dq\frac{e^{-q\zeta}H_{\nu}^{(2)}(qR)}{q_{\text{spp}}-q}\approx i\pi q_{\text{spp}}e^{-q_{\text{spp}}\zeta}H_{\nu}^{(2)}\left(q_{\text{spp}}R\right),
\end{equation}
and finally for the delta function terms

\begin{equation}
\mathcal{I}_{2}^{(2)}(\omega)=-\frac{i\pi}{R}\int_{0}^{\infty}dqqe^{-q\zeta}J_{\nu}(qR)\delta(\omega^{2}-A(\omega)q)=-\frac{i\pi}{RA(\omega)}q_{\text{spp}}e^{-q_{\text{spp}}\zeta}J_{\nu}\left(q_{\text{spp}}R\right).
\end{equation}
We therefore can calculate the aforementioned integral as

\begin{align}
\mathcal{I}^{(2)} & =\mathcal{I}_{1}^{(2)}+\mathcal{I}_{2}^{(2)}\nonumber \\
 & =\frac{\pi}{RA(\omega)}q_{\text{spp}}e^{-q_{\text{spp}}\zeta}\left[Y_{\nu}\left(q_{\text{spp}}R\right)-iJ_{\nu}\left(q_{\text{spp}}R\right)\right].
\end{align}
Having calculated all these integrals, all we have left is to calculate
the Green's tensor by putting together all the results. We shall do
this in the next section of this appendix.

\subsection{Putting the results together\label{subsec:Putting-the-results}}

Here, we collect the results of the previous Appendices and note that
we now have all the results we need to construct the Green's function.
It can be written as

\begin{align}
\bar{\bar{G}}(\boldsymbol{r},\boldsymbol{r}';\omega)=c^{2}\frac{\pi^{2}}{\bar{\varepsilon}}e^{-q_{\text{spp}}(z+z')}\frac{q_{\text{spp}}^{2}}{A(\omega)}\bar{\bar{M}}(q_{\text{ssp}},R,\phi)\label{eq:GrapheneGreensFunction-1}
\end{align}
where the matrix $\bar{\bar{M}}(q,\phi)$ has diagonal elements

\begin{equation}
\begin{cases}
M_{11}=\left[Y_{0}\left(q_{\text{spp}}R\right)-iJ_{0}\left(q_{\text{spp}}R\right)\right]\cos^{2}\phi-\frac{1}{Rq_{\text{ssp}}}\left[Y_{1}\left(q_{\text{spp}}R\right)-iJ_{1}\left(q_{\text{spp}}R\right)\right]\cos2\phi\\
M_{22}=\left[Y_{0}\left(q_{\text{spp}}R\right)-iJ_{0}\left(q_{\text{spp}}R\right)\right]\sin^{2}\phi+\frac{1}{Rq_{\text{ssp}}}\left[Y_{1}\left(q_{\text{spp}}R\right)-iJ_{1}\left(q_{\text{spp}}R\right)\right]\cos2\phi\\
M_{33}=\left[Y_{0}\left(q_{\text{spp}}R\right)-iJ_{0}\left(q_{\text{spp}}R\right)\right]
\end{cases}
\end{equation}
and off-diagonal elements

\begin{equation}
\begin{cases}
M_{12}=M_{21}=-\left[\left[Y_{2}\left(q_{\text{spp}}R\right)-iJ_{2}\left(q_{\text{spp}}R\right)\right]+\frac{2}{\pi q_{\text{spp}}R^{2}}e^{q_{\text{spp}}(z+z')}\frac{A(\omega)}{q_{\text{spp}}}\right]\cos\phi\sin\phi\\
M_{13}=M_{31}=-\left[Y_{1}\left(q_{\text{spp}}R\right)-iJ_{1}\left(q_{\text{spp}}R\right)\right]\cos\phi\\
M_{23}=M_{32}=-\left[Y_{1}\left(q_{\text{spp}}R\right)-iJ_{1}\left(q_{\text{spp}}R\right)\right]\sin\phi
\end{cases},
\end{equation}
where $A(\omega)$ is defined as before by

\begin{equation}
\omega_{\text{spp}}^{2}=A(\omega)q
\end{equation}
In the case of a Drude conductivity, this coefficient has the form

\begin{equation}
A(\omega)\approx\frac{2\alpha cE_{F}}{\hbar\bar{\varepsilon}},
\end{equation}
but more generally, it is defined as

\begin{equation}
A(\omega)\equiv-i\frac{\omega\sigma(\omega)}{2\bar{\varepsilon}\varepsilon_{0}},
\end{equation}
or if we let $\sigma(\omega)=\sigma_{1}(\omega)+i\sigma_{2}(\omega)$,
we can write more explicitly

\begin{align}
A(\omega) & =\frac{1}{2\bar{\varepsilon}\varepsilon_{0}}\left[-i\omega\sigma_{1}(\omega)+\omega\sigma_{2}(\omega)\right]\nonumber \\
 & \approx\frac{\omega\sigma_{2}(\omega)}{2\bar{\varepsilon}\varepsilon_{0}},
\end{align}
if only the imaginary part of the conductivity is considered.

\subsection{Integrals of products of Green's tensors\label{sec:Integrals-of-products}}

In this appendix we derive an important result which we shall use
quite often, regarding the integral of a product of Green's tensors.
Let us consider 
\begin{equation}
\hat{\boldsymbol{A}}(\boldsymbol{r},t)=\mu_{0}\int_{0}^{\infty}d\omega e^{-i\omega t}\int d\boldsymbol{r}^{\prime}\bar{\bar{G}}(\boldsymbol{r},\boldsymbol{r}^{\prime};\omega)\cdot\hat{\boldsymbol{J}}_{N}(\boldsymbol{r}',\omega)+{\rm H.c.}
\end{equation}
and

\begin{equation}
\hat{\boldsymbol{E}}(\boldsymbol{r},t)=-i\mu_{0}\int_{0}^{\infty}d\nu e^{i\nu t}\nu\int d\boldsymbol{r}^{\prime}\hat{\boldsymbol{J}}_{N}^{\dagger}(\boldsymbol{r}^{\prime},\nu)\cdot\bar{\bar{G}}^{\dagger}(\boldsymbol{r},\boldsymbol{r}^{\prime};\nu)+{\rm H.c.}
\end{equation}
and let us make the commutator 
\begin{align}
[\hat{\boldsymbol{A}}(\boldsymbol{r}_{1},t),-\epsilon_{0}\hat{\boldsymbol{E}}(\boldsymbol{r}_{2},t)] & =i\mu_{0}^{2}\epsilon_{0}\int_{0}^{\infty}d\omega e^{-i\omega t}\int_{0}^{\infty}d\nu e^{i\nu t}\nu\int d\boldsymbol{r}^{\prime}\int d\boldsymbol{r}^{\prime\prime}\bar{\bar{G}}(\boldsymbol{r}_{1},\boldsymbol{r}^{\prime};\omega)[\hat{\boldsymbol{J}}_{N}(\boldsymbol{r^{\prime}},\omega),\hat{\boldsymbol{J}}_{N}^{\dagger}(\boldsymbol{r}^{\prime\prime},\nu)\bar{\bar{G}}^{\dagger}(\boldsymbol{r}_{2},\boldsymbol{r}^{\prime\prime};\nu)]\nonumber \\
 & =i\mu_{0}^{2}\epsilon_{0}\int_{0}^{\infty}d\omega e^{-i\omega t}{\cal C}\omega\sigma_{1}(\omega)\int d\boldsymbol{r}^{\prime}\bar{\bar{G}}(\boldsymbol{r}_{1},\boldsymbol{r}^{\prime};\omega)\bar{\bar{G}}^{\dagger}(\boldsymbol{r}_{2},\boldsymbol{r}^{\prime};\omega)\nonumber \\
 & =i\mu_{0}^{2}\epsilon_{0}\int_{0}^{\infty}d\omega e^{-i\omega t}{\cal C}\omega\sigma_{1}(\omega)\int d\boldsymbol{r}^{\prime}\bar{\bar{G}}(\boldsymbol{r}_{1},\boldsymbol{r}^{\prime};\omega)\bar{\bar{G}}^{\ast}(\boldsymbol{r}^{\prime},\boldsymbol{r}_{2};\omega),\label{eq:commutator_symmetry}
\end{align}
where we have used the reciprocity relation $[\bar{\bar{G}}^{\ast}(\boldsymbol{r},\boldsymbol{r}^{\prime};\omega)]^{T}=\bar{\bar{G}}^{\dagger}(\boldsymbol{r},\boldsymbol{r}^{\prime};\omega)=\bar{\bar{G}}^{\ast}(\boldsymbol{r}^{\prime},\boldsymbol{r};\omega)$
. To compute this integral we used Eq. (\ref{eq:LongDeltaCommutator})
. Let us now simplify the previous equation. 
\begin{equation}
-\nabla_{\boldsymbol{r}_{1}}\times\nabla_{\boldsymbol{r}_{1}}\times\bar{\bar{G}}(\boldsymbol{r}_{1},\boldsymbol{r}^{\prime};\omega)+\frac{\omega^{2}}{c^{2}}\epsilon(z)\bar{\bar{G}}(\boldsymbol{r}_{1},\boldsymbol{r}^{\prime};\omega)=-\mathds{1}\delta(\boldsymbol{r}_{1}-\boldsymbol{r}^{\prime}),\label{eq:Green_function_definition-again}
\end{equation}
We start with the Helmholtz equation and multiply from the left by
$\bar{\bar{G}}^{\ast}(\boldsymbol{r},\boldsymbol{r}_{1};\omega)$
and integrating in $\boldsymbol{r}_{1}$ it follows 
\begin{equation}
\int d\boldsymbol{r}_{1}\bar{\bar{G}}^{\ast}(\boldsymbol{r},\boldsymbol{r}_{1};\omega)[-\nabla_{\boldsymbol{r}_{1}}\times\nabla_{\boldsymbol{r}_{1}}\times\bar{\bar{G}}(\boldsymbol{r}_{1},\boldsymbol{r}^{\prime};\omega)]+\frac{\omega^{2}}{c^{2}}\int d\boldsymbol{r}_{1}\epsilon(z)\bar{\bar{G}}^{\ast}(\boldsymbol{r},\boldsymbol{r}_{1};\omega)\bar{\bar{G}}(\boldsymbol{r}_{1},\boldsymbol{r}^{\prime};\omega)=-\bar{\bar{G}}^{\ast}(\boldsymbol{r},\boldsymbol{r}^{\prime};\omega).
\end{equation}
Taking the complex conjugate of the Helmholtz equation (\ref{eq:Green_function_definition-again})
we have

\begin{equation}
-\nabla_{\boldsymbol{r}_{1}}\times\nabla_{\boldsymbol{r}_{1}}\times\bar{\bar{G}}^{\ast}(\boldsymbol{r}_{1},\boldsymbol{r}^{\prime};\omega)+\frac{\omega^{2}}{c^{2}}\epsilon^{\ast}(z)\bar{\bar{G}}^{\ast}(\boldsymbol{r}_{1},\boldsymbol{r^{\prime}};\omega)=-\mathds{1}\delta(\boldsymbol{r}_{1}-\boldsymbol{r}^{\prime}),
\end{equation}
Multiplying from the left by $\bar{\bar{G}}(\boldsymbol{r},\boldsymbol{r}_{1};\omega)$
and integrating over $\boldsymbol{r}_{1}$ yet again, we obtain

\begin{equation}
\int d\boldsymbol{r}_{1}\bar{\bar{G}}(\boldsymbol{r},\boldsymbol{r}_{1};\omega)\left[-\nabla_{\boldsymbol{r}_{1}}\times\nabla_{\boldsymbol{r}_{1}}\times\bar{\bar{G}}^{\ast}(\boldsymbol{r}_{1},\boldsymbol{r^{\prime}};\omega)\right]+\frac{\omega^{2}}{c^{2}}\int d\boldsymbol{r}_{1}\epsilon^{\ast}(z)\bar{\bar{G}}(\boldsymbol{r},\boldsymbol{r}_{1};\omega)\bar{\bar{G}}^{\ast}(\boldsymbol{r}_{1},\boldsymbol{r^{\prime}};\omega)=-\bar{\bar{G}}(\boldsymbol{r},\boldsymbol{r}^{\prime};\omega).
\end{equation}
Making an integration by parts in the first term, we find after subtracting
\begin{equation}
\frac{\omega^{2}}{c^{2}}\int d\boldsymbol{r}_{1}[-\epsilon^{\ast}(z)\bar{\bar{G}}(\boldsymbol{r},\boldsymbol{r}_{1};\omega)\bar{\bar{G}}^{\ast}(\boldsymbol{r}_{1},\boldsymbol{r}^{\prime};\omega)+\epsilon(z)\bar{\bar{G}}^{\ast}(\boldsymbol{r},\boldsymbol{r}_{1};\omega)\bar{\bar{G}}(\boldsymbol{r}_{1},\boldsymbol{r}^{\prime};\omega)]=\bar{\bar{G}}(\boldsymbol{r},\boldsymbol{r}^{\prime};\omega)-\bar{\bar{G}}^{\ast}(\boldsymbol{r},\boldsymbol{r}^{\prime};\omega)
\end{equation}
which can be written as

\begin{equation}
\frac{\omega^{2}}{c^{2}}\int d\boldsymbol{r}_{1}[-\epsilon^{\ast}(z)\bar{\bar{G}}(\boldsymbol{r},\boldsymbol{r}_{1};\omega)\bar{\bar{G}}^{\ast}(\boldsymbol{r}_{1},\boldsymbol{r}^{\prime};\omega)+\epsilon(z)\bar{\bar{G}}(\boldsymbol{r}^{\prime},\boldsymbol{r}_{1};\omega)\bar{\bar{G}}^{\ast}(\boldsymbol{r}_{1},\boldsymbol{r};\omega)]=2\textrm{Im}[\bar{\bar{G}}(\boldsymbol{r},\boldsymbol{r}^{\prime};\omega)]
\end{equation}
or 
\begin{equation}
\frac{\omega^{2}}{c^{2}}\int d\boldsymbol{r}_{1}^{\prime}\textrm{Im}[\epsilon(z)]\bar{\bar{G}}(\boldsymbol{r},\boldsymbol{r}_{1};\omega)\bar{\bar{G}}^{\ast}(\boldsymbol{r}_{1},\boldsymbol{r}^{\prime};\omega)=\textrm{Im}[\bar{\bar{G}}(\boldsymbol{r},\boldsymbol{r}^{\prime};\omega)].\label{eq:Green_useful_relation}
\end{equation}
where we have used the reciprocity relation and the property below

\begin{align}
A_{ik} & =\sum_{j}\bar{\bar{G}}_{ij}^{\ast}(\boldsymbol{r},\boldsymbol{r}_{1};\omega)\bar{\bar{G}}_{jk}(\boldsymbol{r}_{1},\boldsymbol{r}^{\prime};\omega)=\sum_{j}[\bar{\bar{G}}_{ji}^{\ast}(\boldsymbol{r},\boldsymbol{r}_{1};\omega)]^{T}[\bar{\bar{G}}_{kj}(\boldsymbol{r}_{1},\boldsymbol{r}^{\prime};\omega)]^{T},\\
 & =\sum_{j}[\bar{\bar{G}}_{kj}(\boldsymbol{r}_{1},\boldsymbol{r}^{\prime};\omega)]^{T}[\bar{\bar{G}}_{ji}^{\ast}(\boldsymbol{r},\boldsymbol{r}_{1};\omega)]^{T}\\
 & \Rightarrow[\bar{\bar{G}}(\boldsymbol{r}_{1},\boldsymbol{r}^{\prime};\omega)]^{T}[\bar{\bar{G}}^{\ast}(\boldsymbol{r},\boldsymbol{r}_{1};\omega)]^{T}=\bar{\bar{G}}(\boldsymbol{r}^{\prime},\boldsymbol{r}_{1};\omega)\bar{\bar{G}}^{\ast}(\boldsymbol{r}_{1},\boldsymbol{r};\omega)
\end{align}
We also note the following property 
\begin{equation}
G_{ij}(\boldsymbol{\boldsymbol{r}},\boldsymbol{r}^{\prime})=G_{ji}(\boldsymbol{r}^{\prime},\boldsymbol{\boldsymbol{r}})\Leftrightarrow\bar{\bar{G}}(\boldsymbol{\boldsymbol{r}},\boldsymbol{r}^{\prime})=[\bar{\bar{G}}(\boldsymbol{\boldsymbol{r}}^{\prime},\boldsymbol{r})]^{T}.
\end{equation}

\section{Details of the quantization of the plasmon field in graphene\label{sec:Details-of-the}}

\subsection{Transverse and longitudinal $\delta-$functions\label{subsec:Transverse-and-Longitudinal}}

We start this appendix with a small introduction to transverse and
longitudinal $\delta-$functions following \citet{Milonni2019}. Consider
the vector field

\begin{equation}
\boldsymbol{G}(\boldsymbol{r})=\frac{1}{4\pi}\nabla\times\nabla\times\int d^{3}\boldsymbol{r}'\frac{\boldsymbol{F}(\boldsymbol{r}')}{|\boldsymbol{r}-\boldsymbol{r}'|},
\end{equation}
obtained from another vector field $\boldsymbol{F}(\boldsymbol{r}).$We
may write this as

\begin{equation}
4\pi\boldsymbol{G}(\boldsymbol{r})=\nabla\int d^{3}\boldsymbol{r}'\boldsymbol{F}(\boldsymbol{r})'\cdot\nabla\frac{1}{|\boldsymbol{r}-\boldsymbol{r}'|}+4\pi\boldsymbol{F}(\boldsymbol{r}),
\end{equation}
where we have used that $\nabla\times\nabla\times\boldsymbol{C}=\nabla(\nabla\cdot\boldsymbol{C})-\nabla^{2}\boldsymbol{C}$
and also that

\begin{equation}
\nabla^{2}\frac{1}{|\boldsymbol{r}-\boldsymbol{r}'|}=-4\pi\delta(\boldsymbol{r}-\boldsymbol{r}').
\end{equation}
By the chain rule, we have

\begin{align}
4\pi\boldsymbol{G}(\boldsymbol{r}) & =-\nabla\int d^{3}\boldsymbol{r}'\boldsymbol{F}(\boldsymbol{r})'\cdot\nabla'\frac{1}{|\boldsymbol{r}-\boldsymbol{r}'|}+4\pi\boldsymbol{F}(\boldsymbol{r})\nonumber \\
 & =\nabla\int d^{3}\boldsymbol{r}'\frac{\nabla'\cdot\boldsymbol{F}(\boldsymbol{r}')}{|\boldsymbol{r}-\boldsymbol{r}'|}+4\pi\boldsymbol{F}(\boldsymbol{r}),
\end{align}
where in the last step we have integrated by parts and assumed that
$\boldsymbol{F}(\boldsymbol{r})\to0$ as $r\to\infty$. This way we
can write

\begin{equation}
\boldsymbol{F}(\boldsymbol{r})=\boldsymbol{G}(\boldsymbol{r})-\frac{1}{4\pi}\nabla\int d^{3}\boldsymbol{r}'\frac{\nabla'\cdot\boldsymbol{F}(\boldsymbol{r}')}{|\boldsymbol{r}-\boldsymbol{r}'|}.
\end{equation}
Substituting the definition of $\boldsymbol{G}(\boldsymbol{r})$ yet
again, we have

\begin{equation}
\boldsymbol{F}(\boldsymbol{r})=\frac{1}{4\pi}\nabla\times\nabla\times\int d^{3}\boldsymbol{r}'\frac{\boldsymbol{F}(\boldsymbol{r}')}{|\boldsymbol{r}-\boldsymbol{r}'|}-\frac{1}{4\pi}\nabla\int d^{3}\boldsymbol{r}'\frac{\nabla'\cdot\boldsymbol{F}(\boldsymbol{r}')}{|\boldsymbol{r}-\boldsymbol{r}'|},
\end{equation}
and have thus decomposed the field $\boldsymbol{F}$ into a longitudinal
and transverse part

\begin{equation}
\begin{cases}
\boldsymbol{F}^{\perp}\equiv\frac{1}{4\pi}\nabla\times\nabla\times\int d^{3}\boldsymbol{r}'\frac{\boldsymbol{F}(\boldsymbol{r}')}{|\boldsymbol{r}-\boldsymbol{r}'|},\\
\boldsymbol{F}^{\parallel}\equiv-\frac{1}{4\pi}\nabla\int d^{3}\boldsymbol{r}'\frac{\nabla'\cdot\boldsymbol{F}(\boldsymbol{r}')}{|\boldsymbol{r}-\boldsymbol{r}'|},
\end{cases}
\end{equation}
since $\nabla\cdot\boldsymbol{F}^{\perp}=0$ and $\nabla\times\boldsymbol{F}^{\parallel}=0$.
We can also obtain the longitudinal and transverse components of the
field by using

\begin{equation}
\begin{cases}
F_{i}^{\perp}(\boldsymbol{r})=\sum_{j}\int d^{3}\boldsymbol{r}'\delta_{ij}^{\perp}(\boldsymbol{r}-\boldsymbol{r}')F_{j}(\boldsymbol{r}')\\
F_{i}^{\parallel}(\boldsymbol{r})=\sum_{j}\int d^{3}\boldsymbol{r}'\delta_{ij}^{\parallel}(\boldsymbol{r}-\boldsymbol{r}')F_{j}(\boldsymbol{r}')
\end{cases},\label{eq:LongTransdeltaF}
\end{equation}
where we have introduced the definition of the transverse and longitudinal
$\delta-$functions as

\begin{equation}
\delta_{ij}^{\perp}(\boldsymbol{r})\equiv\frac{1}{(2\pi)^{3}}\int d^{3}\boldsymbol{k}\left(\delta_{ij}-\frac{k_{i}k_{j}}{k^{2}}\right)e^{i\boldsymbol{k}\cdot\boldsymbol{r}},
\end{equation}

\begin{equation}
\delta_{ij}^{\parallel}(\boldsymbol{r})\equiv\frac{1}{(2\pi)^{3}}\int d^{3}\boldsymbol{k}\frac{k_{i}k_{j}}{k^{2}}e^{i\boldsymbol{k}\cdot\boldsymbol{r}}.
\end{equation}
We have stated equation (\ref{eq:LongTransdeltaF}) without proof,
but one can be found, for instance in \citet{Milonni2019}. From these
definitions follows

\begin{align}
\delta_{ij}^{\perp}(\boldsymbol{r})+\delta_{ij}^{\parallel}(\boldsymbol{r}) & =\frac{1}{(2\pi)^{3}}\int d^{3}\boldsymbol{k}\delta_{ij}e^{i\boldsymbol{k}\cdot\boldsymbol{r}}\nonumber \\
 & =\delta_{ij}\delta(\boldsymbol{r}),
\end{align}
and also, for each function by carrying out the integrals, that

\begin{equation}
\begin{cases}
\delta_{ij}^{\perp}(\boldsymbol{r})=\frac{2}{3}\delta_{ij}\delta(\boldsymbol{r})-\frac{1}{4\pi r^{3}}\left(\delta_{ij}-\frac{3r_{i}r_{j}}{r^{2}}\right),\\
\delta_{ij}^{\parallel}(\boldsymbol{r})=\frac{1}{3}\delta_{ij}\delta(\boldsymbol{r})+\frac{1}{4\pi r^{3}}\left(\delta_{ij}-\frac{3r_{i}r_{j}}{r^{2}}\right).
\end{cases}\label{eq:deltadefs}
\end{equation}
We can, as an example, carry out the first calculation, where we find

\begin{equation}
\delta_{ij}^{\perp}(\boldsymbol{r})\equiv\frac{1}{(2\pi)^{3}}\int d^{3}\boldsymbol{k}\delta_{ij}e^{i\boldsymbol{k}\cdot\boldsymbol{r}}-\frac{1}{(2\pi)^{3}}\int d^{3}\boldsymbol{k}\frac{k_{i}k_{j}}{k^{2}}e^{i\boldsymbol{k}\cdot\boldsymbol{r}}.
\end{equation}
The first term is immediately recognized as a $\delta$-function,
so we have

\begin{equation}
\delta_{ij}^{\perp}(\boldsymbol{r})=\delta_{ij}\delta(\boldsymbol{r})-\frac{1}{(2\pi)^{3}}\int d^{3}\boldsymbol{k}\frac{k_{i}k_{j}}{k^{2}}e^{i\boldsymbol{k}\cdot\boldsymbol{r}},
\end{equation}
while the second term can be written, for $i\neq j$ as

\begin{align}
\frac{1}{(2\pi)^{3}}\int d^{3}\boldsymbol{k}\frac{k_{i}k_{j}}{k^{2}}e^{i\boldsymbol{k}\cdot\boldsymbol{r}} & =-\frac{1}{(2\pi)^{3}}\frac{\partial^{2}}{\partial r_{i}\partial r_{j}}\int d^{3}\boldsymbol{k}\frac{1}{k^{2}}e^{i\boldsymbol{k}\cdot\boldsymbol{r}}\nonumber \\
 & =-\frac{1}{(2\pi)^{3}}\frac{\partial^{2}}{\partial r_{i}\partial r_{j}}\int_{0}^{\infty}dkk^{2}\int_{0}^{2\pi}d\phi\int_{0}^{\pi}d\theta\sin\theta\frac{1}{k^{2}}e^{i\boldsymbol{k}\cdot\boldsymbol{r}}\nonumber \\
 & =-\frac{1}{(2\pi)^{2}}\frac{\partial^{2}}{\partial r_{i}\partial r_{j}}\int_{0}^{\infty}dk\int_{0}^{\pi}d\theta\sin\theta e^{ikr\cos\theta}\nonumber \\
 & =-\frac{1}{2\pi{}^{2}}\frac{\partial^{2}}{\partial r_{i}\partial r_{j}}\int_{0}^{\infty}dk\frac{\sin(kr)}{kr}\nonumber \\
 & =-\frac{1}{2\pi{}^{2}}\frac{\partial^{2}}{\partial r_{i}\partial r_{j}}\frac{\pi}{2r}\nonumber \\
 & =-\frac{1}{4\pi r^{3}}\frac{3r_{i}r_{j}}{r^{2}},\label{eq:ineqjdeltaT}
\end{align}
and for $i=j$, we have

\begin{align}
\frac{1}{(2\pi)^{3}}\int d^{3}\boldsymbol{k}\frac{k_{i}^{2}}{k^{2}}e^{i\boldsymbol{k}\cdot\boldsymbol{r}} & =-\frac{1}{(2\pi)^{3}}\frac{\partial^{2}}{\partial r_{i}^{2}}\int d^{3}\boldsymbol{k}\frac{1}{k^{2}}e^{i\boldsymbol{k}\cdot\boldsymbol{r}}\nonumber \\
 & =-\frac{1}{2\pi{}^{2}}\frac{\partial^{2}}{\partial r_{i}^{2}}\frac{\pi}{2r}\nonumber \\
 & =\frac{1}{3}\delta(\boldsymbol{r})-\frac{1}{4\pi}\left(\frac{3r_{i}^{2}}{r^{5}}-\frac{1}{r^{3}}\right)\nonumber \\
 & =\frac{1}{3}\delta(\boldsymbol{r})+\frac{1}{4\pi r^{3}}\left(\delta_{ij}-\frac{3r_{i}r_{i}}{r^{2}}\right),\label{eq:ieqjdeltaT}
\end{align}
where we have inserted the second order derivative must be performed
carefully as it needs to be regularized (see \citet{Stewart2008,Frahm1983}),
and thus we have inserted in the third line $\delta(\boldsymbol{r})/3$.
In the last step we have inserted the Kronecker delta $\delta_{ij}$
for convenience, which is possible since we assumed from the start
that $i=j$. We can group the results of equations (\ref{eq:ineqjdeltaT})
and (\ref{eq:ieqjdeltaT}) to recover the result for equation

\begin{equation}
\delta_{ij}^{\perp}(\boldsymbol{r})=\frac{2}{3}\delta_{ij}\delta(\boldsymbol{r})-\frac{1}{4\pi r^{3}}\left(\delta_{ij}-\frac{3r_{i}r_{j}}{r^{2}}\right).
\end{equation}

The procedure for calculating the longitudinal $\delta$-function
is exactly the same, so we choose to omit it.

\subsection{Quantizing the plasmon field\label{subsec:Quantizing-the-Plasmon}}

Here we start the quantization procedure of the Plasmon field. If
we promote the vector potential to an operator we have the relation
with the noise operator \textbf{$\hat{\boldsymbol{J}}_{N}(\boldsymbol{\rho},\omega)$}.

\begin{equation}
\hat{\boldsymbol{A}}(\boldsymbol{r},\omega)=\mu_{0}\int d^{3}\boldsymbol{r}'\bar{\bar{G}}(\boldsymbol{r},\boldsymbol{r}';\omega)\hat{\boldsymbol{J}}_{N}(\boldsymbol{\rho}',\omega)\delta(z'),
\end{equation}
and its transpose

\begin{equation}
\hat{\boldsymbol{A}}^{\dagger}(\boldsymbol{r},\omega)=\mu_{0}\int d^{3}\boldsymbol{r}'\hat{\boldsymbol{J}}_{N}^{\dagger}(\boldsymbol{\rho}',\omega)\bar{\bar{G}}^{\dagger}(\boldsymbol{r},\boldsymbol{r}';\omega)\delta(z').
\end{equation}
In the Weyl gauge, the electric field reads

\begin{equation}
\hat{\boldsymbol{E}}(\boldsymbol{r},\omega)=i\omega\mu_{0}^{2}\int d^{3}\boldsymbol{r}'\bar{\bar{G}}(\boldsymbol{r},\boldsymbol{r}';\omega)\hat{\boldsymbol{J}}_{N}(\boldsymbol{\rho}',\omega)\delta(z'),
\end{equation}
and its transpose

\begin{equation}
\hat{\boldsymbol{E}}^{\dagger}(\boldsymbol{r},\omega)=i\omega\mu_{0}^{2}\int d^{3}\boldsymbol{r}'\hat{\boldsymbol{J}}_{N}^{\dagger}(\boldsymbol{\rho}',\omega)\bar{\bar{G}}^{\dagger}(\boldsymbol{r},\boldsymbol{r}';\omega)\delta(z').
\end{equation}
The noise operator is assumed to obey commutation relations 
\begin{equation}
\left[\hat{J}_{N,i}(\boldsymbol{\rho};\omega),\hat{J}_{N,j}^{\dagger}(\boldsymbol{\rho}';\omega')\right]=\mathcal{C}\alpha(\omega)\delta_{ij}\delta(\boldsymbol{\rho}-\boldsymbol{\rho}')\delta(\omega-\omega'),\label{eq:LongDeltaCommutator}
\end{equation}
while in the Weyl gauge, since the Lagrangian density of the E-M field
is 
\begin{equation}
\mathcal{L}=(\varepsilon_{0}\dot{\boldsymbol{A}}^{2}+(\nabla\times\boldsymbol{A})^{2}/\mu_{0})/2,
\end{equation}
we can find the canonically conjugate variable to $\boldsymbol{A}$
by calculating

\begin{equation}
\partial\mathcal{L}/\partial\dot{\boldsymbol{A}}=\varepsilon_{0}\dot{\boldsymbol{A}}(\boldsymbol{r})=-\varepsilon_{0}\boldsymbol{E}(\boldsymbol{r},\omega).
\end{equation}
This means that when we promote the vector potential and the electric
field to operators, we have the canonical commutation relation

\begin{equation}
\left[\hat{A}_{i}(\boldsymbol{r},\omega),-\varepsilon_{0}\hat{E}_{j}(\boldsymbol{r}',\omega)\right]=i\hbar\delta_{ij}\delta(\boldsymbol{r}-\boldsymbol{r}'),
\end{equation}
or since we are interested only in the longitudinal components of
the operators we substitute the full $\delta$-function with

\begin{equation}
\left[\hat{A}_{i}(\boldsymbol{r},\omega),-\varepsilon_{0}\hat{E}_{j}(\boldsymbol{r}',\omega)\right]=i\hbar\delta_{ij}\delta^{\parallel}(\boldsymbol{r}-\boldsymbol{r}')\label{eq:canonicalLong}
\end{equation}
If we want represent the vector potential in real space and time we
calculate

\begin{equation}
\hat{\boldsymbol{A}}(\boldsymbol{r},t)=\mu_{0}\int_{0}^{\infty}d\omega e^{-i\omega t}\hat{\boldsymbol{A}}(\boldsymbol{r},\omega)+\mu_{0}\int_{0}^{\infty}d\omega e^{i\omega t}\hat{\boldsymbol{A}}^{\dagger}(\boldsymbol{r},\omega),
\end{equation}
and similarly for the electric field. The commutator of the vector
potential operator and the electric field operator will thus be

\begin{align}
\left[\hat{\boldsymbol{A}}(\boldsymbol{r},t),-\varepsilon_{0}\hat{\boldsymbol{E}}(\boldsymbol{r}',t)\right]= & \int_{0}^{\infty}\frac{d\omega}{2\pi}e^{-i\omega t}\int_{0}^{\infty}\frac{d\nu}{2\pi}e^{i\nu t}\int d^{3}\boldsymbol{r}_{1}\int d^{3}\boldsymbol{r}_{2}\left\{ \left(i\nu\mu_{0}^{2}\right)\right.\nonumber \\
 & \left.\left[\bar{\bar{G}}(\boldsymbol{r},\boldsymbol{r}_{1};\omega)\hat{\boldsymbol{J}}_{N}(\boldsymbol{\rho}_{1},\omega)\delta(z_{1}),-\varepsilon_{0}\hat{\boldsymbol{J}}_{N}^{\dagger}(\boldsymbol{\rho}_{2},\nu)\bar{\bar{G}}(\boldsymbol{r}',\boldsymbol{r}_{2};\nu)\delta(z_{2})\right]+\cdots\right\} \nonumber \\
= & \int_{0}^{\infty}\frac{d\omega}{2\pi}e^{-i\omega t}\int_{0}^{\infty}\frac{d\nu}{2\pi}e^{i\nu t}\int d^{3}\boldsymbol{r}_{1}\int d^{3}\boldsymbol{r}_{2}\left\{ i\nu\frac{\mu_{0}}{c^{2}}\delta(z_{1})\delta(z_{2})\right.\nonumber \\
 & \left.\left[\bar{\bar{G}}(\boldsymbol{r},\boldsymbol{\rho}_{1};\omega)\hat{\boldsymbol{J}}_{N}(\boldsymbol{\rho}_{1},\omega),\hat{\boldsymbol{J}}_{N}^{\dagger}(\boldsymbol{\rho}_{2},\nu)\bar{\bar{G}}(\boldsymbol{r}',\boldsymbol{\rho}_{2};\omega)\right]+\cdots\right\} ,
\end{align}
where the Green's functions are evaluated at $z=0$. The remaining
commutator can be simplified as

\begin{align}
\left[\bar{\bar{G}}(\boldsymbol{r},\boldsymbol{\rho}_{1};\omega)\hat{\boldsymbol{J}}_{N}(\boldsymbol{\rho}_{1},\omega),\hat{\boldsymbol{J}}_{N}^{\dagger}(\boldsymbol{\rho}_{2},\nu)\bar{\bar{G}}^{\dagger}(\boldsymbol{r}',\boldsymbol{\rho}_{2};\omega)\right]\nonumber \\
=\bar{\bar{G}}(\boldsymbol{r},\boldsymbol{\rho}_{1};\omega)\left[\hat{\boldsymbol{J}}_{N}(\boldsymbol{\rho}_{1},\omega),\hat{\boldsymbol{J}}_{N}^{\dagger}(\boldsymbol{\rho}_{2},\nu)\right]\bar{\bar{G}}^{\dagger}(\boldsymbol{r}',\boldsymbol{\rho}_{2};\omega)\nonumber \\
=\mathcal{C}\alpha(\omega)\bar{\bar{G}}(\boldsymbol{r},\boldsymbol{\rho}_{1};\omega)\delta(\boldsymbol{\rho}_{1}-\boldsymbol{\rho}_{2})\delta(\omega-\nu)\bar{\bar{G}}^{\dagger}(\boldsymbol{r}',\boldsymbol{\rho}_{2};\omega),
\end{align}
where the constant $\mathcal{C}$ and the function $\alpha(\omega)$
are to be determined such that the commutator relations of equation
(\ref{eq:LongDeltaCommutator}) are obeyed. Substituting into the
integral, we are left with

\begin{align}
\left[\hat{\boldsymbol{A}}(\boldsymbol{r},t),-\varepsilon_{0}\hat{\boldsymbol{E}}(\boldsymbol{r}',t)\right]= & \frac{i\mu_{0}\mathcal{C}}{2\pi c^{2}}\int_{0}^{\infty}\frac{d\omega}{2\pi}\omega\alpha(\omega)\int d^{2}\boldsymbol{\rho}_{1}\bar{\bar{G}}(\boldsymbol{r},\boldsymbol{\rho}_{1};\omega)\bar{\bar{G}}^{\dagger}(\boldsymbol{r}',\boldsymbol{\rho}_{1};\omega)\nonumber \\
 & +\cdots.
\end{align}
The other integral we have so far represented by elipses will contribute
equally to the result, thus we have

\begin{equation}
\left[\hat{\boldsymbol{A}}(\boldsymbol{r},t),-\varepsilon_{0}\hat{\boldsymbol{E}}(\boldsymbol{r}',t)\right]=2\frac{i\mu_{0}\mathcal{C}}{2\pi c^{2}}\int_{0}^{\infty}\frac{d\omega}{2\pi}\omega\alpha(\omega)\int d^{2}\boldsymbol{\rho}_{1}\bar{\bar{G}}(\boldsymbol{r},\boldsymbol{\rho}_{1};\omega)\bar{\bar{G}}^{\dagger}(\boldsymbol{r}',\boldsymbol{\rho}_{1};\omega).
\end{equation}
We may now use, as before, only the part of the Green's function associated
with the SPP field. We thus, focus on performing the integral

\begin{align*}
\int d^{2}\boldsymbol{\rho}_{1}\bar{\bar{G}}(\boldsymbol{r},\boldsymbol{\rho}_{1};\omega)\bar{\bar{G}}^{\dagger}(\boldsymbol{r}',\boldsymbol{\rho}_{1};\omega)= & \frac{1}{4\bar{\varepsilon}^{2}}\int d^{2}\boldsymbol{\rho}_{1}\int d^{2}\boldsymbol{q}\frac{1}{q}\frac{e^{i\boldsymbol{q}\cdot(\boldsymbol{\rho}-\boldsymbol{\rho}_{1})}e^{-qz}}{i\omega\sigma(\omega)q/(2\bar{\varepsilon}\varepsilon_{0}c^{2})+\omega^{2}/c^{2}}\tilde{G}(\boldsymbol{q})\\
 & \int d^{2}\boldsymbol{p}\frac{1}{p}\frac{e^{-i\boldsymbol{p}\cdot(\boldsymbol{\rho}'-\boldsymbol{\rho}_{1})}e^{-pz'}}{-i\omega\sigma^{*}(\omega)p/(2\bar{\varepsilon}\varepsilon_{0}c^{2})+\omega^{2}/c^{2}}\tilde{G}^{\dagger}(\boldsymbol{p})
\end{align*}

\begin{equation}
=\frac{1}{4\bar{\varepsilon}^{2}}\int d^{2}\boldsymbol{q}\int d^{2}\boldsymbol{p}\int d^{2}\boldsymbol{\rho}_{1}\frac{1}{qp}\frac{e^{i\boldsymbol{q}\cdot\boldsymbol{\rho}+i\left(\boldsymbol{p}-\boldsymbol{q}\right)\cdot\boldsymbol{\rho}_{1}-i\boldsymbol{p}\cdot\boldsymbol{\rho}'}e^{-qz}e^{-pz'}}{D(q,\omega)D^{*}(p,\omega)}\tilde{G}(\boldsymbol{q})\tilde{G}^{\dagger}(\boldsymbol{p}),
\end{equation}
where $D(q,\omega)$ is shorthand for the denominator. Using the fact
that

\begin{equation}
\int d^{2}\boldsymbol{\rho}_{1}e^{i\left(\boldsymbol{p}-\boldsymbol{q}\right)\cdot\boldsymbol{\rho}_{1}}=(2\pi)^{2}\delta(\boldsymbol{p}-\boldsymbol{q}),
\end{equation}
we have

\begin{equation}
\int d^{2}\boldsymbol{\rho}_{1}\bar{\bar{G}}(\boldsymbol{r},\boldsymbol{\rho}_{1};\omega)\bar{\bar{G}}^{\dagger}(\boldsymbol{r}',\boldsymbol{\rho}_{1};\omega)=\frac{(2\pi)^{2}}{4\bar{\varepsilon}^{2}}\int d^{2}\boldsymbol{q}\frac{1}{q^{2}}\frac{e^{i\boldsymbol{q}\cdot\left(\boldsymbol{\rho}-\boldsymbol{\rho}'\right)}e^{-q(z+z')}}{\left|i\omega\sigma(\omega)q/(2\bar{\varepsilon}\varepsilon_{0}c^{2})+\omega^{2}/c^{2}\right|^{2}}\tilde{G}(\boldsymbol{q})\tilde{G}^{\dagger}(\boldsymbol{q}).
\end{equation}
The commutator then reads

\begin{align}
\left[\hat{\boldsymbol{A}}(\boldsymbol{r},t),-\varepsilon_{0}\hat{\boldsymbol{E}}(\boldsymbol{r}',t)\right]= & \frac{i\mu_{0}\pi\mathcal{C}}{c^{2}\bar{\varepsilon}^{2}}\int d^{2}\boldsymbol{q}\frac{1}{q^{2}}e^{i\boldsymbol{q}\cdot\left(\boldsymbol{\rho}-\boldsymbol{\rho}'\right)}e^{-q(z+z')}\tilde{G}(\boldsymbol{q})\tilde{G}^{\dagger}(\boldsymbol{q})\nonumber \\
 & \int_{0}^{\infty}\frac{d\omega}{2\pi}\frac{\omega\alpha(\omega)}{\left|i\omega\sigma(\omega)q/(2\bar{\varepsilon}\varepsilon_{0}c^{2})+\omega^{2}/c^{2}\right|^{2}}\nonumber \\
= & \frac{i\pi\mathcal{C}}{\varepsilon_{0}\bar{\varepsilon}^{2}}\int d^{2}\boldsymbol{q}\frac{1}{q^{2}}e^{i\boldsymbol{q}\cdot\left(\boldsymbol{\rho}-\boldsymbol{\rho}'\right)}e^{-q(z+z')}\tilde{G}(\boldsymbol{q})\tilde{G}^{\dagger}(\boldsymbol{q})\nonumber \\
 & \int_{0}^{\infty}\frac{d\omega}{2\pi}\frac{\omega\alpha(\omega)}{\left|i\omega\sigma(\omega)q/(2\bar{\varepsilon}\varepsilon_{0})+\omega^{2}\right|^{2}}.
\end{align}
Now we can move on to the evaluation of the integral over the frequency

\begin{equation}
\mathcal{I}(q)=\int\frac{d\omega}{2\pi}\frac{\omega\alpha(\omega)}{\left|i\omega\sigma(\omega)q/(2\bar{\varepsilon}\varepsilon_{0})+\omega^{2}\right|^{2}},\label{eq:Integralaim}
\end{equation}
which when considering the complex conductivity $\sigma(\omega)=\sigma_{1}(\omega)+i\sigma_{2}(\omega)$
as usual becomes

\begin{equation}
\mathcal{I}(q)=\int_{0}^{\infty}\frac{d\omega}{2\pi}\frac{\omega\alpha(\omega)}{\left|i\omega\sigma_{1}(\omega)q/(2\bar{\varepsilon}\varepsilon_{0}c^{2})-\omega\sigma_{2}(\omega)q/(2\bar{\varepsilon}\varepsilon_{0}c^{2})+\omega^{2}/c^{2}\right|^{2}}.
\end{equation}
We can yet again evaluate this integral in the complex plane. In particular
we may look at an auxilliary problem where we consider a function
$g(z)$ such that its contour integral over a large semi-circle in
the upper half-plane vanishes, that is to say

\begin{equation}
\oint dzg(z)=\lim_{R\to\infty}\left[\int_{-R}^{R}d\omega g(\omega)+\int_{0}^{\pi}d\theta iRe^{i\theta}g(Re^{i\theta})\right]=0,\label{eq:uhpcontour0}
\end{equation}
which is equivalent to stating that $g(z)$ has no poles in the upper
half plane. We therefore know that the integrals are symmetric and
as such, we may evaluate one to find the other. In particular, we
may evaluate pick a $g(z)$ of the following form

\begin{equation}
g(z)=\frac{z}{z^{2}+iqz\sigma(z)/(2\bar{\varepsilon}\varepsilon_{0})},\label{eq:guess}
\end{equation}
noting that if

\begin{equation}
\lim_{R\to\infty}g(Re^{i\theta})=\frac{1}{Re^{i\theta}},
\end{equation}
we have that

\begin{equation}
\int_{-\infty}^{\infty}d\omega g(\omega)=-i\pi,\label{eq:intg}
\end{equation}
which folllows directly from equation (\ref{eq:uhpcontour0}), and
noting that if we assume that $iqz\sigma(z)/(2\bar{\varepsilon}\varepsilon_{0})$
grows more slowly than $z^{2}$ (which it does, virtue of the dispersion
relation we had found), then equation (\ref{eq:guess}) obeys the
desired limit and thus can be evaluated as in equation (\ref{eq:intg}).
Multiplying and dividing by the conjugate of the denominator of $g(\omega),$we
may write

\begin{align}
\int_{-\infty}^{\infty}d\omega\frac{\omega^{3}-iq\omega^{2}\sigma^{*}(\omega)/(2\bar{\varepsilon}\varepsilon_{0})}{|\omega^{2}+iq\omega\sigma(\omega)/(2\bar{\varepsilon}\varepsilon_{0})|^{2}} & =-i\pi\nonumber \\
\Leftrightarrow\int_{-\infty}^{\infty}d\omega\frac{\omega^{3}-iq\omega^{2}\left[\sigma_{1}(\omega)-i\sigma_{2}(\omega)\right]/(2\bar{\varepsilon}\varepsilon_{0})}{|\omega^{2}+iq\omega\sigma(\omega)/(2\bar{\varepsilon}\varepsilon_{0})|^{2}} & =-i\pi,
\end{align}
and if we consider the dispersion relation we found for $\sigma_{2}(\omega)$,
written

\begin{equation}
\frac{\sigma_{2}(\omega)}{\varepsilon_{0}\omega}=\frac{2\bar{\varepsilon}}{q}
\end{equation}
we may write $\sigma_{2}(\omega)=\frac{2\bar{\varepsilon}\varepsilon_{0}\omega}{q}$,
and therefore

\begin{align}
\int_{-\infty}^{\infty}d\omega\frac{\omega^{3}-iq\omega^{2}\left[\sigma_{1}(\omega)-i\sigma_{2}(\omega)\right]/(2\bar{\varepsilon}\varepsilon_{0})}{|\omega^{2}+iq\omega\sigma(\omega)/(2\bar{\varepsilon}\varepsilon_{0})|^{2}} & =\frac{-iq}{2\bar{\varepsilon}\varepsilon_{0}}\int_{-\infty}^{\infty}d\omega\frac{\sigma_{1}(\omega)\omega^{2}}{|\omega^{2}+iq\omega\sigma(\omega)/(2\bar{\varepsilon}\varepsilon_{0})|^{2}}\nonumber \\
 & =-i\pi.
\end{align}
We can now multiply by $-\frac{1}{2\pi i}$ on both sides, to obtain

\[
\int_{-\infty}^{\infty}d\omega\frac{1}{2\pi}\frac{\sigma_{1}(\omega)\omega^{2}}{|\omega^{2}+iq\omega\sigma(\omega)/(2\bar{\varepsilon}\varepsilon_{0})|^{2}}=\frac{\bar{\varepsilon}\varepsilon_{0}}{q},
\]
and so, we can set $\alpha(\omega)=\omega\sigma_{1}(\omega)$, and
solve the integral of equation (\ref{eq:Integralaim}) which (given
that $\sigma_{1}(\omega)$ is an even function) will evaluate to

\begin{equation}
\mathcal{I}(q)=\frac{\bar{\varepsilon}\varepsilon_{0}}{2q}.
\end{equation}
Remembering that $\alpha(\omega)$ is chosen such that the commutator
becomes a longitudinal $\delta-$function, this choice of $\alpha(\omega)$
allows us to write

\begin{align}
\left[\hat{\boldsymbol{A}}(\boldsymbol{r},t),-\varepsilon_{0}\hat{\boldsymbol{E}}(\boldsymbol{r}',t)\right] & =\frac{i\pi\mathcal{C}}{\varepsilon_{0}\bar{\varepsilon}^{2}}\bar{\varepsilon}\varepsilon_{0}\int d^{2}\boldsymbol{q}\frac{e^{i\boldsymbol{q}\cdot\left(\boldsymbol{\rho}-\boldsymbol{\rho}'\right)}}{q^{3}}e^{-q(z+z')}\tilde{G}(\boldsymbol{q})\tilde{G}^{\dagger}(\boldsymbol{q})\nonumber \\
 & =\frac{i\pi\mathcal{C}}{\bar{\varepsilon}}\int d^{2}\boldsymbol{q}\frac{e^{i\boldsymbol{q}\cdot\left(\boldsymbol{\rho}-\boldsymbol{\rho}'\right)}}{q^{3}}e^{-q(z+z')}\tilde{G}(\boldsymbol{q})\tilde{G}^{\dagger}(\boldsymbol{q}).
\end{align}
Noting that using the general expression for $\tilde{G}(\boldsymbol{q}),\tilde{G}^{\dagger}(\boldsymbol{q})$,
one can easily verify (in the electrostatic limit) that

\begin{equation}
\tilde{G}(\boldsymbol{q})\tilde{G}^{\dagger}(\boldsymbol{q})=2q^{2}\tilde{G}(\boldsymbol{q}),
\end{equation}
which gives the commutator

\begin{equation}
\left[\hat{\boldsymbol{A}}(\boldsymbol{r},t),-\varepsilon_{0}\hat{\boldsymbol{E}}(\boldsymbol{r}',t)\right]=\frac{i\pi\mathcal{C}}{\bar{\varepsilon}}\int d^{2}\boldsymbol{q}\frac{e^{i\boldsymbol{q}\cdot\left(\boldsymbol{\rho}-\boldsymbol{\rho}'\right)}}{q}e^{-q(z+z')}\tilde{G}(\boldsymbol{q}).
\end{equation}
We can now consider elements of this commutator, in particular, we
may look at

\begin{equation}
\left[\hat{A}_{i}(\boldsymbol{r},t),-\varepsilon_{0}\hat{E}_{j}(\boldsymbol{r}',t)\right]=\frac{i\pi\mathcal{C}}{\bar{\varepsilon}}\int d^{2}\boldsymbol{q}\frac{q_{i}q_{j}}{q}e^{i\boldsymbol{q}\cdot\left(\boldsymbol{\rho}-\boldsymbol{\rho}'\right)}e^{-q(z+z')},
\end{equation}
for the longitudinal components of the fields. $i=x,y$. Bringing
the exponentials together, by reintroducing $\boldsymbol{k}\equiv\boldsymbol{p}_{1}=(q_{x},q_{y},i\kappa_{1})\approx(q_{x},q_{y},iq)$
and $\boldsymbol{r}=(x,y,z)$ and $\boldsymbol{r}'=(x',y',z')$, we
can write

\begin{align}
\left[\hat{A}_{i}(\boldsymbol{r},t),-\varepsilon_{0}\hat{E}_{j}(\boldsymbol{r}',t)\right] & =\frac{i\pi\mathcal{C}}{\bar{\varepsilon}}\int d^{2}\boldsymbol{q}\frac{q_{i}q_{j}}{q}e^{i\boldsymbol{k}\cdot\left(\boldsymbol{r}-\boldsymbol{r}'\right)},
\end{align}
and extending the integral to the whole $q-$space as can easily be
shown by integrating in the complex plane

\begin{align}
\int d^{3}\boldsymbol{k}\frac{k_{i}k_{j}}{q^{2}+k_{z}^{2}}e^{i\boldsymbol{k}\cdot\left(\boldsymbol{r}-\boldsymbol{r}'\right)} & =\int d^{2}kk_{i}k_{j}e^{iq(\boldsymbol{\rho}-\boldsymbol{\rho}')}\int dk_{z}\frac{e^{-k_{z}(z+z')}}{(q+ik_{z})(q-ik_{z})}\nonumber \\
 & =\int d^{2}kk_{i}k_{j}e^{iq(\boldsymbol{\rho}-\boldsymbol{\rho}')}2\pi i\frac{e^{-q(z+z')}}{2qi}\nonumber \\
 & =\pi\int d^{2}\boldsymbol{q}\frac{q_{i}q_{j}}{q}e^{iq(\boldsymbol{\rho}-\boldsymbol{\rho}')}.
\end{align}
The commutator evaluates to

\[
\left[\hat{A}_{i}(\boldsymbol{r},t),-\varepsilon_{0}\hat{E}_{j}(\boldsymbol{r}',t)\right]=\frac{i\mathcal{C}}{\bar{\varepsilon}}\int d^{3}\boldsymbol{k}\frac{k_{i}k_{j}}{k^{2}}e^{i\boldsymbol{k}\cdot\left(\boldsymbol{r}-\boldsymbol{r}'\right)}.
\]
This is very similar to the for of the longitudinal delta function
$\delta_{ij}^{\parallel}(\boldsymbol{r}-\boldsymbol{r}')$. In fact,
recalling that

\begin{equation}
\delta_{ij}^{\parallel}(\boldsymbol{r})\equiv\frac{1}{(2\pi)^{3}}\int d^{3}\boldsymbol{k}\frac{k_{i}k_{j}}{k^{2}}e^{i\boldsymbol{k}\cdot\boldsymbol{r}},
\end{equation}
we can recover equation (\ref{eq:canonicalLong}) if we let

\begin{equation}
\mathcal{C}=\frac{\bar{\varepsilon}\hbar}{(2\pi)^{3}}.
\end{equation}
Using the form of $\mathcal{C}$ and $\alpha(\omega)$ we discovered,
the commutators for the noise operators can also be found for $i$
and $j=x,y$, we can write

\begin{equation}
\left[\hat{J}_{N,i}(\boldsymbol{\rho};\omega),\hat{J}_{N,j}^{\dagger}(\boldsymbol{\rho}';\omega')\right]=\frac{\bar{\varepsilon}\hbar}{(2\pi)^{3}}\omega\sigma_{1}(\omega)\delta_{ij}\delta(\boldsymbol{\rho}-\boldsymbol{\rho}')\delta(\omega-\omega'),
\end{equation}
and therefore we can define the operators

\begin{equation}
\hat{\text{\textbf{f}}}(\boldsymbol{\rho},\omega)=\frac{\hat{\boldsymbol{J}}_{N}(\boldsymbol{\rho},\omega)}{\sqrt{\frac{\bar{\varepsilon}\hbar}{(2\pi)^{3}}\omega\sigma_{1}(\omega)}},\label{eq:fop}
\end{equation}
which obey bosonic commutation relations

\begin{equation}
\left[\hat{\text{f}}_{i}(\boldsymbol{\rho},\omega),\hat{\text{f}}_{i}^{\dagger}(\boldsymbol{\rho}',\omega')\right]=\delta_{ij}\delta(\boldsymbol{\rho}-\boldsymbol{\rho}')\delta(\omega-\omega').
\end{equation}
The vector potential can thus be written making use of these operators
as

\begin{align}
\hat{\boldsymbol{A}}(\boldsymbol{r},\omega) & =\mu_{0}\sqrt{\frac{\bar{\varepsilon}\hbar}{(2\pi)^{3}}\omega\sigma_{1}(\omega)}\int d^{3}\boldsymbol{r}'\bar{\bar{G}}(\boldsymbol{r},\boldsymbol{r}',\omega)\cdot\hat{\text{\textbf{f}}}(\boldsymbol{\rho}',\omega)\delta(z')\nonumber \\
 & \equiv\beta(\omega)\int d^{2}\boldsymbol{\rho}'\bar{\bar{G}}(\boldsymbol{r},\boldsymbol{\rho}',z'=0;\omega)\cdot\hat{\text{\textbf{f}}}(\boldsymbol{\rho}',\omega).
\end{align}
The electric field operator can then be written at the cost of the
vector potential operator in the Schrodinger picture, by Fourier transforming
at $t=0$, giving the quantized field

\begin{equation}
\hat{\boldsymbol{E}}(\boldsymbol{r})=i\int_{0}^{\infty}d\omega\omega\hat{\boldsymbol{A}}(\boldsymbol{r},\omega)-i\int_{0}^{\infty}d\omega\omega\hat{\boldsymbol{A}}^{\dagger}(\boldsymbol{r},\omega).
\end{equation}
This result concludes the quantization of the SPP field. In the next
few sections we will aim to use these results, especially the bosonic
creation and annihilation operators of equation (\ref{eq:fop}), to
derive the dynamics of two level-systems coupled to graphene plasmons.

\section{Details of the calculation of the dynamics of one qubit coupled to
the plasmonic bath\label{sec:Details-of-the-calculation-of-the-dynamics-of-one-qubit}}

\subsection{Obtaining the dynamics via the Schrodinger equation\label{subsec:Obtaining-the-dynamics}}

In the previous appendices we went over all the building blocks to
set up a mathematical description of our system, in particular the
plasmonic bath. We now move on to a quantitative description of the
interaction between a two-level system and the plasmonic bath. The
Hamiltonian of a two-level system, field and atom-field interaction
is given, in the Schrodinger picture by

\begin{align}
\hat{H}= & \hbar\omega_{0}\sigma_{+}\sigma_{-}+\int d^{2}\boldsymbol{\rho}\int_{0}^{\infty}d\omega\hbar\omega\hat{\text{\textbf{f}}}^{\dagger}(\boldsymbol{\rho},\omega)\cdot\hat{\text{\textbf{f}}}(\boldsymbol{\rho},\omega)\nonumber \\
 & -\sigma_{+}\boldsymbol{d}_{eg}\cdot\int_{0}^{\infty}d\omega\hat{\boldsymbol{E}}(\boldsymbol{r}_{A},\omega)-\sigma_{-}\boldsymbol{d}_{eg}\cdot\int_{0}^{\infty}d\omega\hat{\boldsymbol{E}}(\boldsymbol{r}_{A},\omega).
\end{align}
In the rotating wave approximation, we keep only the resonant terms
making up the electric field. The Hamiltonian will therefore be

\begin{align}
\hat{H}= & \hbar\omega_{0}\sigma_{+}\sigma_{-}+\int d^{2}\boldsymbol{\rho}\int_{0}^{\infty}d\omega\hbar\omega\hat{\text{\textbf{f}}}^{\dagger}(\boldsymbol{\rho},\omega)\cdot\hat{\text{\textbf{f}}}(\boldsymbol{\rho},\omega)\nonumber \\
 & -\sigma_{+}\boldsymbol{d}_{eg}\cdot i\int_{0}^{\infty}d\omega\omega\hat{\boldsymbol{A}}(\boldsymbol{r}_{0},\omega)+i\int_{0}^{\infty}d\omega\omega\hat{\boldsymbol{A}}^{\dagger}(\boldsymbol{r}_{0},\omega)\cdot\boldsymbol{d}_{eg}\sigma_{-},
\end{align}
or at the cost of only the creation and destruction operators

\begin{align}
\hat{H}= & \hbar\omega_{0}\sigma_{+}\sigma_{-}+\int d^{2}\boldsymbol{\rho}\int_{0}^{\infty}d\omega\hbar\omega\hat{\text{\textbf{f}}}^{\dagger}(\boldsymbol{\rho},\omega)\cdot\hat{\text{\textbf{f}}}(\boldsymbol{\rho},\omega)\nonumber \\
 & -\sigma_{+}\boldsymbol{d}_{eg}\cdot i\int_{0}^{\infty}d\omega\omega\beta(\omega)\int d^{2}\boldsymbol{\rho}\bar{\bar{G}}(\boldsymbol{r},\boldsymbol{\rho}',z'=0;\omega)\cdot\hat{\text{\textbf{f}}}(\boldsymbol{\rho}',\omega)\nonumber \\
 & +i\int_{0}^{\infty}d\omega\omega\beta^{*}(\omega)\int d^{2}\boldsymbol{\rho}'\bar{\bar{G}}^{\dagger}(\boldsymbol{r},\boldsymbol{\rho}',z'=0;\omega)\cdot\hat{\text{\textbf{f}}}^{\dagger}(\boldsymbol{\rho}',\omega)\cdot\boldsymbol{d}_{eg}\sigma_{-}.
\end{align}
,We therefore have a zeroth order Hamiltonian of the form

\begin{equation}
\hat{H}_{0}=\hbar\omega_{0}\sigma_{+}\sigma_{-}+\int d^{2}\boldsymbol{\rho}\int_{0}^{\infty}d\omega\hbar\omega\hat{\text{\textbf{f}}}^{\dagger}(\boldsymbol{\rho},\omega)\cdot\hat{\text{\textbf{f}}}(\boldsymbol{\rho},\omega),
\end{equation}
and an interaction term which can be written as

\begin{align}
\hat{V}= & -\sigma_{+}\boldsymbol{d}_{eg}\cdot i\int_{0}^{\infty}d\omega\omega\beta(\omega)\int d^{2}\boldsymbol{\rho}'\bar{\bar{G}}(\boldsymbol{r},\boldsymbol{\rho}',z'=0;\omega)\cdot\hat{\text{\textbf{f}}}(\boldsymbol{\rho}',\omega)\nonumber \\
 & +i\int_{0}^{\infty}d\omega\omega\beta(\omega)\int d^{2}\boldsymbol{\rho}'\bar{\bar{G}}^{\dagger}(\boldsymbol{r},\boldsymbol{\rho}',z'=0;\omega)\cdot\hat{\text{\textbf{f}}}^{\dagger}(\boldsymbol{\rho}',\omega)\cdot\boldsymbol{d}_{eg}\sigma_{-}.
\end{align}
We can now aim to solve the Schrodinger equation

\begin{equation}
i\hbar\frac{\partial}{\partial t}\left|\psi(t)\right\rangle =H\left|\psi(t)\right\rangle .
\end{equation}
To solve the Schrodinger equation we make the following ansatz

\begin{equation}
\left|\psi(t)\right\rangle =C_{e}(t)e^{-i\omega_{0}t}\left|0\right\rangle \left|e\right\rangle +\int d^{2}\boldsymbol{\rho}\int_{0}^{\infty}d\omega e^{-i\omega t}\boldsymbol{C}_{g}(\boldsymbol{\rho},\omega,t)\left|1(\boldsymbol{\rho},\omega,t)\right\rangle \left|g\right\rangle .
\end{equation}
Substituting onto the wave equation, on the left we have

\begin{align}
i\hbar\frac{\partial}{\partial t}\left|\psi(t)\right\rangle =i\hbar\dot{C}_{e}(t)e^{-i\omega_{0}t}\left|0\right\rangle \left|e\right\rangle +\hbar\omega_{0}C_{e}(t)e^{-i\omega_{0}t}\left|0\right\rangle \left|e\right\rangle \nonumber \\
+\int d^{2}\boldsymbol{\rho}\int_{0}^{\infty}d\omega\hbar\omega e^{-i\omega t}\boldsymbol{C}_{g}(\boldsymbol{\rho},\omega,t)\left|1(\boldsymbol{\rho},\omega)\right\rangle \left|g\right\rangle \nonumber \\
+\int d^{2}\boldsymbol{\rho}\int_{0}^{\infty}d\omega i\hbar e^{-i\omega t}\dot{\boldsymbol{C}}_{g}(\boldsymbol{\rho},\omega,t)\left|1(\boldsymbol{\rho},\omega)\right\rangle \left|g\right\rangle .
\end{align}
We now note that we have written $\left|\psi(t)\right\rangle $ as
a function of tensor product states made up of $\left|0\right\rangle ,\left|1(\boldsymbol{\rho},\omega)\right\rangle $
which describe the plasmonic vacuum and the presence of a single plasmon
in the graphene and the states of the atom $\left|e\right\rangle $
and $\left|g\right\rangle $, which are respectively the excited and
ground state for the two-level atom. This way, the operators which
make up the Hamiltonian act in the following ways

\begin{equation}
\begin{cases}
\sigma_{+}\left|i\right\rangle \left|e\right\rangle =0,\\
\sigma_{+}\left|i\right\rangle \left|g\right\rangle =\left|i\right\rangle \left|e\right\rangle ,\\
\sigma_{-}\left|i\right\rangle \left|e\right\rangle =\left|i\right\rangle \left|g\right\rangle ,\\
\sigma_{-}\left|i\right\rangle \left|g\right\rangle =0,
\end{cases}
\end{equation}
\begin{equation}
\begin{cases}
\hat{\text{\textbf{f}}}^{\dagger}(\boldsymbol{\rho},\omega)\left|1(\boldsymbol{\rho},\omega)\right\rangle \left|j\right\rangle =\left|2(\boldsymbol{\rho},\omega)\right\rangle \left|j\right\rangle ,\\
\hat{\text{\textbf{f}}}^{\dagger}(\boldsymbol{\rho},\omega)\left|0\right\rangle \left|j\right\rangle =\left|1(\boldsymbol{\rho},\omega)\right\rangle \left|j\right\rangle ,\\
\hat{\text{\textbf{f}}}(\boldsymbol{\rho},\omega)\left|1(\boldsymbol{\rho},\omega)\right\rangle \left|j\right\rangle =\left|0\right\rangle \left|j\right\rangle ,\\
\hat{\text{\textbf{f}}}(\boldsymbol{\rho},\omega)\left|0\right\rangle \left|j\right\rangle =0,
\end{cases},\label{eq:feffec}
\end{equation}
where $i=0,1$ and $j=g,e$. Note that since the operators are matched
such that $\hat{\text{\textbf{f}}}(\boldsymbol{\rho},\omega)$ is
always paired with $\sigma_{+}$ and vice-versa, the state $\left|\boldsymbol{2}(\boldsymbol{\rho}_{1},\boldsymbol{\rho}_{2},\omega_{1},\omega_{2})\right\rangle \left|j\right\rangle $
is never actually produced. Acting with the zeroth order Hamiltonian
on the state therefore gives

\begin{equation}
\hat{H}_{0}\left|\psi(t)\right\rangle =\hbar\omega_{0}C_{e}(t)e^{-i\omega_{0}t}+\int d^{2}\boldsymbol{\rho}\int_{0}^{\infty}d\omega\hbar\omega\boldsymbol{C}_{g}(\boldsymbol{\rho},\omega,t)e^{-i\omega t}\left|1(\boldsymbol{\rho}',\nu,t)\right\rangle \left|g\right\rangle ,
\end{equation}
where we have also used that $\hat{\text{\textbf{f}}}(\boldsymbol{\rho},\omega)\left|1(\boldsymbol{\rho}',\nu,t)\right\rangle =\left|0\right\rangle \delta(\omega-\nu)\delta(\boldsymbol{\rho}-\boldsymbol{\rho}')$.
Using the same properties the interaction Hamiltonian gives

\begin{align*}
\hat{V}\left|\psi(t)\right\rangle = & -i\int d^{2}\boldsymbol{\rho}\int_{0}^{\infty}d\omega\omega\beta(\omega)\boldsymbol{d}_{eg}\cdot\bar{\bar{G}}(\boldsymbol{r},\boldsymbol{\rho},z=0;\omega)\cdot e^{-i\omega t}\boldsymbol{C}_{g}(\boldsymbol{\rho},\omega,t)\left|0\right\rangle \left|e\right\rangle \\
 & +i\int d^{2}\boldsymbol{\rho}\int_{0}^{\infty}d\omega\omega\beta(\omega)\bar{\bar{G}}^{\dagger}(\boldsymbol{r},\boldsymbol{\rho},z=0;\omega)\cdot\boldsymbol{d}_{eg}C_{e}(t)e^{-i\omega_{0}t}\left|1(\boldsymbol{\rho},\omega,t)\right\rangle \left|g\right\rangle .
\end{align*}
Acting on the left with the bra $\left\langle e\right|\left\langle 0\right|$
on the left and hand side of the Schrodinger equation, we have

\begin{align*}
i\hbar\dot{C}_{e}(t)e^{-i\omega_{0}t}= & -i\int d^{2}\boldsymbol{\rho}\int_{0}^{\infty}d\omega\omega\beta(\omega)\boldsymbol{d}_{eg}\cdot\bar{\bar{G}}(\boldsymbol{r},\boldsymbol{\rho},z=0;\omega)e^{-i\omega t}\cdot\boldsymbol{C}_{g}(\boldsymbol{\rho},\omega,t),
\end{align*}
or

\begin{equation}
\dot{C}_{e}(t)=-\frac{1}{\hbar}\int d^{2}\boldsymbol{\rho}\int_{0}^{\infty}d\omega\omega\beta(\omega)\boldsymbol{d}_{eg}\cdot\bar{\bar{G}}(\boldsymbol{r},\boldsymbol{\rho},z=0;\omega)e^{-i(\omega-\omega_{0})t}\cdot\boldsymbol{C}_{g}(\boldsymbol{\rho},\omega,t),\label{eq:dtCe}
\end{equation}
while on the other hand, acting on the left with $\left\langle g\right|\left\langle 1(\boldsymbol{\rho},\omega)\right|$
we have

\begin{align}
\hbar\omega e^{-i\omega t}\boldsymbol{C}_{g}(\boldsymbol{\rho},\omega,t)+i\hbar e^{-i\omega t}\dot{\boldsymbol{C}}_{g}(\boldsymbol{\rho},\omega,t)= & \hbar\omega\boldsymbol{C}_{g}(\boldsymbol{\rho},\omega,t)e^{-i\omega t}\nonumber \\
 & +i\omega\beta(\omega)\bar{\bar{G}}^{\dagger}(\boldsymbol{r},\boldsymbol{\rho},z=0;\omega)\cdot\boldsymbol{d}_{0}C_{e}(t)e^{-i\omega_{0}t}.
\end{align}

\begin{equation}
\Leftrightarrow\dot{\boldsymbol{C}}_{g}(\boldsymbol{\rho},\omega,t)=\frac{1}{\hbar}\omega\beta(\omega)\boldsymbol{d}_{eg}\cdot\bar{\bar{G}}^{\dagger}(\boldsymbol{r},\boldsymbol{\rho},z=0;\omega)\cdot e^{-i(\omega_{0}-\omega)t}\boldsymbol{C}_{g}(\boldsymbol{\rho},\omega,t).
\end{equation}
Integrating the previous equations with respect to time, choosing
$C_{e}(0)=1$ and $\boldsymbol{C}_{g}(\boldsymbol{\rho},\omega,0)=\boldsymbol{0}$
we have

\begin{equation}
C_{e}(t)=1-\frac{1}{\hbar}\int d^{2}\boldsymbol{\rho}\int_{0}^{\infty}d\omega\omega\beta(\omega)\boldsymbol{d}_{eg}\cdot\bar{\bar{G}}(\boldsymbol{r},\boldsymbol{\rho},z=0;\omega)\cdot\int_{0}^{t}dt_{1}e^{-i(\omega-\omega_{0})t_{1}}\boldsymbol{C}_{g}(\boldsymbol{\rho},\omega,t_{1}),
\end{equation}

\begin{equation}
\boldsymbol{C}_{g}(\boldsymbol{\rho},\omega,t)=\frac{1}{\hbar}\omega\beta(\omega)\bar{\bar{G}}^{\dagger}(\boldsymbol{r},\boldsymbol{\rho},z=0;\omega)\cdot\boldsymbol{d}_{eg}\int_{0}^{t}dt_{1}e^{-i(\omega_{0}-\omega)t_{1}}C_{e}(t_{1}).\label{eq:Cg}
\end{equation}
We can in particular, substitute equation (\ref{eq:Cg}) into equation
(\ref{eq:dtCe}), to obtain the differential equation

\begin{align}
\dot{C}_{e}(t)= & -\frac{1}{\hbar}\int d^{2}\boldsymbol{\rho}\int_{0}^{\infty}d\omega\omega\beta(\omega)\boldsymbol{d}_{eg}\cdot\bar{\bar{G}}(\boldsymbol{r},\boldsymbol{\rho},z=0;\omega)e^{-i(\omega-\omega_{0})t}\nonumber \\
 & \cdot\frac{1}{\hbar}\omega\beta(\omega)\bar{\bar{G}}^{\dagger}(\boldsymbol{r},\boldsymbol{\rho},z=0;\omega)\cdot\boldsymbol{d}_{eg}\int_{0}^{t}dt_{1}e^{-i(\omega_{0}-\omega)t_{1}}C_{e}(t_{1}).
\end{align}
We can simplify the previous expression as

\begin{align}
\dot{C}_{e}(t)= & -\frac{1}{\hbar^{2}}\int_{0}^{\infty}d\omega\omega^{2}\beta^{2}(\omega)\boldsymbol{d}_{eg}\cdot\int d^{2}\boldsymbol{\rho}\bar{\bar{G}}(\boldsymbol{r},\boldsymbol{\rho},z=0;\omega)\nonumber \\
 & \cdot\bar{\bar{G}}^{\dagger}(\boldsymbol{r},\boldsymbol{\rho},z=0;\omega)\cdot\boldsymbol{d}_{eg}e^{-i(\omega-\omega_{0})t}\int_{0}^{t}dt_{1}e^{-i(\omega_{0}-\omega)t_{1}}C_{e}(t_{1}),
\end{align}
and we can further introduce the kernel (see appendix \ref{sec:Integrals-of-products})
$\bar{\mathcal{K}}(\boldsymbol{r},\omega)$ defined by

\begin{align}
\bar{\mathcal{K}}(\boldsymbol{r},\omega) & =\varepsilon_{0}\mu_{0}\omega^{2}\frac{\sigma_{1}(\omega)}{\varepsilon_{0}\omega}\int d^{2}\boldsymbol{\rho}\bar{\bar{G}}(\boldsymbol{r},\boldsymbol{\rho},z=0;\omega)\cdot\bar{\bar{G}}^{\dagger}(\boldsymbol{r},\boldsymbol{\rho},z=0;\omega)\nonumber \\
 & =\frac{\omega^{2}}{c^{2}}\frac{\sigma_{1}(\omega)}{\varepsilon_{0}\omega}\int d^{2}\boldsymbol{\rho}\bar{\bar{G}}(\boldsymbol{r},\boldsymbol{\rho},z=0;\omega)\cdot\bar{\bar{G}}^{\dagger}(\boldsymbol{r},\boldsymbol{\rho},z=0;\omega)\nonumber \\
 & =\text{Im}[\bar{\bar{G}}(\boldsymbol{r},\boldsymbol{r};\omega)],
\end{align}
We shall discuss the evaluation of this integral in due time, but
first we move on to write

\begin{equation}
\dot{C}_{e}(t)=-\frac{\bar{\varepsilon}\mu_{0}}{(2\pi)^{3}\hbar}\int_{0}^{\infty}d\omega\omega^{2}\boldsymbol{d}_{eg}\cdot\bar{\mathcal{K}}(\boldsymbol{r},\omega)\cdot\boldsymbol{d}_{eg}\int_{0}^{t}dt_{1}e^{-i(\omega_{0}-\omega)(t_{1}-t)}C_{e}(t_{1}).
\end{equation}
We can now make the first Markov approximation to write $C_{e}(t_{1})\approx C_{e}(t)$.
We then have

\begin{align}
\dot{C}_{e}(t) & =-\frac{\bar{\varepsilon}\mu_{0}}{(2\pi)^{3}\hbar}\int_{0}^{\infty}d\omega\omega^{2}\boldsymbol{d}_{eg}\cdot\bar{\mathcal{K}}(\boldsymbol{r},\omega)\cdot\boldsymbol{d}_{eg}C_{e}(t)\int_{0}^{t}dt_{1}e^{-i(\omega_{0}-\omega)(t_{1}-t)}\nonumber \\
 & =-\frac{\bar{\varepsilon}\mu_{0}}{(2\pi)^{3}\hbar}\int_{0}^{\infty}d\omega\omega^{2}\boldsymbol{d}_{eg}\cdot\bar{\mathcal{K}}(\boldsymbol{r},\omega)\cdot\boldsymbol{d}_{eg}C_{e}(t).
\end{align}
By the Sokhotski--Plemelj identity the integral is

\begin{equation}
\mathcal{I}_{SP}(\omega)=\pi\delta(\omega-\omega_{0})+i\mathcal{P}\frac{1}{\omega-\omega_{0}}
\end{equation}
Using this result, we have

\begin{equation}
\dot{C_{e}}(t)=iC_{e}(t)\left[-\Delta+i\Gamma\right],
\end{equation}
where we have introduced the Lamb Shift

\begin{equation}
\Delta=\frac{\bar{\varepsilon}\mu_{0}}{(2\pi)^{3}\hbar}\mathcal{P}\int d\omega\omega^{2}\frac{\boldsymbol{d}_{eg}\cdot\text{Im}[\bar{\bar{G}}(\boldsymbol{r},\boldsymbol{r};\omega)]\cdot\boldsymbol{d}_{eg}}{\omega-\omega_{0}},
\end{equation}
and the decay rate

\begin{align}
\Gamma & =-\pi\frac{\bar{\varepsilon}\mu_{0}}{(2\pi)^{3}\hbar}\int d\omega\omega^{2}\boldsymbol{d}_{eg}\cdot\text{Im}[\bar{\bar{G}}(\boldsymbol{r},\boldsymbol{r};\omega)]\cdot\boldsymbol{d}_{eg}\delta(\omega-\omega_{0})\nonumber \\
 & =-\pi\frac{\bar{\varepsilon}}{(2\pi)^{3}\hbar\varepsilon_{0}c^{2}}\omega_{0}^{2}\boldsymbol{d}_{eg}\cdot\text{Im}[\bar{\bar{G}}(\boldsymbol{r},\boldsymbol{r};\omega_{0})]\cdot\boldsymbol{d}_{eg}.
\end{align}
We will now compute this rate using the dispersion relation

\subsubsection{Computation of the decay rate of a single qubit\label{subsec:Computation-of-the}}

We now aim to calculate the decay rate for our two-level system close
to the Graphene sheet. We therefore will make use of the Green's function
we found previously in equation (\ref{eq:GrapheneGreensFunction}),
to calculate

\begin{equation}
\text{Im}[\bar{\bar{G}}(\boldsymbol{r},\boldsymbol{r};\omega_{0})]=c^{2}\frac{\pi^{2}}{\bar{\varepsilon}}e^{-2q_{\text{spp}}z}\frac{q_{\text{spp}}^{2}}{A(\omega_{0})}\text{Im}[\bar{M}(q_{\text{ssp}},0,0)],
\end{equation}
where $A$ is in general, a function of $\omega_{0}$. The imaginary
part of the evaluated matrix gives

\begin{equation}
\begin{cases}
\text{Im}\left[M_{11}\right]=-J_{0}\left(q_{\text{spp}}R\right)+\frac{J_{1}\left(q_{\text{spp}}R\right)}{q_{\text{spp}}R},\\
\text{Im}\left[M_{22}\right]=-\frac{J_{1}\left(q_{\text{spp}}R\right)}{q_{\text{spp}}R},\\
\text{Im}\left[M_{33}\right]=-J_{0}\left(q_{\text{spp}}R\right),
\end{cases}
\end{equation}
and

\begin{equation}
\begin{cases}
\text{Im}\left[M_{12}\right]=\text{Im}\left[M_{21}\right]=0,\\
\text{Im}\left[M_{13}\right]=\text{Im}\left[M_{31}\right]=J_{1}\left(q_{\text{spp}}R\right),\\
\text{Im}\left[M_{23}\right]=\text{Im}\left[M_{32}\right]=0.
\end{cases}.
\end{equation}
Therefore, we have, if we let $a\equiv q_{\text{spp}}R$ for the sake
of simplifying the notation, we find

\[
\Gamma=-\pi\frac{\bar{\varepsilon}}{(2\pi)^{3}\hbar\varepsilon_{0}c^{2}}\omega_{0}^{2}\boldsymbol{d}_{eg}\cdot\text{Im}[\bar{G}_{1}(\boldsymbol{r},\boldsymbol{r};\omega_{0})]\cdot\boldsymbol{d}_{eg}
\]

\begin{align}
\Gamma & =-\frac{\pi\bar{\varepsilon}\mu_{0}}{(2\pi)^{3}\hbar}\omega_{0}^{2}\left(-c^{2}\frac{\pi^{2}}{\bar{\varepsilon}}e^{-2q_{\text{spp}}z}\frac{q_{\text{spp}}^{2}}{A(\omega)}\right)\nonumber \\
 & \times\boldsymbol{d}_{eg}\cdot\left[\lim_{a\to0}\begin{pmatrix}J_{0}\left(a\right)-\frac{J_{1}\left(a\right)}{a} & 0 & -J_{1}\left(a\right)\\
0 & \frac{J_{1}\left(a\right)}{a} & 0\\
-J_{1}\left(a\right) & 0 & J_{0}\left(a\right)
\end{pmatrix}\right]\cdot\boldsymbol{d}_{eg}.
\end{align}
Introducing the notation

\begin{align}
\Gamma & =c^{2}\frac{\mu_{0}}{8\hbar}\omega_{0}^{2}e^{-2q_{\text{spp}}z}\frac{q_{\text{spp}}^{2}}{A(\omega_{0})}\boldsymbol{d}_{eg}\cdot\begin{pmatrix}1/2 & 0 & 0\\
0 & 1/2 & 0\\
0 & 0 & 1
\end{pmatrix}\cdot\boldsymbol{d}_{eg}\nonumber \\
 & =\frac{1}{16\hbar\varepsilon_{0}}\frac{\omega_{0}^{2}}{A(\omega_{0})}e^{-2q_{\text{spp}}z}q_{\text{spp}}^{2}\boldsymbol{d}_{eg}\cdot\begin{pmatrix}1 & 0 & 0\\
0 & 1 & 0\\
0 & 0 & 2
\end{pmatrix}\cdot\boldsymbol{d}_{eg}\nonumber \\
 & =\frac{1}{16\hbar\varepsilon_{0}}q_{\text{spp}}^{3}e^{-2q_{\text{spp}}z}(d_{eg}^{\parallel2}+2d_{eg}^{\perp2}),
\end{align}
and writing

\begin{equation}
q_{\text{spp}}^{3}=\frac{(2\pi)^{3}}{\lambda_{\text{spp}}^{3}},
\end{equation}
we have a result compatible with the literature

\begin{equation}
\Gamma=\frac{1}{16\hbar\varepsilon_{0}}\frac{(2\pi)^{3}}{\lambda_{\text{spp}}^{3}}e^{-2q_{\text{spp}}z}(d_{eg}^{\parallel2}+2d_{eg}^{\perp2})
\end{equation}
or by writing $\Gamma=\gamma/2$, we find the rate

\begin{equation}
\gamma=\frac{1}{8\hbar\varepsilon_{0}}\frac{(2\pi)^{3}}{\lambda_{\text{spp}}^{3}}e^{-2q_{\text{spp}}z}(d_{eg}^{\parallel2}+2d_{eg}^{\perp2}).
\end{equation}
This is the final result for the rate $\gamma$ that characterizes
the spontaneous decay of a two-level system coupled to an SPP field.

\subsection{Deriving the Lindblad equation and computing the dynamics of a single
qubit\label{subsec:Deriving-the-Lindblad}}

We now aim to present a microscopic derivation of the Lindblad equation
for our system. This will allow us to extend these calculations to
the case where both qubits are excited, but we first present a derivation
for a single qubit coupled to the SPP field. As with the case of the
thermal EM field, we start with noting that for interaction Hamiltonians
of the form

\begin{equation}
\hat{V}=\sum_{i}\beta_{i}\otimes\Gamma_{i},\label{eq:NiceHamiltonian}
\end{equation}
it is possible to write the Redfield equation in terms of the reservoir
correlation functions as

\begin{align}
\frac{d\hat{\rho}_{S}}{dt}= & -\frac{1}{\hbar^{2}}\sum_{ij}\int_{t_{0}}^{t}d\tau\left(\beta_{i}(t)\beta_{j}(\tau)\hat{\rho}_{S}(t)-\beta_{j}(\tau)\hat{\rho}_{S}(t)\beta_{i}(t)\right)\left\langle \Gamma_{i}(t)\Gamma_{j}(\tau)\right\rangle _{R}\nonumber \\
 & +\left(\hat{\rho}_{S}(t)\beta_{j}(\tau)\beta_{i}(t)-\beta_{i}(t)\hat{\rho}_{S}(t)\beta_{j}(\tau)\right)\left\langle \Gamma_{j}(\tau)\Gamma_{i}(t)\right\rangle _{R}.\label{eq:Redfield-1}
\end{align}
These reservoir correlation functions are defined as

\begin{equation}
\begin{cases}
\left\langle \Gamma_{i}(t)\Gamma_{j}(\tau)\right\rangle \equiv\text{tr}_{R}\left\{ \hat{\rho}_{R}\Gamma_{i}(t)\Gamma_{j}(\tau)\right\} ,\\
\left\langle \Gamma_{j}(t)\Gamma_{i}(\tau)\right\rangle \equiv\text{tr}_{R}\left\{ \hat{\rho}_{R}\Gamma_{j}(t)\Gamma_{i}(\tau)\right\} .
\end{cases}
\end{equation}
The Hamiltonian for a single qubit are given by equation (\ref{eq:Hamiltonians})
if the indices $\alpha$ are all set to $1$. Since the commutation
relations for SPP creation and destruction operators follow directly
from equation (\ref{eq:feffec}), we can find the operators in the
interaction picture via

\begin{align*}
\hat{\text{\textbf{f}}}_{I}^{\dagger}(\boldsymbol{\rho},\omega) & =e^{i\hat{H}_{0}t/\hbar}\hat{\text{\textbf{f}}}^{\dagger}(\boldsymbol{\rho},\omega)e^{-i\hat{H}_{0}t/\hbar}.
\end{align*}
We make use of the commutator property $Ae^{B}=e^{c}e^{B}A$, where
$\left[A,B\right]=cA$. In this case, we have

\begin{equation}
e^{i\hat{H}_{0}t/\hbar}\hat{\text{\textbf{f}}}^{\dagger}(\boldsymbol{\rho},\omega)e^{-i\hat{H}_{0}t/\hbar}=e^{c}\hat{\text{\textbf{f}}}^{\dagger}(\boldsymbol{\rho},\omega),
\end{equation}
where

\begin{align}
\left[\hat{\text{\textbf{f}}}^{\dagger}(\boldsymbol{\rho},\omega),-i\hat{H}_{0}t/\hbar\right] & =\left[\hat{\text{\textbf{f}}}^{\dagger}(\boldsymbol{\rho},\omega),-it/\hbar\int d^{2}\boldsymbol{\rho}'\int_{0}^{\infty}d\nu\hbar\nu\hat{\text{\textbf{f}}}^{\dagger}(\boldsymbol{\rho}',\nu)\hat{\text{\textbf{f}}}(\boldsymbol{\rho}',\nu)\right]\nonumber \\
 & =-it\int d^{2}\boldsymbol{\rho}'\int_{0}^{\infty}d\nu\nu\left[\hat{\text{\textbf{f}}}^{\dagger}(\boldsymbol{\rho},\omega),\hat{\text{\textbf{f}}}^{\dagger}(\boldsymbol{\rho}',\nu)\hat{\text{\textbf{f}}}(\boldsymbol{\rho}',\nu)\right].
\end{align}
The commutator can be expanded as

\begin{align}
\left[\hat{\text{\textbf{f}}}^{\dagger}(\boldsymbol{\rho},\omega),\hat{\text{\textbf{f}}}^{\dagger}(\boldsymbol{\rho}',\nu)\hat{\text{\textbf{f}}}(\boldsymbol{\rho}',\nu)\right] & =\left[\hat{\text{\textbf{f}}}^{\dagger}(\boldsymbol{\rho},\omega),\hat{\text{\textbf{f}}}^{\dagger}(\boldsymbol{\rho}',\nu)\right]\hat{\text{\textbf{f}}}(\boldsymbol{\rho}',\nu)+\hat{\text{\textbf{f}}}^{\dagger}(\boldsymbol{\rho}',\nu)\left[\hat{\text{\textbf{f}}}^{\dagger}(\boldsymbol{\rho},\omega),\hat{\text{\textbf{f}}}(\boldsymbol{\rho}',\nu)\right]\nonumber \\
 & =-\hat{\text{\textbf{f}}}^{\dagger}(\boldsymbol{\rho}',\nu)\delta(\boldsymbol{\rho}-\boldsymbol{\rho}')\delta(\omega-\nu),
\end{align}
from which we have

\begin{align}
\left[\hat{\text{\textbf{f}}}^{\dagger}(\boldsymbol{\rho},\omega),-i\hat{H}_{0}t/\hbar\right] & =it\int d^{2}\boldsymbol{\rho}'\int_{0}^{\infty}d\nu\nu\hat{\text{\textbf{f}}}^{\dagger}(\boldsymbol{\rho}',\nu)\delta(\boldsymbol{\rho}-\boldsymbol{\rho}')\delta(\omega-\nu)\nonumber \\
 & =i\omega t\hat{\text{\textbf{f}}}^{\dagger}(\boldsymbol{\rho},\omega).
\end{align}
Thus, we can use the privously mentioned identity, with $c=i\omega t$.
The result is the expected

\begin{equation}
\hat{\text{\textbf{f}}}_{I}^{\dagger}(\boldsymbol{\rho},\omega)=\hat{\text{\textbf{f}}}^{\dagger}(\boldsymbol{\rho},\omega)e^{i\omega t}.
\end{equation}
A similar calculation yields

\begin{equation}
\left[\hat{\text{\textbf{f}}}(\boldsymbol{\rho},\omega),-i\hat{H}_{0}t/\hbar\right]=\int d^{2}\boldsymbol{\rho}'\int_{0}^{\infty}d\nu\nu\left[\hat{\text{\textbf{f}}}(\boldsymbol{\rho},\omega),\hat{\text{\textbf{f}}}^{\dagger}(\boldsymbol{\rho}',\nu)\hat{\text{\textbf{f}}}(\boldsymbol{\rho}',\nu)\right],
\end{equation}
where the commutator yields

\begin{align}
\left[\hat{\text{\textbf{f}}}(\boldsymbol{\rho},\omega),\hat{\text{\textbf{f}}}^{\dagger}(\boldsymbol{\rho}',\nu)\hat{\text{\textbf{f}}}(\boldsymbol{\rho}',\nu)\right] & =\left[\hat{\text{\textbf{f}}}(\boldsymbol{\rho},\omega),\hat{\text{\textbf{f}}}^{\dagger}(\boldsymbol{\rho}',\nu)\right]\hat{\text{\textbf{f}}}(\boldsymbol{\rho}',\nu)+\hat{\text{\textbf{f}}}^{\dagger}(\boldsymbol{\rho}',\nu)\left[\hat{\text{\textbf{f}}}(\boldsymbol{\rho},\omega),\hat{\text{\textbf{f}}}(\boldsymbol{\rho}',\nu)\right]\nonumber \\
 & =\hat{\text{\textbf{f}}}(\boldsymbol{\rho}',\nu)\delta(\boldsymbol{\rho}-\boldsymbol{\rho}')\delta(\omega-\nu),
\end{align}
from where follows

\begin{equation}
\left[\hat{\text{\textbf{f}}}(\boldsymbol{\rho},\omega),-i\hat{H}_{0}t/\hbar\right]=-i\omega t\hat{\text{\textbf{f}}}^{\dagger}(\boldsymbol{\rho},\omega),
\end{equation}
which gives the representation

\begin{equation}
\hat{\text{\textbf{f}}}_{I}(\boldsymbol{\rho},\omega)=\hat{\text{\textbf{f}}}(\boldsymbol{\rho},\omega)e^{-i\omega t}.
\end{equation}
As we have seen before, we also have for the qubit raising and lowering
operators

\begin{equation}
\begin{cases}
\sigma_{+}^{I}=\sigma_{+}e^{i\omega_{0}t},\\
\sigma_{-}^{I}=\sigma_{-}e^{-i\omega_{0}t},
\end{cases}
\end{equation}
We can now identify the Hamiltonian of SPP field coupled to the qubit
as one of the form (\ref{eq:NiceHamiltonian}), by making the definitions

\[
\begin{cases}
\hbar\mathcal{G}^{-}(\boldsymbol{r},\boldsymbol{\rho};\omega)\equiv\omega\beta(\omega)\boldsymbol{d}\cdot\bar{\bar{G}}(\boldsymbol{r},\boldsymbol{\rho},z=0;\omega),\\
\hbar\mathcal{G}^{+}(\boldsymbol{r},\boldsymbol{\rho};\omega)\equiv\omega\beta^{*}(\omega)\bar{\bar{G}}^{\dagger}(\boldsymbol{r},\boldsymbol{\rho},z=0;\omega)\cdot\boldsymbol{d}^{\dagger},
\end{cases}
\]
and then

\begin{equation}
\begin{cases}
\beta_{1}=\sigma_{-},\\
\beta_{2}=\sigma_{+},\\
\Gamma_{1}=\Gamma^{\dagger}=i\int_{0}^{\infty}d\omega\int d^{2}\boldsymbol{\rho}\hat{\text{\textbf{f}}}^{\dagger}(\boldsymbol{\rho},\omega)\cdot\mathcal{G}^{+}(\boldsymbol{r},\boldsymbol{\rho};\omega),\\
\Gamma_{2}=\Gamma=-i\int_{0}^{\infty}d\omega\int d^{2}\boldsymbol{\rho}\mathcal{G}^{-}(\boldsymbol{r},\boldsymbol{\rho};\omega)\cdot\hat{\text{\textbf{f}}}(\boldsymbol{\rho},\omega),
\end{cases}
\end{equation}
with which the interaction Hamiltonian takes a form identical to that
we had previously obtained when treating the thermal E-M field

\[
\hat{V}=\hbar\left(\sigma_{-}\Gamma^{\dagger}+\sigma_{+}\Gamma\right).
\]
The procedure to derive the Lindblad equation is therefore very similar
to the case of the thermal E-M field. We can write all the operators
in the interaction picture

\begin{equation}
\begin{cases}
\tilde{\beta}_{1}(t)=\sigma_{-}e^{-i\omega_{0}t},\\
\tilde{\beta}_{2}(t)=\sigma_{+}e^{i\omega_{0}t},\\
\tilde{\Gamma}_{1}(t)=i\int_{0}^{\infty}d\omega\int d^{2}\boldsymbol{\rho}\hat{\text{\textbf{f}}}^{\dagger}(\boldsymbol{\rho},\omega)\cdot\mathcal{G}^{+}(\boldsymbol{r},\boldsymbol{\rho};\omega)e^{i\omega t},\\
\tilde{\Gamma}_{2}(t)=-i\int_{0}^{\infty}d\omega\int d^{2}\boldsymbol{\rho}\mathcal{G}^{-}(\boldsymbol{r},\boldsymbol{\rho};\omega)\cdot\hat{\text{\textbf{f}}}(\boldsymbol{\rho},\omega)e^{-i\omega t}.
\end{cases}
\end{equation}
The summation present in the Redfield equation is over $i=1,2$ and
$j=1,2$ and therefore, we can evaluate all the reservoir correlation
functions in the interaction picture

\begin{align}
\left\langle \tilde{\Gamma}_{1}(t)\tilde{\Gamma}_{1}(\tau)\right\rangle  & =\text{tr}_{R}\left\{ \hat{\rho}_{R}\tilde{\Gamma}(t)\tilde{\Gamma}(\tau)\right\} \nonumber \\
 & =\text{tr}_{R}\left\{ -\hat{\rho}_{R}\int_{0}^{\infty}d\omega\int d^{2}\boldsymbol{\rho}\hat{\text{\textbf{f}}}^{\dagger}(\boldsymbol{\rho},\omega)\cdot\mathcal{G}^{+}(\boldsymbol{r},\boldsymbol{\rho};\omega)e^{i\omega t}\right.\nonumber \\
 & \left.\int_{0}^{\infty}d\omega'\int d^{2}\boldsymbol{\rho}'\hat{\text{\textbf{f}}}^{\dagger}(\boldsymbol{\rho}',\omega')\cdot\mathcal{G}^{+}(\boldsymbol{r},\boldsymbol{\rho}';\omega')e^{i\omega'\tau}\right\} \nonumber \\
 & =0,
\end{align}
in the rotating wave approximation. Similarly

\begin{equation}
\left\langle \tilde{\Gamma}_{2}(t)\tilde{\Gamma}_{2}(\tau)\right\rangle =0.
\end{equation}
The story becomes much more complicated, however, for the remaining
reservoir correlation functions $\left\langle \tilde{\Gamma}_{1}(t)\tilde{\Gamma}_{2}(\tau)\right\rangle $
and $\left\langle \tilde{\Gamma}_{2}(t)\tilde{\Gamma}_{1}(\tau)\right\rangle $.
We will therefore consider first the case of zero temperature, which
is much simpler and will allow us to recover the results obtained
in the Schrodinger picture. Then, we will move on to a description
at finite temperature.

\subsection{Evaluating the correlation functions at temperature $T=0$: Density
matrix $\rho_{R}=\vert0\rangle\langle0\vert$\label{subsec:Evaluating-the-correlation}}

In this appendix we provide the aforementioned zero temperature description
of the The reason for choosing the temperature $T=0$ is that the
density matrix of the reservoir will be given by 
\begin{equation}
\rho_{R}=\vert\psi\rangle\langle\psi\vert=\vert0\rangle\langle0\vert,
\end{equation}
since it has not been perturbed by the decay of the qubit. We can
also work in the subspace of the reservoir with basis $\left\{ \left|0\right\rangle ,\left|\mathds{1}(\boldsymbol{\rho},\omega)\right\rangle \right\} $,
since at most there will be a single SPP present. In this subspace,
the trace will in principle be calculated over two terms, however
we see that one of these terms immediately vanishes due to the form
of the density matrix. This results in

\begin{align*}
\left\langle \tilde{\Gamma}_{1}(t)\tilde{\Gamma}_{2}(\tau)\right\rangle = & \text{tr}_{R}\left\{ \hat{\rho}_{R}\tilde{\Gamma}_{1}(t)\tilde{\Gamma}_{2}(\tau)\right\} \\
 & =\braket{0|0}\left\langle 0\right|\int_{0}^{\infty}d\omega\int d^{2}\boldsymbol{\rho}\hat{\text{\textbf{f}}}^{\dagger}(\boldsymbol{\rho},\omega)\cdot\mathcal{G}^{+}(\boldsymbol{r},\boldsymbol{\rho};\omega)e^{i\omega t}\\
 & \int_{0}^{\infty}d\omega'\int d^{2}\boldsymbol{\rho}'\mathcal{G}^{-}(\boldsymbol{r},\boldsymbol{\rho}';\omega')\cdot\hat{\text{\textbf{f}}}(\boldsymbol{\rho}',\omega')e^{i\omega'\tau}\left|0\right\rangle \\
 & =0.
\end{align*}
The other correlation function is given by

\begin{align}
\left\langle \tilde{\Gamma}_{2}(t)\tilde{\Gamma}_{1}(\tau)\right\rangle  & =\text{tr}_{R}\left\{ \hat{\rho}_{R}\tilde{\Gamma}_{2}(t)\tilde{\Gamma}_{1}(\tau)\right\} \nonumber \\
 & =\braket{0|0}\left\langle 0\right|\int_{0}^{\infty}d\omega\int d^{2}\boldsymbol{\rho}\mathcal{G}^{-}(\boldsymbol{r},\boldsymbol{\rho};\omega)\cdot\hat{\text{\textbf{f}}}(\boldsymbol{\rho},\omega)e^{i\omega t}\int_{0}^{\infty}d\omega'\int d^{2}\boldsymbol{\rho}'\hat{\text{\textbf{f}}}^{\dagger}(\boldsymbol{\rho}',\omega')\cdot\mathcal{G}^{+}(\boldsymbol{r}',\boldsymbol{\rho}';\omega')e^{i\omega'\tau}\left|0\right\rangle \nonumber \\
 & =\int_{0}^{\infty}d\omega\int d^{2}\boldsymbol{\rho}\mathcal{G}^{-}(\boldsymbol{r},\boldsymbol{\rho};\omega)\cdot\left\langle \mathds{1}(\boldsymbol{\rho},\omega)\right|e^{-i\omega t}\int_{0}^{\infty}d\omega'\int d^{2}\boldsymbol{\rho}'\left|\mathds{1}(\boldsymbol{\rho}',\omega')\right\rangle \cdot\mathcal{G}^{+}(\boldsymbol{r}',\boldsymbol{\rho}';\omega')e^{i\omega'\tau}\nonumber \\
 & =\int_{0}^{\infty}d\omega'\int d^{2}\boldsymbol{\rho}'\int_{0}^{\infty}d\omega\int d^{2}\boldsymbol{\rho}\mathcal{G}^{-}(\boldsymbol{r},\boldsymbol{\rho};\omega)\cdot\mathcal{G}^{+}(\boldsymbol{r}',\boldsymbol{\rho}';\omega')\delta(\boldsymbol{\rho}-\boldsymbol{\rho}')\delta(\omega-\omega')e^{i\omega'\tau}e^{-i\omega t}\nonumber \\
 & =\int_{0}^{\infty}d\omega\int d^{2}\boldsymbol{\rho}\mathcal{G}^{-}(\boldsymbol{r},\boldsymbol{\rho};\omega)\cdot\mathcal{G}^{+}(\boldsymbol{r},\boldsymbol{\rho};\omega)e^{i\omega(\tau-t)}.
\end{align}
Similarly

\[
\left\langle \tilde{\Gamma}_{2}(\tau)\tilde{\Gamma}_{1}(t)\right\rangle =\left\langle \tilde{\Gamma}_{2}(t)\tilde{\Gamma}_{1}(\tau)\right\rangle ^{*}.
\]
Substituting into the Redfield equation (\ref{eq:Redfield-2-1}),
all terms but those proportional to this correlation function vanish,
and therefore we are left with

\begin{align}
\frac{d\hat{\rho}_{S}^{I}}{dt}= & -\frac{1}{\hbar^{2}}\int_{t_{0}}^{t}d\tau\left(\tilde{\beta}_{2}(t)\tilde{\beta}_{1}(\tau)\hat{\rho}_{S}^{I}(t)-\tilde{\beta}_{1}(\tau)\hat{\rho}_{S}^{I}(t)\tilde{\beta}_{2}(t)\right)\left\langle \tilde{\Gamma}_{2}(t)\tilde{\Gamma}_{1}(\tau)\right\rangle _{R}\nonumber \\
 & +\left(\hat{\rho}_{S}^{I}(t)\tilde{\beta}_{2}(\tau)\tilde{\beta}_{1}(t)-\tilde{\beta}_{1}(t)\hat{\rho}_{S}^{I}(t)\tilde{\beta}_{2}(\tau)\right)\left\langle \tilde{\Gamma}_{2}(\tau)\tilde{\Gamma}_{1}(t)\right\rangle _{R},
\end{align}
and substituting the operators, we obtain

\begin{align}
\frac{d\hat{\rho}_{S}^{I}}{dt}= & -\frac{1}{\hbar^{2}}\int_{0}^{t}d\tau\left(\sigma_{+}\sigma_{-}e^{i\omega_{0}(t-\tau)}\hat{\rho}_{S}^{I}(t)-\sigma_{-}\hat{\rho}_{S}^{I}(t)\sigma_{+}e^{i\omega_{0}(t-\tau)}\right)\left\langle \tilde{\Gamma}_{1}(t)\tilde{\Gamma}_{2}(\tau)\right\rangle \nonumber \\
 & +\left(\hat{\rho}_{S}^{I}(t)\sigma_{+}\sigma_{-}e^{i\omega_{0}(\tau-t)}-\sigma_{-}\hat{\rho}_{S}^{I}(t)\sigma_{+}e^{i\omega_{0}(\tau-t)}\right)\left\langle \tilde{\Gamma}_{2}(t)\tilde{\Gamma}_{1}(\tau)\right\rangle ^{*}.
\end{align}
If we introduce the notation

\begin{equation}
\boldsymbol{d}_{eg}\cdot\bar{\mathcal{K}}(\boldsymbol{r},\omega)\cdot\boldsymbol{d}_{eg}\equiv\frac{1}{\hbar^{2}}\int d^{2}\boldsymbol{\rho}\mathcal{G}^{-}(\boldsymbol{r},\boldsymbol{\rho};\omega)\cdot\mathcal{G}^{+}(\boldsymbol{r},\boldsymbol{\rho};\omega).
\end{equation}
We can perform the Markov approximation, extending the integral in
time to infinity and writing $\hat{\rho}_{S}^{I}(t_{1})\approx\hat{\rho}_{S}^{I}(t)$.
This yields

\begin{align}
\frac{d\hat{\rho}_{S}^{I}}{dt}= & -\left(\sigma_{+}\sigma_{-}\hat{\rho}_{S}^{I}(t)-\sigma_{-}\hat{\rho}_{S}^{I}(t)\sigma_{+}\right)\int_{0}^{\infty}d\omega\boldsymbol{d}_{eg}\cdot\bar{\mathcal{K}}(\boldsymbol{r},\omega)\cdot\boldsymbol{d}_{eg}\int_{0}^{\infty}dt_{1}e^{i(\omega-\omega_{0})(t_{1}-t)}\nonumber \\
 & -\left(\hat{\rho}_{S}^{I}(t)\sigma_{+}\sigma_{-}-\sigma_{-}\hat{\rho}_{S}^{I}(t)\sigma_{+}\right)\int_{0}^{\infty}d\omega\boldsymbol{d}_{eg}\cdot\bar{\mathcal{K}}(\boldsymbol{r},\omega)\cdot\boldsymbol{d}_{eg}\int_{0}^{\infty}dt_{1}e^{-i(\omega-\omega_{0})(t_{1}-t)},
\end{align}
and we obtain

\begin{align}
\frac{d\hat{\rho}_{S}^{I}}{dt}= & -\left(\sigma_{+}\sigma_{-}\hat{\rho}_{S}^{I}(t)-\sigma_{-}\hat{\rho}_{S}^{I}(t)\sigma_{+}\right)\int_{0}^{\infty}d\omega\bar{\mathcal{K}}(\boldsymbol{r},\omega)\mathcal{I}(\omega)\nonumber \\
 & -\left(\hat{\rho}_{S}^{I}(t)\sigma_{+}\sigma_{-}-\sigma_{-}\hat{\rho}_{S}^{I}(t)\sigma_{+}\right)\int_{0}^{\infty}d\omega\bar{\mathcal{K}}(\boldsymbol{r},\omega)\mathcal{I}^{*}(\omega),
\end{align}
where

\begin{equation}
\mathcal{I}(\omega)=\pi\delta(\omega-\omega_{\alpha'})-i\mathcal{P}\frac{1}{\omega-\omega_{\alpha'}}.
\end{equation}
We can therefore define

\begin{equation}
\int_{0}^{\infty}d\omega\boldsymbol{d}_{eg}\cdot\bar{\mathcal{K}}(\boldsymbol{r},\omega)\cdot\boldsymbol{d}_{eg}\mathcal{I}(\omega)\equiv\Gamma+i\Delta,
\end{equation}
which lets us write

\begin{align}
\frac{d\hat{\rho}_{S}^{I}}{dt}= & -\left(\sigma_{+}\sigma_{-}\hat{\rho}_{S}^{I}(t)-\sigma_{-}\hat{\rho}_{S}^{I}(t)\sigma_{+}\right)(\Gamma+i\Delta)\nonumber \\
 & -\left(\hat{\rho}_{S}^{I}(t)\sigma_{+}\sigma_{-}-\sigma_{-}\hat{\rho}_{S}^{I}(t)\sigma_{+}\right)(\Gamma-i\Delta).
\end{align}
We can write this more compactly as

\begin{align*}
\frac{d\hat{\rho}_{S}^{I}}{dt}= & -\left(\sigma_{+}\sigma_{-}\hat{\rho}_{S}^{I}(t)-\sigma_{-}\hat{\rho}_{S}^{I}(t)\sigma_{+}+\hat{\rho}_{S}^{I}(t)\sigma_{+}\sigma_{-}-\sigma_{-}\hat{\rho}_{S}^{I}(t)\sigma_{+}\right)\Gamma\\
 & -\left(\sigma_{+}\sigma_{-}\hat{\rho}_{S}^{I}(t)-\sigma_{-}\hat{\rho}_{S}^{I}(t)\sigma_{+}-\hat{\rho}_{S}^{I}(t)\sigma_{+}\sigma_{-}+\sigma_{-}\hat{\rho}_{S}^{I}(t)\sigma_{+}\right)i\Delta\\
= & -\Gamma\left(\left\{ \sigma_{+}\sigma_{-},\hat{\rho}_{S}^{I}(t)\right\} -2\sigma_{-}\hat{\rho}_{S}^{I}(t)\sigma_{+}\right)-i\Delta\left(\left[\sigma_{+}\sigma_{-},\hat{\rho}_{S}^{I}(t)\right]\right).
\end{align*}
We now want to return to the Shcrodinger picture, and therefore use
the definition of the operator in the interaction picture

\begin{equation}
\hat{\rho}_{S}^{I}(t)=e^{iH_{0}t/\hbar}\hat{\rho}_{S}e^{-iH_{0}t/\hbar},
\end{equation}
with which we can calculate the derivative 
\begin{equation}
\frac{d\hat{\rho}_{S}^{I}}{dt}=e^{iH_{0}t/\hbar}\frac{d\hat{\rho}_{S}}{dt}e^{-iH_{0}t/\hbar}+\frac{i}{\hbar}e^{iH_{0}t/\hbar}\left[H_{0},\hat{\rho}_{S}(t)\right]e^{-iH_{0}t/\hbar},
\end{equation}
and can further rewrite

\begin{equation}
\frac{d\hat{\rho}_{S}}{dt}=e^{-iH_{0}t/\hbar}\frac{d\hat{\rho}_{S}^{I}}{dt}e^{iH_{0}t/\hbar}-\frac{i}{\hbar}\left[H_{0},\hat{\rho}_{S}(t)\right].
\end{equation}
We can now convert from the interaction picture to the Schrodinger
picture, by writing

\begin{align}
\frac{d\hat{\rho}_{S}}{dt}= & -\Gamma\left(e^{-iH_{0}t/\hbar}\left\{ \sigma_{+}\sigma_{-},\hat{\rho}_{S}^{I}(t)\right\} e^{iH_{0}t/\hbar}-e^{-iH_{0}t/\hbar}2\sigma_{-}\hat{\rho}_{S}^{I}(t)\sigma_{+}e^{iH_{0}t/\hbar}\right)\nonumber \\
 & -i\Delta\left(e^{-iH_{0}t/\hbar}\left[\sigma_{+}\sigma_{-},\hat{\rho}_{S}^{I}(t)\right]e^{iH_{0}t/\hbar}\right)\nonumber \\
 & -\frac{i}{\hbar}\left[H_{0},\hat{\rho}_{S}(t)\right]\nonumber \\
 & =\Gamma\left(2\sigma_{-}\hat{\rho}_{S}(t)\sigma_{+}-\left\{ \sigma_{+}\sigma_{-},\hat{\rho}_{S}(t)\right\} \right)-i\Delta\left(\left[\sigma_{+}\sigma_{-},\hat{\rho}_{S}(t)\right]\right)-\frac{i}{\hbar}\left[H_{0},\hat{\rho}_{S}(t)\right].
\end{align}
We can now define

\begin{equation}
\begin{cases}
\hat{L}=\sqrt{\Gamma}\sigma_{-},\\
\hat{L}^{\dagger}=\sqrt{\Gamma}\sigma_{+},
\end{cases}
\end{equation}
which yields

\begin{align}
\frac{d\hat{\rho}_{S}}{dt} & =2\hat{L}\hat{\rho}_{S}(t)\hat{L}^{\dagger}-\left\{ \hat{L}^{\dagger}\hat{L},\hat{\rho}_{S}(t)\right\} -\frac{i}{\hbar}\left[(\omega+\Delta)\sigma_{+}\sigma_{-},\hat{\rho}_{S}(t)\right].
\end{align}
This is an equation in Lindblad form, and we now check that the dynamics
reproduces those found via Schrodinger's equation. We write the previous
equation in the convenient form

\[
\frac{d\hat{\rho}_{S}}{dt}=\Gamma\left(2\sigma_{-}\hat{\rho}_{S}(t)\sigma_{+}-\left\{ \sigma_{+}\sigma_{-},\hat{\rho}_{S}(t)\right\} \right)-\frac{i}{\hbar}\left[\hbar(\omega_{0}+\Delta)\sigma_{+}\sigma_{-},\hat{\rho}_{S}(t)\right].
\]
We now expand the density matrix as

\begin{align}
\hat{\rho}_{S} & =\sum_{ij}\rho_{ij}(t)\left|i\right\rangle \left\langle j\right|,
\end{align}
but look only at the populations $P_{i}=\rho_{ii}$. To do so, we
substitute this onto the differential equation above and act on the
left and right with the bra and ket corresponding to the state $\left|e\right\rangle $.
This yields

\begin{align}
\dot{P}_{e}(t)= & \Gamma\left(\left\langle e\right|2\sigma_{-}\hat{\rho}_{S}(t)\sigma_{+}\left|e\right\rangle -\left\langle e\right|\sigma_{+}\sigma_{-}\hat{\rho}_{S}(t)\left|e\right\rangle -\left\langle e\right|\hat{\rho}_{S}(t)\sigma_{+}\sigma_{-}\left|e\right\rangle \right)\nonumber \\
 & -\frac{i}{\hbar}(\omega_{0}+\Delta)\left(\left\langle e\right|\sigma_{+}\sigma_{-}\hat{\rho}_{S}(t)\left|e\right\rangle -\left\langle e\right|\hat{\rho}_{S}(t)\sigma_{+}\sigma_{-}\left|e\right\rangle \right)\nonumber \\
= & \Gamma\left(-P_{e}-P_{e}\right)-\frac{i}{\hbar}(\omega_{0}+\Delta)(P_{e}-P_{e})\nonumber \\
= & -2\Gamma P_{e}(t).
\end{align}
While for the ground state

\begin{align}
\dot{P}_{g}(t)= & \Gamma\left(\left\langle g\right|2\sigma_{-}\hat{\rho}_{S}(t)\sigma_{+}\left|g\right\rangle -\left\langle g\right|\sigma_{+}\sigma_{-}\hat{\rho}_{S}(t)\left|g\right\rangle -\left\langle g\right|\hat{\rho}_{S}(t)\sigma_{+}\sigma_{-}\left|g\right\rangle \right)\nonumber \\
 & -\frac{i}{\hbar}(\omega_{0}+\Delta)\left(\left\langle g\right|\sigma_{+}\sigma_{-}\hat{\rho}_{S}(t)\left|g\right\rangle -\left\langle g\right|\hat{\rho}_{S}(t)\sigma_{+}\sigma_{-}\left|g\right\rangle \right)\nonumber \\
= & 2\Gamma P_{e}(t).
\end{align}
This yields the same dynamics as we had obtained via Schrodinger's
equation, namely

\begin{equation}
P_{e}(t)=P_{e}(0)e^{-2\Gamma t}.
\end{equation}
and for the ground state

\begin{align}
P_{g}(t) & =2\Gamma\int_{0}^{t}P_{e}(t')\nonumber \\
 & =2\Gamma P_{e}(0)\frac{e^{-2\Gamma t}-1}{-2\Gamma}\nonumber \\
 & =P_{e}(0)\left(1-e^{-2\Gamma t}\right)
\end{align}
Setting $\gamma=2\Gamma$ we obtain

\begin{equation}
P_{g}(t)=P_{e}(0)\left(1-e^{-\gamma t}\right).
\end{equation}
Note that $\gamma$ has exactly of the same form as derived in appendix
\ref{subsec:Computation-of-the}.

\subsection{Evaluating the correlation functions at a finite temperature $T$\label{subsec:Evaluating-the-correlation-1}}

The Lindblad approach has the advantage of being able to derive these
results for a finite temperature aswell, where the SPP field will
be described by a bosonic distribution

\begin{equation}
\bar{n}_{\text{spp}}(\omega)=\frac{1}{e^{\hbar\omega/k_{B}T}-1}.
\end{equation}
We will use the following results found in \citet{scheel2009macroscopic}.

\begin{equation}
\begin{cases}
\left\langle \hat{\text{\textbf{f}}}(\boldsymbol{\rho},\omega)\hat{\text{\textbf{f}}}(\boldsymbol{\rho}',\omega')\right\rangle _{R}=0,\\
\left\langle \hat{\text{\textbf{f}}}^{\dagger}(\boldsymbol{\rho},\omega)\hat{\text{\textbf{f}}}^{\dagger}(\boldsymbol{\rho}',\omega')\right\rangle _{R}=0,\\
\left\langle \hat{\text{\textbf{f}}}^{\dagger}(\boldsymbol{\rho},\omega)\hat{\text{\textbf{f}}}(\boldsymbol{\rho}',\omega')\right\rangle _{R}=\bar{n}_{\text{spp}}(\omega)\delta(\boldsymbol{\rho}-\boldsymbol{\rho}')\delta(\omega-\omega'),\\
\left\langle \hat{\text{\textbf{f}}}(\boldsymbol{\rho},\omega)\hat{\text{\textbf{f}}}^{\dagger}(\boldsymbol{\rho}',\omega')\right\rangle _{R}=\left[\bar{n}_{\text{spp}}(\omega)+1\right]\delta(\boldsymbol{\rho}-\boldsymbol{\rho}')\delta(\omega-\omega').
\end{cases}.
\end{equation}
The derivation of the Lindblad equation proceeds exactly as described
in section \ref{subsec:Detailed-derivation-of}, and as before we
have only to derive the reservoir correlation functions

\begin{align}
\left\langle \tilde{\Gamma}_{1}(t)\tilde{\Gamma}_{1}(\tau)\right\rangle  & =\int_{0}^{\infty}d\omega\int d^{2}\boldsymbol{\rho}\mathcal{G}^{-}(\boldsymbol{r},\boldsymbol{\rho}';\omega')\mathcal{G}^{+}(\boldsymbol{r},\boldsymbol{\rho};\omega)e^{-i\omega'\tau}e^{i\omega t}\left\langle \hat{\text{\textbf{f}}}^{\dagger}(\boldsymbol{\rho},\omega)\hat{\text{\textbf{f}}}^{\dagger}(\boldsymbol{\rho}',\omega')\right\rangle _{R}\nonumber \\
 & =0.
\end{align}
Similarly, using $\left\langle \hat{\text{\textbf{f}}}(\boldsymbol{\rho},\omega)\hat{\text{\textbf{f}}}(\boldsymbol{\rho}',\omega')\right\rangle =0$,
we find

\begin{equation}
\left\langle \tilde{\Gamma}_{2}(t)\tilde{\Gamma}_{2}(\tau)\right\rangle =0.
\end{equation}
Therefore, we have to calculate only

\begin{align}
\left\langle \tilde{\Gamma}_{2}(t)\tilde{\Gamma}_{1}(\tau)\right\rangle  & =\int_{0}^{\infty}d\omega\int d^{2}\boldsymbol{\rho}\int_{0}^{\infty}d\omega'\int d^{2}\boldsymbol{\rho}'\mathcal{G}^{-}(\boldsymbol{r},\boldsymbol{\rho};\omega)\mathcal{G}^{+}(\boldsymbol{r},\boldsymbol{\rho}';\omega')e^{-i\omega t}e^{i\omega'\tau}\left\langle \hat{\text{\textbf{f}}}(\boldsymbol{\rho},\omega)\hat{\text{\textbf{f}}}^{\dagger}(\boldsymbol{\rho}',\omega')\right\rangle _{R}\nonumber \\
 & =\int_{0}^{\infty}d\omega\int d^{2}\boldsymbol{\rho}\mathcal{G}^{-}(\boldsymbol{r},\boldsymbol{\rho};\omega)\mathcal{G}^{+}(\boldsymbol{r},\boldsymbol{\rho};\omega)\left[\bar{n}_{\text{spp}}(\omega)+1\right]e^{i\omega(\tau-t)}.
\end{align}
In the same manner, we can find

\begin{align}
\left\langle \tilde{\Gamma}_{1}(t)\tilde{\Gamma}_{2}(\tau)\right\rangle  & =\int_{0}^{\infty}d\omega\int d^{2}\boldsymbol{\rho}\int_{0}^{\infty}d\omega'\int d^{2}\boldsymbol{\rho}'\mathcal{G}^{-}(\boldsymbol{r},\boldsymbol{\rho};\omega)\mathcal{G}^{+}(\boldsymbol{r},\boldsymbol{\rho}';\omega')e^{i\omega t}e^{-i\omega'\tau}\left\langle \hat{\text{\textbf{f}}}^{\dagger}(\boldsymbol{\rho},\omega)\hat{\text{\textbf{f}}}(\boldsymbol{\rho}',\omega')\right\rangle _{R}\nonumber \\
 & =\int_{0}^{\infty}d\omega\int d^{2}\boldsymbol{\rho}\mathcal{G}^{-}(\boldsymbol{r},\boldsymbol{\rho};\omega)\mathcal{G}^{+}(\boldsymbol{r},\boldsymbol{\rho};\omega)\bar{n}_{\text{spp}}(\omega)e^{-i\omega(\tau-t)}.
\end{align}
We also have

\begin{equation}
\begin{cases}
\left\langle \tilde{\Gamma}_{2}(\tau)\tilde{\Gamma}_{1}(t)\right\rangle _{R}=\left\langle \tilde{\Gamma}_{2}(t)\tilde{\Gamma}_{1}(\tau)\right\rangle _{R}^{*},\\
\left\langle \tilde{\Gamma}_{1}(\tau)\tilde{\Gamma}_{2}(t)\right\rangle _{R}=\left\langle \tilde{\Gamma}_{1}(t)\tilde{\Gamma}_{2}(\tau)\right\rangle _{R}^{*}.
\end{cases}
\end{equation}

\begin{align}
\frac{d\hat{\rho}_{S}}{dt}= & -\frac{1}{\hbar^{2}}\sum_{ij}\int_{t_{0}}^{t}d\tau\left(\beta_{i}(t)\beta_{j}(\tau)\hat{\rho}_{S}(t)-\beta_{j}(\tau)\hat{\rho}_{S}(t)\beta_{i}(t)\right)\left\langle \Gamma_{i}(t)\Gamma_{j}(\tau)\right\rangle _{R}\nonumber \\
 & +\left(\hat{\rho}_{S}(t)\beta_{j}(\tau)\beta_{i}(t)-\beta_{i}(t)\hat{\rho}_{S}(t)\beta_{j}(\tau)\right)\left\langle \Gamma_{j}(\tau)\Gamma_{i}(t)\right\rangle _{R}.\label{eq:Redfield-2-1}
\end{align}
Therefore, the Redfield equation reads,

\begin{align}
\frac{d\hat{\rho}_{S}^{I}}{dt}= & -\frac{1}{\hbar^{2}}\int_{0}^{t}d\tau\left(\tilde{\beta}_{2}(t)\tilde{\beta}_{1}(\tau)\hat{\rho}_{S}^{I}(t)-\tilde{\beta}_{1}(\tau)\hat{\rho}_{S}^{I}(t)\tilde{\beta}_{2}(t)\right)\left\langle \tilde{\Gamma}_{2}(t)\tilde{\Gamma}_{1}(\tau)\right\rangle _{R}\nonumber \\
 & +\left(\hat{\rho}_{S}^{I}(t)\tilde{\beta}_{2}(\tau)\tilde{\beta}_{1}(t)-\tilde{\beta}_{1}(t)\hat{\rho}_{S}^{I}(t)\tilde{\beta}_{2}(\tau)\right)\left\langle \tilde{\Gamma}_{2}(\tau)\tilde{\Gamma}_{1}(t)\right\rangle _{R}\nonumber \\
 & +\left(\tilde{\beta}_{1}(t)\tilde{\beta}_{2}(\tau)\hat{\rho}_{S}^{I}(t)-\tilde{\beta}_{2}(\tau)\hat{\rho}_{S}^{I}(t)\tilde{\beta}_{1}(t)\right)\left\langle \tilde{\Gamma}_{1}(t)\tilde{\Gamma}_{2}(\tau)\right\rangle _{R}\nonumber \\
 & +\left(\hat{\rho}_{S}^{I}(t)\tilde{\beta}_{1}(\tau)\tilde{\beta}_{2}(t)-\tilde{\beta}_{2}(t)\hat{\rho}_{S}^{I}(t)\tilde{\beta}_{1}(\tau)\right)\left\langle \tilde{\Gamma}_{1}(\tau)\tilde{\Gamma}_{2}(t)\right\rangle _{R},
\end{align}
and using the previous definitions for the operators $\beta_{i}$
and $\Gamma_{i}$ we write it as 
\begin{align}
\frac{d\hat{\rho}_{S}^{I}}{dt}= & -\frac{1}{\hbar^{2}}\int_{0}^{t}d\tau\left(\sigma_{+}\sigma_{-}\hat{\rho}_{S}^{I}(t)-\sigma_{-}\hat{\rho}_{S}^{I}(t)\sigma_{+}\right)e^{-i\omega_{0}(\tau-t)}\left\langle \tilde{\Gamma}_{2}(t)\tilde{\Gamma}_{1}(\tau)\right\rangle _{R}\nonumber \\
 & +\left(\hat{\rho}_{S}^{I}(t)\sigma_{+}\sigma_{-}-\sigma_{-}\hat{\rho}_{S}^{I}(t)\sigma_{+}\right)e^{i\omega_{0}(\tau-t)}\left\langle \tilde{\Gamma}_{2}(t)\tilde{\Gamma}_{1}(\tau)\right\rangle _{R}^{*}\nonumber \\
 & +\left(\sigma_{-}\sigma_{+}\hat{\rho}_{S}^{I}(t)-\sigma_{+}\hat{\rho}_{S}^{I}(t)\sigma_{-}\right)e^{i\omega_{0}(\tau-t)}\left\langle \tilde{\Gamma}_{1}(t)\tilde{\Gamma}_{2}(\tau)\right\rangle _{R}\nonumber \\
 & +\left(\hat{\rho}_{S}^{I}(t)\sigma_{-}\sigma_{+}-\sigma_{+}\hat{\rho}_{S}^{I}(t)\sigma_{-}\right)e^{-i\omega_{0}(\tau-t)}\left\langle \tilde{\Gamma}_{1}(t)\tilde{\Gamma}_{2}(\tau)\right\rangle _{R}^{*}.
\end{align}
Substituting the previously evaluated reservoir correlation functions,
and using the notation

\begin{equation}
\boldsymbol{d}_{eg}\cdot\bar{\mathcal{K}}(\boldsymbol{r},\omega)\cdot\boldsymbol{d}_{eg}\equiv\frac{1}{\hbar^{2}}\int d^{2}\boldsymbol{\rho}\mathcal{G}^{-}(\boldsymbol{r},\boldsymbol{\rho};\omega)\mathcal{G}^{+}(\boldsymbol{r},\boldsymbol{\rho};\omega),
\end{equation}
we write

\begin{align}
\frac{d\hat{\rho}_{S}^{I}}{dt}= & -\int_{0}^{\infty}d\omega\left(\sigma_{+}\sigma_{-}\hat{\rho}_{S}^{I}(t)-\sigma_{-}\hat{\rho}_{S}^{I}(t)\sigma_{+}\right)\bar{\mathcal{K}}(\boldsymbol{r},\omega)\left[\bar{n}_{\text{spp}}(\omega)+1\right]\int_{0}^{t}d\tau e^{i(\omega-\omega_{0})(\tau-t)}\nonumber \\
 & +\left(\hat{\rho}_{S}^{I}(t)\sigma_{+}\sigma_{-}-\sigma_{-}\hat{\rho}_{S}^{I}(t)\sigma_{+}\right)\bar{\mathcal{K}}(\boldsymbol{r},\omega)\left[\bar{n}_{\text{spp}}(\omega)+1\right]\int_{0}^{t}d\tau e^{-i(\omega-\omega_{0})(\tau-t)}\nonumber \\
 & +\left(\sigma_{-}\sigma_{+}\hat{\rho}_{S}^{I}(t)-\sigma_{+}\hat{\rho}_{S}^{I}(t)\sigma_{-}\right)\bar{\mathcal{K}}(\boldsymbol{r},\omega)\bar{n}_{\text{spp}}(\omega)\int_{0}^{t}d\tau e^{-i(\omega-\omega_{0})(\tau-t)}\nonumber \\
 & +\left(\hat{\rho}_{S}^{I}(t)\sigma_{-}\sigma_{+}-\sigma_{+}\hat{\rho}_{S}^{I}(t)\sigma_{-}\right)\bar{\mathcal{K}}(\boldsymbol{r},\omega)\bar{n}_{\text{spp}}(\omega)\int_{0}^{t}d\tau e^{i(\omega-\omega_{0})(\tau-t)}.
\end{align}
Letting the integral go to infinity and using the Sokhotski-Plemelj,
we can define

\begin{equation}
\begin{cases}
\Gamma\equiv\pi\int_{0}^{\infty}d\omega\boldsymbol{d}_{eg}\cdot\bar{\mathcal{K}}(\boldsymbol{r},\omega)\cdot\boldsymbol{d}_{eg}\delta(\omega-\omega_{0}),\\
\Delta\equiv\mathcal{P}\int_{0}^{\infty}d\omega\frac{\boldsymbol{d}_{eg}\cdot\bar{\mathcal{K}}(\boldsymbol{r},\omega)\cdot\boldsymbol{d}_{eg}}{\omega_{0}-\omega},\\
\Delta'\equiv\mathcal{P}\int_{0}^{\infty}d\omega\frac{\boldsymbol{d}_{eg}\cdot\bar{\mathcal{K}}(\boldsymbol{r},\omega)\cdot\boldsymbol{d}_{eg}\bar{n}_{\text{spp}}(\omega)}{\omega_{0}-\omega}.
\end{cases}
\end{equation}

to obtain

\begin{align}
\frac{d\hat{\rho}_{S}^{I}}{dt}= & \left(\sigma_{-}\hat{\rho}_{S}^{I}(t)\sigma_{+}-\sigma_{+}\sigma_{-}\hat{\rho}_{S}^{I}(t)\right)\left[\Gamma\left[\bar{n}_{\text{spp}}(\omega_{0})+1\right]+i(\Delta+\Delta')\right]\nonumber \\
 & +\left(\sigma_{-}\hat{\rho}_{S}^{I}(t)\sigma_{+}-\hat{\rho}_{S}^{I}(t)\sigma_{+}\sigma_{-}\right)\left[\Gamma\left[\bar{n}_{\text{spp}}(\omega_{0})+1\right]-i(\Delta+\Delta')\right]\nonumber \\
 & \text{+}\left(\sigma_{+}\hat{\rho}_{S}^{I}(t)\sigma_{-}-\sigma_{-}\sigma_{+}\hat{\rho}_{S}^{I}(t)\right)\left[\Gamma\bar{n}_{\text{spp}}(\omega_{0})-i\Delta'\right]\nonumber \\
 & +\left(\sigma_{+}\hat{\rho}_{S}^{I}(t)\sigma_{-}-\hat{\rho}_{S}^{I}(t)\sigma_{-}\sigma_{+}\right)\left[\Gamma\bar{n}_{\text{spp}}(\omega_{0})+i\Delta'\right].
\end{align}
Grouping together all the terms in , we obtain

\begin{align}
\frac{d\hat{\rho}_{S}^{I}}{dt}= & \Gamma\left[\bar{n}_{\text{spp}}(\omega_{0})+1\right]\left(\sigma_{-}\hat{\rho}_{S}^{I}(t)\sigma_{+}-\sigma_{+}\sigma_{-}\hat{\rho}_{S}^{I}(t)+\sigma_{-}\hat{\rho}_{S}^{I}(t)\sigma_{+}-\hat{\rho}_{S}^{I}(t)\sigma_{+}\sigma_{-}\right)\nonumber \\
 & +i(\Delta+\Delta')\left(\sigma_{-}\hat{\rho}_{S}^{I}(t)\sigma_{+}-\sigma_{+}\sigma_{-}\hat{\rho}_{S}^{I}(t)-\sigma_{-}\hat{\rho}_{S}^{I}(t)\sigma_{+}+\hat{\rho}_{S}^{I}(t)\sigma_{+}\sigma_{-}\right)\nonumber \\
 & +\Gamma\bar{n}_{\text{spp}}(\omega_{0})\left(\sigma_{+}\hat{\rho}_{S}^{I}(t)\sigma_{-}-\sigma_{-}\sigma_{+}\hat{\rho}_{S}^{I}(t)+\sigma_{+}\hat{\rho}_{S}^{I}(t)\sigma_{-}-\hat{\rho}_{S}^{I}(t)\sigma_{-}\sigma_{+}\right)\nonumber \\
 & -i\Delta'\left(\sigma_{+}\hat{\rho}_{S}^{I}(t)\sigma_{-}-\sigma_{-}\sigma_{+}\hat{\rho}_{S}^{I}(t)-\sigma_{+}\hat{\rho}_{S}^{I}(t)\sigma_{-}+\hat{\rho}_{S}^{I}(t)\sigma_{-}\sigma_{+}\right),
\end{align}
which can be compacted as

\begin{align}
\frac{d\hat{\rho}_{S}^{I}}{dt}= & \Gamma\left[\bar{n}_{\text{spp}}(\omega_{0})+1\right]\left(2\sigma_{-}\hat{\rho}_{S}^{I}(t)\sigma_{+}-\left\{ \sigma_{+}\sigma_{-},\hat{\rho}_{S}^{I}(t)\right\} \right)\nonumber \\
 & -i(\Delta+\Delta')\left[\sigma_{+}\sigma_{-},\hat{\rho}_{S}^{I}(t)\right]\nonumber \\
 & +\Gamma\bar{n}_{\text{spp}}(\omega_{0})\left(2\sigma_{+}\hat{\rho}_{S}^{I}(t)\sigma_{-}-\left\{ \sigma_{-}\sigma_{+},\hat{\rho}_{S}^{I}(t)\right\} \right)\nonumber \\
 & +i\Delta'\left[\sigma_{-}\sigma_{+},\hat{\rho}_{S}^{I}(t)\right].
\end{align}
We now note that the $\sigma_{-}\sigma_{+}$ can be written as

\begin{equation}
\sigma_{-}\sigma_{+}=\sigma_{+}\sigma_{-}+\left[\sigma_{-},\sigma_{+}\right],
\end{equation}
where this commutator can be calculated employing

\begin{equation}
\begin{cases}
\sigma_{-}=\frac{1}{2}(\sigma_{x}-i\sigma_{y}),\\
\sigma_{+}=\frac{1}{2}(\sigma_{x}+i\sigma_{y}),
\end{cases}
\end{equation}
and the commutator relations of the Pauli matrices

\begin{equation}
\left[\sigma_{i},\sigma_{j}\right]=2i\varepsilon_{ijk}\sigma_{k}.
\end{equation}
This yields

\begin{align}
\left[\sigma_{-},\sigma_{+}\right] & =\frac{1}{4}\left[\sigma_{x}-i\sigma_{y},\sigma_{x}+i\sigma_{y}\right]\nonumber \\
 & =\frac{1}{4}\left(\left[\sigma_{x},\sigma_{x}\right]+i\left[\sigma_{x},\sigma_{y}\right]-i\left[\sigma_{y},\sigma_{x}\right]+\left[\sigma_{y},\sigma_{y}\right]\right)\nonumber \\
 & =\frac{i}{2}\left[\sigma_{x},\sigma_{y}\right]\nonumber \\
 & =-\sigma_{z}.
\end{align}
Substituting this into our dynamical equation, we obtain

\begin{align}
\frac{d\hat{\rho}_{S}^{I}}{dt}= & \Gamma\left[\bar{n}_{\text{spp}}(\omega_{0})+1\right]\left(2\sigma_{-}\hat{\rho}_{S}^{I}(t)\sigma_{+}-\left\{ \sigma_{+}\sigma_{-},\hat{\rho}_{S}^{I}(t)\right\} \right)\nonumber \\
 & -\Gamma\bar{n}_{\text{spp}}(\omega_{0})\left(2\sigma_{+}\hat{\rho}_{S}^{I}(t)\sigma_{-}-\left\{ \sigma_{-}\sigma_{+},\hat{\rho}_{S}^{I}(t)\right\} \right)\nonumber \\
 & -i(\Delta+\Delta')\left[\sigma_{+}\sigma_{-},\hat{\rho}_{S}^{I}(t)\right]\nonumber \\
 & +i\Delta'\left[\sigma_{+}\sigma_{-}-\sigma_{z},\hat{\rho}_{S}^{I}(t)\right].
\end{align}
We can also use the fact that

\begin{equation}
\sigma_{+}\sigma_{-}=\frac{1}{2}\left(\mathds{1}+\sigma_{z}\right),
\end{equation}
or rather, that

\begin{equation}
\sigma_{z}=2\sigma_{+}\sigma_{-}-\mathds{1}.
\end{equation}
This yields

\begin{align}
\frac{d\hat{\rho}_{S}^{I}}{dt}= & \Gamma\left[\bar{n}_{\text{spp}}(\omega_{0})+1\right]\left(2\sigma_{-}\hat{\rho}_{S}^{I}(t)\sigma_{+}-\left\{ \sigma_{+}\sigma_{-},\hat{\rho}_{S}^{I}(t)\right\} \right)\nonumber \\
 & -\Gamma\bar{n}_{\text{spp}}(\omega_{0})\left(2\sigma_{+}\hat{\rho}_{S}^{I}(t)\sigma_{-}-\left\{ \sigma_{-}\sigma_{+},\hat{\rho}_{S}^{I}(t)\right\} \right)\nonumber \\
 & -i(\Delta+\Delta')\left[\sigma_{+}\sigma_{-},\hat{\rho}_{S}^{I}(t)\right]\nonumber \\
 & +i\Delta'\left[\sigma_{+}\sigma_{-}-2\sigma_{+}\sigma_{-}+\mathds{1},\hat{\rho}_{S}^{I}(t)\right],
\end{align}
and since $\left[\mathds{1},\hat{\rho}_{S}^{I}(t)\right]=0$, we obtain

\begin{align}
\frac{d\hat{\rho}_{S}^{I}}{dt}= & \Gamma\left[\bar{n}_{\text{spp}}(\omega_{0})+1\right]\left(2\sigma_{-}\hat{\rho}_{S}^{I}(t)\sigma_{+}-\left\{ \sigma_{+}\sigma_{-},\hat{\rho}_{S}^{I}(t)\right\} \right)\nonumber \\
 & +\Gamma\bar{n}_{\text{spp}}(\omega_{0})\left(2\sigma_{+}\hat{\rho}_{S}^{I}(t)\sigma_{-}-\left\{ \sigma_{-}\sigma_{+},\hat{\rho}_{S}^{I}(t)\right\} \right)\nonumber \\
 & -i(\Delta+\Delta')\left[\sigma_{+}\sigma_{-},\hat{\rho}_{S}^{I}(t)\right]\nonumber \\
 & -i\Delta'\left[\sigma_{+}\sigma_{-},\hat{\rho}_{S}^{I}(t)\right],
\end{align}
which finally yields

\begin{align}
\frac{d\hat{\rho}_{S}^{I}}{dt}= & \Gamma\left[\bar{n}_{\text{spp}}(\omega_{0})+1\right]\left(2\sigma_{-}\hat{\rho}_{S}^{I}(t)\sigma_{+}-\left\{ \sigma_{+}\sigma_{-},\hat{\rho}_{S}^{I}(t)\right\} \right)\nonumber \\
 & +\Gamma\bar{n}_{\text{spp}}(\omega_{0})\left(2\sigma_{+}\hat{\rho}_{S}^{I}(t)\sigma_{-}-\left\{ \sigma_{-}\sigma_{+},\hat{\rho}_{S}^{I}(t)\right\} \right)\nonumber \\
 & -i(\Delta+2\Delta')\left[\sigma_{+}\sigma_{-},\hat{\rho}_{S}^{I}(t)\right].
\end{align}
We can convert this into the Schrodinger picture, and the procedure
is similar to that carried out in detail for a single qubit. In the
end, we simply get rid of the $I$ index and add a term corresponding
to the coherent evolution of the system. This procedure results in

\begin{align}
\frac{d\hat{\rho}_{S}}{dt}= & \Gamma\left[\bar{n}_{\text{spp}}(\omega_{0})+1\right]\left(2\sigma_{-}\hat{\rho}_{S}(t)\sigma_{+}-\left\{ \sigma_{+}\sigma_{-},\hat{\rho}_{S}(t)\right\} \right)\nonumber \\
 & +\Gamma\bar{n}_{\text{spp}}(\omega_{0})\left(2\sigma_{+}\hat{\rho}_{S}(t)\sigma_{-}-\left\{ \sigma_{-}\sigma_{+},\hat{\rho}_{S}(t)\right\} \right)\nonumber \\
 & -i(\Delta+2\Delta')\left[\sigma_{+}\sigma_{-},\hat{\rho}_{S}(t)\right]-\frac{i}{\hbar}\left[H_{0},\hat{\rho}_{S}(t)\right].
\end{align}
Since $H_{0}=\hbar\omega_{0}\sigma_{+}\sigma_{-}$, we can write compactly

\begin{align}
\frac{d\hat{\rho}_{S}}{dt}= & \Gamma\left[\bar{n}_{\text{spp}}(\omega_{0})+1\right]\left(2\sigma_{-}\hat{\rho}_{S}(t)\sigma_{+}-\left\{ \sigma_{+}\sigma_{-},\hat{\rho}_{S}(t)\right\} \right)\nonumber \\
 & +\Gamma\bar{n}_{\text{spp}}(\omega_{0})\left(2\sigma_{+}\hat{\rho}_{S}(t)\sigma_{-}-\left\{ \sigma_{-}\sigma_{+},\hat{\rho}_{S}(t)\right\} \right)\nonumber \\
 & -i(\omega_{0}+\Delta+2\Delta')\left[\sigma_{+}\sigma_{-},\hat{\rho}_{S}(t)\right].\label{eq:LindbladOp}
\end{align}
Defining the operators

\begin{equation}
\begin{cases}
\hat{L}_{1}=\sqrt{\Gamma\bar{n}_{\text{spp}}(\omega_{0})}\sigma_{\text{+}},\\
\hat{L}_{2}=\sqrt{\Gamma\left[\bar{n}_{\text{spp}}(\omega_{0})+1\right]}\sigma_{-},
\end{cases}
\end{equation}
we can write the equation in Lindblad form

\begin{align}
\frac{d\hat{\rho}_{S}}{dt}= & \left(2\hat{L}_{2}\hat{\rho}_{S}(t)\hat{L}_{2}^{\dagger}-\left\{ \hat{L}_{2}^{\dagger}\hat{L}_{2},\hat{\rho}_{S}(t)\right\} \right)\nonumber \\
 & +\left(2\hat{L}_{1}\hat{\rho}_{S}(t)\hat{L}_{1}^{\dagger}-\left\{ \hat{L}_{1}^{\dagger}\hat{L}_{1},\hat{\rho}_{S}(t)\right\} \right)\nonumber \\
 & -i(\omega_{0}+\Delta+2\Delta')\left[\sigma_{+}\sigma_{-},\hat{\rho}_{S}(t)\right],
\end{align}
or introducing a sum over $k=1,2$, the result

\begin{equation}
\frac{d\hat{\rho}_{S}}{dt}=-\frac{i}{\hbar}\left[\hbar\omega_{0}'\sigma_{+}\sigma_{-},\hat{\rho}_{S}(t)\right]+\sum_{k=1,2}\left(2\hat{L}_{k}\hat{\rho}_{S}(t)\hat{L}_{k}^{\dagger}-\left\{ \hat{L}_{k}^{\dagger}\hat{L}_{k},\hat{\rho}_{S}(t)\right\} \right),\label{eq:LindbladfiniteT}
\end{equation}
with $\omega_{0}'=\omega_{0}+\Delta+\Delta'$. We can convert this
into a dynamical equation for the probability of finding each state
by proceeding analogously to the case of a single qubit. We write

\begin{align}
\hat{\rho}_{S} & =\rho_{ee}\left|e\right\rangle \left\langle e\right|+\rho_{eg}\left|e\right\rangle \left\langle g\right|+\rho_{ge}\left|g\right\rangle \left\langle e\right|+\rho_{gg}\left|g\right\rangle \left\langle g\right|\nonumber \\
 & =\begin{bmatrix}\rho_{ee} & \rho_{eg}\\
\rho_{ge} & \rho_{gg}
\end{bmatrix}.
\end{align}
We can write the Lindblad jump operators in matrix form

\begin{equation}
\begin{cases}
\hat{L}_{1} & =\sqrt{\Gamma\bar{n}_{\text{spp}}(\omega_{0})}\begin{bmatrix}0 & 1\\
0 & 0
\end{bmatrix},\\
\hat{L}_{2} & =\sqrt{\Gamma\left[\bar{n}_{\text{spp}}(\omega_{0})+1\right]}\begin{bmatrix}0 & 0\\
1 & 0
\end{bmatrix}.
\end{cases}
\end{equation}
We can now substitute these objects into the Lindblad equation and
setting $\gamma=2\Gamma$, we find

\begin{equation}
\begin{bmatrix}\dot{\rho}_{ee} & \dot{\rho}_{eg}\\
\dot{\rho}_{ge} & \dot{\rho}_{gg}
\end{bmatrix}=-i4\omega_{0}\begin{bmatrix}0 & \rho_{eg}\\
\rho_{ge} & 0
\end{bmatrix}+\begin{bmatrix}-\gamma\left[\bar{n}_{\text{spp}}(\omega_{0})+1\right]\rho_{ee}+\gamma\bar{n}_{\text{spp}}(\omega_{0})\rho_{gg} & -\frac{\gamma}{2}\left(\left[\bar{n}_{\text{spp}}(\omega_{0})+1\right]+\bar{n}_{\text{spp}}(\omega_{0})\right)\rho_{eg}\\
-\frac{\gamma}{2}\left(\left[\bar{n}_{\text{spp}}(\omega_{0})+1\right]+\bar{n}_{\text{spp}}(\omega_{0})\right)\rho_{ge} & \gamma\left[\bar{n}_{\text{spp}}(\omega_{0})+1\right]\rho_{ee}-\gamma\bar{n}_{\text{spp}}(\omega_{0})\rho_{gg}
\end{bmatrix}.
\end{equation}
Substituting into equation (\ref{eq:LindbladOp}) and looking at the
diagonal elements which correspond to the populations, we obtain a
coupled set of differential equations for the probabilities

\begin{equation}
\begin{cases}
\dot{P}_{e}=-\gamma\left[\bar{n}_{\text{spp}}(\omega_{0})+1\right]P_{e}+\gamma\bar{n}_{\text{spp}}(\omega_{0})P_{g},\\
\dot{P}_{g}=\gamma\left[\bar{n}_{\text{spp}}(\omega_{0})+1\right]P_{e}-\gamma\bar{n}_{\text{spp}}(\omega_{0})P_{g}.
\end{cases}
\end{equation}
This is slightly more complicated then the case at zero temperature
since the qubits may also be excited by thermal fluctuations of the
SPP field. These differential equations have a general solution given
by

\begin{align}
P_{e}(t)= & C_{1}\frac{\gamma\left(\bar{n}_{\text{spp}}(\omega_{0})+\left[\bar{n}_{\text{spp}}(\omega_{0})+1\right]e^{-\gamma(2\bar{n}_{\text{spp}}(\omega_{0})+1)t}\right)}{\gamma\left[2\bar{n}_{\text{spp}}(\omega_{0})+1\right]}\nonumber \\
 & +C_{2}\frac{\gamma\bar{n}_{\text{spp}}(\omega_{0})\left(-1+e^{-\gamma(2\bar{n}_{\text{spp}}(\omega_{0})+1)t}\right)}{\gamma\left[2\bar{n}_{\text{spp}}(\omega_{0})+1\right]},
\end{align}

\begin{align}
P_{g}(t)= & -C_{1}\frac{\gamma\left[\bar{n}_{\text{spp}}(\omega_{0})+1\right]\left(-1+e^{-\gamma(2\bar{n}_{\text{spp}}(\omega_{0})+1)t}\right)}{2\Gamma\left[2\bar{n}_{\text{spp}}(\omega_{0})+1\right]}\nonumber \\
 & +C_{2}\frac{\gamma\left(\left[\bar{n}_{\text{spp}}(\omega_{0})+1\right]+\bar{n}_{\text{spp}}(\omega_{0})e^{-\gamma(2\bar{n}_{\text{spp}}(\omega_{0})+1)t}\right)}{\gamma\left[2\bar{n}_{\text{spp}}(\omega_{0})+1\right]}.
\end{align}
If we choose initial conditions such that $P_{g}(0)=0$ and $P_{e}(0)=1$,
we can simplifiy the previous equations since we find

\begin{equation}
C_{2}\frac{\left(\left[\bar{n}_{\text{spp}}(\omega_{0})+1\right]+\bar{n}_{\text{spp}}(\omega_{0})\right)}{\left[2\bar{n}_{\text{spp}}(\omega_{0})+1\right]}=0,
\end{equation}
which gives $C_{2}=0$. Therefore, we can write

\begin{equation}
P_{e}(0)=C_{1}\frac{2\bar{n}_{\text{spp}}(\omega_{0})+1}{2\bar{n}_{\text{spp}}(\omega_{0})+1}=1,
\end{equation}
which gives $C_{1}=1$. We therefore have

\begin{equation}
P_{e}(t)=\frac{\left(\bar{n}_{\text{spp}}(\omega_{0})+\left[\bar{n}_{\text{spp}}(\omega_{0})+1\right]e^{-\gamma(2\bar{n}_{\text{spp}}(\omega_{0})+1)t}\right)}{2\bar{n}_{\text{spp}}(\omega_{0})+1}.
\end{equation}
We can even attempt to further simplify this, using

\begin{equation}
\bar{n}_{\text{spp}}(\omega)=\frac{1}{e^{\hbar\omega/k_{B}T}-1},
\end{equation}
which yields

\begin{align*}
\bar{n}_{\text{spp}}(\omega)+1 & =\frac{1}{e^{\hbar\omega/k_{B}T}-1}+1\\
 & =\frac{e^{\hbar\omega/k_{B}T}}{e^{\hbar\omega/k_{B}T}-1},
\end{align*}
and

\begin{align}
2\bar{n}_{\text{spp}}(\omega)+1 & =\frac{2}{e^{\hbar\omega/k_{B}T}-1}+1\nonumber \\
 & =\frac{e^{\hbar\omega/k_{B}T}+1}{e^{\hbar\omega/k_{B}T}-1}.
\end{align}
We can see that for $\hbar\omega_{0}\gg k_{B}T$, both terms approach
$1$, while $\bar{n}_{\text{spp}}(\omega)\to0$. In this limit we
recover

\begin{equation}
P_{e}(t)\approx e^{-\gamma t},
\end{equation}
which is the result for zero temperature. The results thus appear
to be consistent. In the other limit for large temperature, we find
$\bar{n}_{\text{spp}}(\omega)+1\approx\bar{n}_{\text{spp}}(\omega)$.
And thus, we find

\begin{align}
P_{e}(t) & =\frac{\bar{n}_{\text{spp}}(\omega_{0})\left(1+e^{-2\gamma\bar{n}_{\text{spp}}(\omega_{0})t}\right)}{2\bar{n}_{\text{spp}}(\omega_{0})}\nonumber \\
 & =\frac{1}{2}\left(1+e^{-2\gamma\bar{n}_{\text{spp}}(\omega_{0})t}\right),
\end{align}
while

\begin{align}
P_{g}(t) & =-\frac{\left[\bar{n}_{\text{spp}}(\omega_{0})+1\right]\left(-1+e^{-\gamma(2\bar{n}_{\text{spp}}(\omega_{0})+1)t}\right)}{\left[2\bar{n}_{\text{spp}}(\omega_{0})+1\right]}\nonumber \\
 & \approx-\frac{1}{2}\left(-1+e^{-2\gamma\bar{n}_{\text{spp}}(\omega_{0})t}\right)\nonumber \\
 & \approx\frac{1}{2}\left(1-e^{-2\gamma\bar{n}_{\text{spp}}(\omega_{0})t}\right),
\end{align}
and both states will tend to equilibrium at equal probability.

\section{Thermal averages of $\hat{\boldsymbol{\text{f}}}$-operators \label{sec:Thermal-averages-of}}

In this appendix we present a small derivation of the thermal averages
of the surface-plasmon-polariton creation and anihilation operators
and their product, based on discretizing the integrals that they are
made up of. Let us consider the Hamiltonian of the free electromagnetic
field described by the operators $\hat{\mathbf{f}}(\boldsymbol{r},\omega)=\left\{ f_{x}(\mathbf{r},\omega),f_{y}(\mathbf{r},\omega),f_{z}(\mathbf{r},\omega)\right\} $

\begin{align}
H_{F} & =\int d^{3}\boldsymbol{r}\int d\omega\hbar\omega\hat{\mathbf{f}}^{\dagger}(\boldsymbol{r},\omega)\cdot\hat{\mathbf{f}}(\boldsymbol{r},\omega)\nonumber \\
 & =\sum_{i}\int d^{3}\boldsymbol{r}\int d\omega\hbar\omega f_{i}^{\dagger}(\boldsymbol{r},\omega)f_{i}(\boldsymbol{r},\omega).
\end{align}
Let us discretize the integrals over $\mathbf{r}$ and $\omega$ to
write the Hamiltonian as 
\begin{equation}
H_{F}=\sum_{i}\sum_{n}\Delta_{\boldsymbol{r}}\sum_{\ell}\Delta_{\omega}\hbar\omega_{\ell}f_{i}^{\dagger}(\boldsymbol{r}_{n},\omega_{\ell})f_{i}(\boldsymbol{r}_{n},\omega_{\ell}).
\end{equation}
Now let us define the operators 
\begin{align}
\tilde{f}_{i,n,\ell} & =\sqrt{\Delta_{\boldsymbol{r}}\Delta_{\omega}}f_{i}(\boldsymbol{r}_{n},\omega_{\ell}),\\
\tilde{f}_{i,n,\ell}^{\dagger} & =\sqrt{\Delta_{\boldsymbol{r}}\Delta_{\omega}}f_{i}^{\dagger}(\boldsymbol{r}_{n},\omega_{\ell}).
\end{align}
Let us assume that 
\begin{equation}
\left[\tilde{f}_{i,n,\ell},\tilde{f}_{i^{\prime},n^{\prime},\ell^{\prime}}^{\dagger}\right]=\delta_{i,i^{\prime}}\delta_{n,n^{\prime}}\delta_{\ell,\ell^{\prime}}
\end{equation}
This leads us to 
\begin{align}
\left[f_{i}(\boldsymbol{r}_{n},\omega_{\ell}),f_{i^{\prime}}^{\dagger}(\boldsymbol{r}_{n^{\prime}},\omega_{\ell^{\prime}})\right] & =\frac{1}{\Delta_{\boldsymbol{r}}\Delta_{\omega}}\left[\tilde{f}_{i,n,\ell},\tilde{f}_{i^{\prime},n^{\prime},\ell^{\prime}}^{\dagger}\right]\nonumber \\
 & =\frac{1}{\Delta_{\boldsymbol{r}}\Delta_{\omega}}\delta_{i,i^{\prime}}\delta_{n,n^{\prime}}\delta_{\ell,\ell^{\prime}}\nonumber \\
 & =\delta_{i,i^{\prime}}\frac{1}{\Delta_{\boldsymbol{r}}}\delta_{n,n^{\prime}}\frac{1}{\Delta_{\omega}}\delta_{\ell,\ell^{\prime}}
\end{align}
In the limit of $\Delta_{\mathbf{r}}\rightarrow0$ and $\Delta_{\omega}\rightarrow0$,
we obtain 
\begin{align}
\lim_{\Delta_{\mathbf{r}}\rightarrow0}\frac{1}{\Delta_{\mathbf{r}}}\delta_{n,n^{\prime}} & =\delta(\boldsymbol{r}_{n}-\boldsymbol{r}_{n^{\prime}})\\
\lim_{\Delta_{\omega}\rightarrow0}\frac{1}{\Delta_{\omega}}\delta_{\ell,\ell^{\prime}} & =\delta(\omega_{\ell}-\omega_{\ell^{\prime}})
\end{align}
such that 
\begin{equation}
\left[f_{i}(\boldsymbol{r}_{n},\omega_{\ell}),f_{i^{\prime}}^{\dagger}(\boldsymbol{r}_{n^{\prime}},\omega_{\ell^{\prime}})\right]=\delta_{i,i^{\prime}}\delta(\boldsymbol{r}_{n}-\boldsymbol{r}_{n^{\prime}})\delta(\omega_{\ell}-\omega_{\ell^{\prime}})
\end{equation}
as intended. Therefore, we can write 
\begin{equation}
H_{F}=\sum_{i}\sum_{n}\sum_{\ell}\hbar\omega_{\ell}\tilde{f}_{i,n,\ell}^{\dagger}\tilde{f}_{i,n,\ell}.
\end{equation}
An eigenstate of this Hamiltonian is characterized by the occupation
number of each state $\left(i,n,\ell\right)$, $n_{\lambda,i,n,\ell}$.
We represent such state as 
\begin{equation}
\left|\left\{ n_{i_{j},n_{j},\ell_{j}}\right\} \right\rangle =\left|n_{i_{1},n_{1},\ell_{1}},...,n_{i_{N},n_{N},\ell_{N}}\right\rangle ,
\end{equation}
which has $n_{i_{j},n_{j},\ell_{j}}$ particles in state $\left(i_{j},n_{j},\ell_{j}\right)$.
Now let us compute 
\begin{equation}
\left\langle f_{i^{\prime}}^{\dagger}(\boldsymbol{r}_{n^{\prime}},\omega_{\ell^{\prime}})f_{i}(\boldsymbol{r}_{n},\omega_{\ell})\right\rangle =\frac{1}{\Delta_{\boldsymbol{r}}\Delta_{\omega}}\left\langle \tilde{f}_{\lambda^{\prime}i^{\prime},n^{\prime},\ell'}^{\dagger}\tilde{f}_{\lambda^{\prime}i^{\prime},n^{\prime},\ell'}\right\rangle .
\end{equation}
Now we evaluate $\left\langle \tilde{f}_{i,n,\ell}^{\dagger}\tilde{f}_{i,n,\ell}\right\rangle $.
For simplicity we group all the state labels into a superindex $I=\left(i,n,\ell\right)$.
Therefore, we want to evaluate 
\begin{align}
\left\langle \tilde{f}_{I}^{\dagger}\tilde{f}_{I}\right\rangle  & =\frac{\sum_{\left\{ n_{I^{\prime}}\right\} }\left\langle \left\{ n_{I^{\prime}}\right\} \right|e^{-\beta\sum_{J^{\prime}}\hbar\omega_{J^{\prime}}\tilde{f}_{J^{\prime}}^{\dagger}\tilde{f}_{J^{\prime}}}\tilde{f}_{I}^{\dagger}\tilde{f}_{I}\left|\left\{ n_{I^{\prime}}\right\} \right\rangle }{\sum_{\left\{ n_{I^{\prime}}\right\} }\left\langle \left\{ n_{I^{\prime}}\right\} \right|e^{-\beta\sum_{J^{\prime}}\hbar\omega_{J^{\prime}}\tilde{f}_{J^{\prime}}^{\dagger}\tilde{f}_{J^{\prime}}}\left|\left\{ n_{I^{\prime}}\right\} \right\rangle },
\end{align}
where $\beta=1/k_{B}T$. Now we write, assuming that the system is
populated by $N$ bosons, that

\begin{align}
 & \sum_{\left\{ n_{I^{\prime}}\right\} }\left\langle \left\{ n_{I^{\prime}}\right\} \right|e^{-\beta\sum_{J^{\prime}}\hbar\omega_{J^{\prime}}\tilde{f}_{J^{\prime}}^{\dagger}\tilde{f}_{J^{\prime}}}\tilde{f}_{I}^{\dagger}\tilde{f}_{I}\left|\left\{ n_{I^{\prime}}\right\} \right\rangle \nonumber \\
= & \sum_{n_{I_{1}}=0}^{+\infty}...\sum_{n_{I}=0}^{+\infty}...\sum_{n_{I_{N}}=0}^{+\infty}\left\langle n_{I_{1}},...,n_{I},...,n_{I_{N}}\right|e^{-\beta\hbar\omega_{I_{1}}\tilde{f}_{I_{1}}^{\dagger}\tilde{f}_{I_{1}}}...e^{-\beta\hbar\omega_{I}\tilde{f}_{I}^{\dagger}\tilde{f}_{I}}\tilde{f}_{I}^{\dagger}\tilde{f}_{I}...e^{-\beta\hbar\omega_{I_{1}}\tilde{f}_{I_{N}}^{\dagger}\tilde{f}_{I_{N}}}\left|n_{I_{1}},...,n_{I},...,n_{I_{N}}\right\rangle \nonumber \\
= & \sum_{n_{I_{1}}=0}^{+\infty}...\sum_{n_{I}=0}^{+\infty}...\sum_{n_{I_{N}}=0}^{+\infty}e^{-\beta\hbar\omega_{I_{1}}n_{I_{1}}}...e^{-\beta\hbar\omega_{I}n_{I}}n_{I}...e^{-\beta\hbar\omega_{I_{1}}n_{I_{N}}}\nonumber \\
= & \sum_{n_{I}=0}^{+\infty}e^{-\beta\hbar\omega_{I}n_{I}}n_{I}\prod_{n_{I^{\prime}}\neq n_{I}}\sum_{n_{I^{\prime}}=0}^{+\infty}e^{-\beta\hbar\omega_{I^{\prime}}n_{I^{\prime}}}.
\end{align}
Likewise

\begin{align}
 & \sum_{\left\{ n_{I^{\prime}}\right\} }\left\langle \left\{ n_{I^{\prime}}\right\} \right|e^{-\beta\sum_{J^{\prime}}\hbar\omega_{J^{\prime}}\tilde{f}_{J^{\prime}}^{\dagger}\tilde{f}_{J^{\prime}}}\left|\left\{ n_{I^{\prime}}\right\} \right\rangle \nonumber \\
= & \sum_{n_{I_{1}}=0}^{+\infty}...\sum_{n_{I}=0}^{+\infty}...\sum_{n_{I_{N}}=0}^{+\infty}\left\langle n_{I_{1}},...,n_{I},...,n_{I_{N}}\right|e^{-\beta\hbar\omega_{I_{1}}\tilde{f}_{I_{1}}^{\dagger}\tilde{f}_{I_{1}}}...e^{-\beta\hbar\omega_{I}\tilde{f}_{I}^{\dagger}\tilde{f}_{I}}...e^{-\beta\hbar\omega_{I_{1}}\tilde{f}_{I_{N}}^{\dagger}\tilde{f}_{I_{N}}}\left|n_{I_{1}},...,n_{I},...,n_{I_{N}}\right\rangle \nonumber \\
= & \sum_{n_{I_{1}}=0}^{+\infty}...\sum_{n_{I}=0}^{+\infty}...\sum_{n_{I_{N}}=0}^{+\infty}e^{-\beta\hbar\omega_{I_{1}}n_{I_{1}}}...e^{-\beta\hbar\omega_{I}n_{I}}...e^{-\beta\hbar\omega_{I_{1}}n_{I_{N}}}\nonumber \\
= & \sum_{n_{I}=0}^{+\infty}e^{-\beta\hbar\omega_{I}n_{I}}\prod_{n_{I^{\prime}}\neq n_{I}}\sum_{n_{I^{\prime}}=0}^{+\infty}e^{-\beta\hbar\omega_{I^{\prime}}n_{I^{\prime}}}.
\end{align}
Therefore 
\begin{align}
\left\langle \tilde{f}_{I}^{\dagger}\tilde{f}_{I}\right\rangle  & =\frac{\sum_{n_{I}=0}^{+\infty}e^{-\beta\hbar\omega_{I}n_{I}}n_{I}\prod_{n_{I^{\prime}}\neq n_{I}}\sum_{n_{I^{\prime}}=0}^{+\infty}e^{-\beta\hbar\omega_{I^{\prime}}n_{I^{\prime}}}}{\sum_{n_{I}=0}^{+\infty}e^{-\beta\hbar\omega_{I}n_{I}}\prod_{n_{I^{\prime}}\neq n_{I}}\sum_{n_{I^{\prime}}=0}^{+\infty}e^{-\beta\hbar\omega_{I^{\prime}}n_{I^{\prime}}}}\nonumber \\
 & =\frac{\sum_{n_{I}=0}^{+\infty}e^{-\beta\hbar\omega_{I}n_{I}}n_{I}}{\sum_{n_{I}=0}^{+\infty}e^{-\beta\hbar\omega_{I}n_{I}}}\nonumber \\
 & =\frac{1}{\sum_{n_{I}=0}^{+\infty}e^{-\beta\hbar\omega_{I}n_{I}}}\left(-\frac{\partial}{\partial\beta\hbar\omega_{I}}\sum_{n_{I}=0}^{+\infty}e^{-\beta\hbar\omega_{I}n_{I}}\right).
\end{align}
Since 
\begin{align}
\sum_{n_{I}=0}^{+\infty}e^{-\beta\hbar\omega_{I}n_{I}} & =1+\left(e^{-\beta\hbar\omega_{I}}\right)+\left(e^{-\beta\hbar\omega_{I}}\right)^{2}+...\nonumber \\
 & =\frac{1}{1-e^{-\beta\hbar\omega_{I}}}.
\end{align}
Now 
\begin{equation}
\frac{\partial}{\partial x}\left(\frac{1}{1-e^{-x}}\right)=-\frac{e^{-x}}{\left(1-e^{-x}\right)^{2}}.
\end{equation}
Therefore 
\begin{align}
\left\langle \tilde{f}_{I}^{\dagger}\tilde{f}_{I}\right\rangle  & =\left(1-e^{-\beta\hbar\omega_{I}}\right)\left(\frac{e^{-\beta\hbar\omega_{I}}}{\left(1-e^{-\beta\hbar\omega_{I}}\right)^{2}}\right)\nonumber \\
 & =\frac{e^{-\beta\hbar\omega_{I}}}{1-e^{-\beta\hbar\omega_{I}}}\nonumber \\
 & =\frac{1}{e^{\beta\hbar\omega_{I}}-1}\nonumber \\
 & =\bar{n}_{\text{spp}}(\omega_{I}).
\end{align}
If $I\neq J$, then we have 
\begin{equation}
\left\langle \tilde{f}_{I}^{\dagger}\tilde{f}_{J}\right\rangle =0,I\neq J.
\end{equation}
Therefore, we can write 
\begin{equation}
\left\langle \tilde{f}_{i^{\prime},n^{\prime},\ell^{\prime}}^{\dagger}\tilde{f}_{i,n,\ell}\right\rangle =\delta_{i,i^{\prime}}\delta_{n,n^{\prime}}\delta_{\ell,\ell^{\prime}}\bar{n}_{\text{spp}}(\omega_{\ell}).
\end{equation}
Therefore 
\begin{equation}
\left\langle f_{i^{\prime}}^{\dagger}(\boldsymbol{r}_{n^{\prime}},\omega_{\ell^{\prime}})f_{i}(\mathbf{r}_{n},\omega_{\ell})\right\rangle =\frac{1}{\Delta_{\mathbf{r}}\Delta_{\omega}}\delta_{i,i^{\prime}}\delta_{n,n^{\prime}}\delta_{\ell,\ell^{\prime}}\bar{n}_{\text{spp}}(\omega_{\ell})
\end{equation}
As before, in the limit of $\Delta_{\mathbf{r}},\Delta_{\omega}\rightarrow0$,
we obtain 
\begin{equation}
\left\langle f_{i^{\prime}}^{\dagger}(\mathbf{r}_{n^{\prime}},\omega_{\ell^{\prime}})f_{i}(\mathbf{r}_{n},\omega_{\ell})\right\rangle =\delta_{i,i^{\prime}}\delta(\boldsymbol{r}_{n}-\boldsymbol{r}_{n^{\prime}})\delta(\omega_{\ell}-\omega_{\ell^{\prime}})\bar{n}_{\text{spp}}(\omega_{\ell}),
\end{equation}
or 
\begin{equation}
\left\langle f_{i^{\prime}}^{\dagger}(\boldsymbol{r}^{\prime},\omega^{\prime})f_{i}(\boldsymbol{r},\omega)\right\rangle =\delta_{i,i^{\prime}}\delta(\boldsymbol{r}-\boldsymbol{r}^{\prime})\delta(\omega-\omega^{\prime})\bar{n}_{\text{spp}}(\omega).
\end{equation}

This is the result presented in the main text without further justification,
but this brief derivation shows that it is indeed valid.

\section{Details of the calculation of the dynamics of two qubits coupled
to the plasmonic bath\label{sec:Details-of-the-1}}

\subsection{Obtaining the dynamics via the Schrodinger equation\label{subsec:Obtaining-the-dynamics-1}}

The Hamiltonian for this system can be constructed analogously to
the previous results. We consider the Hamiltonian corresponding to
each atom $H_{0,1}$ and $H_{0,2}$ as well as the SPP field $H_{\text{spp}}.$
In addition, we must consider the interaction terms between the plasmon
field and both qubits in the dipole approximation $H_{d,1}$ and $H_{d,2}$.
The Hamiltonian is written as

\begin{equation}
H=\sum_{\alpha=1,2}H_{0,\alpha}+H_{\text{spp}}+\sum_{\alpha=1,2}H_{d,\alpha}.
\end{equation}
To simplify the following equations, we introduce the notation

\begin{equation}
\begin{cases}
\hbar\mathcal{G}_{\alpha}^{-}(\boldsymbol{r},\boldsymbol{\rho};\omega)\equiv\omega\beta(\omega)\boldsymbol{d}_{\alpha}\cdot\bar{\bar{G}}(\boldsymbol{r},\boldsymbol{\rho},z=0;\omega),\\
\hbar\mathcal{G}_{\alpha}^{+}(\boldsymbol{r},\boldsymbol{\rho};\omega)\equiv\omega\beta^{*}(\omega)\bar{\bar{G}}^{\dagger}(\boldsymbol{r},\boldsymbol{\rho},z=0;\omega)\cdot\boldsymbol{d}_{\alpha}^{\dagger}.
\end{cases}
\end{equation}
The components of the Hamiltonian are therefore, in the rotating wave
approximation, given by

\begin{equation}
\begin{cases}
H_{0,\alpha}=\hbar\omega_{0}\mathds{1_{\text{spp}}}\sigma_{\alpha}^{+}\sigma_{\alpha}^{-},\\
H_{\text{spp}}=\int d^{2}\boldsymbol{\rho}\int_{0}^{\infty}d\omega\hbar\omega\hat{\text{\textbf{f}}}^{\dagger}(\boldsymbol{\rho},\omega)\hat{\text{\textbf{f}}}(\boldsymbol{\rho},\omega),\\
H_{d,\alpha}=i\hbar\int_{0}^{\infty}d\omega\int d^{2}\boldsymbol{\rho}\left[\hat{\text{\textbf{f}}}^{\dagger}(\boldsymbol{\rho},\omega)\cdot\mathcal{G}_{\alpha}^{+}(\boldsymbol{r}_{\alpha},\boldsymbol{\rho};\omega)\sigma_{\alpha}^{-}-\mathcal{G}_{\alpha}^{-}(\boldsymbol{r}_{\alpha},\boldsymbol{\rho};\omega)\cdot\hat{\text{\textbf{f}}}(\boldsymbol{\rho},\omega)\sigma_{\alpha}^{+}\right].
\end{cases}.\label{eq:Hamiltonians}
\end{equation}
As in the previous notes, we want to solve the Schrodinger equation

\begin{equation}
i\hbar\frac{\partial}{\partial t}\left|\psi(t)\right\rangle =H\left|\psi(t)\right\rangle .
\end{equation}
We considet that the system is coupled to a single excitation of the
field. This means that initially one of the atoms (label it as atom
$\alpha$) will be in the excited state while the other will be in
the ground state and there will be no SPPs in the graphene. We write
an ansatz for the wave-function as

\begin{equation}
\left|\psi(t)\right\rangle =\sum_{\alpha=1,2}C_{\alpha}(t)e^{-i\omega_{\alpha}t}\left|\alpha\right\rangle \left|0\right\rangle +\int d^{2}\boldsymbol{\rho}\int_{0}^{\infty}d\omega\boldsymbol{C}_{g}(\boldsymbol{\rho},t;\omega)e^{-i\omega t}\cdot\left|g\right\rangle \left|\mathds{1}(\boldsymbol{\rho},\omega)\right\rangle .
\end{equation}
We now substitute this into the Schrodinger equation on both sides.
The left-hand side, involving the time derivative, yields

\begin{align}
i\hbar\frac{\partial}{\partial t}\left|\psi(t)\right\rangle = & \sum_{\alpha'=1,2}\left[\hbar\omega_{\alpha'}C_{\alpha'}(t)e^{-i\omega_{\alpha'}t}+i\hbar\dot{C}_{\alpha'}(t)e^{-i\omega_{\alpha'}t}\right]\left|\alpha'\right\rangle \left|0\right\rangle \nonumber \\
 & +\int d^{2}\boldsymbol{\rho}\int_{0}^{\infty}d\omega\left[\hbar\omega\boldsymbol{C}_{g}(\boldsymbol{\rho},t;\omega)e^{-i\omega t}+i\hbar\dot{\boldsymbol{C}}_{g}(\boldsymbol{\rho},t;\omega)e^{-i\omega t}\right]\cdot\left|g\right\rangle \left|\mathds{1}(\boldsymbol{\rho};\omega)\right\rangle .
\end{align}
On the other hand, for the right hand side, we have

\begin{align*}
\hat{H}\left|\psi(t)\right\rangle = & \sum_{\alpha'=1,2}\hbar\omega_{\alpha'}C_{\alpha'}(t)e^{-i\omega_{\alpha'}t}\\
 & +\int d^{2}\boldsymbol{\rho}\int_{0}^{\infty}d\omega\hbar\omega\boldsymbol{C}_{g}(t)e^{-i\omega t}\left|g\right\rangle \left|\mathds{1}(\boldsymbol{\rho};\omega)\right\rangle \\
 & +i\hbar\sum_{\alpha'=1,2}\int_{0}^{\infty}d\omega\int d^{2}\boldsymbol{\rho}_{\alpha'}(t)e^{-i\omega_{\alpha'}t}\left|g\right\rangle \left|\mathds{1}(\boldsymbol{\rho};\omega)\right\rangle \cdot\mathcal{G}_{\alpha'}^{+}(\boldsymbol{r}_{\alpha'},\boldsymbol{\rho};\omega)\\
 & -i\hbar\sum_{\alpha'=1,2}\int_{0}^{\infty}d\omega\int d^{2}\boldsymbol{\rho}\mathcal{G}_{\alpha'}^{-}(\boldsymbol{r}_{\alpha'},\boldsymbol{\rho};\omega)\cdot\boldsymbol{C}_{g}(\boldsymbol{\rho},t;\omega)e^{-i\omega t}\left|\alpha'\right\rangle \left|0\right\rangle .
\end{align*}
Thus, we see that, equating both sides, many terms cancel, leaving
us with

\begin{align}
i\hbar\sum_{\alpha'=1,2}\dot{C}_{\alpha'}(t)e^{-i\omega_{\alpha'}t}\left|\alpha'\right\rangle \left|0\right\rangle +i\hbar\int d^{2}\boldsymbol{\rho}\int_{0}^{\infty}d\omega\dot{\boldsymbol{C}}_{g}(\boldsymbol{\rho},t;\omega)e^{-i\omega t}\cdot\left|g\right\rangle \left|\mathds{1}(\boldsymbol{\rho};\omega)\right\rangle \nonumber \\
=i\hbar\sum_{\alpha'=1,2}\int_{0}^{\infty}d\omega\int d^{2}\boldsymbol{\rho}_{\alpha'}(t)e^{-i\omega_{\alpha}t}\left|g\right\rangle \left|\mathds{1}(\boldsymbol{\rho};\omega)\right\rangle \cdot\mathcal{G}_{\alpha'}^{+}(\boldsymbol{r}_{\alpha'},\boldsymbol{\rho};\omega)\nonumber \\
-i\hbar\sum_{\alpha'=1,2}\int_{0}^{\infty}d\omega\int d^{2}\boldsymbol{\rho}\mathcal{G}_{\alpha'}^{-}(\boldsymbol{r}_{\alpha'},\boldsymbol{\rho};\omega)\cdot\boldsymbol{C}_{g}(\boldsymbol{\rho},t;\omega)\left|\alpha'\right\rangle \left|0\right\rangle .
\end{align}
Acting on the left on both sides of the equation with the state $\left\langle \alpha\right|\left\langle 0\right|$,
we get

\begin{equation}
\dot{C}_{\alpha}(t)e^{-i\omega_{\alpha}t}=-\int_{0}^{\infty}d\omega\int d^{2}\boldsymbol{\rho}\mathcal{G}_{\alpha}^{-}(\boldsymbol{r}_{\alpha},\boldsymbol{\rho};\omega)\cdot\boldsymbol{C}_{g}(\boldsymbol{\rho},t;\omega)e^{-i\omega t},
\end{equation}
while if we act only on the state with $\left\langle g\right|\left\langle \mathds{1}(\boldsymbol{\rho};\omega)\right|$,
we obtain

\begin{equation}
\dot{\boldsymbol{C}}_{g}(\boldsymbol{\rho},t;\omega)e^{-i\omega t}=\sum_{\alpha'=1,2}C_{\alpha'}(t)e^{-i\omega_{\alpha'}t}\mathcal{G}_{\alpha'}^{+}(\boldsymbol{r}_{\alpha'},\boldsymbol{\rho};\omega).
\end{equation}
We can now integrate this final expression in time, and via performing
the first Markov approximation, isolate the integral over the exponentials
by writing $C_{\alpha'}(t_{1})\approx C_{\alpha}(t)$.

\begin{equation}
\boldsymbol{C}_{g}(\boldsymbol{\rho},t;\omega)=\sum_{\alpha'=1,2}C_{\alpha'}(t)\mathcal{G}_{\alpha'}^{+}(\boldsymbol{r}_{\alpha'},\boldsymbol{\rho};\omega)\int_{0}^{t}dt_{1}e^{i\left(\omega-\omega_{\alpha'}\right)t_{1}}.
\end{equation}
We now take this expression and substitute it back into the equation
for $C_{\alpha}(t)$ to obtain

\begin{equation}
\dot{C}_{\alpha}(t)=-\int_{0}^{\infty}d\omega\int d^{2}\boldsymbol{\rho}\mathcal{G}_{\alpha}^{-}(\boldsymbol{r}_{\alpha},\boldsymbol{\rho};\omega)e^{-i(\omega-\omega_{\alpha})t}\cdot\left[\sum_{\alpha'=1,2}C_{\alpha'}(t)\mathcal{G}_{\alpha'}^{+}(\boldsymbol{r}_{\alpha'},\boldsymbol{\rho};\omega)\int_{0}^{t}dt_{1}e^{-i(\omega_{\alpha'}-\omega)t_{1}}\right].
\end{equation}
Grouping the terms together everything together and introducing the
notation

\[
\boldsymbol{d}_{\alpha}\cdot\bar{\mathcal{K}}_{\alpha\alpha'}(\omega)\cdot\boldsymbol{d}_{\alpha'}=\int d^{2}\boldsymbol{\rho}\mathcal{G}_{\alpha}^{-}(\boldsymbol{r}_{\alpha},\boldsymbol{\rho};\omega)\mathcal{G}_{\alpha'}^{+}(\boldsymbol{r}_{\alpha'},\boldsymbol{\rho};\omega),
\]
we obtain

\begin{equation}
\dot{C}_{\alpha}(t)=-\int_{0}^{\infty}d\omega\sum_{\alpha'=1,2}\boldsymbol{d}_{\alpha}\cdot\bar{\mathcal{K}}_{\alpha\alpha'}(\omega)\cdot\boldsymbol{d}_{\alpha'}\int_{0}^{t}dt_{1}e^{-i(\omega_{\alpha'}-\omega)t_{1}-i(\omega-\omega_{\alpha})t}.
\end{equation}
For simplicity we admit that the two-level atoms are identical, i.e.,
$\omega_{1}=\omega_{2}$. This allows us to write

\begin{equation}
\dot{C}_{\alpha}(t)=-\int_{0}^{\infty}d\omega\sum_{\alpha'=1,2}C_{\alpha'}(t)\boldsymbol{d}_{\alpha}\cdot\bar{\mathcal{K}}_{\alpha\alpha'}(\omega)\cdot\boldsymbol{d}_{\alpha'}\int_{0}^{t}dt_{1}e^{-i(\omega_{\alpha'}-\omega)\left(t_{1}-t\right)},
\end{equation}
The integral over time by making the change of variables $\tau=t-t_{1}$
as well as using the Sokhotski-Plemelj identity, where in the third
line we extended the region of integration via the second Markov approximation,
and then applied the Sokhtski-Plemelj theorem. We are left with the
differential equation

\begin{align}
\dot{C}_{\alpha}(t) & =-\sum_{\alpha'=1,2}C_{\alpha'}(t)\left[-i\mathcal{P}\int_{0}^{\infty}d\omega\frac{\boldsymbol{d}_{\alpha}\cdot\bar{\mathcal{K}}_{\alpha\alpha'}(\omega)\cdot\boldsymbol{d}_{\alpha'}}{\omega-\omega_{\alpha}}+\pi\boldsymbol{d}_{\alpha}\cdot\bar{\mathcal{K}}_{\alpha\alpha'}(\omega_{\alpha})\cdot\boldsymbol{d}_{\alpha'}\right]\nonumber \\
 & \equiv-\sum_{\alpha'=1,2}C_{\alpha'}(t)\left[-i\Delta_{\alpha\alpha'}+\Gamma_{\alpha\alpha'}\right].
\end{align}
Explicitly, for our two-level system, we have

\begin{equation}
\begin{cases}
\dot{C}_{1}(t)=-C_{1}(t)\left[\Gamma_{11}-i\Delta_{11}\right]-C_{2}(t)\left[\Gamma_{12}-i\Delta_{12}\right],\\
\dot{C}_{2}(t)=-C_{1}(t)\left[\Gamma_{21}-i\Delta_{21}\right]-C_{2}(t)\left[\Gamma_{22}-i\Delta_{22}\right].
\end{cases}.
\end{equation}
We can construct the states

\[
\begin{cases}
\left|+\right\rangle =\frac{1}{\sqrt{2}}\left(\left|1\right\rangle +\left|2\right\rangle \right),\\
\left|-\right\rangle =\frac{1}{\sqrt{2}}\left(\left|1\right\rangle -\left|2\right\rangle \right).
\end{cases},
\]
with amplitudes

\begin{equation}
\begin{cases}
C_{+}(t)=\frac{1}{\sqrt{2}}\left[C_{1}(t)+C_{2}(t)\right],\\
C_{-}(t)=\frac{1}{\sqrt{2}}\left[C_{1}(t)-C_{2}(t)\right].
\end{cases}
\end{equation}
If we now admit that $\Delta_{11}=\Delta_{22}\equiv\Delta,$$\Gamma_{11}=\Gamma_{22}\equiv\Gamma$
and $\Delta_{21}=\Delta_{12}\equiv g_{12},\Gamma_{12}=\Gamma_{21}\equiv\Gamma_{12}$,
we have 
\begin{equation}
\begin{cases}
\dot{C}_{1}(t)=-C_{1}(t)\left[\Gamma-i\Delta\right]-C_{2}(t)\left[\Gamma_{12}-ig_{12}\right],\\
\dot{C}_{2}(t)=-C_{1}(t)\left[\Gamma_{12}-ig_{12}\right]-C_{2}(t)\left[\Gamma-i\Delta\right].
\end{cases}.
\end{equation}

\begin{equation}
\begin{cases}
\dot{C}_{+}(t)=C_{+}(t)\left[i\left(\Delta+g_{12}\right)-\left(\Gamma+\Gamma_{12}\right)\right],\\
\dot{C}_{-}(t)=C_{-}(t)\left[i\left(\Delta-g_{12}\right)-\left(\Gamma-\Gamma_{12}\right)\right].
\end{cases}
\end{equation}
We note also that the decay rates are given by the dyadic Green's
function as

\begin{align}
\Gamma_{\alpha\alpha'} & =\pi\boldsymbol{d}_{\alpha}\cdot\bar{\mathcal{K}}_{\alpha\alpha'}(\omega)\cdot\boldsymbol{d}_{\alpha'}\nonumber \\
 & =\frac{\pi\bar{\varepsilon}}{(2\pi)^{3}\hbar\varepsilon_{0}c^{2}}\omega_{\alpha}^{2}\boldsymbol{d}_{\alpha}\cdot\text{Im}[\bar{\bar{G}}(\boldsymbol{r}_{\alpha},\boldsymbol{r}_{\alpha'};\omega_{\alpha})]\cdot\boldsymbol{d}_{\alpha\text{'}}.
\end{align}
This yields the solutions

\begin{equation}
\begin{cases}
C_{+}(t)=C_{+}^{0}e^{i\left(\Delta+g_{12}\right)t-\left(\Gamma+\Gamma_{12}\right)t}\\
C_{-}(t)=C_{-}^{0}e^{i\left(\Delta-g_{12}\right)t-\left(\Gamma-\Gamma_{12}\right)t}
\end{cases}
\end{equation}
and the probability is given by

\begin{equation}
\begin{cases}
P_{+}(t)=P_{+}^{0}e^{-\left(\gamma+\gamma_{12}\right)t},\\
P_{-}(t)=P_{-}^{0}e^{-\left(\gamma-\gamma_{12}\right)t}.
\end{cases}
\end{equation}
We can now imagine that initially $C_{1}(t=0)=1$ and $C_{2}(t=0)=0$
in order to look at the dynamics of the system. In particular, we
can write

\[
\begin{cases}
C_{1}(t)=\frac{1}{\sqrt{2}}\left[C_{+}^{0}e^{-i\left(\Delta+g_{12}\right)t-\left(\Gamma+\Gamma_{12}\right)t}+C_{-}^{0}e^{-i\left(\Delta-g_{12}\right)t-\left(\Gamma-\Gamma_{12}\right)t}\right],\\
C_{2}(t)=\frac{1}{\sqrt{2}}\left[C_{+}^{0}e^{-i\left(\Delta+g_{12}\right)t-\left(\Gamma+\Gamma_{12}\right)t}-C_{-}^{0}e^{-i\left(\Delta-g_{12}\right)t-\left(\Gamma-\Gamma_{12}\right)t}\right].
\end{cases}
\]
or using the initial conditions

\begin{equation}
C_{2}(t=0)=0\Rightarrow C_{+}^{0}=C_{-}^{0}\equiv C_{1}(t=0).
\end{equation}
Therefore, grouping the necessary terms together, yields

\begin{equation}
\begin{cases}
C_{1}(t)=\frac{1}{\sqrt{2}}C_{1}(0)e^{-i\Delta-\Gamma}\left(e^{-i\left(g_{12}-i\Gamma_{12}\right)t}+e^{i\left(g_{12}-i\Gamma_{12}\right)t}\right),\\
C_{2}(t)=\frac{1}{\sqrt{2}}C_{2}(0)e^{-i\Delta-\Gamma}\left(e^{-i\left(g_{12}-i\Gamma_{12}\right)t}-e^{i\left(g_{12}-i\Gamma_{12}\right)t}\right).
\end{cases}
\end{equation}
and the corresponding probabilities are

\begin{equation}
\begin{cases}
P_{1}(t) & =P_{1}^{0}e^{-\gamma}\left(\cos\left(2g_{12}t\right)+\cosh\left(\gamma_{12}t\right)\right),\\
P_{2}(t) & =P_{2}^{0}e^{-\gamma}\left(\cos\left(2g_{12}t\right)-\cosh\left(\gamma_{12}t\right)\right).
\end{cases}
\end{equation}
Plots of the corresponding probabilities can be found in the main
text in figures \ref{fig:Schrodinger-a} and \ref{fig:Schrodinger-b}
in the main text.

\subsection{Deriving the Lindblad equation for two qubits\label{subsec:Deriving-the-Lindblad-1}}

To obtain the dynamics of two coupled qubits the procedure is yet
again very similar to the previous calculations. We start with the
Hamiltonian, as usual, which can be written as

\begin{equation}
\hat{H}=\sum_{\alpha=1,2}\hat{H}_{\alpha}+\hat{H}_{\text{spp}}+\hat{V},
\end{equation}
with each component given in equation (\ref{eq:Hamiltonians}). We
then have the Redfield equation and with

\begin{equation}
\begin{cases}
\beta_{\alpha,1}=\sigma_{\alpha}^{-},\\
\beta_{\alpha,2}=\sigma_{\alpha}^{-},\\
\Gamma_{\alpha,1}=\Gamma_{\alpha}^{\dagger}=i\int_{0}^{\infty}d\omega\int d^{2}\boldsymbol{\rho}\hat{\text{\textbf{f}}}^{\dagger}(\boldsymbol{\rho},\omega)\cdot\mathcal{G}^{+}(\boldsymbol{r}_{\alpha},\boldsymbol{\rho};\omega),\\
\Gamma_{\alpha,2}=\Gamma_{\alpha}=-i\int_{0}^{\infty}d\omega\int d^{2}\boldsymbol{\rho}\mathcal{G}^{-}(\boldsymbol{r}_{\alpha},\boldsymbol{\rho};\omega)\cdot\hat{\text{\textbf{f}}}(\boldsymbol{\rho},\omega).
\end{cases},
\end{equation}
which we can write as

\begin{align}
\frac{d\hat{\rho}_{S}}{dt}= & -\frac{1}{\hbar^{2}}\sum_{\alpha\beta}\sum_{ij}\int_{t_{0}}^{t}d\tau\left(\beta_{\alpha,i}(t)\beta_{\beta,j}(\tau)\hat{\rho}_{S}(t)-\beta_{\beta,j}(\tau)\hat{\rho}_{S}(t)\beta_{\alpha,i}(t)\right)\left\langle \Gamma_{\alpha,i}(t)\Gamma_{\beta,j}(\tau)\right\rangle _{R}\nonumber \\
 & +\left(\hat{\rho}_{S}(t)\beta_{\beta,j}(\tau)\beta_{\alpha,i}(t)-\beta_{\alpha,i}(t)\hat{\rho}_{S}(t)\beta_{\beta,j}(\tau)\right)\left\langle \Gamma_{\beta,j}(\tau)\Gamma_{\alpha,i}(t)\right\rangle _{R}.\label{eq:Redfield-2}
\end{align}
By analogy with the previous results, the reservoir correlation functions
are

\begin{equation}
\begin{cases}
\left\langle \Gamma_{\alpha,1}(t)\Gamma_{\beta,1}(\tau)\right\rangle _{R}=\left\langle \Gamma_{\alpha,2}(t)\Gamma_{\beta,2}(\tau)\right\rangle _{R}=0,\\
\left\langle \Gamma_{\alpha,1}(t)\Gamma_{\beta,2}(\tau)\right\rangle _{R}=\int_{0}^{\infty}d\omega\int d^{2}\boldsymbol{\rho}\mathcal{G}^{-}(\boldsymbol{r}_{\beta},\boldsymbol{\rho};\omega)\mathcal{G}^{+}(\boldsymbol{r}_{\alpha},\boldsymbol{\rho};\omega)\bar{n}_{\text{spp}}(\omega)e^{-i\omega(\tau-t)},\\
\left\langle \Gamma_{\alpha,2}(t)\Gamma_{\beta,1}(\tau)\right\rangle _{R}=\int_{0}^{\infty}d\omega\int d^{2}\boldsymbol{\rho}\mathcal{G}^{-}(\boldsymbol{r}_{\alpha},\boldsymbol{\rho};\omega)\mathcal{G}^{+}(\boldsymbol{r}_{\beta},\boldsymbol{\rho};\omega)\left[\bar{n}_{\text{spp}}(\omega)+1\right]e^{i\omega(\tau-t)},
\end{cases}
\end{equation}
and if we define, for ease of notation the kernel

\begin{equation}
\boldsymbol{d}_{\alpha}\cdot\bar{\mathcal{K}}_{\alpha\beta}(\boldsymbol{r}_{\alpha},\boldsymbol{r}_{\beta},\omega)\cdot\boldsymbol{d}_{\beta}\equiv\frac{1}{\hbar^{2}}\int d^{2}\boldsymbol{\rho}\mathcal{G}^{-}(\boldsymbol{r}_{\alpha},\boldsymbol{\rho};\omega)\cdot\mathcal{G}^{+}(\boldsymbol{r}_{\beta},\boldsymbol{\rho};\omega),
\end{equation}
and the integral

\begin{equation}
\mathcal{I}_{SP}(\omega)=\int_{0}^{\infty}d\tau e^{i(\omega-\omega_{0})(\tau-t)}
\end{equation}
we obtain, evaluating the sums of equation (\ref{eq:Redfield-2})
and in the Markov approximation, the result

\begin{align}
\frac{d\hat{\rho}_{S}^{I}}{dt}= & -\int_{0}^{\infty}d\omega\sum_{\alpha\beta}\left(\sigma_{\alpha}^{+}\sigma_{\beta}^{-}\hat{\rho}_{S}^{I}(t)-\sigma_{\beta}^{-}\hat{\rho}_{S}^{I}(t)\sigma_{\alpha}^{+}\right)\boldsymbol{d}_{\alpha}\cdot\bar{\mathcal{K}}_{\alpha\beta}(\boldsymbol{r}_{\alpha},\boldsymbol{r}_{\beta},\omega)\cdot\boldsymbol{d}_{\beta}\left[\bar{n}_{\text{spp}}(\omega)+1\right]\mathcal{I}(\omega)\nonumber \\
 & +\left(\hat{\rho}_{S}^{I}(t)\sigma_{\beta}^{\text{+}}\sigma_{\alpha}^{-}-\sigma_{\alpha}^{-}\hat{\rho}_{S}^{I}(t)\sigma_{\beta}^{+}\right)\boldsymbol{d}_{\beta}\cdot\bar{\mathcal{K}}_{\beta\alpha}(\boldsymbol{r}_{\beta},\boldsymbol{r}_{\alpha},\omega)\cdot\boldsymbol{d}_{\alpha}\left[\bar{n}_{\text{spp}}(\omega)+1\right]\mathcal{I}^{*}(\omega)\nonumber \\
 & +\left(\sigma_{\alpha}^{-}\sigma_{\beta}^{+}\hat{\rho}_{S}^{I}(t)-\sigma_{\beta}^{+}\hat{\rho}_{S}^{I}(t)\sigma_{\alpha}^{-}\right)\boldsymbol{d}_{\alpha}\bar{\mathcal{K}}_{\alpha\beta}(\boldsymbol{r}_{\alpha},\boldsymbol{r}_{\beta},\omega)\cdot\boldsymbol{d}_{\beta}\bar{n}_{\text{spp}}(\omega)\mathcal{I}^{*}(\omega)\nonumber \\
 & +\left(\hat{\rho}_{S}^{I}(t)\sigma_{\beta}^{-}\sigma_{\alpha}^{+}-\sigma_{\alpha}^{+}\hat{\rho}_{S}^{I}(t)\sigma_{\beta}^{-}\right)\boldsymbol{d}_{\beta}\cdot\bar{\mathcal{K}}_{\beta\alpha}(\boldsymbol{r}_{\beta},\boldsymbol{r}_{\alpha},\omega)\cdot\boldsymbol{d}_{\alpha}\bar{n}_{\text{spp}}(\omega)\mathcal{I}(\omega).
\end{align}
Due to the summation, this equation will contain terms in $\alpha=\beta=1,2$
coming from the evolution of each qubit as well as cross terms $\alpha=1,\beta=2$
or $\alpha=2,\beta=1$ coming from the interaction between the qubits.
The equation will be written as

\begin{align}
\frac{d\hat{\rho}_{S}^{I}}{dt}= & -\int_{0}^{\infty}d\omega\sum_{\alpha\beta}\left(\sigma_{\alpha}^{+}\sigma_{\beta}^{-}\hat{\rho}_{S}^{I}(t)-\sigma_{\beta}^{-}\hat{\rho}_{S}^{I}(t)\sigma_{\alpha}^{+}\right)\boldsymbol{d}_{\alpha}\cdot\bar{\mathcal{K}}_{\alpha\beta}(\boldsymbol{r}_{\alpha},\boldsymbol{r}_{\beta},\omega)\cdot\boldsymbol{d}_{\beta}\left[\bar{n}_{\text{spp}}(\omega)+1\right]\mathcal{I}(\omega)\nonumber \\
 & +\left(\hat{\rho}_{S}^{I}(t)\sigma_{\beta}^{\text{+}}\sigma_{\alpha}^{-}-\sigma_{\alpha}^{-}\hat{\rho}_{S}^{I}(t)\sigma_{\beta}^{+}\right)\boldsymbol{d}_{\beta}\cdot\bar{\mathcal{K}}_{\beta\alpha}(\boldsymbol{r}_{\beta},\boldsymbol{r}_{\alpha},\omega)\cdot\boldsymbol{d}_{\alpha}\left[\bar{n}_{\text{spp}}(\omega)+1\right]\mathcal{I}^{*}(\omega)\nonumber \\
 & +\left(\sigma_{\alpha}^{-}\sigma_{\beta}^{+}\hat{\rho}_{S}^{I}(t)-\sigma_{\beta}^{+}\hat{\rho}_{S}^{I}(t)\sigma_{\alpha}^{-}\right)\boldsymbol{d}_{\alpha}\bar{\mathcal{K}}_{\alpha\beta}(\boldsymbol{r}_{\alpha},\boldsymbol{r}_{\beta},\omega)\cdot\boldsymbol{d}_{\beta}\bar{n}_{\text{spp}}(\omega)\mathcal{I}^{*}(\omega)\nonumber \\
 & +\left(\hat{\rho}_{S}^{I}(t)\sigma_{\beta}^{-}\sigma_{\alpha}^{+}-\sigma_{\alpha}^{+}\hat{\rho}_{S}^{I}(t)\sigma_{\beta}^{-}\right)\boldsymbol{d}_{\beta}\cdot\bar{\mathcal{K}}_{\beta\alpha}(\boldsymbol{r}_{\beta},\boldsymbol{r}_{\alpha},\omega)\cdot\boldsymbol{d}_{\alpha}\bar{n}_{\text{spp}}(\omega)\mathcal{I}(\omega).
\end{align}
Notice that we can split the sums into terms where $\alpha=\beta$
and where $\alpha\neq\beta$. We wrtite the following

\begin{align}
\frac{d\hat{\rho}_{S}^{I}}{dt}= & -\int_{0}^{\infty}d\omega\sum_{\alpha}\left(\sigma_{\alpha}^{+}\sigma_{\alpha}^{-}\hat{\rho}_{S}^{I}(t)-\sigma_{\alpha}^{-}\hat{\rho}_{S}^{I}(t)\sigma_{\alpha}^{+}\right)\boldsymbol{d}_{\alpha}\cdot\bar{\mathcal{K}}_{\alpha\alpha}(\boldsymbol{r}_{\alpha},\boldsymbol{r}_{\alpha},\omega)\cdot\boldsymbol{d}_{\alpha}\left[\bar{n}_{\text{spp}}(\omega)+1\right]\mathcal{I}(\omega)\nonumber \\
 & +\left(\hat{\rho}_{S}^{I}(t)\sigma_{\alpha}^{\text{+}}\sigma_{\alpha}^{-}-\sigma_{\alpha}^{-}\hat{\rho}_{S}^{I}(t)\sigma_{\alpha}^{+}\right)\boldsymbol{d}_{\alpha}\cdot\bar{\mathcal{K}}_{\alpha\alpha}(\boldsymbol{r}_{\alpha},\boldsymbol{r}_{\alpha},\omega)\cdot\boldsymbol{d}_{\alpha}\left[\bar{n}_{\text{spp}}(\omega)+1\right]\mathcal{I}^{*}(\omega)\nonumber \\
 & +\left(\sigma_{\alpha}^{-}\sigma_{\alpha}^{+}\hat{\rho}_{S}^{I}(t)-\sigma_{\alpha}^{+}\hat{\rho}_{S}^{I}(t)\sigma_{\alpha}^{-}\right)\boldsymbol{d}_{\alpha}\bar{\mathcal{K}}_{\alpha\alpha}(\boldsymbol{r}_{\alpha},\boldsymbol{r}_{\alpha},\omega)\cdot\boldsymbol{d}_{\alpha}\bar{n}_{\text{spp}}(\omega)\mathcal{I}^{*}(\omega)\nonumber \\
 & +\left(\hat{\rho}_{S}^{I}(t)\sigma_{\alpha}^{-}\sigma_{\alpha}^{+}-\sigma_{\alpha}^{+}\hat{\rho}_{S}^{I}(t)\sigma_{\alpha}^{-}\right)\boldsymbol{d}_{\alpha}\cdot\bar{\mathcal{K}}_{\alpha\alpha}(\boldsymbol{r}_{\alpha},\boldsymbol{r}_{\alpha},\omega)\cdot\boldsymbol{d}_{\alpha}\bar{n}_{\text{spp}}(\omega)\mathcal{I}(\omega)\nonumber \\
 & -\int_{0}^{\infty}d\omega\sum_{\alpha\neq\beta}\left(\sigma_{\alpha}^{+}\sigma_{\beta}^{-}\hat{\rho}_{S}^{I}(t)-\sigma_{\beta}^{-}\hat{\rho}_{S}^{I}(t)\sigma_{\alpha}^{+}\right)\boldsymbol{d}_{\alpha}\cdot\bar{\mathcal{K}}_{\alpha\beta}(\boldsymbol{r}_{\alpha},\boldsymbol{r}_{\beta},\omega)\cdot\boldsymbol{d}_{\beta}\left[\bar{n}_{\text{spp}}(\omega)+1\right]\mathcal{I}(\omega)\nonumber \\
 & +\left(\hat{\rho}_{S}^{I}(t)\sigma_{\beta}^{\text{+}}\sigma_{\alpha}^{-}-\sigma_{\alpha}^{-}\hat{\rho}_{S}^{I}(t)\sigma_{\beta}^{+}\right)\boldsymbol{d}_{\beta}\cdot\bar{\mathcal{K}}_{\beta\alpha}(\boldsymbol{r}_{\beta},\boldsymbol{r}_{\alpha},\omega)\cdot\boldsymbol{d}_{\alpha}\left[\bar{n}_{\text{spp}}(\omega)+1\right]\mathcal{I}^{*}(\omega)\nonumber \\
 & +\left(\sigma_{\alpha}^{-}\sigma_{\beta}^{+}\hat{\rho}_{S}^{I}(t)-\sigma_{\beta}^{+}\hat{\rho}_{S}^{I}(t)\sigma_{\alpha}^{-}\right)\boldsymbol{d}_{\alpha}\bar{\mathcal{K}}_{\alpha\beta}(\boldsymbol{r}_{\alpha},\boldsymbol{r}_{\beta},\omega)\cdot\boldsymbol{d}_{\beta}\bar{n}_{\text{spp}}(\omega)\mathcal{I}^{*}(\omega)\nonumber \\
 & +\left(\hat{\rho}_{S}^{I}(t)\sigma_{\beta}^{-}\sigma_{\alpha}^{+}-\sigma_{\alpha}^{+}\hat{\rho}_{S}^{I}(t)\sigma_{\beta}^{-}\right)\boldsymbol{d}_{\beta}\cdot\bar{\mathcal{K}}_{\beta\alpha}(\boldsymbol{r}_{\beta},\boldsymbol{r}_{\alpha},\omega)\cdot\boldsymbol{d}_{\alpha}\bar{n}_{\text{spp}}(\omega)\mathcal{I}(\omega).
\end{align}
Performing the same analysis we had done for the case of the single
qubit, we find that the first summation becomes, with the jump operators

\begin{equation}
\begin{cases}
\hat{L}_{\alpha,1}=\sqrt{\Gamma\bar{n}_{\text{spp}}(\omega_{0})}\sigma_{\alpha}^{+},\\
\hat{L}_{\alpha,2}=\sqrt{\Gamma\left[\bar{n}_{\text{spp}}(\omega_{0})+1\right]}\sigma_{\alpha}^{-}.
\end{cases}
\end{equation}

\begin{align}
 & \sum_{\alpha}\left(2L_{\alpha,2}\hat{\rho}_{S}(t)L_{\alpha,2}^{\dagger}-\left\{ L_{\alpha,2}^{\dagger}L_{\alpha,2},\hat{\rho}_{S}(t)\right\} \right)\nonumber \\
 & +\left(2L_{\alpha,1}\hat{\rho}_{S}(t)L_{\alpha,1}^{\dagger}-\left\{ L_{\alpha,1}^{\dagger}L_{\alpha,1},\hat{\rho}_{S}(t)\right\} \right)-i(\omega_{0}+\Delta_{\alpha}+2\Delta'_{\alpha})\left[\sigma_{\alpha}^{+}\sigma_{\alpha}^{-},\hat{\rho}_{S}(t)\right],
\end{align}
whereas the for the second expression we can split it apart and write

\begin{align}
\sum_{\alpha\neq\beta}\int_{0}^{\infty}d\omega\left(\sigma_{\alpha}^{+}\sigma_{\beta}^{-}\hat{\rho}_{S}^{I}(t)-\sigma_{\beta}^{-}\hat{\rho}_{S}^{I}(t)\sigma_{\alpha}^{+}\right)\boldsymbol{d}_{\alpha}\cdot\bar{\mathcal{K}}_{\alpha\beta}(\boldsymbol{r}_{\alpha},\boldsymbol{r}_{\beta},\omega)\cdot\boldsymbol{d}_{\beta}\left[\bar{n}_{\text{spp}}(\omega)+1\right]\mathcal{I}(\omega)\nonumber \\
+\left(\hat{\rho}_{S}^{I}(t)\sigma_{\beta}^{\text{+}}\sigma_{\alpha}^{-}-\sigma_{\alpha}^{-}\hat{\rho}_{S}^{I}(t)\sigma_{\beta}^{+}\right)\boldsymbol{d}_{\beta}\cdot\bar{\mathcal{K}}_{\beta\alpha}(\boldsymbol{r}_{\beta},\boldsymbol{r}_{\alpha},\omega)\cdot\boldsymbol{d}_{\alpha}\left[\bar{n}_{\text{spp}}(\omega)+1\right]\mathcal{I}^{*}(\omega)\nonumber \\
+\left(\sigma_{\alpha}^{-}\sigma_{\beta}^{+}\hat{\rho}_{S}^{I}(t)-\sigma_{\beta}^{+}\hat{\rho}_{S}^{I}(t)\sigma_{\alpha}^{-}\right)\boldsymbol{d}_{\alpha}\cdot\bar{\mathcal{K}}_{\alpha\beta}(\boldsymbol{r}_{\alpha},\boldsymbol{r}_{\beta},\omega)\cdot\boldsymbol{d}_{\beta}\bar{n}_{\text{spp}}(\omega)\mathcal{I}^{*}(\omega)\\
+\left(\hat{\rho}_{S}^{I}(t)\sigma_{\beta}^{-}\sigma_{\alpha}^{+}-\sigma_{\alpha}^{+}\hat{\rho}_{S}^{I}(t)\sigma_{\beta}^{-}\right)\boldsymbol{d}_{\beta}\cdot\bar{\mathcal{K}}_{\beta\alpha}(\boldsymbol{r}_{\beta},\boldsymbol{r}_{\alpha},\omega)\cdot\boldsymbol{d}_{\alpha}\bar{n}_{\text{spp}}(\omega)\mathcal{I}(\omega).\nonumber 
\end{align}
We can make our usual definitions

\begin{equation}
\begin{cases}
\Gamma_{\alpha\beta}\equiv\pi\int_{0}^{\infty}d\omega\boldsymbol{d}_{\alpha}\cdot\bar{\mathcal{K}}(\boldsymbol{r}_{\alpha},\boldsymbol{r}_{\beta},\omega)\cdot\boldsymbol{d}_{\beta}\delta(\omega-\omega_{0}),\\
\Delta_{\alpha\beta}\equiv\mathcal{P}\int_{0}^{\infty}d\omega\frac{\boldsymbol{d}_{\alpha}\cdot\bar{\mathcal{K}}(\boldsymbol{r}_{\alpha},\boldsymbol{r}_{\beta},\omega)\cdot\boldsymbol{d}_{\beta}}{\omega_{0}-\omega},\\
\Delta'_{\alpha\beta}\equiv\mathcal{P}\int_{0}^{\infty}d\omega\frac{\boldsymbol{d}_{\alpha}\cdot\bar{\mathcal{K}}(\boldsymbol{r}_{\alpha},\boldsymbol{r}_{\beta},\omega)\cdot\boldsymbol{d}_{\beta}\bar{n}_{\text{spp}}(\omega)}{\omega_{0}-\omega}.
\end{cases}
\end{equation}
and write

\begin{align}
-\sum_{\alpha\neq\beta}\left(\sigma_{\alpha}^{+}\sigma_{\beta}^{-}\hat{\rho}_{S}^{I}(t)-\sigma_{\beta}^{-}\hat{\rho}_{S}^{I}(t)\sigma_{\alpha}^{+}\right)\left[\Gamma_{\alpha\beta}\left[\bar{n}_{\text{spp}}(\omega_{0})+1\right]+i\left(\Delta_{\alpha\beta}+\Delta'_{\alpha\beta}\right)\right]\nonumber \\
+\left(\hat{\rho}_{S}^{I}(t)\sigma_{\beta}^{\text{+}}\sigma_{\alpha}^{-}-\sigma_{\alpha}^{-}\hat{\rho}_{S}^{I}(t)\sigma_{\beta}^{+}\right)\left[\Gamma_{\beta\alpha}\left[\bar{n}_{\text{spp}}(\omega_{0})+1\right]-i\left(\Delta_{\beta\alpha}+\Delta'_{\beta\alpha}\right)\right]\nonumber \\
+\left(\sigma_{\alpha}^{-}\sigma_{\beta}^{+}\hat{\rho}_{S}^{I}(t)-\sigma_{\beta}^{+}\hat{\rho}_{S}^{I}(t)\sigma_{\alpha}^{-}\right)\left[\Gamma_{\alpha\beta}\left(\bar{n}_{\text{spp}}(\omega_{0})+1\right)-i\Delta'_{\alpha\beta}\right]\nonumber \\
+\left(\hat{\rho}_{S}^{I}(t)\sigma_{\beta}^{-}\sigma_{\alpha}^{+}-\sigma_{\alpha}^{+}\hat{\rho}_{S}^{I}(t)\sigma_{\beta}^{-}\right)\left[\Gamma_{\beta\alpha}\left(\bar{n}_{\text{spp}}(\omega_{0})+1\right)+i\Delta'_{\beta\alpha}\right].
\end{align}
We can group together common terms noting that we can particularize
for our two qubit system by defining $\Gamma_{\alpha\beta}=\Gamma_{\beta\alpha}\equiv\Gamma_{12}$
as well as $\Delta_{\alpha\beta}=\Delta_{\beta\alpha}\equiv g_{12}$
and $\Delta_{\alpha\beta}'=\Delta_{\beta\alpha}'\equiv\Delta_{12}'$.
The summation yields

\begin{align}
-\left(\sigma_{1}^{+}\sigma_{2}^{-}\hat{\rho}_{S}^{I}(t)-\sigma_{2}^{-}\hat{\rho}_{S}^{I}(t)\sigma_{1}^{+}+\sigma_{2}^{+}\sigma_{1}^{-}\hat{\rho}_{S}^{I}(t)-\sigma_{1}^{-}\hat{\rho}_{S}^{I}(t)\sigma_{2}^{+}\right)\left[\Gamma_{12}\left[\bar{n}_{\text{spp}}(\omega_{0})+1\right]+i\left(\Delta_{12}+\Delta'_{12}\right)\right]\nonumber \\
-\left(\hat{\rho}_{S}^{I}(t)\sigma_{2}^{\text{+}}\sigma_{1}^{-}-\sigma_{1}^{-}\hat{\rho}_{S}^{I}(t)\sigma_{2}^{+}+\hat{\rho}_{S}^{I}(t)\sigma_{1}^{\text{+}}\sigma_{2}^{-}-\sigma_{2}^{-}\hat{\rho}_{S}^{I}(t)\sigma_{1}^{+}\right)\left[\Gamma_{12}\left[\bar{n}_{\text{spp}}(\omega_{0})+1\right]-i\left(\Delta_{12}+\Delta'_{12}\right)\right]\nonumber \\
-\left(\sigma_{1}^{-}\sigma_{2}^{+}\hat{\rho}_{S}^{I}(t)-\sigma_{2}^{+}\hat{\rho}_{S}^{I}(t)\sigma_{1}^{-}+\sigma_{2}^{-}\sigma_{1}^{+}\hat{\rho}_{S}^{I}(t)-\sigma_{1}^{+}\hat{\rho}_{S}^{I}(t)\sigma_{2}^{-}\right)\left[\Gamma_{12}\bar{n}_{\text{spp}}(\omega_{0})-i\Delta'_{12}\right]\nonumber \\
-\left(\hat{\rho}_{S}^{I}(t)\sigma_{2}^{-}\sigma_{1}^{+}-\sigma_{1}^{+}\hat{\rho}_{S}^{I}(t)\sigma_{2}^{-}+\hat{\rho}_{S}^{I}(t)\sigma_{1}^{-}\sigma_{2}^{+}-\sigma_{2}^{+}\hat{\rho}_{S}^{I}(t)\sigma_{1}^{-}\right)\left[\Gamma_{12}\bar{n}_{\text{spp}}(\omega_{0})+i\Delta'_{12}\right],
\end{align}
and what follows is

\begin{align}
\Gamma_{12}\left[\bar{n}_{\text{spp}}(\omega_{0})+1\right]\left(2\sigma_{2}^{-}\hat{\rho}_{S}^{I}(t)\sigma_{1}^{+}+2\sigma_{1}^{-}\hat{\rho}_{S}^{I}(t)\sigma_{2}^{+}-\left\{ \sigma_{1}^{+}\sigma_{2}^{-},\hat{\rho}_{S}^{I}(t)\right\} -\left\{ \sigma_{2}^{+}\sigma_{1}^{-},\hat{\rho}_{S}^{I}(t)\right\} \right)\nonumber \\
+\Gamma_{12}\bar{n}_{\text{spp}}(\omega_{0})\left(2\sigma_{2}^{+}\hat{\rho}_{S}^{I}(t)\sigma_{1}^{-}+2\sigma_{1}^{+}\hat{\rho}_{S}^{I}(t)\sigma_{2}^{-}-\left\{ \sigma_{1}^{-}\sigma_{2}^{+},\hat{\rho}_{S}^{I}(t)\right\} -\left\{ \sigma_{2}^{-}\sigma_{1}^{+},\hat{\rho}_{S}^{I}(t)\right\} \right)\nonumber \\
+i\left(g_{12}+\Delta'_{12}\right)\left(\left[\sigma_{1}^{+}\sigma_{2}^{-}+\sigma_{2}^{+}\sigma_{1}^{-},\hat{\rho}_{S}^{I}(t)\right]\right)\nonumber \\
-i\Delta'_{12}\left(\left[\sigma_{1}^{-}\sigma_{2}^{+}+\sigma_{2}^{-}\sigma_{1}^{+},\hat{\rho}_{S}^{I}(t)\right]\right),
\end{align}
since the rasing and lowering operators act in different qubits we
can write the commutator

\begin{equation}
\left[\sigma_{i}^{+},\sigma_{j}^{-}\right]=0,
\end{equation}
for $i\neq j$, using which we obtain

\begin{align}
\Gamma_{12}\left[\bar{n}_{\text{spp}}(\omega_{0})+1\right]\left(2\sigma_{2}^{-}\hat{\rho}_{S}^{I}(t)\sigma_{1}^{+}+2\sigma_{1}^{-}\hat{\rho}_{S}^{I}(t)\sigma_{2}^{+}-\left\{ \sigma_{1}^{+}\sigma_{2}^{-},\hat{\rho}_{S}^{I}(t)\right\} -\left\{ \sigma_{2}^{+}\sigma_{1}^{-},\hat{\rho}_{S}^{I}(t)\right\} \right)\nonumber \\
+\Gamma_{12}\bar{n}_{\text{spp}}(\omega_{0})\left(2\sigma_{2}^{+}\hat{\rho}_{S}^{I}(t)\sigma_{1}^{-}+2\sigma_{1}^{+}\hat{\rho}_{S}^{I}(t)\sigma_{2}^{-}-\left\{ \sigma_{1}^{-}\sigma_{2}^{+},\hat{\rho}_{S}^{I}(t)\right\} -\left\{ \sigma_{2}^{-}\sigma_{1}^{+},\hat{\rho}_{S}^{I}(t)\right\} \right)\nonumber \\
+ig_{12}\left(\left[\sigma_{1}^{+}\sigma_{2}^{-}+\sigma_{2}^{+}\sigma_{1}^{-},\hat{\rho}_{S}^{I}(t)\right]\right).
\end{align}
We can now define additional jump operators

\begin{equation}
\begin{cases}
\hat{K}_{\alpha,1}=\sqrt{\Gamma_{12}\bar{n}_{\text{spp}}(\omega_{0})}\sigma_{\alpha}^{+},\\
\hat{K}_{\alpha,2}=\sqrt{\Gamma_{12}\left[\bar{n}_{\text{spp}}(\omega_{0})+1\right]}\sigma_{\alpha}^{-}.
\end{cases}
\end{equation}
Converting back to the Schrodinger picture, we have, with the previously
defined jump operators

\begin{align}
\sum_{\alpha\neq\beta}\left(2\hat{K}_{\alpha,2}\hat{\rho}_{S}(t)\hat{K}_{\beta,2}^{\dagger}-\left\{ \hat{K}_{\alpha,2}^{\dagger}\hat{K}_{\beta,2},\hat{\rho}_{S}(t)\right\} \right)\nonumber \\
+\left(2\hat{K}_{\alpha,1}\hat{\rho}_{S}(t)\hat{K}_{\beta,1}^{\dagger}-\left\{ \hat{K}_{\alpha,1}^{\dagger}\hat{K}_{\beta,1},\hat{\rho}_{S}(t)\right\} \right)-i(\omega_{0}+g_{12})\left[\sigma_{\alpha}^{+}\sigma_{\beta}^{-},\hat{\rho}_{S}(t)\right],
\end{align}
To all intents and purposes we can now write all terms including the
coherent shift $g_{12}$ and the operators $\hat{L}_{\alpha,k}$ and
$\hat{K}_{\alpha,k}$ in the master equation, and therefore we have
as a final result

\begin{align}
\frac{d\hat{\rho}_{S}}{dt}= & -\frac{i}{\hbar}\left[\hbar\omega_{12}\sum_{\alpha\neq\beta}\sigma_{\alpha}^{+}\sigma_{\beta}^{-},\hat{\rho}_{S}(t)\right]-\frac{i}{\hbar}\left[\hbar\omega_{0}'\sum_{\alpha}\sigma_{\alpha}^{+}\sigma_{\alpha}^{-},\hat{\rho}_{S}(t)\right]\nonumber \\
 & +\sum_{\alpha}\left(\sum_{k}2\hat{L}_{\alpha,k}\hat{\rho}_{S}(t)\hat{L}_{\alpha,k}^{\dagger}-\left\{ \hat{L}_{\alpha,k}^{\dagger}\hat{L}_{\alpha,k},\hat{\rho}_{S}(t)\right\} \right)\nonumber \\
 & +\sum_{\alpha\neq\beta}\left(\sum_{k}2\hat{K}_{\alpha,k}\hat{\rho}_{S}(t)\hat{K}_{\beta,k}^{\dagger}-\left\{ \hat{K}_{\alpha,k}^{\dagger}\hat{K}_{\beta,k},\hat{\rho}_{S}(t)\right\} \right).
\end{align}
with $\omega_{0}'=\omega_{0}+\Delta_{\alpha}+2\Delta_{\alpha}'$ and
$\omega_{12}=\omega_{0}+g_{12}$. This brings our equation to Lindblad
form. To bring this into an equation for the probability we can proceed
as before, expanding the density matrix in the respective terms

\begin{align}
\hat{\rho}_{S} & =\sum_{ij}\sum_{kl}\rho_{ij,kl}\left|ij\right\rangle \left\langle kl\right|,\nonumber \\
 & =\begin{bmatrix}\rho_{ee,ee} & \rho_{ee,eg} & \rho_{ee,ge} & \rho_{ee,gg}\\
\rho_{eg,ee} & \rho_{eg,eg} & \rho_{eg,ge} & \rho_{eg,gg}\\
\rho_{ge,ee} & \rho_{ge,eg} & \rho_{ge,ge} & \rho_{ge,gg}\\
\rho_{gg,ee} & \rho_{gg,eg} & \rho_{gg,ge} & \rho_{gggg}
\end{bmatrix}.
\end{align}
We define for simplicity the following super-operators

\begin{equation}
\mathbb{I}_{\alpha\beta}\left[\hat{\rho}_{S}(t)\right]=-\frac{i}{\hbar}\left[\hbar\omega_{12}\sigma_{\alpha}^{+}\sigma_{\beta}^{-},\hat{\rho}_{S}(t)\right],
\end{equation}
corresponding to the term in the dynamical equation that gives the
interaction between the two-level systems

\begin{equation}
\mathbb{C}_{\alpha}\left[\hat{\rho}_{S}(t)\right]=-\frac{i}{\hbar}\left[\hbar\omega_{0}'\sigma_{\alpha}^{+}\sigma_{\alpha}^{-},\hat{\rho}_{S}(t)\right],
\end{equation}
corresponding to the term that gives the coherent evolution of each
qubit,

\begin{equation}
\mathbb{L}_{\alpha}\left[\hat{\rho}_{S}(t)\right]=\sum_{k}2\hat{L}_{\alpha,k}\hat{\rho}_{S}(t)\hat{L}_{\alpha,k}^{\dagger}-\left\{ \hat{L}_{\alpha,k}^{\dagger}\hat{L}_{\alpha,k},\hat{\rho}_{S}(t)\right\} ,
\end{equation}
which gives the incoherent evolution of both qubits due to the exchange
of plasmons with themselves and finally

\begin{equation}
\mathbb{K}_{\alpha\beta}\left[\hat{\rho}_{S}(t)\right]=\sum_{k}2\hat{K}_{\alpha,k}\hat{\rho}_{S}(t)\hat{K}_{\beta,k}^{\dagger}-\left\{ \hat{K}_{\alpha,k}^{\dagger}\hat{K}_{\beta,k},\hat{\rho}_{S}(t)\right\} ,
\end{equation}
which gives the incoherent evolution of both qubits due to the exchange
of plasmons between eachother. The information regarding this exchange
is contained in the jump operators $K_{\alpha,k}$. We have therefore

\begin{equation}
\frac{d\hat{\rho}_{S}}{dt}=\left[\sum_{\alpha}\mathbb{C}_{\alpha}+\sum_{\alpha\neq\beta}\mathbb{I}_{\alpha\beta}+\sum_{\alpha}\mathbb{L}_{\alpha}+\sum_{\alpha\neq\beta}\mathbb{K}_{\alpha\beta}\right]\left[\hat{\rho}_{S}(t)\right].
\end{equation}
We make these definitions in order to simplify the following calculation.
Note that we can write the jump operators in matrix form since we
have

\begin{equation}
\sigma_{1}^{+}=\sigma^{+}\otimes\mathds{1}=\begin{bmatrix}0 & 0 & 1 & 0\\
0 & 0 & 0 & 1\\
0 & 0 & 0 & 0\\
0 & 0 & 0 & 0
\end{bmatrix}.
\end{equation}

\begin{equation}
\sigma_{2}^{+}=\mathds{1}\otimes\sigma^{+}=\begin{bmatrix}0 & 1 & 0 & 0\\
0 & 0 & 0 & 0\\
0 & 0 & 0 & 1\\
0 & 0 & 0 & 0
\end{bmatrix}.
\end{equation}
We write each term of the dynamical equation in matrix form using
these operators. We are left with

\begin{equation}
\sum_{\alpha}\mathbb{C}_{\alpha}=i\omega_{0}'\begin{bmatrix}0 & -\rho_{eg,ee} & -\rho_{ge,ee} & -2\rho_{gg,ee}\\
\rho_{eg,ee} & 0 & 0 & -\rho_{gg,eg}\\
\rho_{ge,ee} & 0 & 0 & -\rho_{gg,ge}\\
2\rho_{gg,ee} & \rho_{gg,eg} & \rho_{gg,ge} & 0
\end{bmatrix},
\end{equation}

\begin{equation}
\sum_{\alpha\neq\beta}\mathbb{I}_{\alpha\beta}=i\omega_{12}\begin{bmatrix}0 & \rho_{eg,ee} & \rho_{eg,ee} & 0\\
-\rho_{ge,ee} & 0 & \rho_{eg,eg}-\rho_{ge,ge} & -\rho_{gg,ge}\\
-\rho_{eg,ee} & \rho_{ge,ge}-\rho_{eg,eg} & 0 & -\rho_{gg,eg}\\
0 & \rho_{gg,ge} & \rho_{gg,eg} & 0
\end{bmatrix}.
\end{equation}
The whole matrix can easily be calculated for the remaining terms
as well $\sum_{\alpha}\mathbb{L}_{\alpha}$ and $\sum_{\alpha\neq\beta}\mathbb{K}_{\alpha\beta}$,
however, since we are intrested only in the populations, that is to
say, the diagonal elements of the density matrix, we find the neither
$\mathbb{C}_{\alpha}$ nor $\mathbb{I}_{\alpha\beta}$ contribute,
since they both have zeros along the diagonals. The diagonal contributions
of $\mathbb{L_{\alpha}}$ are, if organized in a vector

\begin{equation}
\text{diag}\left[\sum_{\alpha}\mathbb{L}_{\alpha}\right]=\gamma\begin{bmatrix}-2\rho_{ee,ee}\left[\bar{n}_{\text{spp}}(\omega_{0})+1\right]+\bar{n}_{\text{spp}}(\omega_{0})(\rho_{eg,eg}+\rho_{ge,ge})\\
\rho_{ee,ee}\left[\bar{n}_{\text{spp}}(\omega_{0})+1\right]-\rho_{eg,eg}\left[\bar{n}_{\text{spp}}(\omega_{0})+1\right]-\bar{n}_{\text{spp}}(\omega_{0})\rho_{eg,eg}+\bar{n}_{\text{spp}}(\omega_{0})\rho_{gg,gg}\\
\rho_{ee,ee}\left[\bar{n}_{\text{spp}}(\omega_{0})+1\right]-\rho_{ge,ge}\left[\bar{n}_{\text{spp}}(\omega_{0})+1\right]-\bar{n}_{\text{spp}}(\omega_{0})\rho_{ge,ge}+\bar{n}_{\text{spp}}(\omega_{0})\rho_{gg,gg}\\
-2\rho_{gg,gg}\bar{n}_{\text{spp}}(\omega_{0})+\left[\bar{n}_{\text{spp}}(\omega_{0})+1\right](\rho_{eg,eg}+\rho_{ge,ge})
\end{bmatrix},
\end{equation}
and the contributions from $\mathbb{K}_{\alpha\beta}$ read

\begin{equation}
\text{diag}\left[\sum_{\alpha\neq\beta}\mathbb{K}_{\alpha\beta}\right]=\gamma_{12}\begin{bmatrix}2\bar{n}_{\text{spp}}(\omega_{0})\rho_{ge,eg}\\
-\bar{n}_{\text{spp}}(\omega_{0})\rho_{ge,eg}-\left[\bar{n}_{\text{spp}}(\omega_{0})+1\right]\rho_{ge,eg}\\
-\bar{n}_{\text{spp}}(\omega_{0})\rho_{ge,eg}-\left[\bar{n}_{\text{spp}}(\omega_{0})+1\right]\rho_{ge,eg}\\
2\left[\bar{n}_{\text{spp}}(\omega_{0})+1\right]\rho_{ge,eg}
\end{bmatrix}.
\end{equation}
As expected, the coherent evolutions do not contribute to the populations
since they are responsible at most for a phase shift, however in evaluating
the evolution of the diagonal terms of the density matrix we also
find a dependance on off-diagonal terms. This suggests that a change
of basis is in order, and in fact, we can look at the basis we already
used for the Schrodinger equation approach. In this case, we have
the states $\left|3\right\rangle =\left|ee\right\rangle ,\left|+\right\rangle =\left(\left|eg\right\rangle +\left|ge\right\rangle \right)/\sqrt{2},\left|-\right\rangle =\left(\left|eg\right\rangle -\left|ge\right\rangle \right)/\sqrt{2}$
and $\left|0\right\rangle =\left|gg\right\rangle $. We can construct
a change of basis matrix by writing

\begin{equation}
U=\begin{bmatrix}1 & 0 & 0 & 0\\
0 & \frac{1}{\sqrt{2}} & \frac{1}{\sqrt{2}} & 0\\
0 & \frac{1}{\sqrt{2}} & -\frac{1}{\sqrt{2}} & 0\\
0 & 0 & 0 & 1
\end{bmatrix},
\end{equation}
and construct the raising and lowering operators in this basis by
writing

\begin{equation}
\sigma'{}_{\alpha}^{\pm}=U^{\dagger}\sigma_{\alpha}^{\pm}U.
\end{equation}
We find, for instance

\begin{equation}
\sigma'{}_{1}^{+}=\begin{bmatrix}0 & \frac{1}{\sqrt{2}} & -\frac{1}{\sqrt{2}} & 0\\
0 & 0 & 0 & \frac{1}{\sqrt{2}}\\
0 & 0 & 0 & \frac{1}{\sqrt{2}}\\
0 & 0 & 0 & 0
\end{bmatrix},
\end{equation}

\begin{equation}
\sigma'{}_{2}^{+}=\begin{bmatrix}0 & \frac{1}{\sqrt{2}} & \frac{1}{\sqrt{2}} & 0\\
0 & 0 & 0 & \frac{1}{\sqrt{2}}\\
0 & 0 & 0 & -\frac{1}{\sqrt{2}}\\
0 & 0 & 0 & 0
\end{bmatrix}.
\end{equation}
In this basis the density matrix also takes the shape

\begin{equation}
\hat{\rho}_{S}=\begin{bmatrix}\rho_{3,3} & \rho_{+,+} & \rho_{+,-} & \rho_{3,0}\\
\rho_{+,3} & \rho_{+,+} & \rho_{+,-} & \rho_{+,0}\\
\rho_{-,3} & \rho_{-,+} & \rho_{-,-} & \rho_{-,0}\\
\rho_{0,3} & \rho_{0,+} & \rho_{0,-} & \rho_{0,0}
\end{bmatrix},
\end{equation}
which means that we can calculate the several components of the dynamical
equation, speciffically, the populations by taking the diagonal elements.
As before, there is no contribution from the coherent evolution given
by $\mathbb{C}_{\alpha}$ or $\mathbb{I}_{\alpha\beta}$, however
we can look at the diagonal elements of the remaining parts

\begin{equation}
\text{diag}\left[\sum_{\alpha}\mathbb{L}_{\alpha}\right]=\gamma\begin{bmatrix}-2\rho_{3,3}\left[\bar{n}_{\text{spp}}(\omega_{0})+1\right]-\bar{n}_{\text{spp}}(\omega_{0})\rho_{3,3}+\bar{n}_{\text{spp}}(\omega_{0})\left(\rho_{-,-}+\rho_{+,+}\right)\\
\rho_{3,3}\left[\bar{n}_{\text{spp}}(\omega_{0})+1\right]+\bar{n}_{\text{spp}}(\omega_{0})\rho_{0,0}-\bar{n}_{\text{spp}}(\omega_{0})\rho_{+.+}-\left[\bar{n}_{\text{spp}}(\omega_{0})+1\right]\rho_{+,+}\\
\rho_{3,3}\left[\bar{n}_{\text{spp}}(\omega_{0})+1\right]+\bar{n}_{\text{spp}}(\omega_{0})\rho_{0,0}-\bar{n}_{\text{spp}}(\omega_{0})\rho_{-,-}-\left[\bar{n}_{\text{spp}}(\omega_{0})+1\right]\rho_{-,-}\\
-2\rho_{0,0}\bar{n}_{\text{spp}}(\omega_{0})+\left[\bar{n}_{\text{spp}}(\omega_{0})+1\right]\left(\rho_{-,-}+\rho_{+,+}\right)
\end{bmatrix},
\end{equation}

\begin{equation}
\text{diag}\left[\sum_{\alpha}\mathbb{K}_{\alpha}\right]=\gamma_{12}\begin{bmatrix}\bar{n}_{\text{spp}}(\omega_{0})\left(\rho_{+,+}-\rho_{-,-}\right)\\
\left[\bar{n}_{\text{spp}}(\omega_{0})+1\right]\rho_{3,3}+\bar{n}_{\text{spp}}(\omega_{0})\rho_{0,0}-\bar{n}_{\text{spp}}(\omega_{0})\rho_{+,+}-\left[\bar{n}_{\text{spp}}(\omega_{0})+1\right]\rho_{+,+}\\
-\left[\bar{n}_{\text{spp}}(\omega_{0})+1\right]\rho_{3,3}+\bar{n}_{\text{spp}}(\omega_{0})\rho_{0,0}+\bar{n}_{\text{spp}}(\omega_{0})\rho_{-,-}+\left[\bar{n}_{\text{spp}}(\omega_{0})+1\right]\rho_{-,-}\\
-\left[\bar{n}_{\text{spp}}(\omega_{0})+1\right]\left(\rho_{-,-}-\rho_{+,+}\right)
\end{bmatrix}.
\end{equation}
We see that the diagonal elements of the density matrix in this basis
are self-contained, that is to say, they depend only on each other,
as opposed to the natural basis we analyzed previously. This allows
us to evaluate the dynamics of the density matrix looking only at
the diagonal elements. Solving the dynamical equation can be done
numerically, and we obtain the figures presented in the main text.
Note that if we let the temperature go to 0, which means $\bar{n}_{\text{spp}}(\omega_{0})\to0$,
we find

\begin{equation}
\text{diag}\left[\sum_{\alpha}\mathbb{L}_{\alpha}\right]=\gamma\begin{bmatrix}-2\rho_{3,3}\\
\rho_{3,3}-\rho_{+,+}\\
\rho_{3,3}-\rho_{-,-}\\
\rho_{-,-}+\rho_{+,+}
\end{bmatrix},
\end{equation}

\begin{equation}
\text{diag}\left[\sum_{\alpha}\mathbb{K}_{\alpha}\right]=\gamma_{12}\begin{bmatrix}0\\
\rho_{3,3}-\rho_{+,+}\\
-\rho_{3,3}+\rho_{-,-}\\
\rho_{+,+}-\rho_{-,-}
\end{bmatrix},
\end{equation}
and therefore, writing the populations as $\rho_{3,3}\equiv P_{3}$,
$\rho_{\pm,\pm}\equiv P_{\pm}$ and $\rho_{0,0}\equiv P_{0}$, we
find the dinamical equation

\begin{equation}
\begin{bmatrix}\dot{P}_{3}\\
\dot{P}_{\text{+}}\\
\dot{P}_{-}\\
\dot{P}_{0}
\end{bmatrix}=\gamma\begin{bmatrix}-2P_{3}\\
P_{3}-P_{+}\\
P_{3}-P_{-}\\
P_{-}+P_{+}
\end{bmatrix}+\gamma_{12}\begin{bmatrix}0\\
P_{3}-P_{+}\\
-P_{3}+P_{-}\\
P_{+}-P_{-}
\end{bmatrix},
\end{equation}
which paired with the initial conditions $P_{3}(0)=0$, $P_{+}(0)=1/2$,
$P_{-}(0)=1/2$, $P_{0}(0)=0$ gives the dynamics

\begin{equation}
\begin{cases}
P_{3}(t)=0,\\
P_{+}(t)=\frac{1}{2}e^{-(\gamma+\gamma_{12})t},\\
P_{-}(t)=\frac{1}{2}e^{-(\gamma-\gamma_{12})t},\\
P_{0}(t)=1-\frac{e^{-(\gamma-\gamma_{12})t}+e^{-(\gamma+\gamma_{12})t}}{2}.
\end{cases}
\end{equation}
which matches exactly those obtained via the Schrodinger equation.


%

\end{document}